\documentclass[11pt,a4paper]{article}
\pdfoutput=1
\usepackage{jheppub}
\allowdisplaybreaks
\usepackage{xcolor}
\usepackage{amsmath}
\usepackage{graphicx}
\usepackage{esint}
\usepackage{slashed}
\usepackage{multirow}
\usepackage{gensymb}
\usepackage{makecell}
\usepackage{caption}
\usepackage{subcaption}
\usepackage{comment}
\usepackage[normalem]{ulem}
\usepackage{float}
\usepackage{pifont}
\newcommand{\xmark}{\ding{55}}
\usepackage[capitalise]{cleveref}
\usepackage[utf8]{inputenc}
\usepackage[greek,english]{babel}
\usepackage{bm}

\newcommand{\nuDM}{{\boldmath \texorpdfstring{$\nu$}{\textnu}-DM}}

\usepackage{babel}
\newcommand{\genericT}{\ensuremath{T}}
\newcommand{\met}{\slashed{E}_\genericT}

\begin{document}
\preprint{CETUP-2023-025}
\title{New Constraints on Neutrino-Dark Matter Interactions: A Comprehensive Analysis}

\author{P. S. Bhupal Dev$^{1,2}$, Doojin Kim$^{3,4}$, Deepak Sathyan$^{4,5}$, Kuver Sinha$^6$, Yongchao Zhang$^{7,8}$}

\affiliation{$^1$Department of Physics and McDonnell Center for the Space Sciences, Washington University, St. Louis, MO 63130, USA}
\affiliation{$^2$PRISMA$^{++}$ Cluster of Excellence \& Mainz Institute for Theoretical Physics, 
Johannes Gutenberg-Universit\"{a}t Mainz, 55099 Mainz, Germany}

\affiliation{$^3$Department of Physics, University of South Dakota, Vermillion, SD 57069, USA}

\affiliation{$^4$Mitchell Institute for Fundamental Physics and Astronomy, Department of Physics and Astronomy, Texas A\&M University, College Station, TX 77843, USA}

\affiliation{$^5$Maryland Center for Fundamental Physics, Department of Physics, University of Maryland, College Park, MD 20742, USA}

\affiliation{$^6$Department of Physics and Astronomy, University of Oklahoma, Norman, OK 73019, USA}

\affiliation{$^7$School of Physics, Southeast University, Nanjing 211189, China}

\affiliation{$^8$Center for High Energy Physics, Peking University, Beijing 100871, China}

\emailAdd{bdev@wustl.edu, doojin.kim@usd.edu, dsathyan@tamu.edu, kuver.sinha@ou.edu, zhangyongchao@seu.edu.cn}

\abstract{We present a comprehensive analysis of the interactions of neutrinos with the dark sector within the simplified model framework. We first derive the exact analytic formulas for the differential scattering cross sections of neutrinos with scalar, fermion, and vector dark matter (DM) for light dark sector models with mediators of different types. We then implement the full catalog of constraints on the parameter space of the neutrino-DM and neutrino-mediator couplings and masses, including cosmological and astrophysical bounds coming from Big Bang Nucleosynthesis, Cosmic Microwave Background, DM and neutrino self-interactions, DM collisional damping, and astrophysical neutrino sources, as well as laboratory constraints from 3-body meson decays and invisible $Z$ decays. 
We find that most of the benchmarks in the DM mass-coupling plane adopted in previous studies to get an observable neutrino-DM interaction effect are actually ruled out by a combination of the above-mentioned constraints, especially the laboratory ones which are robust against astrophysical uncertainties and independent of the cosmological history. 
To illustrate the consequences of our new results, we take the galactic supernova neutrinos in the MeV energy range as a concrete example and highlight the difficulties in finding any observable effect of neutrino-DM interactions. Finally, we identify new benchmark points potentially promising for future observational prospects of the attenuation of the galactic supernova neutrino flux and comment on their implications for the detection prospects in future large-volume neutrino experiments such as JUNO, Hyper-K, and DUNE. We also comment on the ultraviolet-embedding of the effective neutrino-DM couplings. 
}


\maketitle

\section{Introduction}

Neutrinos and dark matter (DM) provide the two most compelling pieces of empirical evidence for the existence of new physics beyond the Standard Model (BSM). They are often treated as uncorrelated sectors while studying BSM phenomenology, but the fact is that the physics of both the neutrino and DM sectors remains largely unknown. Therefore, it is intriguing to consider the possibility that these two sectors are intimately connected to each other. In other words, {\it can DM have sizable interactions with neutrinos?} 

From the experimental point of view, this seems like a difficult question, since both DM and neutrinos are ``invisible'' in detectors. In fact, the most stringent experimental constraints on DM come from their assumed interactions with ``visible'' Standard Model (SM) particles, such as baryons, electrons, and photons~\cite{Cirelli:2024ssz}. It is conceivable that all the direct detection and collider constraints, as well as the majority of indirect detection constraints involving visible final states, can be evaded (or at least relaxed) if DM dominantly couples only to neutrinos and no other SM species at leading order. This is theoretically possible, for instance, if the DM couplings to charged particles in the SM are suppressed either by mixing, higher-order effects, or higher-dimensional operators (see Appendix~\ref{sec:UV} for concrete examples). 

However, it turns out that while secluded, the interactions between neutrinos and DM are not entirely hidden. In fact, there are several important cosmological and astrophysical consequences if neutrino-DM interactions are sizable. For instance, light ($\lesssim 10$ MeV) DM can transfer entropy into the neutrino sector at its thermal decoupling, affecting the expansion rate of the Universe and dramatically altering Big Bang Nucleosynthesis (BBN) and Cosmic Microwave Background (CMB) observables~\cite{Serpico:2004nm,Boehm:2013jpa, Escudero:2018mvt, Sabti:2019mhn, Giovanetti:2024orj, Pal:2024yom}. A neutrino-DM interaction can also lead to diffusion damping of cosmological perturbations on small scales, suppressing the CMB matter power spectrum and smoothing out the CMB temperature anisotropies~\cite{Boehm:2003xr,Mangano:2006mp, Wilkinson:2014ksa,Escudero:2015yka,Diacoumis:2017hff,DiValentino:2017oaw,Mosbech:2020ahp, Brax:2023tvn}. It could even enhance the primordial gravitational wave amplitude on small scales~\cite{Ghosh:2017jdy}. At late times, DM can also impede the formation of small-scale structure, known as ``collisional damping''~\cite{Boehm:2000gq,Boehm:2004th}, with measurable effects in the Lyman-$\alpha$ forest~\cite{vandenAarssen:2012vpm, Wilkinson:2014ksa, Hooper:2021rjc} and in Milky Way satellites and subhalos~\cite{Hooper:2007tu,Boehm:2014vja,Bertoni:2014mva,Schewtschenko:2014fca,Schewtschenko:2015rno,Akita:2023yga, Heston:2024ljf}. The reduction in halo formation and number of galaxies in those halos may result in a smaller number of ionizing photons in the intergalactic medium (IGM), which reduces the amount of ionized hydrogen in the IGM, and consequently, affects the reionization history~\cite{Dey:2022ini, Mosbech:2022uud}. Similarly, one can place astrophysical constraints on neutrino-DM interactions from core-collapse supernovae~\cite{Mangano:2006mp,Fayet:2006sa, Sun:2025gyj}, diffuse supernova neutrino background~\cite{Farzan:2014gza}, cosmic neutrino background interactions with ultralight scalars~\cite{Cheek:2025kks,Lambiase:2025twn}, spectral modifications of high-energy astrophysical neutrinos~\cite{Reynoso:2016hjr,Arguelles:2017atb,Kelly:2018tyg,Pandey:2018wvh,Koren:2019wwi,Murase:2019xqi,Karmakar:2020yzn,McMullen:2021ikf, Cline:2022qld, Ferrer:2022kei, Cline:2023tkp, Fujiwara:2023lsv}, and neutrino signals from DM self-annihilation/decay in indirect detection experiments~\cite{Beacom:2006tt, Yuksel:2007ac, Palomares-Ruiz:2007trf, Palomares-Ruiz:2007egs,Covi:2009xn,Moline:2014xua,ElAisati:2017ppn,Arguelles:2022nbl,IceCube:2023ies}. Standard neutrino oscillations can also be modified in a medium of DM~\cite{deSalas:2016svi,Choi:2019zxy,Cheek:2025kks}. The basic idea is that in cosmological and astrophysical settings, the neutrinos encounter a large column depth of DM before reaching the Earth, thereby providing an ideal testing ground for neutrino-DM interactions. 

The key purpose of this paper is to perform a comprehensive analysis of the neutrino-DM interactions. To this end, we $(i)$ summarize all DM-neutrino constraints existing in the literature, compare the results of different groups including ours, and point out potential discrepancies and caveats; $(ii)$  incorporate new laboratory constraints on DM-neutrino interactions, including a subtle cancellation effect that weakens the meson and $Z$ decay constraints for light mediators; and $(iii)$  consider the prospects of probing neutrino-DM interactions in a future galactic supernova event by identifying suitable benchmarks that evade all existing constraints and by calculating the associated neutrino fluxes at future neutrino experiments.

From the theoretical viewpoint, the neutrino-DM interactions can be described in a model-independent way via a simplified model framework~\cite{Olivares-DelCampo:2017feq, Blennow:2019fhy}. Since we do not know the properties of the DM, such as its mass and spin, we consider all possible renormalizable interaction terms involving a scalar, fermion, or vector DM. Also, the energy scale at which these new interaction terms are generated is not known unless an ultraviolet (UV)-complete model is specified (see Appendix~\ref{sec:UV} for some examples); therefore, we will use a simplified-model approach by introducing a single mediator of arbitrary mass which couples to both neutrinos and DM, and again the mediator can be either a scalar, fermion, or vector. Thus, the only free parameters in this phenomenological framework are the DM mass, mediator mass, and their couplings to each other and/or neutrinos. With this setup, in Section~\ref{sec:nuDMInteractions} we derive exact analytic expressions for all relevant differential scattering cross sections in terms of these free parameters. 

Since the mediator couples directly to both neutrinos and DM, it introduces additional constraints. Depending on the type of mediator and DM under consideration, there will be astrophysical constraints from DM self-interactions~\cite{Adhikari:2022sbh}, as well as cosmological and laboratory constraints from neutrino self-interactions~\cite{Berryman:2022hds}. In addition, there are often stringent laboratory constraints on the mediators coupling to neutrinos, such as from the double-beta ($\beta\beta$) decays, invisible $Z$ decays, and rare meson and charged lepton decays~\cite{Lessa:2007up, Berryman:2018ogk, deGouvea:2019qaz, Brdar:2020nbj, Dev:2024ygx}. We carefully reevaluate the meson, tau, and $Z$ decay constraints, taking into account the subtle cancellation effect between the tree and loop level diagrams, which is especially important in the light mediator case to cancel the infrared divergence.  Our implementation of the full catalog of constraints for each combination of DM-mediators is described in Section~\ref{sec:bounds}. 
This work presents the first comprehensive compilation of cosmological, astrophysical, and laboratory constraints on neutrino-DM interactions within the phenomenologically relevant parameter space.
The goal of this exercise is to determine whether there exists any viable parameter space for an observable effect of sizable neutrino-DM interactions. In this context, we point out that the cosmological and astrophysical limits are usually the most stringent for masses below MeV scale, whereas the laboratory constraints are often the most stringent above MeV scale. Thus a combination of these limits rules out a wide range of parameter space for DM and mediator masses up to the GeV scale. 

As a concrete physical example where sizable neutrino-DM interactions can play an important role, we consider the case of supernova neutrinos in Section~\ref{sec:supernova}. It is well known from simulations~\cite{Janka:2025tvf},  validated by the SN1987A data~\cite{Kamiokande-II:1987idp, Bionta:1987qt,Alekseev:1988gp}, that neutrinos carry away roughly 99\% of the energy in a core-collapse supernova explosion. If it happens in the local vicinity of our galaxy (within $\sim$10 kpc), these supernova neutrinos, with energy up to a few tens of MeV, will be detected with high statistics in next-generation large-volume neutrino detectors such as DUNE~\cite{DUNE:2020ypp}, Hyper-K~\cite{Hyper-Kamiokande:2018ofw} and JUNO~\cite{JUNO:2015sjr}. 
Since these neutrinos travel astrophysical distances and must go through the galactic DM halo before reaching the Earth, they are ideal candidates for testing the hypothesis of neutrino-DM interactions. In particular, a significant neutrino-DM interaction could attenuate the expected neutrino flux, shift the neutrino spectrum to lower energies, or cause a time-delayed signal due to neutrino-DM scattering~\cite{Franarin:2018gfk,Carpio:2022sml, Balantekin:2023jlg, Chauhan:2025hoz}. 
All of these effects are potentially measurable in the next-generation neutrino experiments. Thus, it is important to check if there exists any viable parameter space for this to happen while satisfying all existing constraints mentioned above. As we will find out, it seems rather challenging to probe new regions of parameter space with the interactions between supernova neutrinos and DM in the aforementioned neutrino experiments in many cases, for a cored DM density profile such as the Einasto profile~\cite{Navarro:2003ew}. On the other hand, if the DM density profile is NFW-type~\cite{Navarro:1995iw} or has a spike structure~\cite{Gondolo:1999ef}, the density could be orders of magnitude larger at the galactic center. In such special cases and if the supernova event happens in the vicinity of the galactic center, we show that the supernova neutrinos can probe some regions of the parameter space of neutrino-DM interactions that are still unconstrained. 

The rest of the paper is organized as follows: We begin by specifying the neutrino-DM interaction models considered in this work in Section~\ref{sec:nuDMInteractions}. The comprehensive astrophysical, cosmological, and laboratory limits are detailed in Section~\ref{sec:bounds}, and we emphasize which limits are new in this paper. In Section~\ref{sec:supernova}, we calculate the attenuated supernova neutrino fluxes due to neutrino-DM scattering. The effects on the neutrino signals at future neutrino experiments are presented for some benchmark points in the parameter space of one particular model, which satisfy all the existing constraints. Our goal in this part of the paper is to show that this method of probing the parameter space of neutrino-DM interactions is useful and complementary to other methods.  We present our conclusions and further discussion in Section~\ref{sec:con}. The UV-completions for the couplings of neutrinos and DM are collected in Appendix~\ref{sec:UV}. The calculation details for the neutrino-DM scattering, DM-DM scattering, neutrino-neutrino scattering, DM annihilation to neutrinos, and the mediator widths are listed in the Appendices~\ref{Appendix:xs:nuDM},  \ref{app:dmself}, \ref{Appendix:xs:nu-nu}, \ref{app:thermalrelic}, and \ref{appendix:width}, respectively. 

\section{\nuDM~scattering cross sections}
\label{sec:nuDMInteractions}

In this section, we provide the formulae of $\nu$-DM scattering cross sections, considering all possible combinations of scalar, fermionic, and vectorial mediators and DM. We also provide a comparison of our results with those in the existing literature~\cite{Olivares-DelCampo:2017feq, Arguelles:2017atb}.

For concreteness, we make the following simplifications and assumptions: 
\begin{itemize}
    \item The DM is non-relativistic with velocity $v \sim 10^{-3}c$. In the calculations of $\nu$-DM scattering cross sections below, we neglect the kinetic energy of DM in the initial state and assume the DM is at rest in the initial state. This is a valid assumption as long as the incoming neutrino energy is much larger than the DM kinetic energy.
    \item All the $\nu$-DM couplings considered below are flavor-diagonal (or equivalently, flavor-universal). If the couplings are lepton flavor-changing, they might contribute to neutrino flavor transitions in some way and there might be additional stringent constraints, which we do not consider here.
\end{itemize}
For convenience, we summarize our key results in Table~\ref{tab:models}, with the analytic expressions for all the $\nu$-DM scattering amplitude squares given in Appendix~\ref{Appendix:xs:nuDM}.

\begin{table}[!t]
\centering
\caption{Summary of the DM scenarios (2nd column) considered in this paper, with a scalar, fermion, or vector mediator (1st column) for $\nu$-DM scattering. The relevant Lagrangians are given in this section (3rd column), while the corresponding amplitude squares are collected in Appendix~\ref{Appendix:xs:nuDM} (5th column). Note that in some models the $\nu$-DM and $\bar\nu$-DM scatterings proceed in different channels (cf. 4th column). In the last two columns, we compare our results with those from Refs.~\cite{Olivares-DelCampo:2017feq, Arguelles:2017atb}.
The stars indicate our results are larger than those in Refs.~\cite{Olivares-DelCampo:2017feq,Arguelles:2017atb} by a factor of 2. The dagger means our result is consistent with the code from Ref.~\cite{Arguelles:2017atb}, but not the same as the formula in their paper. See text for more details.
}
\label{tab:models}
\vspace{5pt}
 \begin{tabular}{|c | c | c | c | c | c | c | c |} 
 \hline\hline
 Med. & DM & Lagrangian & Channels & Amp. sq. & Ref.~\cite{Olivares-DelCampo:2017feq} & Ref.~\cite{Arguelles:2017atb} \\ \hline
 \parbox[t]{2mm}{\multirow{6}{*}{\rotatebox[origin=c]{90}{Scalar}}} & Complex scalar  
 & (\ref{eqn:L:scalar:scalar}) & $t$ & (\ref{eqn:ampsq:scalar:scalar}) & $-$ & \checkmark *
 \\ \cline{2-7}
 & \multirow{2}*{Dirac fermion}   
  & \multirow{2}*{(\ref{eqn:L:scalar:Dirac})}   & $\nu$-DM: $u$ & (\ref{eqn:ampsq:scalar:Dirac:u}) & \checkmark & $-$  \\ 
  & &  & $\bar\nu$-DM: $s$ & (\ref{eqn:ampsq:scalar:Dirac:s}) & $-$ & $-$ \\ \cline{2-7}
  & Majorana fermion  & (\ref{eqn:L:scalar:Dirac}) & $s$, $u$ & (\ref{eqn:ampsq:scalar:Majorana}) & \xmark & $-$  \\ \cline{2-7}
  & Dirac fermion & (\ref{eqn:L:scalar:Dirac:t}) & $t$ & (\ref{eqn:ampsq:scalar:Dirac:t}) & $-$ & \checkmark * \\ \cline{2-7}
  & Complex vector & (\ref{eqn:L:scalar:vector}) & $t$ & (\ref{eqn:ampsq:scalar:vector}) & $-$ & $-$ \\ \hline\hline
  
 \parbox[t]{2mm}{\multirow{4}{*}{\rotatebox[origin=c]{90}{Fermion}}} & real scalar & (\ref{eqn:L:fermion:scalar_complex}) & $s$, $u$ & (\ref{eqn:ampsq:fermion:scalar_real}) &  \checkmark * & \checkmark * \\ \cline{2-7}
 & complex scalar  & (\ref{eqn:L:fermion:scalar_complex}) & \makecell{$\nu$-DM: $s$ \\ $\bar\nu$-DM: $u$} & \makecell{(\ref{eqn:ampsq:fermion:scalar_complex:s}) \\ (\ref{eqn:ampsq:fermion:scalar_complex:u})} & \makecell{$-$ \\ \checkmark} & $-$ \\ \cline{2-7}
 & vector & (\ref{eqn:L:fermion:vector}) & $s$, $u$ & (\ref{eqn:ampsq:fermion:vector}) &  \xmark & $-$ \\ \hline\hline
 
 \parbox[t]{2mm}{\multirow{8}{*}{\rotatebox[origin=c]{90}{Vector}}} & complex scalar & (\ref{eqn:L:vector:scalar_complex:t}) & $t$ & (\ref{eqn:ampsq:vector:scalar_complex:t}) & \checkmark & $-$ \\ \cline{2-7}
 & Dirac fermion & (\ref{eqn:L:vector:Dirac:t}) & $t$ & (\ref{eqn:ampsq:vector:Dirac:t}) & \checkmark & \checkmark $\dagger$ \\ \cline{2-7}
 & \multirow{2}*{Dirac fermion}   
  & \multirow{2}*{(\ref{eqn:L:vector:Dirac:u})}   & $\nu$-DM: $u$ & (\ref{eqn:ampsq:vector:Dirac:u}) & $-$ & $-$  \\ 
  & &  & $\bar\nu$-DM: $s$ & (\ref{eqn:ampsq:vector:Dirac:s}) & $-$ & $-$ \\ \cline{2-7}
  & Majorana fermion & (\ref{eqn:L:vector:Majorana:t}) & $t$ & (\ref{eqn:ampsq:vector:Majorana:t}) & \checkmark & $-$ \\ \cline{2-7}
  & Majorana fermion  & (\ref{eqn:L:vector:Dirac:u}) & $s$, $u$ & (\ref{eqn:ampsq:vector:Majorana:su}) & $-$ & $-$ \\ \cline{2-7}
  & real vector & (\ref{eqn:L:vector:vector_real:t}) & $t$ & (\ref{eqn:ampsq:vector:vector_real:t}) & \xmark & $-$  \\ \cline{2-7}
  & complex vector & (\ref{eqn:L:vector:vector_complex:t}) & $t$ & (\ref{eqn:ampsq:vector:vector_complex:t}) & $-$ & $-$  \\ \hline\hline
\end{tabular}
\end{table}

\subsection{Model Lagrangians}
Let us start with DM scenarios involving a real scalar mediator denoted by $\phi$ with mass being $m_\phi$. 
\begin{itemize}
    \item If the DM particle $\chi$ is a complex scalar, the Lagrangian can be written as
\begin{eqnarray}
\label{eqn:L:scalar:scalar}
{\cal L} = - \phi \bar\nu \left( g_{\nu\, s} + i g_{\nu\, p} \gamma_5 \right) \nu
- \mu\, \phi\, \chi^\dagger \chi  \,,
\end{eqnarray}
where $\mu$ is a dimensionful quantity parameterizing the interaction strength between the mediator and DM, and $g_{\nu\,s}$ and $g_{\nu\,p}$ are respectively the scalar and pseudo-scalar couplings. With our amplitude square in Eq.~(\ref{eqn:ampsq:scalar:scalar}), we can reproduce Eq.~(12) in Ref.~\cite{Arguelles:2017atb}, if we replace the couplings $g_{\nu\, s} \to g$, $g_{\nu\, p} \to 0$, and the dimensionful parameter $\mu \to g'$. For the case of real scalar DM, the Lagrangian is written as 
\begin{eqnarray}
\label{eqn:L:realscalar:scalar}
{\cal L} = - \phi \bar\nu \left( g_{\nu\, s} + i g_{\nu\, p} \gamma_5 \right) \nu
-\mu \phi \chi^2  \,,
\end{eqnarray}
and the corresponding amplitude square is larger by a factor of $2^2 = 4$. We parameterize $\mu$ in terms of a dimensionless coupling between dark matter and a mediator $g_\chi$ and a scale, and set the scale of $\mu$ to be 1 MeV throughout our analysis.

\item For the case of Dirac fermion DM $\chi$, the Lagrangian for the $\phi$-$\chi$-$\nu$ coupling can be written as
\begin{eqnarray}
\label{eqn:L:scalar:Dirac}
{\cal L} = - \phi \overline{\chi} (g_s + i g_p \gamma_5) P_L \nu ~+~ {\rm H.c.} 
\end{eqnarray}
where we assume only the left-handed neutrinos have interactions of this sort, for which $g_s$ and $g_p$ respectively parameterize the scalar and pseudo-scalar coupling strengths. 
Then the $\nu$-DM and $\bar\nu$-DM scatterings can only proceed via $u$ and $s$ channels, respectively [cf. Eqs.~(\ref{eqn:ampsq:scalar:Dirac:u}) and (\ref{eqn:ampsq:scalar:Dirac:s})].  For the case of Majorana fermion, the $\nu$-DM scattering can proceed via both $s$ and $u$ channels, and the contributions should be summed up [cf. Eq.~(\ref{eqn:ampsq:scalar:Majorana})]. 

\hspace{0.5cm} If the couplings are in the form of $\phi$-$\chi$-$\chi$ and $\phi$-$\nu$-$\nu$, the corresponding Lagrangian terms are
\begin{eqnarray}
\label{eqn:L:scalar:Dirac:t}
{\cal L} = - \phi \bar\nu \left( g_{\nu\, s} + i g_{\nu\, p} \gamma_5 \right) \nu 
- \phi \bar\chi \left( g_{\chi\, s} + i g_{\chi\, p} \gamma_5 \right) \chi \,,
\end{eqnarray}
where we include both the scalar and pseudo-scalar couplings as usual, without loss of generality, for both neutrinos and DM. The spin sum-averaged amplitude square is given in Eq.~(\ref{eqn:ampsq:scalar:Dirac:t}); for the case of Majorana fermion DM, there is an extra factor of $2^2=4$.

\item For the case of complex vector DM $\chi_\mu$, the interaction Lagrangian contains
\begin{eqnarray}
\label{eqn:L:scalar:vector}
{\cal L} = - \phi \bar\nu \left( g_{\nu\, s} + i g_{\nu\, p} \gamma_5 \right) \nu 
- \mu \, \phi \, \chi^{\dagger \mu} \chi_\mu \,,
\end{eqnarray}
and the corresponding amplitude square is given in Eq.~(\ref{eqn:ampsq:scalar:vector}). For the case of real vector DM, the Lagrangian can be written as
\begin{eqnarray}
\label{eqn:L:scalar:realvector}
{\cal L} = - \phi \bar\nu \left( g_{\nu\, s} + i g_{\nu\, p} \gamma_5 \right) \nu 
- \mu \, \phi \, \chi^{\mu} \chi_\mu \,,
\end{eqnarray}
giving again an extra factor of $2^2=4$. 
\end{itemize}

For the scenarios with a fermionic mediator $N$ with mass being $m_N$, let us consider the following models.
\begin{itemize}
    \item For the case of real scalar DM $\chi$, if the mediator is a Dirac or Majorana fermion $N$, the Lagrangian can be written as
\begin{eqnarray}
\label{eqn:L:fermion:scalar_complex}
{\cal L} = - \chi \overline{N} (g_s + i g_p \gamma_5) P_L \nu ~+~ {\rm H.c.}  
\end{eqnarray}
The $\nu$-DM scattering can proceed in both $s$- and $u$-channels, and the corresponding amplitude square can be found in Eq.~(\ref{eqn:ampsq:fermion:scalar_real}). If DM is a complex scalar, the scatterings of DM with neutrino and antineutrino are in the $s$- and $u$-channels, respectively [cf. Eq.~(\ref{eqn:ampsq:fermion:scalar_complex:su})].

\item The active neutrinos are left-handed fermions in the SM. However, they may have a small fraction of right-handed couplings, e.g. from mixing with right-handed neutrinos. The current limits on heavy-light neutrino mixing can be found e.g. in Refs.~\cite{Bolton:2019pcu, neutriomixing}. For simplicity, we neglect the right-handed component of active neutrinos in this paper. Then the couplings of vector DM $\chi$ with neutrinos can be written as
\begin{eqnarray}
\label{eqn:L:fermion:vector}
{\cal L} = -  g \chi^\mu \overline{N} \gamma_\mu P_L \nu ~+~ {\rm H.c.}
\end{eqnarray}
The corresponding amplitude square is given in Eq.~(\ref{eqn:ampsq:fermion:vector}). 
\end{itemize}

Let us now move on to the cases with a (neutral) gauge boson mediator $Z'$ with mass $m_{Z'}$.
\begin{itemize}
    \item The couplings of $Z'$ with neutrinos can in principle have both left- and right-handed components, as just aforementioned. For simplicity, we neglect the right-handed components of active neutrinos. Then in the case of complex scalar DM $\chi$, the relevant couplings are:
\begin{eqnarray}
\label{eqn:L:vector:scalar_complex:t}
{\cal L} =  Z_\mu^\prime \left[ g_\nu \overline{\nu} \gamma^\mu P_L \nu + g_\chi \left( \chi^{\dagger} \left( \partial^\mu \chi \right) - \left( \partial^\mu \chi^\dagger \right) \chi \right) \right] \,.
\end{eqnarray}
The corresponding amplitude square is given in Eq.~(\ref{eqn:ampsq:vector:scalar_complex:t}).

\item For the case of Dirac fermion DM, the couplings in the $t$-channel are
\begin{eqnarray}
\label{eqn:L:vector:Dirac:t}
{\cal L} = Z_\mu^\prime \Big[ g_\nu \overline{\nu} \gamma^\mu P_L \nu  + \bar\chi \left( g_{\chi L} \gamma^\mu P_L + g_{\chi R} \gamma^\mu P_R \right) \chi \Big] \,.
\end{eqnarray}
One should note there are some special cases in this scenario: $g_{\chi R} = 0$ and $g_{\chi L}=0$ correspond respectively to the cases with $Z'$ coupling only to the left- and right-handed fermion DM, while $g_{\chi L} = \pm g_{\chi R}$ are the cases with DM has only vector $g_{\chi V} = (g_{\chi L} + g_{\chi R})/2$ and axial-vector $g_{\chi A} = - (g_{\chi L} - g_{\chi R})/2$ couplings. 
The amplitude square is given in Eq.~(\ref{eqn:ampsq:vector:Dirac:t}). If DM is a Majorana particle, its coupling with the $Z'$ mediator can only be in the axial-vector form of $\gamma_\mu \gamma_5$, and the Lagrangian is
\begin{eqnarray}
\label{eqn:L:vector:Majorana:t}
{\cal L} = Z_\mu^\prime \Big[ g_\nu \overline{\nu} \gamma^\mu P_L \nu + g_{\chi}  \bar\chi \gamma^\mu \gamma_5 \chi \Big] \,.
\end{eqnarray}
The corresponding amplitude square is presented in Eq.~(\ref{eqn:ampsq:vector:Majorana:t}).

For the $s$- and $u$-channel processes, we need a coupling in the form of 
\begin{eqnarray}
\label{eqn:L:vector:Dirac:u}
{\cal L} = g Z_\mu^\prime \overline{\chi} \gamma^\mu P_L  \nu ~+~ {\rm H.c.} 
\end{eqnarray} 
The amplitude squares for the neutrino and antineutrino are given in Eqs.~(\ref{eqn:ampsq:vector:Dirac:u}) and (\ref{eqn:ampsq:vector:Dirac:s}), respectively. For the case of Majorana fermion DM, there are both $s$- and $u$-channel diagrams, and their contributions should be summed up [cf. Eq.~(\ref{eqn:ampsq:vector:Majorana:su})]. 

\item  For the case of real vector DM, the relevant couplings are:
\begin{eqnarray}
\label{eqn:L:vector:vector_real:t}
{\cal L} = g_\nu Z_\mu^\prime \overline{\nu} \gamma^\mu P_L \nu  
+ \left( \frac12 g_\chi \chi^\mu \left( \partial_\mu \chi^\nu \right) Z_\nu^{\prime} ~+~ {\rm H.c.} \right) \,,
\end{eqnarray}
and the corresponding amplitude square is given in Eq.~(\ref{eqn:ampsq:vector:vector_real:t}). 
If the DM particle is a complex vector, the couplings are 
\begin{eqnarray}
\label{eqn:L:vector:vector_complex:t}
{\cal L} = g_\nu  Z_\mu^\prime \overline{\nu} \gamma^\mu P_L \nu  
+ \left( g_\chi \chi^{\dagger \mu} \left( \partial^\nu \chi_\mu \right) Z_\nu^{\prime} ~+~ {\rm H.c.} \right) \,,
\end{eqnarray}
and the resultant amplitude square can be found in Eq.~(\ref{eqn:ampsq:vector:vector_complex:t}).
\end{itemize}

\subsection{Comparison with previous results} 
\begin{figure}[!t]
    \centering
    \includegraphics[width=0.48\textwidth]{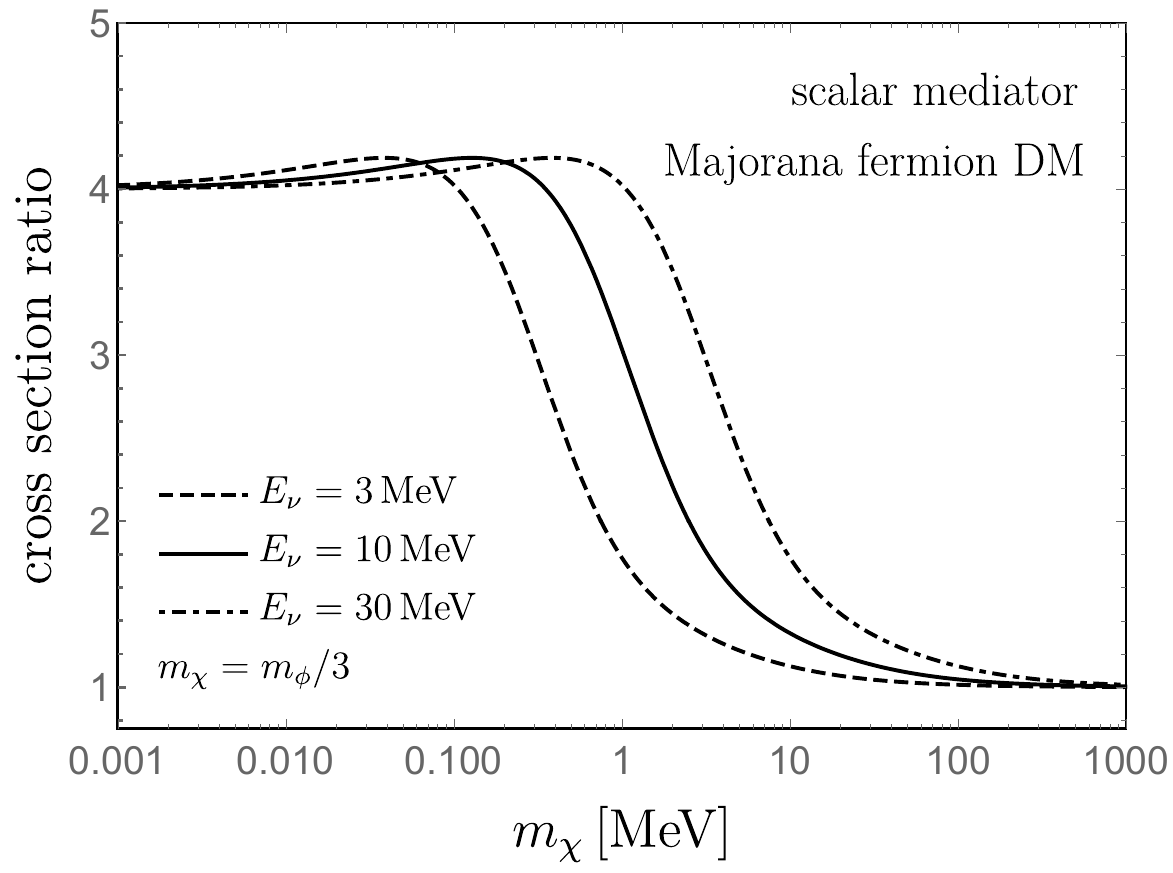}
    \includegraphics[width=0.48\textwidth]{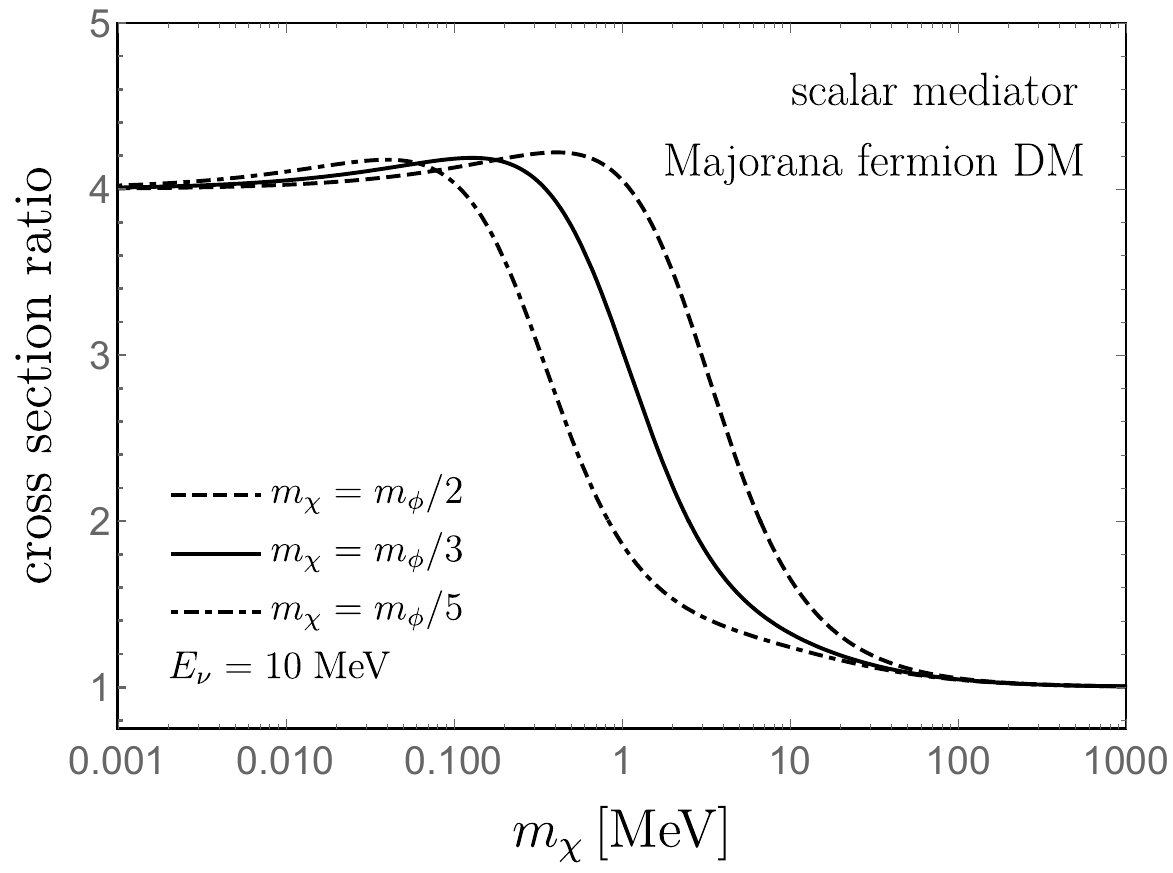}
    \caption{Cross section ratios of those in Ref.~\cite{Olivares-DelCampo:2017feq} with respect to our calculations for the case of scalar mediator and Majorana fermion DM (cf. Table~\ref{tab:models}). The left panel is for the cases of $E_\nu = 3$ MeV, 10 MeV, and 30 MeV with $m_\chi = m_\phi/3$, while the right panel is for the cases of $m_\phi/m_\chi = 2$, 3, and $5$ with $E_\nu = 10$ MeV. 
    }
    \label{fig:xs:comparison:1}
 \end{figure}
 \begin{figure}[!t]
    \centering
    \includegraphics[width=0.48\textwidth]{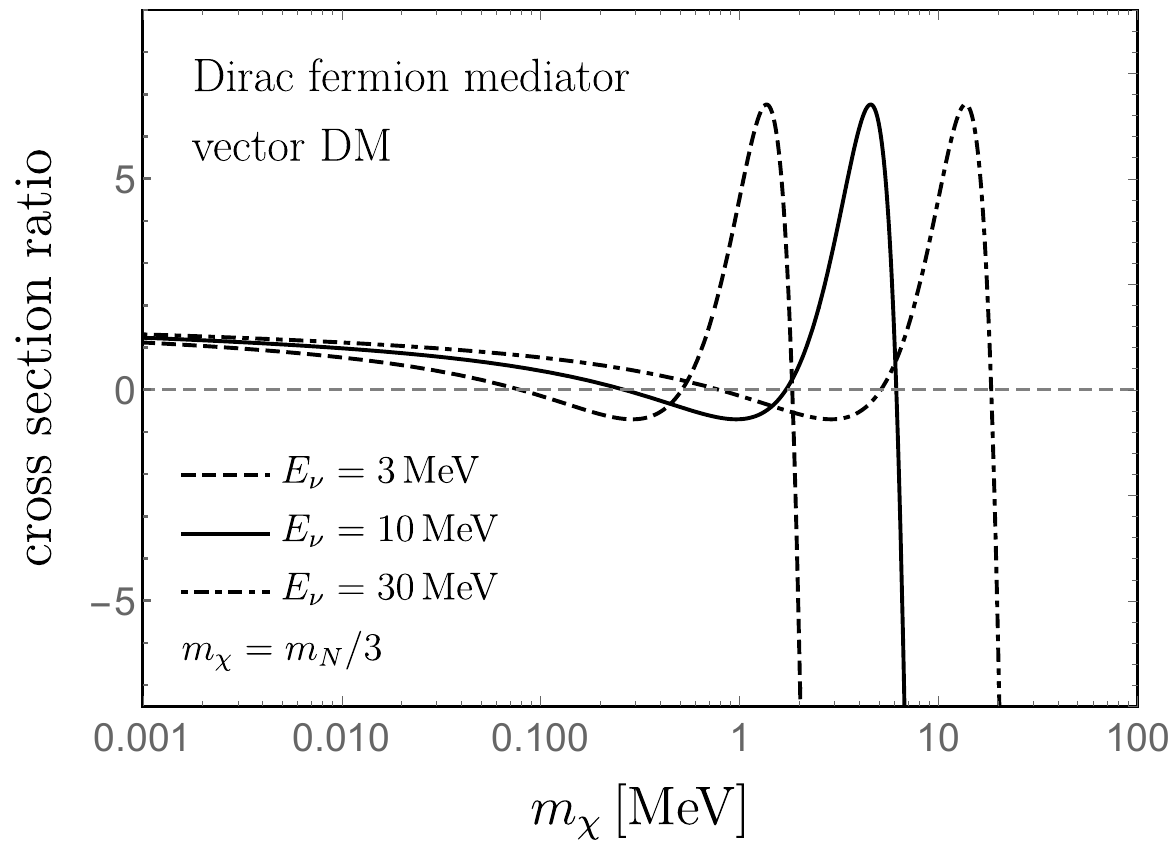}
    \includegraphics[width=0.49\textwidth]{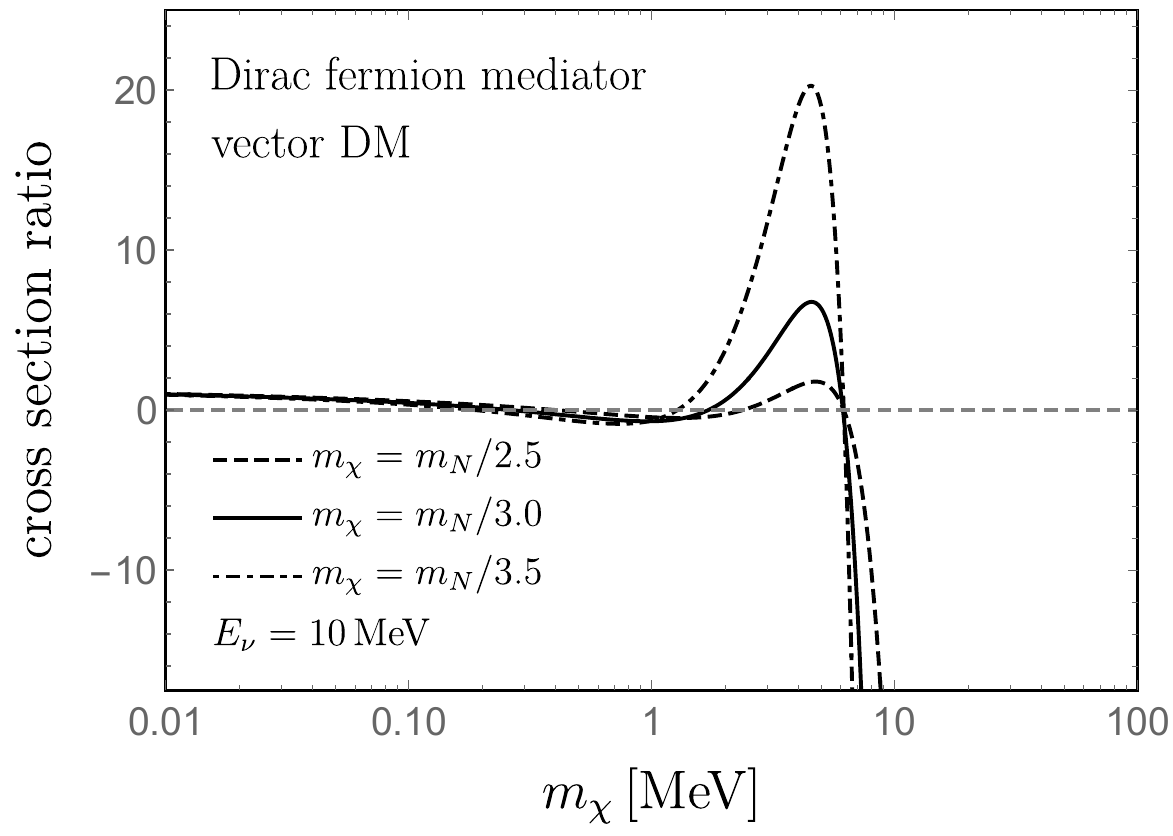}
    \caption{Cross section ratios of those in Ref.~\cite{Olivares-DelCampo:2017feq} with respect to our calculations for the case of Dirac fermion mediator and vector DM (cf. Table~\ref{tab:models}). The left panel is for the cases of $E_\nu = 3$ MeV, 10 MeV, and 30 MeV with $m_\chi = m_N/3$, while the right panel is for the cases of $m_N/m_\chi = 2.5$, 3.0, and $3.5$ with $E_\nu = 10$ MeV. }
    \label{fig:xs:comparison:2}
 \end{figure}
 \begin{figure}[!t]
    \centering
    \includegraphics[width=0.48\textwidth]{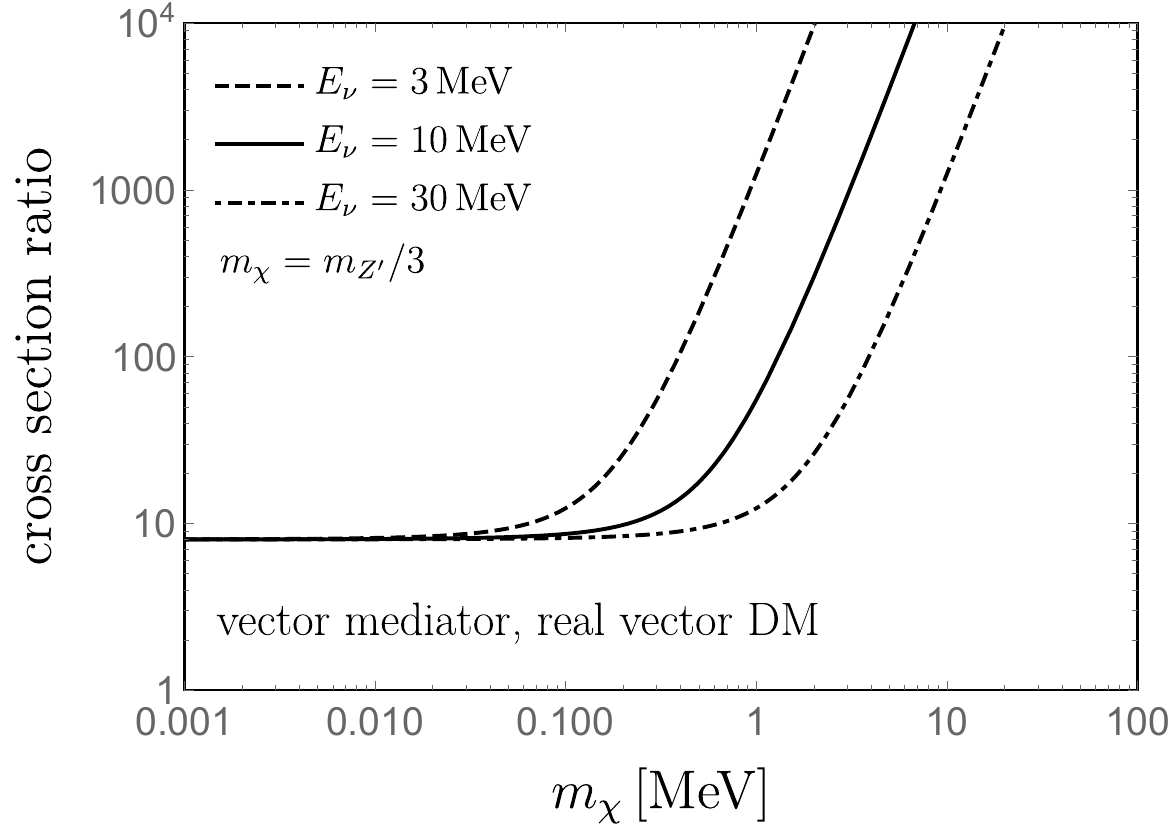}
    \includegraphics[width=0.48\textwidth]{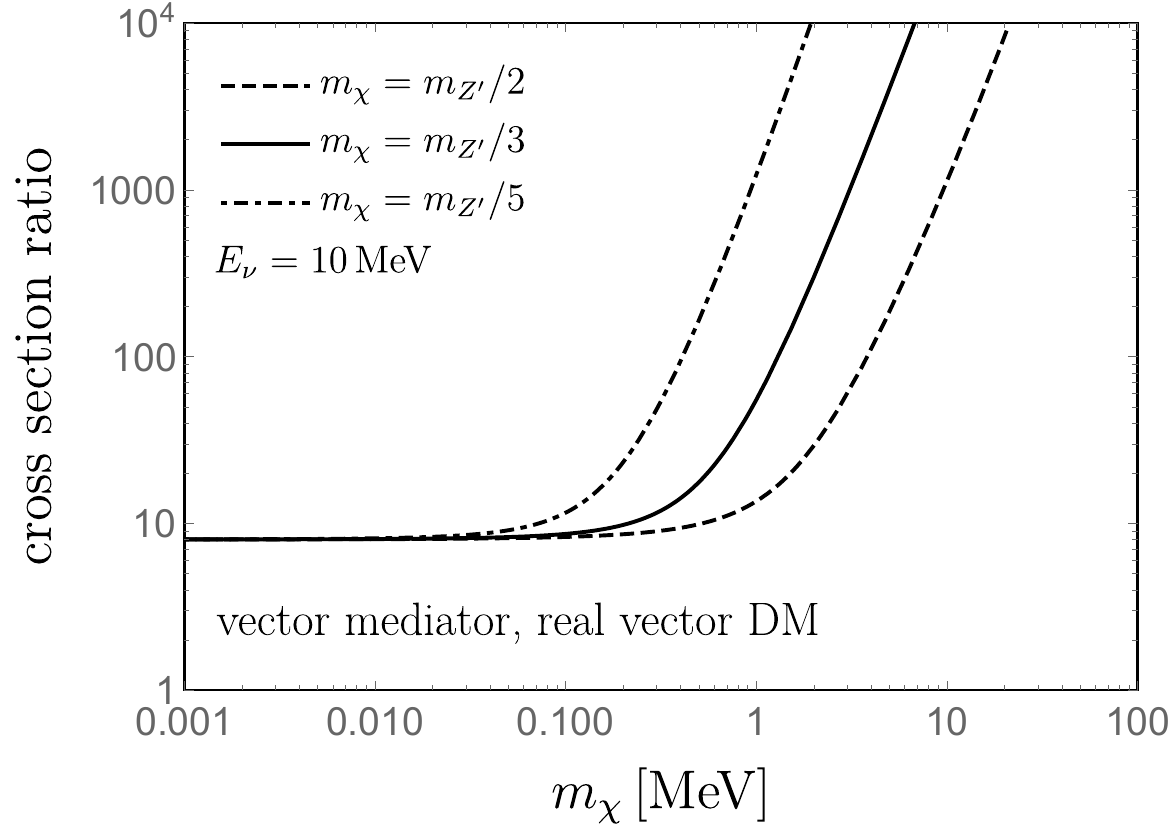}
    \caption{Cross section ratios of those in Ref.~\cite{Olivares-DelCampo:2017feq} with respect to our calculations for the case of vector mediator and real vector DM (cf. Table~\ref{tab:models}). The left panel is for the cases of $E_\nu = 3$ MeV, 10 MeV, and 30 MeV with $m_\chi = m_{Z'}/3$, while the right panel is for the cases of $m_{Z'}/m_\chi = 2$, 3, and $5$ with $E_\nu = 10$ MeV.}
    \label{fig:xs:comparison:3}
\end{figure}

The comparison of our cross sections for $\nu$-DM scattering with those from Refs.~\cite{Olivares-DelCampo:2017feq, Arguelles:2017atb} in all the models are collected in the last two columns of Table~\ref{tab:models}. As shown in this table, in most of the cases our results are consistent with Refs.~\cite{Olivares-DelCampo:2017feq, Arguelles:2017atb}, up to a factor of 2. However, for three models in the table, our calculations are different from those given in Ref.~\cite{Olivares-DelCampo:2017feq}, which are labeled by the cross marks and are described below:  
\begin{itemize}
\item The comparison of our calculations with those of Ref.~\cite{Olivares-DelCampo:2017feq} for the case of scalar mediator and Majorana fermion DM in the $s$- and $u$-channels is shown in Fig.~\ref{fig:xs:comparison:1}. The ratio of the $\nu$-DM scattering cross section from Ref.~\cite{Olivares-DelCampo:2017feq} over ours in this work is drawn as a function of the DM mass, for fixed neutrino energy $E_\nu$ and DM-mediator mass ratio. The lines in the left panel are for the cases of neutrino energy of $E_\nu = 3$ MeV (dashed), 10 MeV (solid), and 30 MeV (dot-dashed) with the fixed DM and mediator mass ratio of $m_\chi = m_\phi/3$, while the right panel is for the cases of DM-mediator mass ratios $m_\phi / m_\chi = 2$ (dashed), 3 (solid), and 5 (dot-dashed) with fixed neutrino energy $E_\nu = 10$ MeV. It is clear that the result from Ref.~\cite{Olivares-DelCampo:2017feq} is larger than ours by roughly a factor of $4$ when DM mass $m_\chi \lesssim 1$ MeV, and approaches to be the same as ours when the DM is heavy. 
    
\item Next, the comparison of the cross sections for the case of Dirac fermion mediator $N$ and vector DM $\chi$ is presented in Fig.~\ref{fig:xs:comparison:2}. The left panel is for the benchmark scenarios of $E_\nu = 3$ MeV (dashed), 10 MeV (solid), and 30 MeV (dot-dashed) with $m_\chi = m_N/3$, while the right panel is for $m_N / m_\chi = 2.5$ (dashed), 3.0 (solid), and 3.5 (dot-dashed) with fixed neutrino energy $E_\nu = 10$ MeV. Our result is close to that in Ref.~\cite{Olivares-DelCampo:2017feq} when DM mass is small, i.e. $m_\chi \lesssim {\cal O} (10 \; {\rm keV})$. However, when $m_\chi \gtrsim {\cal O} (100 \; {\rm keV})$, the $\nu$-DM scattering cross section in Ref.~\cite{Olivares-DelCampo:2017feq} becomes unphysically negative and differs significantly from ours.
    
\item Finally, the comparison of our result with Ref.~\cite{Olivares-DelCampo:2017feq} for the case of vector mediator and real vector DM is shown in Fig.~\ref{fig:xs:comparison:3}, where the parameter setups and labels in the left and right panels are the same as those in Fig.~\ref{fig:xs:comparison:1}. When the DM mass is small, the difference of cross sections is at the order of ${\cal O}(10)$. However, when DM is heavy, i.e. $m_\chi \gtrsim {\cal O}(100 \; {\rm keV})$, the result in Ref.~\cite{Olivares-DelCampo:2017feq} differs more significantly from ours.
\end{itemize}

\section{Bounds on the \nuDM~models}
\label{sec:bounds}

In this section, we discuss the existing bounds on the $\nu$-DM models given in Section~\ref{sec:nuDMInteractions}, including bounds on the neutrino and DM self-interactions, as well as their interactions with the mediators. We summarize the existing limits and highlight the new limits obtained here. We organize the bounds into the following categories: (i) astrophysical bounds, such as those coming from SN1987A, high-energy neutrino sources, and bullet cluster constraints; (ii) cosmological bounds, such as those from CMB, BBN, collisional damping and thermal relic density; and (iii) laboratory bounds, such as from meson, tau and $Z$ decays, and $\beta\beta$ decays. We should emphasize here that this is the minimal set of constraints just taking into account the DM interactions with neutrinos; including DM couplings to other SM fermions (as in specific UV-complete models) will typically result in additional constraints.

\subsection{Astrophysical bounds}

\begin{figure}[t!]
    \centering
  \includegraphics[width=0.8\textwidth]{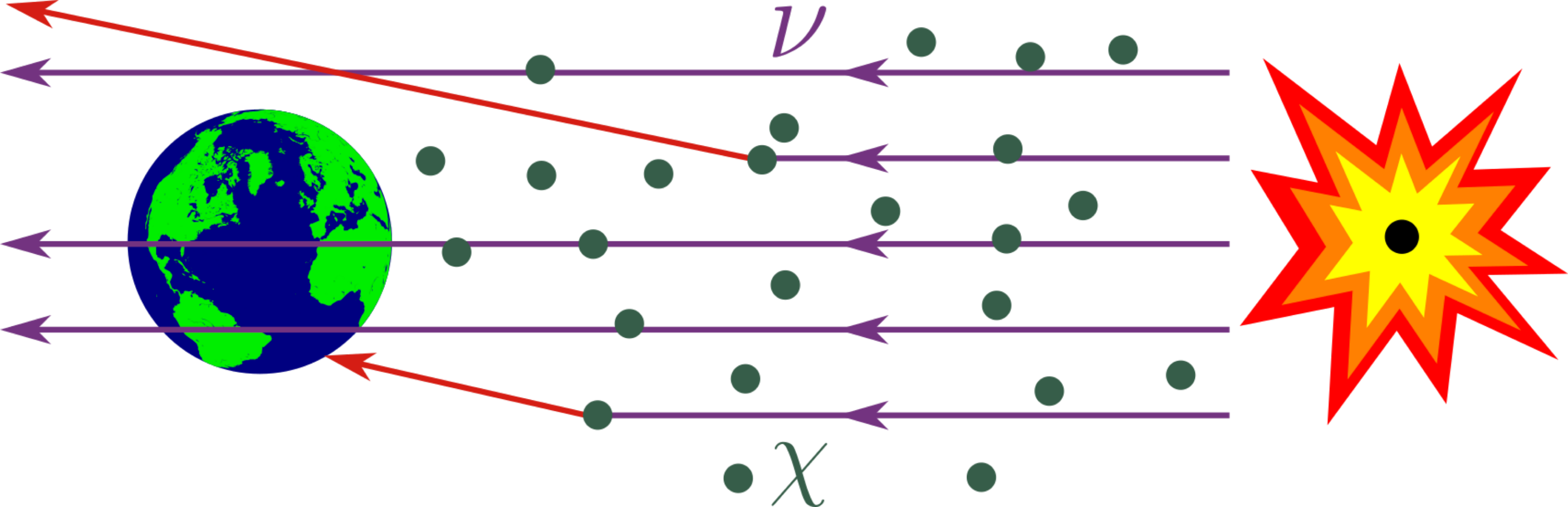}
    \caption{A schematic representation of neutrinos from astrophysical sources such as supernova explosion scattering in the galactic DM halo.}
    \label{fig:schematic}
\end{figure}

This subsection is on the astrophysical limits on the couplings of neutrinos and DM particles. A schematic representation of the scattering of neutrinos from astrophysical sources with DM particles is illustrated in Fig.~\ref{fig:schematic}. 
We would like to emphasize that the {\it bullet cluster limits} on the coupling and mass parameters in the DM models are {\it new} in this paper. These limits are based on the cross sections for DM-DM scattering in Appendix~\ref{app:dmself}. The $\nu$-$\nu$ scattering cross sections are relevant to limits from supernovae and IceCube, as well as those from CMB in the next subsection. Although these calculations are not used for our limits in this paper, for the sake of completeness, they are given in  Appendix~\ref{Appendix:xs:nu-nu}. They may be of interest for future studies or to readers exploring related contexts.

\subsubsection{Supernova 1987A}

Non-standard interactions (NSIs) of neutrinos have far-reaching effects on supernovae, e.g. affecting neutrino evolution in the supernova core, preventing the shock revival, halting the supernova explosion, and inducing neutrino flavor transformation~\cite{Gelmini:1982rr,Goldman:1982bd,Kolb:1987qy,Manohar:1987ec,Choi:1987sd,Dicus:1988jh,Choi:1989hi,Bilenky:1999dn,Kachelriess:2000qc,Tomas:2001dh,Davoudiasl:2005fd,Blennow:2008er,Zhou:2011rc,Galais:2011jh,Sher:2011mx,Heurtier:2016otg,Das:2017iuj,Dighe:2017sur,Brune:2018sab,Yang:2018yvk,Shalgar:2019rqe,Reddy:2021rln,Fiorillo:2022cdq,Chang:2022aas,Akita:2022etk,Akita:2023iwq,Telalovic:2024cot}. 
However, for sufficiently large NSIs, neutrinos stream as a fireball, and the effects of neutrino NSIs on the supernova core are expected to be small for some regions of the parameter space~\cite{Fiorillo:2023ytr,Fiorillo:2023cas,Fiorillo:2024upk}. 
Future observations of the diffuse supernova neutrino background (DSNB) could also constrain neutrino self-interactions and neutrino-DM interactions~\cite{Farzan:2014gza, Akita:2022etk, Balantekin:2023jlg}.\footnote{The Super-K collaboration has recently reported a $2.3\sigma$ excess, which could be possibly from DSNB~\cite{kearns_talk}.} The supernova neutrinos may also scatter with C$\nu$B due to the neutrino NSIs~\cite{Kolb:1987qy,Bilenky:1999dn,Shalgar:2019rqe}. The corresponding limits are more stringent if the mediator mass is below the keV scale.  A (nearly) massless particle $\phi$ coupling to neutrinos could induce some exotic processes such as $\nu + \nu \to \phi$ and $\nu + \bar\nu \to \phi + \phi$, and the SN1987A limit on the coupling is $|g_\nu| \lesssim {\cal O} (10^{-7})$ depending on the neutrino flavors involved~\cite{Kachelriess:2000qc,Farzan:2002wx}. 

Observed neutrinos from SN1987A also place a constraint on the strength of $\nu$-DM interactions. This is discussed in detail in Section~\ref{sec:supernova}; for completeness, we summarize the main results here. A critical quantity in setting limits is the opacity $\tau$ defined in Eq.~(\ref{opacitydef}), which is given by a product of the $\nu$-DM scattering cross section $\sigma_{\nu \chi}$ and the DM column density $\eta$ [defined in Eq.~(\ref{eq:ColumnDensity})] encountered by the neutrino en route to Earth.  If the interaction strength were sufficiently large to produce an opacity $\tau \gtrsim 1$, a significant portion of the line-of-sight neutrino flux would have been deflected away en route to Earth. Since we have observed neutrinos from SN1987A, we know that DM is not a fully opaque medium for neutrinos as they move through it. We can therefore place a bound on the mediator mass and the coupling for each model, ruling out $\tau \gtrsim 1$, where $\tau$ is calculated using the integrated column density of DM from the Earth to the Large Magellanic Cloud (LMC) hosting SN1987A. We use a truncated NFW profile up to 2.7 kpc for the LMC subhalo~\cite{Regis:2021glv}.

There are additional constraints on SN1987A neutrino interactions when considering neutrino annihilation into DM in the supernova core~\cite{Manzari:2023gkt,Cappiello:2025tws}. Given the complexity of this calculation for all the models listed here, these bounds are not given in this paper and will be explored in a future study.

\subsubsection{High-energy astrophysical neutrinos}

With neutrino self-interactions or $\nu$-DM interactions, high-energy astrophysical neutrinos scatter off the C$\nu$B or DM, and in some scenarios induce an absorption line in IceCube neutrino spectra~\cite{Ibe:2014pja, Araki:2014ona}. Thus, the neutrino NSI are constrained by the spectral information of the IceCube data~\cite{Ng:2014pca,Ioka:2014kca,Araki:2014ona,Araki:2015mya,Mazumdar:2020ibx,Bustamante:2020mep,Hyde:2023eph, delaVega:2024pbk} (see also Refs.~\cite{Murase:2019xqi,Creque-Sarbinowski:2020qhz,Esteban:2021tub} for future prospects). The most stringent limits of this type are from IceCube High Energy Starting Events~\cite{Bustamante:2020mep}. 

The IceCube Collaboration has also observed neutrinos from identified neutrino point sources, such as TXS 0506+056~\cite{IceCube:2018cha} and NGC 1068~\cite{IceCube:2022der}. Such neutrinos are also affected by the scattering off the C$\nu$B, and therefore, the survival of the emitted neutrinos en route to Earth can also be used to set limits on neutrino NSI~\cite{Doring:2023vmk}.\footnote{We use the updated limits from Ref.~\cite{Doring:2023vmk} instead of the old ones from Ref.~\cite{Kelly:2018tyg}.} 

Additionally, as shown in \cite{Cline:2023tkp,Fujiwara:2023lsv}, high-energy neutrinos observed at IceCube from active galactic nuclei (AGNs) like TXS 0506+056 and NGC 1068 can place upper limits on neutrino-DM interactions since the observed neutrinos propagated through the host AGN's DM halo, modeled by a spike profile. The cited studies provided upper limits on the neutrino-DM cross section as a function of the DM mass and the integrated number density of DM along the line-of-sight to IceCube. 
In particular, for NGC 1068, the bound is given by~\cite{Cline:2023tkp}
\begin{equation}
    \sigma_{\nu\chi} < 0.34 \frac{m_\chi}{\Sigma_\chi},
\end{equation}
obtained using a flux attenuation scenario, where $\sigma_{\nu\chi}$ is assumed to be constant with respect to the neutrino energy and $\Sigma_\chi \sim 6 \times 10^{31} $ GeV/cm$^2$ is the column density for 1 GeV DM.
We have shown the NGC 1068 constraint for each model in the parameter space considered, with the neutrino energy set to a benchmark value of 10 TeV. As mentioned in \cite{Cline:2023tkp}, the NGC 1068 limit is stronger than that of TXS 0506+056, so we only show the stronger limit.

The recent KM3NeT observation of the highest-energy muon event~\cite{KM3NeT:2025npi} has also been used to derive stringent limits on $\nu$-DM interactions, assuming certain candidate point sources~\cite{Bertolez-Martinez:2025trs, Mondol:2025uuw}. However, since neither the source of the KM3NeT event nor its neutrino origin is confirmed yet,\footnote{In particular, the standard neutrino origin of the observed KM3NeT event is in $2.5-3.6\sigma$ tension with non-observation at IceCube with much larger effective area times exposure time~\cite{Li:2025tqf, KM3NeT:2025ccp}. See Refs.~\cite{Farzan:2025ydi,Dev:2025czz} for a potential DM origin of the KM3NeT event in order to explain this tension. See also Ref.~\cite{Brdar:2025azm} for a nonstandard neutrino explanation.} we do not include these limits here.

\subsubsection{Bullet cluster constraints on self-interacting dark matter}
\label{subsection:bulletcluster}

DM self-interactions could have an impact on colliding galaxy clusters, as they transfer momentum between the DM halos of the host galaxies, causing them to lag behind the collisionless DM hypothesis. Non-observation of such separation in the bullet cluster is used to
set limits on the self-interacting dark matter (SIDM), e.g. in Refs.~\cite{Markevitch:2003at,Bradac:2006er,Robertson:2016xjh, Tulin:2017ara}. The limit is roughly $\sigma / m \leq 1\; {\rm cm}^2/{\rm g}$, with an uncertainty of roughly a factor of 2~\cite{Cirelli:2024ssz}. The SIDM limits are only applicable\footnote{There are loop level self-interactions in the DM-neutrino-mediator models which are highly suppressed.} to those models with DM-DM-mediator type of interactions in Table~\ref{tab:models}. 

\subsubsection{Dark Matter annihilation to neutrinos}

Thermal dark matter in the Milky Way halo can annihilate to neutrinos, with constraints on the annihilation cross section by experiments such as Super-K, Borexino, and XENONnT~\cite{Arguelles:2019ouk,BetancourtKamenetskaia:2025rwk}. We adapt these constraints for our models, showing in the summary plots the region of parameter space that is excluded due to too large an annihilation rate. These bounds follow scale similarly as the thermal relic density contour for each model, described in more details in~\cref{subsubsec:thermalrelic}.

\subsection{Cosmological bounds}

This subsection is devoted to the cosmological limits on the couplings of neutrinos and DM particles. Here the new results include the following: (i) In Section~\ref{subsubsec:BBN}, we clarify some misconceptions on the BBN constraints;  (ii) We consider in Section~\ref{subsubsec:CD} the collisional damping limits for all the DM models in this paper, which were only applied to some of the models in the literature (cf. Refs.~\cite{ Boehm:2000gq,Boehm:2001hm,Boehm:2004th,Olivares-DelCampo:2017feq}); (iii) In Section~\ref{subsubsec:thermalrelic} we obtain the thermal relic density curves for all the DM models in the parameter space of couplings and masses, derived from the annihilation of DM into neutrinos, with the cross sections given in Appendix~\ref{app:thermalrelic}. 

\subsubsection{CMB}

Neutrino NSI affect the time of neutrino free streaming and its interactions with the photon-baryon fluid in the early Universe. The effects of NSI can be seen in the phase shift and amplitude of the matter power spectrum. This bound exists not only for neutrino self-interactions, but also for interactions of neutrinos with any light dark sector; see e.g. Refs.~\cite{Bell:2005dr,Cyr-Racine:2013jua,Archidiacono:2013dua, Lancaster:2017ksf, Oldengott:2017fhy, Kreisch:2019yzn,Park:2019ibn, Escudero:2019gvw, RoyChoudhury:2020dmd, Brinckmann:2020bcn, Taule:2022jrz, Camarena:2024zck,Forastieri:2019cuf,Kreisch:2022zxp,He:2023oke,Camarena:2023cku, Trojanowski:2025oro}.
The neutrino self-interactions or NSI with dark sector may also help alleviate some tensions in the cosmological datasets, including the $H_0$ tension and $\sum m_\nu$ tension, see e.g. Refs.~\cite{Cyr-Racine:2013jua,Oldengott:2014qra,Lancaster:2017ksf,Oldengott:2017fhy,Escudero:2019gvw,Kreisch:2019yzn,Barenboim:2019tux,Ghosh:2019tab,Das:2020xke,RoyChoudhury:2020dmd,Brinckmann:2020bcn,Lyu:2020lps,Mazumdar:2020ibx,Das:2020xke,Huang:2021dba,RoyChoudhury:2022rva,Das:2023npl,Venzor:2023aka,Camarena:2023cku,He:2023oke,Poudou:2025qcx,He:2025jwp, Das:2025asx}. The most recent CMB limit on neutrino NSI is from Ref.~\cite{Camarena:2024zck}, which excludes the effective four-neutrino interactions with the strength of 
\begin{equation}
G_{\rm eff} \simeq \frac{g_\nu^2}{m_{\rm med}^2} < 5.6 \times 10^{-5} \; {\rm MeV}^{-2} \,.
\end{equation}
The CMB matter power spectrum bound can in principle be evaded by $\nu$-DM scattering into electrons. This happens, for instance, in the scotogenic model: $\nu\chi^0\to \chi^+e^-$, where $\chi^+$ is a dark particle which forms a $SU(2)_L$ doublet with the neutral DM particle $\chi^0$~\cite{Tao:1996vb,Ma:2006km}. This is actually a well-motivated UV-completion of our $\phi$-$\nu$-$\chi$ coupling, which is also directly connected to the generation of neutrino masses. More comments on the UV-completions can be found in Appendix~\ref{sec:UV}.

Additionally, there are constraints on the effective number of neutrino species $N_{\rm eff}$ on the couplings of light mediators to neutrinos~\cite{Kamada:2015era,Kamada:2018zxi,Escudero:2019gzq,Li:2023puz,Esseili:2023ldf,Li:2023kuz,Ghosh:2023ilw,Foroughi-Abari:2025mhj,Ghosh:2024cxi} (see also Refs.~\cite{Knapen:2017xzo,Wang:2023csv} for the $N_{\rm eff}$ constraints on the couplings of mediators to DM particles).  The precise measurement of $N_{\rm eff}$ by the Planck data~\cite{Planck:2018vyg} has excluded the couplings of scalar and vector mediators to neutrinos up to ${\cal O} (10^{-10})$~\cite{Li:2023puz}. With the unprecedented precision of $N_{\rm eff}$ at the Simons Observatory~\cite{SimonsObservatory:2018koc,SimonsObservatory:2019qwx}, CMB-S4~\cite{CMB-S4:2016ple,Abazajian:2019eic} and CMB-HD~\cite{CMB-HD:2022bsz}, the effective couplings of mediators to neutrinos can be improved by up to three to four orders of magnitude~\cite{Li:2023puz}.

Finally, there are limits on neutrino-DM interactions from CMB measurements such as the Lyman-$\alpha$ forest~\cite{Wilkinson:2014ksa}, structure formation~\cite{Crumrine:2024sdn}, and small-scale CMB data~\cite{Giare:2023qqn}. We adopt the bounds in Ref.~\cite{Wilkinson:2014ksa} for each model, in a similar style to the bound obtained by collisional damping as described in Sec.~\ref{subsubsec:CD}. The following upper limits on the neutrino-DM scattering cross section are given based on temperature dependence:
\begin{equation}
\sigma_{\nu\chi} < \sigma_n \left( \frac{m_\chi}{\rm GeV} \right) \left(  \frac{T_\nu}{T_\nu^0}\right) \,, \quad \text{with } n = 0,\,2,
\end{equation}
where $T_\nu^0 = 6.1$ K is the cosmic neutrino temperature today~\cite{Akita:2023yga}.
For the indices $n = 0,\, 2$, while allowing $N_{\rm eff}$ to vary, the corresponding limits are, respectively~\cite{Wilkinson:2014ksa}, 
\begin{equation}
\sigma_0 = 2\times 10^{-28} \; {\rm cm}^2 \,, \quad
\sigma_2 = 2\times10^{-39} \; {\rm cm}^2 \,.
\end{equation}
We find that in all models, $n$ = 2 for the parameter space considered. Yet this bound requires a more thoroughly analysis to properly determine the ruled out parameter space for each model since this interpolation is an estimate; therefore, we show a contour on each summary but do not shade the region above the contour.

\subsubsection{BBN}
\label{subsubsec:BBN} 

If the mediator is very light, it will contribute significantly to the effective number $N_{\rm eff}$ of extra relativistic degrees of freedom in the early Universe. The state-of-the-art theoretical prediction for $N_{\rm eff}$ in standard cosmology is $3.0440\pm 0.0002$~\cite{Akita:2020szl,Froustey:2020mcq,Bennett:2020zkv}, whereas the latest Planck data gives $N_{\rm eff} = 2.99 \pm 0.17$~\cite{Planck:2018vyg}, in agreement with the theoretical prediction.
This information, along with the precision measurements of the primordial abundances of light elements, can be used to put stringent constraints on the light mediator mass and couplings~\cite{Huang:2017egl,Blinov:2019gcj,Venzor:2020ova,Grohs:2020xxd} (see also Ref.~\cite{Beacom:2004yd}). 
We adopt the BBN limits from Ref.~\cite{Blinov:2019gcj}, which exclude the mediator mass lower than ${\cal O} ({\rm MeV})$. However, it has been argued that the BBN limits can in principle be avoided or relaxed, if the BSM interactions of neutrinos with other SM particles are activated after BBN and decouple before CMB~\cite{Berbig:2020wve,Wang:2023csv}, or in the dark sink models, with large entropy dump occurring at the QCD phase transition~\cite{Bhattiprolu:2023akk}, or in models with low reheating temperatures~\cite{Berlin:2018ztp}.\footnote{Several recent experimental proposals (see e.g., Refs.~\cite{Dutta:2024kuj, Gao:2024irf, BetancourtKamenetskaia:2025rwk, Chen:2025cvl}) explore the direct detection of sub-MeV DM, setting aside the cosmological bounds. While we agree that the direct DM searches should be pursued irrespective of the cosmological bounds which may be model-dependent, as we argue here, one has to carefully consider the other constraints discussed here (especially the robust laboratory ones) as well to see if sub-MeV DM with sizable coupling to SM particles still remains viable.} 

In the presence of both mediator and DM particles at the MeV scale, there will be additional limits from BBN. For sufficiently light DM, the process $\nu \bar\nu \to \chi \bar\chi$ will keep the DM $\chi$ in equilibrium with SM particles in the early Universe, thus making DM contribute significantly to $N_{\rm eff}$.\footnote{The process $\nu \bar\nu \to \chi \bar\chi$ also offers a mechanism to produce the DM particles in the early Universe via the freezing-in of neutrinos. However, the couplings are required to be rather small, i.e. $g_\nu g_\chi \sim 10^{-12}$, not relevant to the parameter space we are interested in here~\cite{Hufnagel:2021pso}.} Let us make a rough estimate of the corresponding BBN limit. Neglecting the DM mass, the thermally-averaged interaction rate goes as $\Gamma_{\rm int}\sim g_\nu^4 T^5 / m_{\rm med}^4$. Comparing it with the Hubble rate $H\sim T^2/m_{\rm Pl}$ with $m_{\rm Pl}$ being the Planck mass, the ratio is 
\begin{equation}
\frac{\Gamma_{\rm int}}{H}\sim
\left( \frac{g_\nu}{10^{-2}} \right)^4
\left( \frac{m_{\rm med}}{100\; {\rm MeV}} \right)^{-4}
\left( \frac{T}{1\; {\rm MeV}} \right)^3 \,.
\end{equation}
It seems that in large regions of parameter space for the scenarios we are considering, sub-MeV DM will be in equilibrium with neutrinos at $T = 1$ MeV and therefore highly disfavored by BBN, unless the coupling is sufficiently small and/or the mediator is sufficiently heavy.

\begin{figure}[!t]
    \centering
    \includegraphics[width=0.48\textwidth]{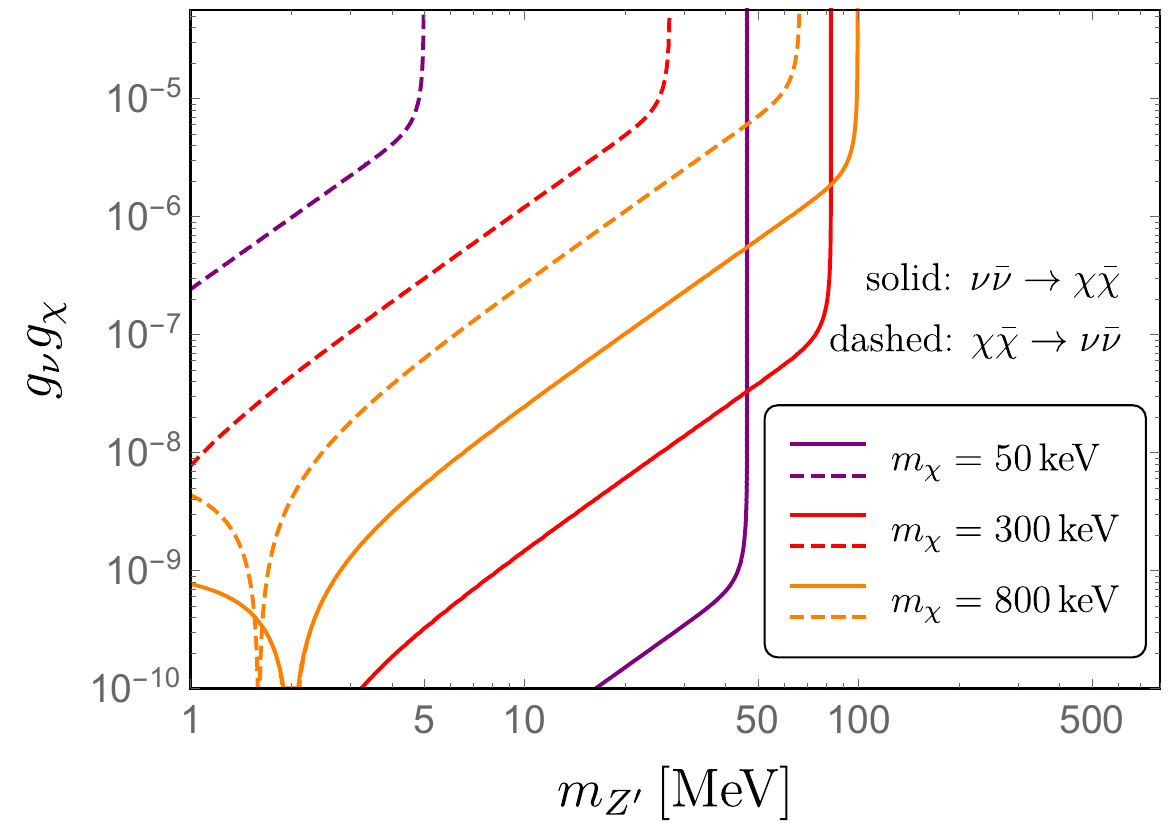}
    \includegraphics[width=0.48\textwidth]{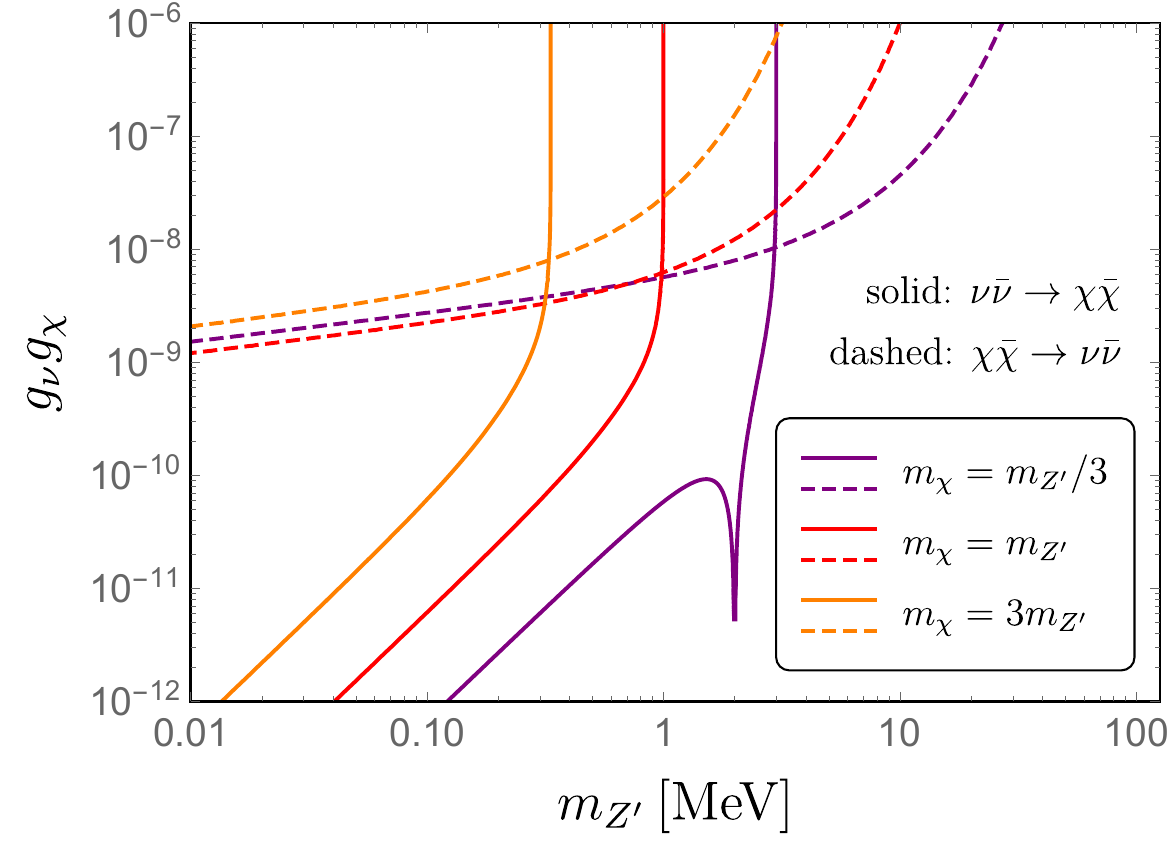}
    \caption{BBN limit on the mediator mass $m_{Z'}$ and the coupling $g_\nu g_\chi$, in the model of vector mediator and complex vector DM, as a result of the processes $\nu \bar\nu \to \chi \bar\chi$ (solid lines) and $\chi\bar\chi \to \nu\bar\nu$ (dashed lines)  at $T = 1$ MeV. In the left panel, the DM mass is fixed to be $m_\chi = 50$ keV (purple), 300 keV (red), and 800 keV (orange). In the right panel, the DM-mediator mass ratio is set to be $m_\chi / m_{Z'} = 1/3$ (purple), $1$ (red), and $3$ (orange). The regions above or to the left of the curves are disfavored by BBN.}
    \label{fig:BBN}
\end{figure}
For illustration purposes, we perform a more accurate estimate for the case of a vector mediator and complex vector DM.
With the couplings in Eq.~(\ref{eqn:L:vector:vector_complex:t}), the scattering cross section $\sigma (\nu \bar\nu \to \chi \chi^\dagger)$ is given in Eq.~(\ref{eqn:xs:nunu_to_DMDM}).
Given $E_\nu \sim T$ with $T$ being the temperature, the corresponding rate is $\Gamma_{\nu \chi} = \sigma (\nu \bar\nu \to \chi \bar\chi) n_\nu (T)$, with $n_\nu (T)$ being the number density of neutrinos at temperature $T$.
The Hubble rate is $H \simeq 1.66 \sqrt{g_\ast} T^2/m_{\rm Pl}$, where $g_\ast = 10.75$ is the effective degree of freedom of SM particles at $T=1$ MeV. 
The BBN limit from the requirement of $\Gamma / H < 1$ for the cases of fixed DM mass of $m_\chi = 50$ keV, 300 keV, and 800 keV are shown, respectively, as the solid purple, red, and orange lines in the left panel of Fig.~\ref{fig:BBN}. For  comparison, we also estimate  the rate for DM annihilating into neutrinos $\Gamma_{\chi \nu} = \sigma (\chi \bar\chi \to \nu \bar\nu) n_\chi (m,\,T)$, where the cross section $\sigma (\chi \bar\chi \to \nu \bar\nu)$ can be found in Eq.~(\ref{eqn:DM:annihilation:vector:vector:c}), and $n_\chi(T)$ is the DM number density at temperature $T$: 
\begin{equation}
n_\chi (m,\, T) = \frac{g_{\rm dof} T^3}{2\pi^2} \int_0^\infty \frac{{\rm d}y}{e^{\sqrt{y^2 + m^2/T^2}} \mp 1} \,,
\end{equation}
with $g_{\rm dof}$ the degree of freedom of DM, and $-1$ for bosonic DM and $+1$ for fermionic DM. For simplicity, we have integrated over the three-momentum ${\bf p}$ of DM (here $y \equiv |{\bf p}|/T$). The resulting limits on $m_{Z'}$ and $g_\nu g_\chi$ from comparing $\Gamma_{\chi \nu}$ with the Hubble rate $H$ are presented as the dashed lines in the left panel of Fig.~\ref{fig:BBN}. In evaluating the cross sections $\sigma(\nu\bar{\nu} \to \chi\bar{\chi})$ and $\sigma(\chi\bar{\chi}\to \nu\bar{\nu})$, we have taken into account the widths of $Z' \to \nu_i \bar\nu_i,\, \chi \bar\chi$, where all the three generations of neutrinos are included $i = 1,\,2,\,3$. It should be noted that the partial width $\Gamma (Z' \to \chi \bar\chi) \propto g_\chi^2 m_{Z'}^5/m_{\chi}^4$ when the DM mass $m_\chi \ll m_{Z'}$ [cf. Eq.~(\ref{eqn:width:MedV:DMV})]. For sufficiently large coupling $g_\chi$, the denominator of Eqs.~(\ref{eqn:xs:nunu_to_DMDM}) and (\ref{eqn:DM:annihilation:vector:vector:c}) becomes dominated by the $m_{Z'}^2 \Gamma_{Z'}^2$ term, then the cross sections scale as 
\begin{equation}
\sigma (\nu \bar\nu \to \chi \bar\chi,\; \chi \bar\chi \to \nu \bar\nu) \propto \frac{g_\nu^2 g_\chi^2}{m_{Z'}^2 \Gamma_{Z'}^2} \propto \frac{g_\nu^2}{g_\chi^2} \frac{m_{\chi}^8}{m_{Z'}^{12}} \,.
\end{equation}
For simplicity, we have set $g_\nu = g_\chi$, and then the BBN limits become independent of the couplings. This generates the BBN lower limit on the mediator mass, which is indicated by the vertical parts of the solid and dashed lines in the left panel of Fig.~\ref{fig:BBN}. 
The BBN limits for the cases of $m_\chi / m_{Z'} = 1/3$, 1, and 3 are shown, respectively, as the purple, red, and orange lines in the right panel of Fig.~\ref{fig:BBN}. The solid and dashed lines are again for the processes $\nu \bar\nu \to \chi \bar\chi$ and $\chi \bar\chi \to \nu \bar\nu$, respectively. As expected, for neutrino annihilating into DM, the mediator mass is required to be larger than the MeV scale, unless the coupling $g_\nu g_\chi$ is very small, say $g_\nu g_\chi \lesssim 10^{-9}$, such that DM cannot be produced efficiently from annihilation of neutrinos. The dip of the solid purple line is due to the resonance effect in the process $\nu \bar\nu \to \chi \bar\chi$. 

As can be seen from Fig.~\ref{fig:BBN}, the exact BBN limit on the mediator mass depends on the value of the DM mass chosen. In our summary plots below, we choose $m_\chi=m_{\rm med}/3$ for which the limit becomes $m_{\rm med}\gtrsim 3$ MeV for $g_\nu g_\chi\gtrsim 10^{-10}$.

\subsubsection{Collisional damping}
\label{subsubsec:CD}

The $\nu$-DM interactions could potentially erase primordial DM fluctuations, suppressing the formation of large-scale structures (LSS) in the Universe, or even smaller-scale structures such as satellite galaxies. This is the so-called collisional damping effect~\cite{ Boehm:2000gq,Boehm:2001hm,Boehm:2004th}.
Such an effect depends to some extent on the neutrino temperature. For simplicity, we can parameterize the limits on $\nu$-DM scattering cross section $\sigma_{\nu\chi}$ in the following way:
\begin{equation}
\sigma_{\nu\chi} < \sigma_n \left( \frac{m_\chi}{\rm MeV} \right) \left( \frac{T_\nu}{T_0} \right)^n \,, \quad \text{with } n = 0,\,2,\,4 \,,
\end{equation}
with $T_0 = 6.1$ K the neutrino temperature today. It is found that for the indices $n = 0,\, 2,\, 4$, the corresponding limits are, respectively~\cite{Akita:2023yga}, 
\begin{equation}
\sigma_0 = 10^{-29} \; {\rm cm}^2 \,, \quad
\sigma_2 = 10^{-40} \; {\rm cm}^2 \,, \quad 
\sigma_4 = 10^{-51} \; {\rm cm}^2 \,.
\end{equation}
Then we make an interpolation of the cross section limits logarithmically, which can be simply written as 
\begin{equation}
\sigma_{\nu\chi} < 10^{-29 - \frac{11}{2}n} \left( \frac{m_\chi}{\rm MeV} \right) \left( \frac{T_\nu}{T_0} \right)^n \; {\rm cm}^2 \,,
\end{equation}
with $n$ now an arbitrary real index which is model-dependent, and $T_\nu = E_\nu / 3.15$ with $E_\nu$ the measured neutrino energy. In our calculations, the index $n$ is obtained from interpolating the data of $\{ \log_{10} (E_\nu/{\rm MeV}),\, \log_{10} (\sigma_{\nu\chi}/{\rm cm}^2) \}$. It is found that the index $n$ depends potentially on the neutrino energy $E_\nu$, as well as other parameters such as the DM and mediator masses involved. 

\subsubsection{Thermal relic density}
\label{subsubsec:thermalrelic}

Interactions between DM and neutrinos provide an efficient annihilation channel for DM into neutrinos in the early Universe~\cite{Palomares-Ruiz:2007trf, Kanemura:2025byi}. For each model, we plot a contour on the mediator mass and coupling plane (fixing DM mass to be $m_\chi / m_{\rm med} = 1/3$), where the average annihilation rate $\langle \sigma_A v \rangle$ equals the rate needed to match the observed thermal relic density of DM at freeze-out, $\sim (2-3) \times 10^{-26}$ cm$^3/$s~\cite{Steigman:2012nb}. In our numerical calculations, we take into account the mild dependence of $\langle \sigma_A v \rangle$ on the DM mass, up to roughly a factor of 2~\cite{Steigman:2012nb}. The formulas for the DM annihilation cross sections $\langle \sigma_A v \rangle$ for different models are given in Appendix~\ref{app:thermalrelic}. A region of parameter space producing a smaller annihilation rate would cause an over-abundance of thermal DM, which could be mitigated by other annihilation channels for the DM candidate. On the other hand, the region of parameter space producing a larger annihilation rate would cause an under-abundance of DM relic density. This scenario may require different assumptions on the DM model, e.g. asymmetric DM, to properly account for the observed relic density. However, it should be noted that the thermal relic density of DM is not strictly a constraint on the models discussed in this paper, rather just a reference line for the $\nu$-DM interactions. There exists a large variety of other DM scenarios, for instance, the freeze-in DM~\cite{Hall:2009bx}, dark sink models~\cite{Bhattiprolu:2023akk,Bhattiprolu:2024dmh}, and non-thermal DM candidates such as axion-like particles~\cite{Hui:2016ltb}, which could achieve the correct relic density without relying on the thermal annihilation.

\subsection{Laboratory bounds}

In this subsection, we update some of the laboratory limits on neutrinos and DM in the literature. The most important new laboratory limits obtained in this paper are the updated meson decay limits on the couplings of neutrinos and DM particles in Section~\ref{subsubsec:mesonlimits}. We generalize the calculations for the neutrinophilic couplings derived in our recent paper~\cite{Dev:2024ygx} to the $\nu$-DM couplings in all models considered here, including the contributions from the 1-loop diagrams and cancellation of infrared divergences in some of the cases. The limits from tau and $Z$ boson decays are relatively weaker. In this subsection, we also collect a comprehensive list of the laboratory limits for completeness, although most of them are rather weak or not relevant to the parameter space in this paper.

\subsubsection{Double-beta decays}

Electron neutrino NSI can be searched for in  $\beta \beta$ decays. For instance, the lepton number violating (LNV) coupling $J \nu_e \nu_e$ of Majoron $J$ to electron neutrinos can induce the exotic neutrinoless $\beta\beta$ decays~\cite{Georgi:1981pg} 
\begin{equation}
(Z,\, A) \to (Z+2,\,A) + e^- + e^- + J \,,
\end{equation}
If the Majoron mass $m_J$ is below the $Q$-value of the nuclear $\beta\beta$ transition (typically a few MeV), the $\beta \beta$ decay searches can put a bound on the corresponding coupling strength~\cite{Blum:2018ljv,Brune:2018sab}. However, we do not show this bound in our summary plots, as the models considered here are lepton-number-conserving in the vertex involving the mediator. 

Even if the couplings of the mediator to electron neutrinos are lepton number conserving, e.g. those in Eqs.~(\ref{eqn:L:scalar:scalar}) and (\ref{eqn:L:vector:scalar_complex:t}), they still suffer from the limits from $\beta\beta$ decay searches (cf. Figs.~1 and 3 in Ref.~\cite{Deppisch:2020sqh}). Because such couplings can generate the effective couplings in the form of $G_S(\bar\nu_e \nu_\alpha)(\bar\nu_e \nu_\beta)$ with the flavor indices $\alpha,\,\beta = e,\, \mu,\,\tau$.\footnote{For LNV couplings such as $J \nu_\alpha \nu_\beta$, the effective operators $(\nu_e \nu_e)(\nu_\alpha \nu_\beta)$ and $(\nu_e \nu_e)(\bar\nu_\alpha \bar\nu_\beta)$ can be generated~\cite{Deppisch:2020sqh}.} The corresponding coefficient is $G_S = g_\nu^2 / (t + m_\phi^2)$, with $t \sim p_F^2$ and $p_F \simeq 100$ MeV the Fermi momentum. Such effective coupling can induce the exotic two-neutrino $\beta\beta$ decays
\begin{equation}
(Z,\, A) \to (Z+2,\,A) + e^- + e^- + \bar\nu_\alpha + \bar\nu_\beta \,.
\end{equation}
The most stringent limit is from the data of two-neutrino $\beta\beta$ decay of $^{136}$Xe, i.e. $T_{1/2} > 2.17 \times 10^{21}$ year~\cite{Barabash:2019nnr, Pritychenko:2024ase}, which leads to the limit of $G_S < 3.2 \times 10^8 \; G_F$. The resultant constraint on $g_\nu$ is rather weak: $g_\nu > 6.1$ for $m_{\phi} \lesssim 100$ MeV. 

Although the current two-neutrino $\beta\beta$ limits are rather weak, this can be naturally generalized to the case of the neutrino-DM couplings, e.g. those in Eqs.~(\ref{eqn:L:scalar:Dirac}) and (\ref{eqn:L:vector:Dirac:u}). 
For sufficiently heavy mediators, the effective operators for the $\beta\beta$ decays are $(\bar\nu_e \chi)(\bar\nu_e \chi)$. If the DM particle $\chi$ is very light, e.g. below the ${\cal O} ({\rm MeV})$ scale, the two-DM $\beta\beta$ decays are allowed: 
\begin{equation}
(Z,\, A) \to (Z+2,\,A) + e^- + e^- + \bar\chi + \bar\chi \,.
\end{equation}
Then the two-neutrino $\beta\beta$ limits can be applied to constrain the $\nu$-DM couplings. However, as for the case of neutrino NSI above, these limits on $\nu$-DM interactions are also very weak. 

\subsubsection{Meson and tau decays}
\label{subsubsec:mesonlimits}

Given couplings to neutrinos, or interactions with neutrino and DM particle, a light mediator $\phi$ or $Z'$ can be produced in meson decays, e.g. ${\sf M}^\pm \to \ell^\pm + \nu/\chi + \phi$, with ${\sf M} = \pi,\; K,\; D,\; B$~\cite{Barger:1981vd, Gelmini:1982rr,Glashow:1985cm,Lessa:2007up,
Pasquini:2015fjv,Berryman:2018ogk} (see also Refs.~\cite{Farzan:2011tz,Carlson:2012pc,Krnjaic:2019rsv,Dutta:2021cip,Bickendorf:2022buy,Dutta:2023fnl}), and ${\sf M}^\pm \to \ell^\pm + \nu/\chi + Z'$~\cite{Laha:2013xua,Bakhti:2017jhm} (see also Refs.~\cite{Barger:2011mt,Carlson:2012pc,Krnjaic:2019rsv,Dutta:2021cip,Dutta:2023fnl}). The precision measurements of meson decays such as ${\sf M}^\pm \rightarrow \ell^\pm \nu$ can thus put a bound on the neutrinophilic interactions. 
However, for the couplings of scalar to neutrinos (and DM), the tree-level processes ${\sf M}^\pm \to \ell^\pm + \nu/\chi + \phi$ suffer from infrared (IR) divergences, i.e. the corresponding partial width goes to infinity in the limit of $m_\phi \to 0$ (see e.g. Refs.~\cite{Carlson:2012pc, Berryman:2018ogk, deGouvea:2019qaz, Dutta:2021cip}). The IR divergent part of the meson decay rate goes as~\cite{Dev:2024ygx} 
\begin{equation}
\label{eqn:IRD}
(\Delta \Gamma)_{\rm IR}\propto m_\ell^2 \log \frac{m_\phi^2}{m_{\sf M}^2} \,.
\end{equation}
Such IR divergences can be canceled out by including the $\phi$-induced 1-loop correction to the SM process ${\sf M}^\pm \to \ell^\pm + \nu$. In particular, letting ${\cal M}^{(0)}$ and ${\cal M}^{(1)}$ denote, respectively, the tree and 1-loop level amplitudes for ${\sf M}^\pm \to \ell^\pm + \nu$, the interference term is at the same order of $g_\nu$ as the tree-level three-body decays ${\sf M}^\pm \to \ell^\pm + \nu/\chi + \phi$, i.e.
\begin{equation}
{\rm Re} \left[ {\cal M}^{(0)\ast} {\cal M}^{(1)} \right] \propto g_\nu^2 \,.
\end{equation}
The IR divergences from these two contributions cancel out with each other~\cite{Pasquini:2015fjv,Brdar:2020nbj,Bickendorf:2022buy,Dev:2024ygx}. The IR divergences and their cancellations here are actually very general features of interacting quantum field theories featuring massless fields within the general context of scattering amplitudes~\cite{Peskin:1995ev, Agarwal:2021ais}, which is expected as a natural consequence of the Kinoshita-Lee-Nauenberg (KLN) theorem~\cite{Kinoshita:1962ur, Lee:1964is}. More details can be found in the Appendix of our recent paper~\cite{Dev:2024ygx}. One should notice that for the pseudoscalar couplings of $\phi$ to neutrinos, neutrino and DM, and charged leptons, e.g. $\phi \bar\nu i\gamma_5 \nu$, $\phi \bar\ell i\gamma_5 \ell$, $\phi \bar\nu i\gamma_5 \chi$, there is no IR divergence~\cite{Rai:2021vvq}, which can be understood due to a shift symmetry.

In analogy to the scalar mediator case above, for the vector mediator $Z'$, there is also the IR divergence in the form of $m_\ell^2 \log (m_{Z'}^2/m_{\sf M
}^2)$. However, in the presence of $Z'$ there is an additional enhancement to the decay width proportional to $m_{\sf M}^2/m_{Z'}^2$, which quadratically diverges as $m_{Z'}$ approaches zero. In the limit of $m_{Z'} \to 0$, the $m_{\sf M}^2/m_{Z'}^2$ terms are more important than the corresponding IR parts, and the latter can be safely neglected~\cite{Barger:2011mt,Carlson:2012pc,Laha:2013xua,Bakhti:2017jhm,Dutta:2021cip}. This is due to the additional physical degree of freedom, i.e. the longitudinal mode of the $Z'$. 

\begin{table}[!t]
    \caption{\label{tab:meson} 
    Meson data for the limits in Figs.~\ref{fig:meson_scalar_summary} and \ref{fig:meson_Zp_summary}, including the central values and the $1\sigma$ uncertainties for the lifetimes, BRs, decay constants $f_{\sf M}$ and the relevant CKM matrix elements $V_{ij}$ for the $\pi$, $K$, $D$ and $B$ mesons. The upper bounds are at the 90\% C.L.. Taken from PDG~\cite{ParticleDataGroup:2024cfk} unless otherwise specified. See text and Ref.~\cite{Dev:2024ygx} for more details. }
\resizebox{\textwidth}{!}{
\begin{tabular}{|c|c|c|c|c|}
\hline\hline
channel & lifetime [sec] & BR & $f_{\sf M}$ [MeV] & $V_{ij}$ \\ \hline
$\pi \to e \nu$ & \multirow{2}*{$(2.6033 \pm 0.0005) \times 10^{-8}$} & $(1.230 \pm 0.004) \times 10^{-4}$ & \multirow{2}*{$(130.2 \pm 0.8)$~\cite{fpi}} & \multirow{2}*{$0.97367 \pm 0.00032$} \\ \cline{1-1} \cline{3-3}
$\pi \to \mu \nu$ & & $0.9998770 \pm 0.0000004$ & & \\ \hline
$K \to e \nu$ & \multirow{2}*{$(1.2380 \pm 0.0020) \times 10^{-8}$} & $(1.582 \pm 0.007) \times 10^{-5}$ & \multirow{2}*{$(155.7 \pm 0.7)$~\cite{fpi}} & \multirow{2}*{$0.22431 \pm 0.00085$} \\ \cline{1-1} \cline{3-3}
$K \to \mu \nu$ & & $0.6356 \pm 0.0011$ &  &  \\ \hline
$D \to e \nu$ & \multirow{2}*{$(1.033 \pm 0.005) \times 10^{-12}$} & $<9.7 \times 10^{-7}$ & \multirow{2}*{$212.0 \pm 0.7$} & \multirow{2}*{$0.22487 \pm 0.00068$} \\ \cline{1-1} \cline{3-3}
$D \to \mu \nu$ &  & $(3.74 \pm 0.17) \times 10^{-4}$ &  &  \\ \hline
$B \to e \nu$ & \multirow{2}*{$(1.638 \pm 0.004) \times 10^{-12}$} & $<9.8 \times 10^{-7}$ & \multirow{2}*{$190.0 \pm 1.3$} & \multirow{2}*{$0.003732 \pm 0.00009$} \\ \cline{1-1} \cline{3-3}
$B \to \mu \nu$ &  & $<8.6 \times 10^{-7}$ &  &  \\ \hline\hline
\end{tabular}}
\end{table}

\begin{figure}[t]
     \centering
     \begin{subfigure}[b]{0.48\textwidth}
        \centering
        \includegraphics[width=1\textwidth]{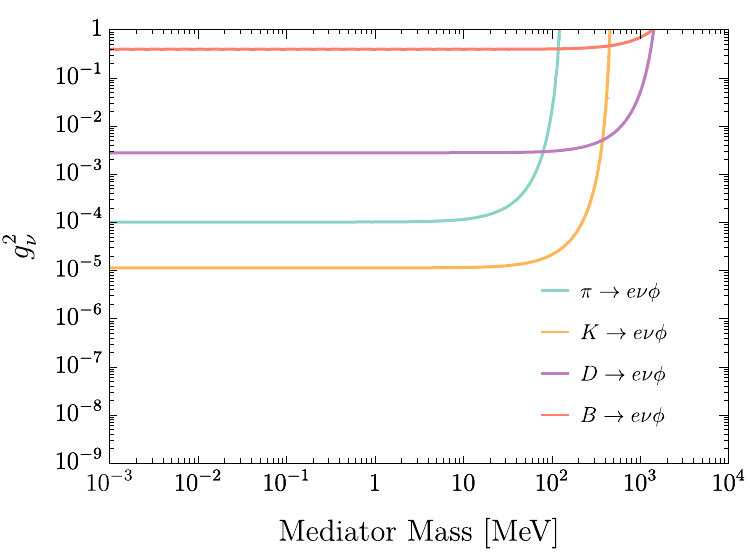}
        \caption{${\sf M} \rightarrow e \nu  \phi$ bounds}
        \label{subfig:meson_scalar_electron}
    \end{subfigure}
     \begin{subfigure}[b]{0.48\textwidth}
        \centering
        \includegraphics[width=1\textwidth]{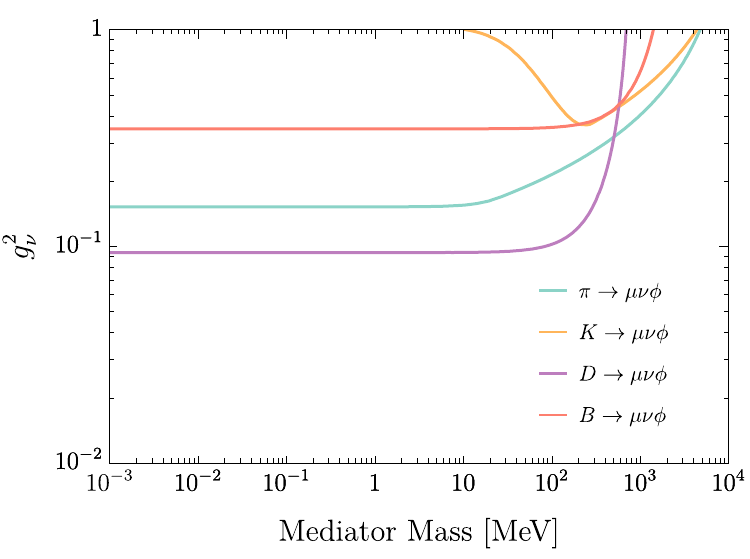}
        \caption{${\sf M} \rightarrow \mu \nu \phi$ bounds}
        \label{subfig:meson_scalar_muon}
    \end{subfigure}
    \caption{
    Limits on the coupling of the scalar mediator $\phi$ to neutrinos from the $\pi$, $K$, $D$, and $B$ meson decays ${\sf M}^\pm \to \ell^\pm + \nu + \phi$, based on the calculations in Ref.~\cite{Dev:2024ygx}. 
    The left and right panels are for charged lepton flavors $\ell = e$ and $\mu$, respectively.
    }
    \label{fig:meson_scalar_summary}
\end{figure}

\begin{figure}[t]
     \centering
     \begin{subfigure}[b]{0.48\textwidth}
        \centering
        \includegraphics[width=1\textwidth]{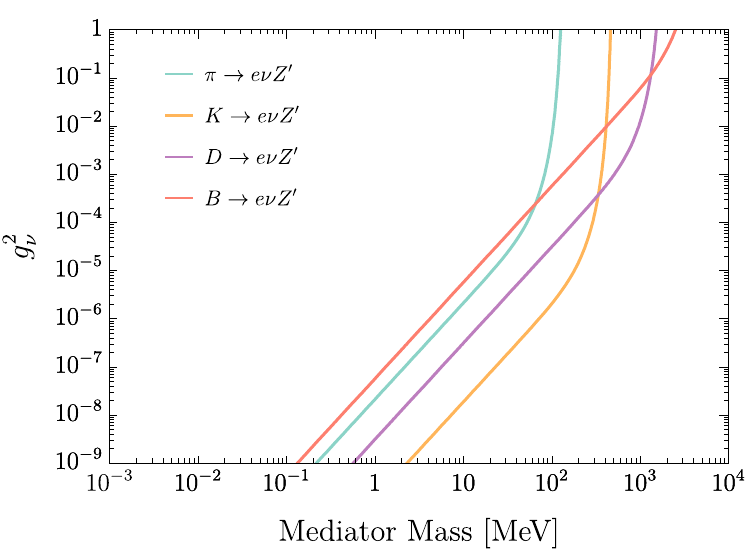}
        \caption{${\sf M} \rightarrow e \nu Z'$ bounds}
        \label{subfig:meson_Zp_electron}
    \end{subfigure}
     \begin{subfigure}[b]{0.48\textwidth}
        \centering
        \includegraphics[width=1\textwidth]{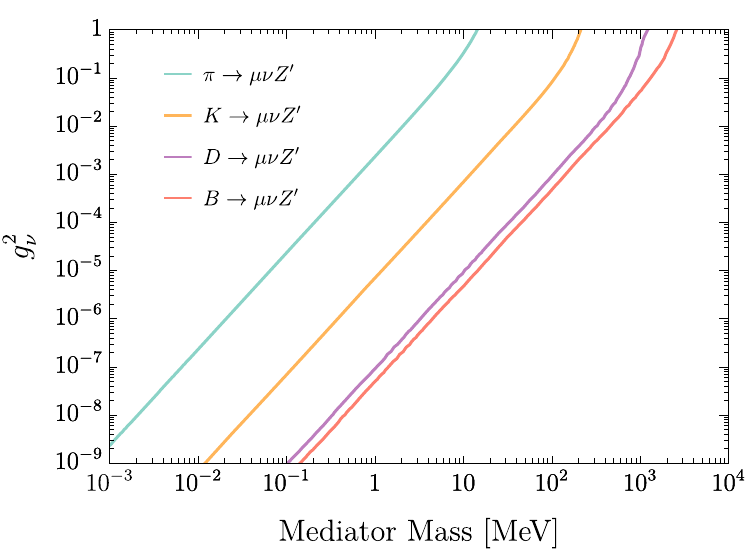}
        \caption{${\sf M} \rightarrow \mu \nu Z'$ bounds}
        \label{subfig:meson_Zp_muon}
    \end{subfigure}
    \caption{
    Limits on the coupling of the vector mediator $Z'$ to neutrinos from the $\pi$, $K$, $D$, and $B$ meson decays ${\sf M}^\pm \to \ell^\pm + \nu + Z'$, based on the calculations in Ref.~\cite{Dev:2024ygx}. 
    The left and right panels are for charged lepton flavors $\ell = e$ and $\mu$, respectively.
    }
    \label{fig:meson_Zp_summary}
\end{figure}

For our interests in this paper, we focus on the following meson and tau decay limits. 
\begin{itemize}
    \item We extend our work in Ref.~\cite{Dev:2024ygx} to include limits from $D$ and $B$ mesons, accounting for their associated lifetimes, masses, decay constants, the relevant Cabibbo-Kobayashi-Maskawa (CKM) matrix elements and their corresponding uncertainties in Table~\ref{tab:meson}. As shown in Figs.~\ref{fig:meson_scalar_summary} and \ref{fig:meson_Zp_summary}, we find that the $D$ and $B$ meson constraints can be competitive with pion and kaon constraints. The strongest constraints for mediator masses up to 1 GeV from $K$ and $D$ mesons are the only ones shown in the summary plots below.
    
    \item For $\phi$-$\nu$-$\bar\nu$ couplings, we adopt the updated limits for the scalar mediator case and $\pi$, $K$ mesons without the IR divergences from Ref.~\cite{Dev:2024ygx}. In particular, the meson partial width data from PDG~\cite{ParticleDataGroup:2024cfk} are used to set the limits. The uncertainty in partial widths are dominated by the meson decay constants $f_{\pi,\,K}$, which are determined by the lattice calculations~\cite{FlavourLatticeAveragingGroupFLAG:2024oxs}. The spectra of charged leptons from $\pi$ and $K$ decays measured by PIENU~\cite{PIENU:2021clt} and NA62~\cite{NA62:2021bji} can improve significantly the meson limits for specific ranges of $m_\phi$.
    \item To obtain the limits on the $\phi$-$\nu$-DM couplings, we generalize the calculations in Ref.~\cite{Dev:2024ygx} with $m_\chi \neq 0$. Then the phase space integration with three nonzero masses $m_\ell$, $m_\phi$, and $m_\chi$ gets much more complicated. We follow Ref.~\cite{Asatrian:2012tp} to do the numerical phase space integration. Setting $m_{\rm DM} = m_\phi/3$, it turns out that the corresponding limits are quite similar to that for the coupling of $\phi$ to neutrinos.

    \item As just aforementioned, for the $Z'$ boson case of ${\sf M}^\pm \to \ell^\pm + \nu / \chi + Z'$, in the limit of small $m_{Z'}$ the partial widths are dominated by the $m_{\sf M}^2/m_{Z'}^2$ terms. Therefore, for simplicity, we neglect the 1-loop contributions. The meson limits on the exotic decay channels ${\sf M}^\pm \to \ell^\pm + \nu + Z'$ are shown in Fig.~\ref{fig:meson_Zp_summary}, with ${\sf M} = \pi,\; K,\; D,\; B$ and $\ell = e,\; \mu$. It is clear from this figure that the most stringent constraint is from the channel $K^\pm \to e^\pm + \nu + Z'$. The meson limits on the decay channels ${\sf M}^\pm \to \ell^\pm + \chi + Z'$ involving the DM $\chi$ are quite similar to that on ${\sf M}^\pm \to \ell^\pm + \nu + Z'$, when we set the DM mass to be one third of the vector mediator mass, i.e. $m_{\rm DM} = m_{Z'}/3$.
\end{itemize}

The following decay channels are also of great interest; however, the corresponding limits are expected to be weaker than the processes above for the mediator mass range that we consider in this paper. 
\begin{itemize}
    \item The exotic three-body tau decays, e.g. $\tau^\pm \to \pi^\pm + \nu/\chi + \phi$. These processes are similar to the meson decays ${\sf M}^\pm \to \ell^\pm + \nu/\chi + \phi$ above, but the corresponding constraints are limited by the precision of the tau data~\cite{Dev:2024ygx}. 
    \item The four-body tau decays, e.g. $\tau^\pm \to \ell^\pm + \nu_\ell + \nu_\tau + \phi$, which are, however, highly suppressed by the phase space~\cite{deGouvea:2019qaz,Brdar:2020nbj}. 
\end{itemize}

\subsubsection{Invisible $Z$ decay}
\label{subsubsec:Zlimits}

Neutrino NSI can also be constrained from the measurement of the invisible $Z$ decay width $\Gamma (Z\rightarrow {\rm inv})$. This is a well-known effect, which has been studied for the scalar case $Z \to \nu + \bar\nu + \phi$~\cite{Berryman:2018ogk,deGouvea:2019qaz, Brdar:2020nbj,Dev:2024ygx} and vector boson case $Z \to \nu + \bar\nu + Z'$~\cite{Laha:2013xua}. 
There are also IR divergences in the limit of $m_{\phi,\; Z'} \to 0$ for the widths of the tree-level three-body processes, and we have to include the 1-loop contributions to cancel the IR divergence, as for the meson and tau cases above. However, for the $Z$ boson decay, there are 1-loop $\phi$ or $Z'$ corrections to both the $Z$-$\nu$-$\bar\nu$ vertex and the neutrino self-energy, and the IR divergences are canceled out for both the $\phi$ and $Z'$ cases~\cite{Dev:2024ygx, Brdar:2020nbj}. When the 1-loop corrections are included, the limits on $Z \to \nu + \bar\nu + \phi$ and $Z \to \nu + \bar\nu + Z'$ are $g_\nu < 1.4$ and $0.54$ for $m_{\phi,\,Z^\prime} \lesssim 10$ GeV, respectively~\cite{Dev:2024ygx}. For the couplings in the form of $Z$-$\nu$-$\chi$, the corresponding limit is to some extent stronger when the 1-loop contributions are taken into account, which turns out to be $g_{\nu\chi}<0.43$ for mediator mass below roughly 10 GeV. 
As for the meson case above, the 1-loop correction of $Z'$ is subdominant to the $m_{Z}^2/m_{Z'}^2$ terms when $Z'$ is light in the $Z'$-$\nu$-$\chi$ case. This is also the case for the $\nu$-$\chi$-$N$ coupling, with $N$ being a fermion mediator and $\chi$ being vector DM, as shown in Fig.~\ref{subfig:VF}.

There should also be limits from $W$ boson decays, e.g. $W \to \ell + \nu / \chi + \phi$, which have been explored in Refs.~\cite{Agashe:2023itp, Agashe:2024owh}. While the precision of $\Gamma (W \to \ell + \nu)$ is significantly weaker than that for the invisible $Z$ data~\cite{ParticleDataGroup:2024cfk}, the shape analysis of the kinematic observables used to measure $m_W$ by ATLAS and CMS can be equally sensitive to neutrinophilic mediators, depending on experimental control over the systematic uncertainties in the measurement \cite{Agashe:2023itp,Agashe:2024owh}. Since a thorough experimental analysis is needed to determine the sensitivity, and the projections show comparable sensitivity to measurements of $\Gamma (Z\rightarrow {\rm inv})$ for sub-GeV mediator masses, we do not show the $W$ boson sensitivity in this paper.

\subsubsection{Other laboratory limits}

Here we list other laboratory limits that are relatively weaker in the parameter space of interest in this paper and are thus not shown in our figures.
\begin{itemize}
    \item A light mediator $X$ can be produced in the tritium beta decay process, i.e. $^3{\rm H} \to {^3{\rm He}} + e^- + \bar\nu_e + X$~\cite{Arcadi:2018xdd}. However, the mediator mass is required to be very small, well below the parameter space that we are interested in.
    \item The mediator can be radiatively produced from muon decays, e.g. $\mu \to e + \phi$~\cite{Santamaria:1985xa}. For the Majoron case with LNV couplings, the light scalar contributes at 1-loop level to the muon decays $\mu \to e + \nu + \bar\nu$ and $\mu \to e + \phi + \phi$~\cite{Goldman:1982bd,Santamaria:1986kg}. 
    As in the tau case above, the mediator can also induce the exotic four-body decays of muon, e.g. $\mu\to e + \nu + \bar\nu + \phi$, which will change the Michel spectrum of electron and modify the Fermi constant $G_F$~\cite{Bickendorf:2022buy}.
    \item If the scalar mass $m_\phi > m_{\sf M}$, the $\phi$-$\nu$-$\bar\nu$ couplings induce the four-body decay of meson ${\sf M}^\pm \to \ell^\pm + \nu + \phi^\ast \to \ell^\pm + \nu + \nu + \bar\nu$, with the neutrino pair $\nu \bar\nu$ from the off-shell scalar $\phi^\ast$~\cite{Bardin:1970wq,Bilenky:1999dn,Kelly:2019wow}. Updating the calculation in Ref.~\cite{Kelly:2019wow}, it is found that the limits from the four-body meson decays are mainly from $\pi,\; K \to e \nu \nu \bar\nu$, and rather weak, only down to $g_\nu > 0.42$. Integrating out the heavy mediator, we can obtain the effective four-neutrino interactions, e.g. $(\bar\nu \nu) (\bar\nu \nu)$~\cite{Gavela:2008ra}. Such interactions induce the four-body decays $Z \to \nu \bar\nu \nu \bar\nu$ and $W \to \ell \nu \nu \bar\nu$~\cite{Bilenky:1992xn,Bilenky:1999dn}. However, suppressed by the four-body phase space, these limits are also rather weak.
    \item The neutrinophilic scalar $\phi$ could induce the process $\nu N \to \ell \phi X$~\cite{Berryman:2018ogk,Kelly:2019wow,deGouvea:2019qaz,Kelly:2021mcd,Adhikary:2024tvl,Liu:2024ywd}. The charged lepton seems ``wrong-signed'' if the coupling is of the LNV type. The current constraints are mainly from MINOS and NOMAD, which are of ${\cal O}(1)$, depending on the neutrino flavors involved~\cite{Berryman:2018ogk}. The sensitivities of such processes can be significantly improved up to ${\cal O} (0.01)$ at future high-precision experiments, such as FLArE~\cite{Batell:2021blf} and DUNE~\cite{DUNE:2015lol}, or even higher at a future muon collider~\cite{Adhikary:2024tvl,Liu:2024ywd}.
    \item The neutrino NSI can induce the trident-like process $\nu + N \to \ell + \nu + \bar\nu + X$, whose limits are, however, rather weak~\cite{Bardin:1970wq}.
    \item In the presence of LNV, the coupling of $\phi$ (or $Z'$) to neutrinos can induce 1-loop contributions to neutrinoless $\beta\beta$ decays, which is, however, suppressed by the loop factor and potentially heavy particles in the loop~\cite{Rodejohann:2019quz}.
    \item The couplings of $\phi$ and $Z'$ with neutrinos could induce 1-loop couplings of $\phi$ and $Z'$ to the quarks and charged leptons~\cite{Chikashige:1980ui,Bauer:2020itv}, which would contribute to neutrino-electron and neutrino-nucleus scattering~\cite{Laha:2013xua,Suliga:2020jfa}, effects on solar neutrino evolution~\cite{Ansarifard:2024zxm}, and radiative emission of neutrino pair from atomic transition~\cite{Ge:2021lur}, and thus get constrained by, e.g. Borexino~\cite{Borexino:2008gab}, COHERENT~\cite{COHERENT:2017ipa}, and Yb-based experiment~\cite{Song:2015xaa}. However, these processes are highly suppressed by the loop factor and the heavy $W$ and $Z$ particles in the loop, and are therefore neglected. 
    \item If the mediator is very light, the neutrinos can decay, e.g. via the process $\nu_j \to \nu_i + \phi$, which could have effects on the neutrino oscillations experiments~\cite{Berezhiani:1987gf,Barger:1998xk,Barger:1999bg,Beacom:2002cb,Abrahao:2015rba,Choubey:2017eyg,Huang:2018nxj,Calatayud-Cadenillas:2024wdw}, supernova neutrinos~\cite{Frieman:1987as,Berezhiani:1989za,Lindner:2001th,Tomas:2001dh}, astrophysical high-energy neutrinos~\cite{Bustamante:2016ciw,Bustamante:2020niz}, cosmic neutrino background (C$\nu$B)~\cite{Huang:2024tbo}, and other astrophysical and cosmological observations~\cite{Cowsik:1977vz,Gelmini:1983ea,Bell:2005dr,Baerwald:2012kc} (see also Ref.~\cite{ParticleDataGroup:2024cfk}).
    \item Given the coupling to the scalar $\phi$, neutrinos might form bound states as long as $g_\nu^2 m_\nu /( 8\pi m_\phi) > 0.84$~\cite{Smirnov:2022sfo}. A large number of neutrinos might also form neutrino clusters, e.g. for the C$\nu$B~\cite{Stephenson:1993rx,Smirnov:2022sfo}, which might have additional astrophysical constraints in case the C$\nu$B was ever detected.  
    \item In neutrinophilic scenarios, the collider signatures are effectively limited to mono-$\gamma+\met$ (or mono-${\rm jet}+\met$) channels, but the signal cross sections in the experimentally accessible regions are too small compared to the large $\nu\bar{\nu}+\gamma$ background, rendering present collider limits rather weak. This is consistent with current BaBar~\cite{BaBar:2017tiz} and ATLAS~\cite{ATLAS:2020uiq} mono-$\gamma$ results and with Belle II projections/discussions of missing-energy searches~\cite{Liang:2023lwh}, which indicate little to no sensitivity for light neutrinophilic dark sectors in our benchmarks.
\end{itemize}

\subsection{Summary Plots}

\begin{table}[!t]
\centering
\caption{Summary of constraint plots,  models and relevant figures.}
\label{tab:vector}
\vspace{3pt}
 \begin{tabular}{|c | c | c | c | } 
 \hline\hline
 DM & Mediator & Channel & Figure \\ \hline
 
 fermion (Dirac)   & scalar & $t$ & Fig.~\ref{fig:exclusion_fermionDM_scalarMed_t} \\ \hline
  fermion (Dirac)   & scalar & $s$, $u$ & Fig.~\ref{subfig:DS}
  \\ \hline
   fermion (Majorana)   & scalar & $s$, $u$ & Fig.~\ref{subfig:MS}
  \\ \hline
  scalar   & scalar & $t$ & Fig.~\ref{subfig:SS} \\ \hline
 vector   & scalar & $t$ & Fig.~\ref{subfig:VS} \\ \hline
   scalar (real)   & fermion & $s$, $u$ & Fig.~\ref{subfig:RSF} \\ \hline
    scalar (complex)   & fermion & $s$, $u$ & Fig.~\ref{subfig:CSF} \\ \hline
   vector   & fermion & $s$, $u$ & Fig.~\ref{subfig:VF} \\ \hline
scalar (complex)   & vector & $t$  & Fig.~\ref{subfig:SV} \\ \hline
fermion (Dirac)   & vector & $t$  & Fig.~\ref{subfig:DFV} \\ \hline
fermion (Majorana)   & vector & $t$  & Fig.~\ref{subfig:MFV} \\ \hline
vector (complex)   & vector & $t$  & Fig.~\ref{subfig:VV} \\ 
\hline\hline   
\end{tabular}\label{resultstable}
\end{table}

\begin{table}[!t]
\centering
\caption{Summary of constraints and for which category of models and to which couplings they apply. In all models, the pion decay bound is weaker than the kaon or $D$ meson decay bounds and is hence not shown. This can be seen from Figs.~\ref{fig:meson_scalar_summary} and \ref{fig:meson_Zp_summary}.}
\begin{tabular}{|c|c|c|c|}
\hline \hline
 & Bound & $t$-channel & $s\,\&\,u$-channel \\
\hline 
\multirow{5}{6em}{Astrophysical} & SN 1987A: $\tau = 1$  & $g_\nu g_\chi$ & $g_{\nu\chi}^2$ \\
& IceCube & $g_\nu^2$ & -- \\ 
& NGC 1068 & $g_\nu g_\chi$ & $g_{\nu\chi}^2$ \\ 
& Bullet Cluster (SIDM) & $g_\chi^2$ & -- \\ 
& DM Annihilation & $g_\nu g_\chi$ & $g_{\nu\chi}^2$ \\
\hline
\multirow{5}{6em}{Cosmological} & CMB: NSI & $g_{\nu}^2$ & -- \\ 
& CMB: $\nu$DM  & $g_\nu g_\chi$ & $g_{\nu\chi}^2$ \\ 
& CMB: $N_{\rm eff}$  & $g_\nu g_\chi$ & $g_{\nu\chi}^2$ \\ 
& BBN & $g_\nu g_\chi$ & $g_{\nu\chi}^2$ \\
& Collisional Damping (CD) & $g_\nu g_\chi$ & $g_{\nu\chi}^2$ \\ 
& Thermal Relic (TR) & $g_\nu g_\chi$ & $g_{\nu\chi}^2$ \\ 
\hline
\multirow{3}{6em}{Laboratory}
& Kaon decay & $g_\nu^2$ & $g_{\nu\chi}^2$ \\ 
& $D$ meson decay & $g_\nu^2$ & $g_{\nu\chi}^2$ \\ 
& $Z$ decay & $g_\nu^2$ & $g_{\nu\chi}^2$ \\
\hline\hline
\end{tabular}
\label{tab:bounds}
\end{table}

\begin{figure}
    \centering
    \includegraphics[width=0.9\textwidth]{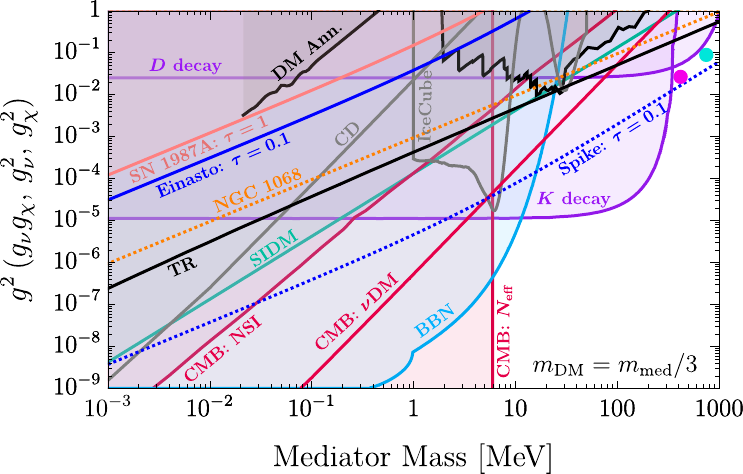}
    \caption{Constraints on the mediator mass and the couplings $g_\nu g_\chi$, $g_\nu^2$, and $g_\chi^2$ for the Dirac fermion DM and scalar mediator in the $t$-channel model. The DM mass is fixed at $m_{\rm DM} = m_{\rm med}/3$. The shaded regions are excluded. Constraints displayed include Supernova 1987A constraints on neutrino NSI and $\nu$-DM interaction with opacity $\tau>1$, limits from IceCube observations of high energy events and NGC 1068, Bullet Cluster constraints on SIDM, CMB limits on neutrino NSI, $\nu$-DM interactions and light mediators via $N_{\rm eff}$, BBN, collisional damping, indirect detection of DM annihilation to neutrinos, and charged kaon and $D$ meson decays.
    The applicability of the constraints on different couplings $g_\nu g_\chi$, $g_\nu^2$, and $g_\chi^2$ can be found in Table~\ref{tab:bounds}. 
    Some of the constraints require a more careful analysis (CMB: $\nu$DM, CD); therefore, the corresponding regions are not shaded. Constraints that require the presence of a spike profile are denoted by dotted lines.
    Opacity contours $\tau = 0.1$ for a local supernova positioned at a distance of $d_{\rm{SN}} = 10$ kpc and galactic coordinates $(\ell, b) = (0, 0)$ are presented for the Einasto profile (blue solid) and in the presence of a DM spike (blue dotted). The parameter space producing the observed thermal relic density (TR) due to DM annihilation into neutrinos is indicated by the solid black line. Two benchmark points labeled by the magenta and cyan points are adopted to estimate the effect of the galactic DM halo on neutrinos from a galactic supernova event in Section~\ref{sec:supernova}. See text for more details.
    } 
    \label{fig:exclusion_fermionDM_scalarMed_t}
\end{figure}

\begin{figure}
     \centering
     \begin{subfigure}[b]{0.49\textwidth}
        \centering
        \includegraphics[width=\linewidth]{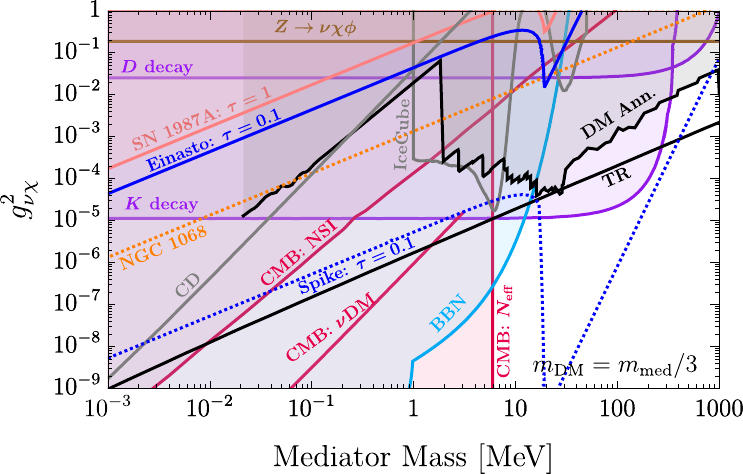}
        \caption{Dirac fermion DM, scalar mediator}
        \label{subfig:DS}
    \end{subfigure}
    \begin{subfigure}[b]{0.49\textwidth}
        \centering
        \includegraphics[width=\linewidth]{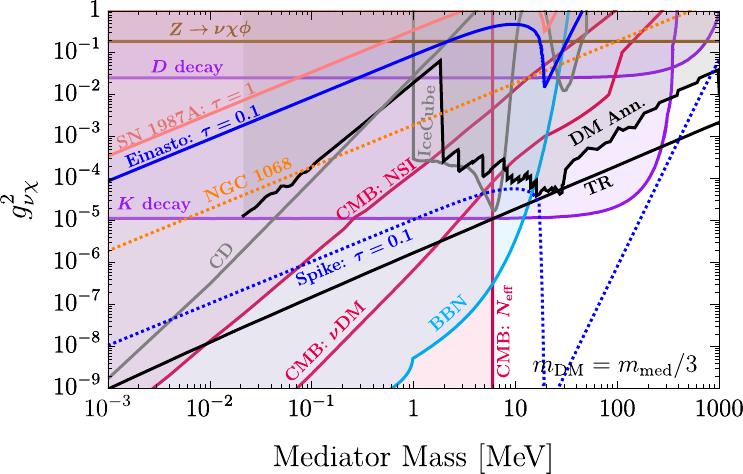}
        \caption{Majorana fermion DM, scalar mediator}
         \label{subfig:MS}
    \end{subfigure}
     \begin{subfigure}[b]{0.49\textwidth}
        \centering
        \includegraphics[width=\linewidth]{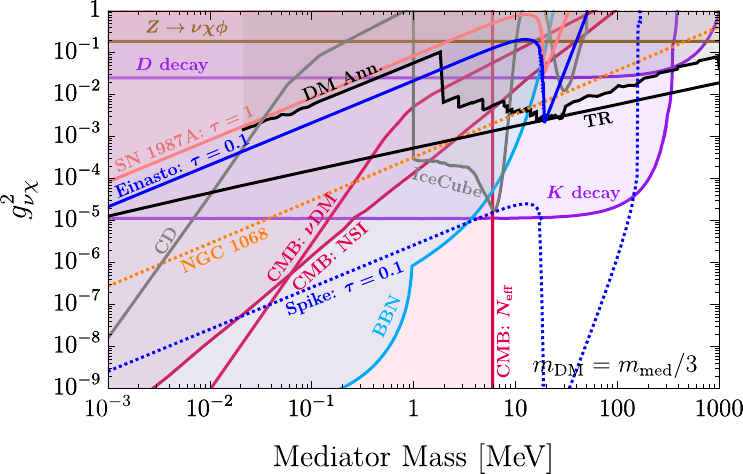}
        \caption{Real Scalar DM, fermion mediator}
        \label{subfig:RSF}
    \end{subfigure}
    \begin{subfigure}[b]{0.49\textwidth}
        \centering
        \includegraphics[width=\linewidth]{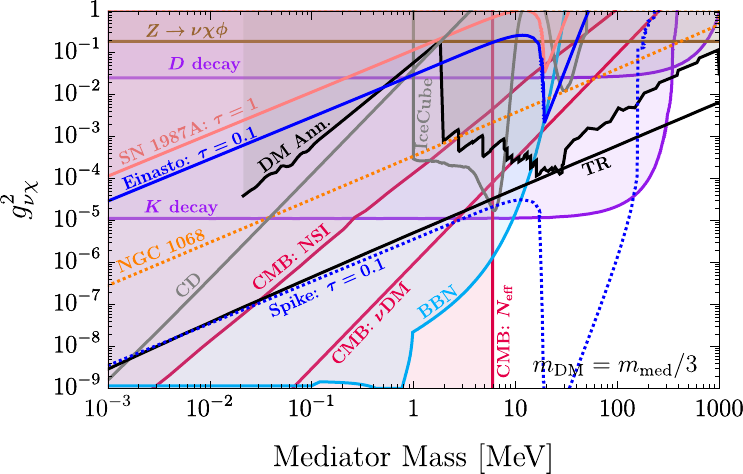}
        \caption{Complex Scalar DM, fermion mediator}
        \label{subfig:CSF}
    \end{subfigure}
    \begin{subfigure}[b]{0.49\textwidth}
        \centering
    \includegraphics[width=\linewidth]{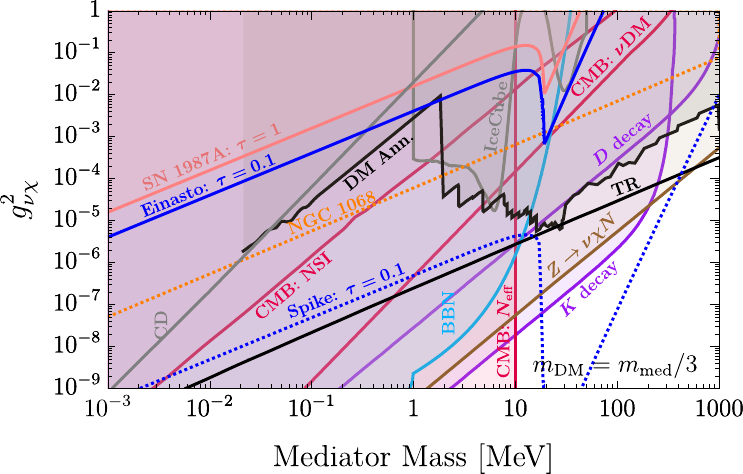}
    \caption{Vector DM, fermionic mediator}
    \label{subfig:VF}
    \end{subfigure}
    \caption{
    The same as Fig.~\ref{fig:exclusion_fermionDM_scalarMed_t}, but for the coupling $g_{\nu\chi}^2$ and for different DM and mediator types as mentioned below each subfigure.
    Also shown in these panels is the invisible $Z$ decay constraint. We note that the resonance structure in the sensitivities from opacities of SN 1987A, Einasto, and Spike profiles shown in the above sensitivity plots is an artifact of determining the opacity for a fixed neutrino energy. Since supernovae produce a spectrum of MeV energy neutrinos, 
    }
    \label{fig:exclusion_2}
\end{figure}

\begin{figure}
\centering
        \begin{subfigure}[b]{0.49\textwidth}
        \centering
        \includegraphics[width=\linewidth]{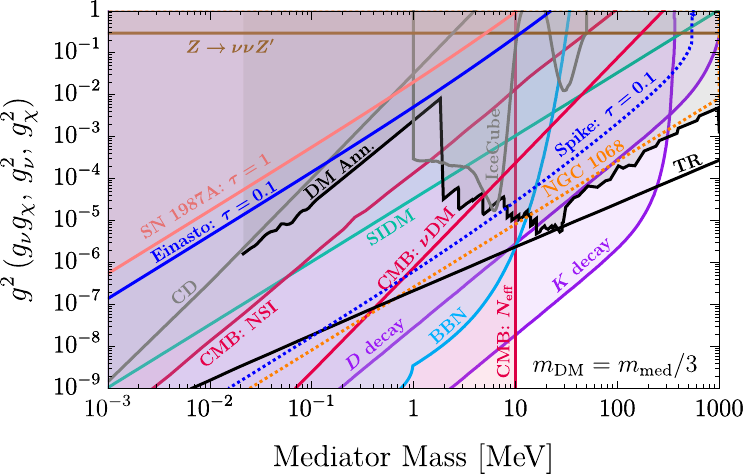}
        \caption{Dirac Fermion DM, vector mediator}
        \label{subfig:DFV}
    \end{subfigure}
    \begin{subfigure}[b]{0.49\textwidth}
        \centering
        \includegraphics[width=\linewidth]{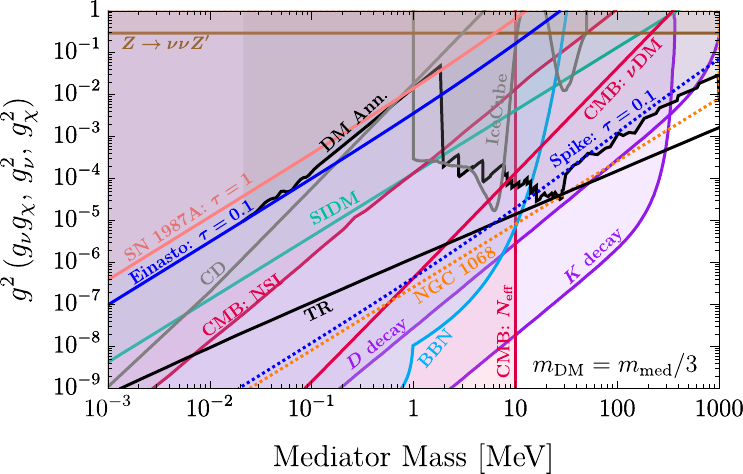}
        \caption{Majorana Fermion DM, vector mediator}
         \label{subfig:MFV}
    \end{subfigure}
     \begin{subfigure}[b]{0.49\textwidth}
        \centering      \includegraphics[width=\linewidth]{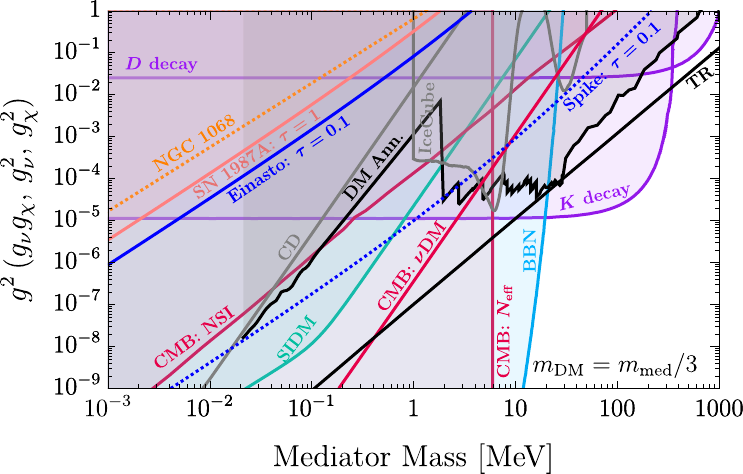}
        \caption{Scalar DM, scalar mediator}
        \label{subfig:SS}
    \end{subfigure}
    \begin{subfigure}[b]{0.49\textwidth}
        \centering       \includegraphics[width=\linewidth]{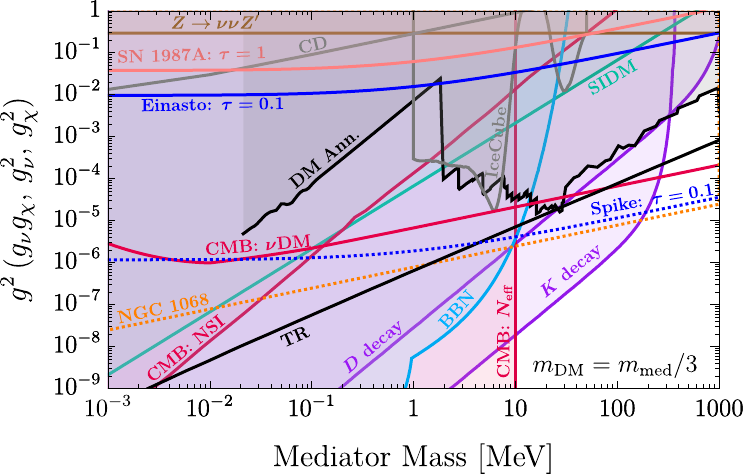}
        \caption{Scalar DM, vector mediator.}
         \label{subfig:SV}
        \end{subfigure}
        \begin{subfigure}[b]{0.49\textwidth}
        \centering       \includegraphics[width=\linewidth]{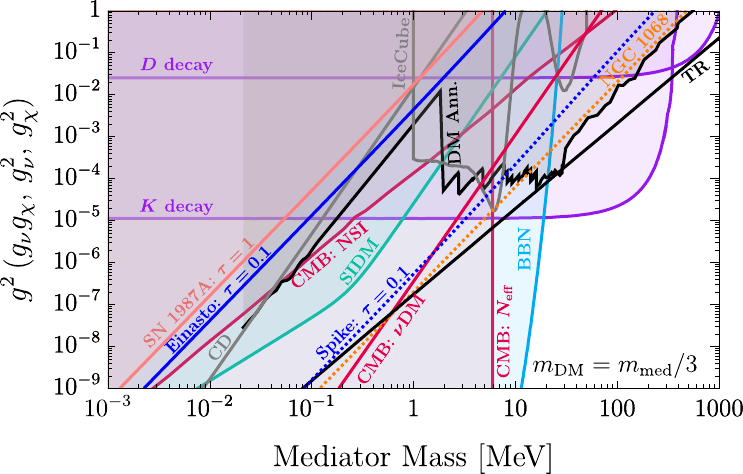}
        \caption{Vector DM, scalar mediator}
              \label{subfig:VS}
    \end{subfigure}
    \begin{subfigure}[b]{0.49\textwidth}
        \centering
        \includegraphics[width=\linewidth]{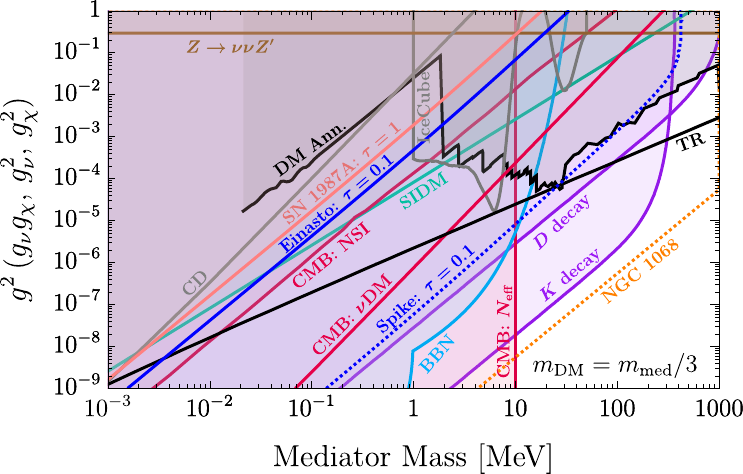}
        \caption{Vector DM, vector mediator}
        \label{subfig:VV}
    \end{subfigure}
    
    \caption{
    The same as Fig.~\ref{fig:exclusion_fermionDM_scalarMed_t}, but for different DM and mediator types as mentioned below each subfigure. 
    Also shown in some panels is the invisible $Z$ decay constraint.
    }   \label{fig:exclusion_3}
\end{figure}

Given the various constraints relevant to neutrino-DM interactions as described above, we are now ready to present our summary plots in Figs.~\ref{fig:exclusion_fermionDM_scalarMed_t}, \ref{fig:exclusion_2} and \ref{fig:exclusion_3}. These results are for various combinations of scalar, fermion, and vector DM and mediators, in the $s$-, $t$-, and $u$-channels where relevant. A guide to the constraint plots is given in Table~\ref{resultstable} for the reader's convenience. 
\begin{itemize}
    \item In all the figures, we have set the DM mass $m_{\rm DM} = m_{\rm med}/3$. 
    \item Some of the contours or shaded regions (CMB NSI, IceCube, BBN, SIDM, DM annihilation) are limits previously explored in the literature, and others (meson/$Z$ decays, SN1987A with $\tau = 1$, Einasto and spike for $d_{\rm SN}=10$ kpc and $\tau = 0.1$, NGC 1068, CMB $\nu$DM, CD) are reinterpretations of existing constraints or new constraints altogether. In many summary plots, the new or reinterpreted constraints are the strongest for $\geq$ MeV mediator masses, highlighting the importance of the new constraints on this parameter space.
    \item In these figures, the limits from SN1987A, IceCube/NGC 1068, bullet clusters (labeled as ``SIDM''), CMB, BBN, CD, and meson/$Z$ decays are shown as the pink, gray/orange, green, red, light blue, gray, and purple/brown shaded regions. The black lines with the label ``TR'' indicate the parameter space that can generate the observed DM relic density due to the thermal annihilation process $\chi \bar\chi \to \nu \bar\nu$.
    \item In the models given in Figs.~\ref{fig:exclusion_fermionDM_scalarMed_t} and \ref{fig:exclusion_3}, 
    the $\nu$-DM scattering are in the $t$-channel, and therefore, we have both the couplings $g_\nu$ and $g_\chi$. For these models, there are limits on neutrino self-interactions $g_\nu^2$ from CMB, BBN, and meson and $Z$ decays, the constraints on DM self-interactions $g_\chi^2$ from bullet clusters, and the limits on $\nu$-DM interactions $g_\nu g_\chi$ from CD. In some models, the mediator is a scalar $\phi$, and the limit of $Z \to \nu + \bar\nu + \phi$ on the coupling $g_\nu$ is weaker than 1 (cf. Section~\ref{subsubsec:Zlimits}); therefore, the corresponding $Z$ limit is not shown in these figures. 
    For these bounds, $g_\nu$ and $g_\chi$ need not be equal, and thus a bound on $g_\nu^2$ does not necessarily rule out the same parameter space for $g_\nu g_\chi$. Unless otherwise stated, we will consider the scenario $g_\nu=g_\chi$ for simplicity. The couplings that each bound probes are listed in Table~\ref{tab:bounds}.
    \item In the models for Fig.~\ref{fig:exclusion_2}, the couplings are in the form of med-$\nu$-$\chi$, and therefore, we have only the coupling $g_{\nu\chi}$. For these cases, there are limits on $g_{\nu\chi}^2$ from BBN, SN1987A, NGC 1068, CMB, CD, and meson and $Z$ decays.
    \item In all these figures, the most stringent astrophysical and cosmological limits are those from BBN, CMB, bullet cluster, and indirect detection of DM annihilation. Above roughly the MeV scale, the constraints are dominated by those from the precision meson decays ${\sf M}^\pm \to \ell^\pm + \nu$. As stated in Section~\ref{subsubsec:mesonlimits}, if the mediator or DM is the $Z'$ boson, the decay ${\sf M}^\pm \to \ell^\pm + \nu/\chi + Z'$ is largely enhanced by the ratio $m_{\sf M}^2 / m_{Z'}^2$ when $Z'$ is very light, as demonstrated for the vector mediator cases. In these cases, to generate the observed DM relic density, the DM and mediator masses are required to be above the scale of ${\cal O} (100\; {\rm MeV})$. As a comparison, for the models  without a vector DM or mediator, there is viable parameter space with DM and mediators down to the scale of ${\cal O} ({\rm MeV})$, which can generate the correct DM relic density.
    \item Dotted contours (NGC 1068 and Spike: $\tau = 0.1$) are dependent on the neutrinos passing through a DM spike profile, either in the MW or another galaxy.
\end{itemize}

\section{Probing \nuDM~interactions with galactic supernovae}
\label{sec:supernova}

In this section, we describe the potential interactions between neutrinos and DM, as well as their observational impact, in particular the DM opacity and the attenuation of neutrinos from galactic supernovae. We then explore phenomenological implications, focusing on new regions of parameter space beyond the constraints discussed in the previous section that can be probed in future neutrino experiments. 

\subsection{DM density profiles and opacity}

We outline the procedure to obtain contours in the parameter space for each DM model where the opacity of $\nu$-DM interactions is significant for $E_\nu \sim 10$ MeV. Determining the opacity contours requires the following components:
\begin{enumerate}
    \item A model describing the $\nu$-DM interactions with specific coupling choices $(g_\nu, g_\chi, g_{\nu \chi}, \mu)$, DM mass $m_\chi$, mediator mass $m_{\rm{med}}$, and neutrino energy $E_\nu$. The cross sections for $\nu$-DM interactions are detailed in Section~\ref{sec:nuDMInteractions}, and the existing bounds on the couplings are discussed in Section~\ref{sec:bounds}.
    \item The integrated column density $\eta$ of DM, which depends on a chosen DM profile and the location of the supernova.
\end{enumerate}

\begin{figure}[t]
    \centering
    \includegraphics[width=0.7\textwidth]{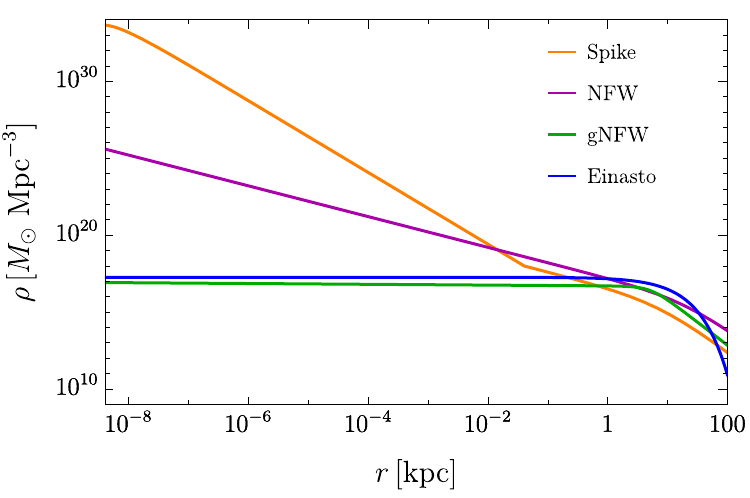}
    \caption{The DM density for the NFW, gNFW, Einasto, and spike profiles, as functions of $r$.}
    \label{fig:profiles}
\end{figure}

For a given model involving $\nu$-DM interactions, the contours of opacity $\tau$ can be derived for various DM profiles and supernova locations.
In our study, for illustration purposes, we consider the following DM halo profiles.
\begin{itemize}
\item The Navarro-Frenk-White (NFW) profile~\cite{Navarro:1995iw,Navarro:1996gj}.
This profile follows a cuspy behavior with two parameters: 
\begin{equation}
    \rho_\chi^{\rm NFW}(r)= \frac{\rho_s}{\left(\frac{r}{r_s} \right)\left(1+\frac{r}{r_s}\right)^2}\,,
\end{equation}
where the scale radius parameter $r_s$ is set to be 8.1 kpc for our analysis, while the characteristic density parameter $\rho_s$ is chosen such that the local DM density at the location of the solar system measured from the galactic center $R_\odot \approx 8.0$ kpc is $\rho_\chi(R_\odot) \approx 0.4$ GeV/cm$^3$~\cite{Lin:2019yux}. 

\item The generalized NFW (gNFW) profile~\cite{Hooper:2010mq,Ou:2023adg}. This is a generalization of the NFW profile above, and allows the profile to range from completely cored to having a density power-law slope as steep as $-3$ at radii larger than the scale radius $r_s$. This profile is given by
\begin{equation}
     \rho_\chi^{\rm gNFW}(r)= \frac{\rho'_s}{\left(\frac{r}{r_s} \right)^\beta\left(1+\frac{r}{r_s}\right)^{3-\beta}}\,,
\end{equation}
where $\beta$ = 0.0258 and $\rho'_s$ is a normalization parameter. $\beta=1$ recovers the standard NFW profile. 

\item The Einasto profile~\cite{Navarro:2003ew}. Unlike cuspy profiles, the logarithmic slope of the profile at highly resolved radii was observed to be shallower in decreasing radius rather than converging to a given value. The Einasto profile was proposed to better describe the observed data, and the functional form is given by
\begin{eqnarray}
\label{eqn:Ein}
    \rho_\chi^{\rm Ein}(r) &=& \rho_{-2} \exp \left\{ -\frac{2}{\alpha}\left[ \left( \frac{r}{r_{-2}}\right)^\alpha -1 \right] \right\} \,,
\end{eqnarray}
where $r_{-2}$ is a scale radius at which the logarithmic slope of the density profile takes an ``isothermal'' value of $-2$, i.e., $\left.{{\rm d} \log \rho_\chi^{\rm Ein}}/{{\rm d} \log r }\right|_{r=r_{-2}}=-2$, and $\rho_{-2}$ is the characteristic density at $r=r_{-2}$.
$\alpha = 0.91$ is a shape parameter determining how fast or slow the slope varies with radius. 

\item The spike DM halo profile~\cite{Gondolo:1999ef}. Particularly in the vicinity of supermassive black holes (SMBHs), the adiabatic growth can significantly enhance DM density, forming a sharp ``spike''~\cite{Gondolo:1999ef}.\footnote{This can be motivated simply from angular momentum conservation. With more mass accreting onto the SMBH, the minimum of the effective potential shifts towards the center, thus giving rise to a spike profile. But the observational evidence for a spike profile at the center of Milky Way is still under debate, and no agreement has been reached in the community~\cite{Ullio:2001fb, Lacroix:2018zmg, Shen:2023kkm}. See Ref.~\cite{Chattopadhyay:2026kbm} for a recent discussion of indirect detection constraints on the existence of a spike profile in Milky Way.} Among various spike profiles, e.g. those in Refs.~\cite{Kim:2017qaw,Lacroix:2018zmg,Shen:2023kkm}, in this study, we consider the NFW spike DM halo profile:
\begin{equation}
    \rho_\chi^{\rm spike}(r)= 
    \begin{cases}
        0, & R <2R_s \,, \\
        \rho_{\rm spike}\left(\frac{r}{R_{\rm spike}} \right)^{-\gamma_{\rm spike}}, & 2R_s \leq r < R_{\rm spike} \,, \\ 
        \rho_{\chi}^{\rm NFW}(r), & r \geq R_{\rm spike} \,,
    \end{cases} 
\end{equation}
where $R_s = 2 G_{N} m_{\rm SMBH}$ is the Schwarzschild radius of the galactic center SMBH (Sagittarius A$^*$) of mass $m_{\rm SMBH} = 4.3 \times 10^6 \, M_\odot$, $R_{\rm spike}$ is the radial extension of the spike, and $\gamma_{\rm spike}$ is a slope parameter. In our study, we obtain $R_{\rm spike} \sim 40$ pc with $\gamma_{\rm spike} = 7/3$~\cite{Lacroix:2018zmg}, following the procedures to obtain the profile normalization found in Refs.~\cite{Lacroix:2016qpq,Fujiwara:2023lsv}. Beyond $R_{\rm spike}$, the DM distribution follows the NFW halo profile as shown in Fig.~\ref{fig:profiles}.
\end{itemize}
The comparison of the NFW, gNFW, Einasto, and spike DM halo profiles discussed above is presented in Fig.~\ref{fig:profiles}, as functions of the distance $r$ from the galactic center. The recent Gaia data tends to be better explained with a cored profile like the Einasto profile over a cuspy NFW profile~\cite{Ou:2023adg, Lim:2023lss}; therefore, we show the opacity contours in Figs.~\ref{fig:exclusion_fermionDM_scalarMed_t}--\ref{fig:exclusion_3} for the Einasto profile, along with an extreme case of a spike profile for comparison. We take a galactic supernova distance benchmark of $d_{\rm{SN}} = 10$ kpc and galactic coordinates $(\ell, b) = (0, 0)$, representing a source directly behind the galactic center.
By design, the spike profile may further enhance the opacity compared to other profiles.

\begin{figure}[t]
     \centering
     \begin{subfigure}[b]{0.6\textwidth}
        \centering
        \includegraphics[width=1\textwidth]{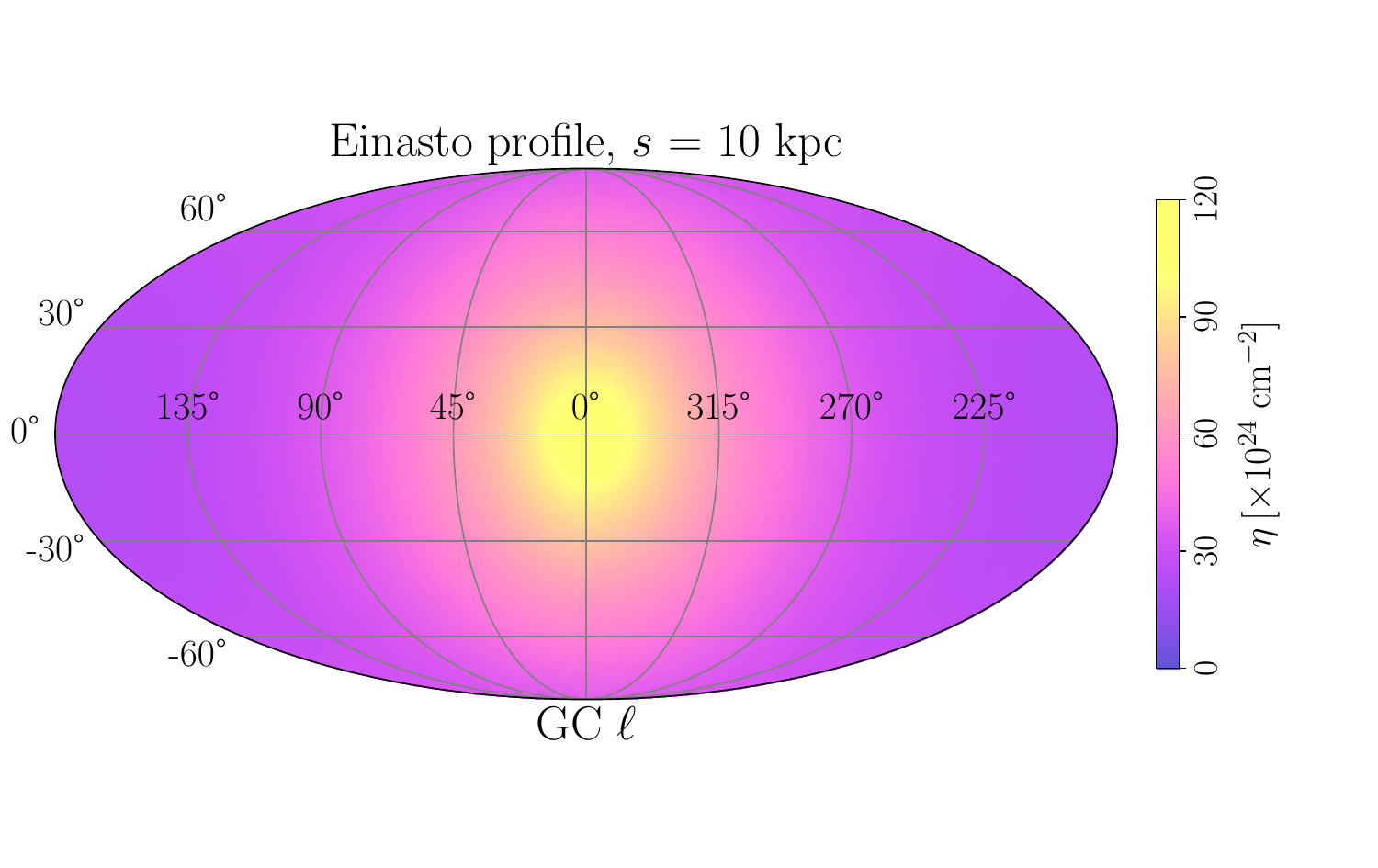}
        \caption{Einasto profile skymap for $d_{\rm SN} = 10$ kpc}
        \label{subfig:Einasto}
    \end{subfigure}
     \begin{subfigure}[b]{0.39\textwidth}
        \centering
        \includegraphics[width=1\textwidth]{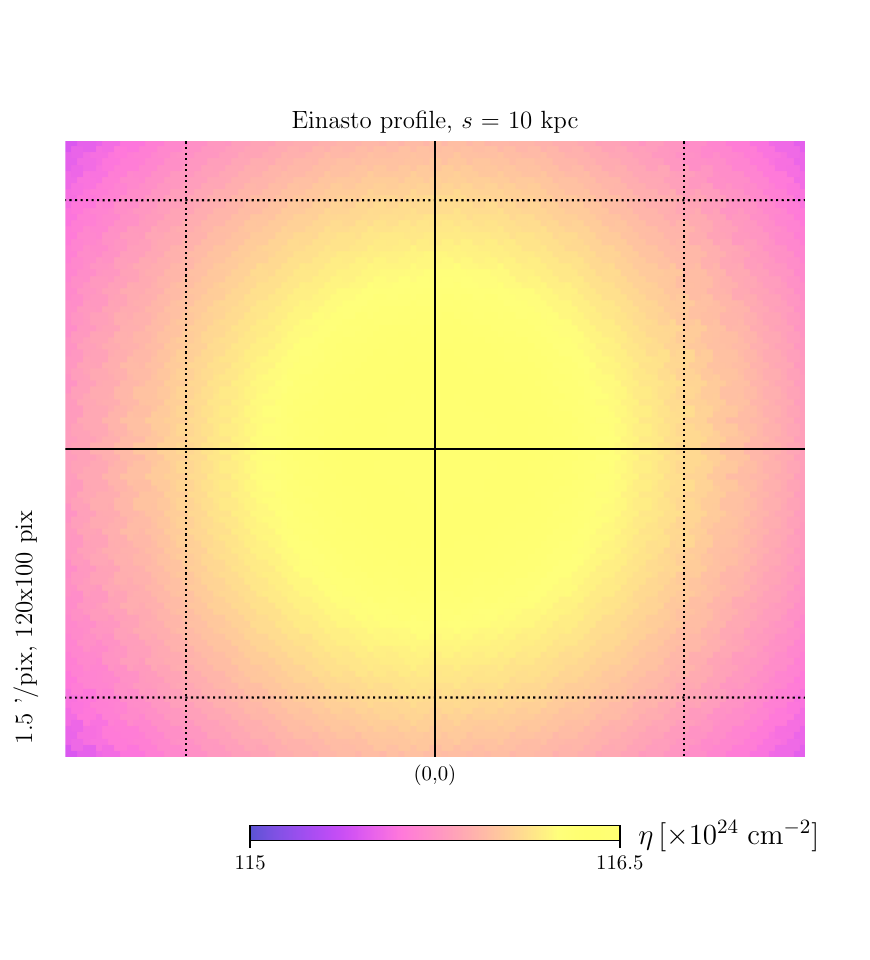}
        \caption{Einasto profile zoomed}
        \label{subfig:EinastoZoom}
    \end{subfigure}
    \caption{Skymap of the integrated column density $\eta$ of DM encountered by neutrinos from a supernova 10 kpc away from the Earth in the direction of $(\ell,b)$, with DM mass $m_{\chi} = 1$ MeV, for the Einasto profile. The panel (a) is for the whole region of $\ell$ and $b$, while the panel (b) is for the zoomed-in region at the center. The bright yellow regions on both skymaps represent $\eta \geq 10^{26}$~cm$^{-2}$. The spacing between grid lines on the zoomed plot in panel (b) corresponds to 1 degree. The column density along the line-of-sight near the the galactic center of the cored profile is relatively uniform, with $\eta$ only varying by $\sim 1\%$ for $1\degree$ away from the galactic center.
    }
    \label{fig:Skymaps:Einasto}
\end{figure}

\begin{figure}[t!]
     \centering
     \begin{subfigure}[b]{0.6\textwidth}
        \centering
        \includegraphics[width=1\textwidth]{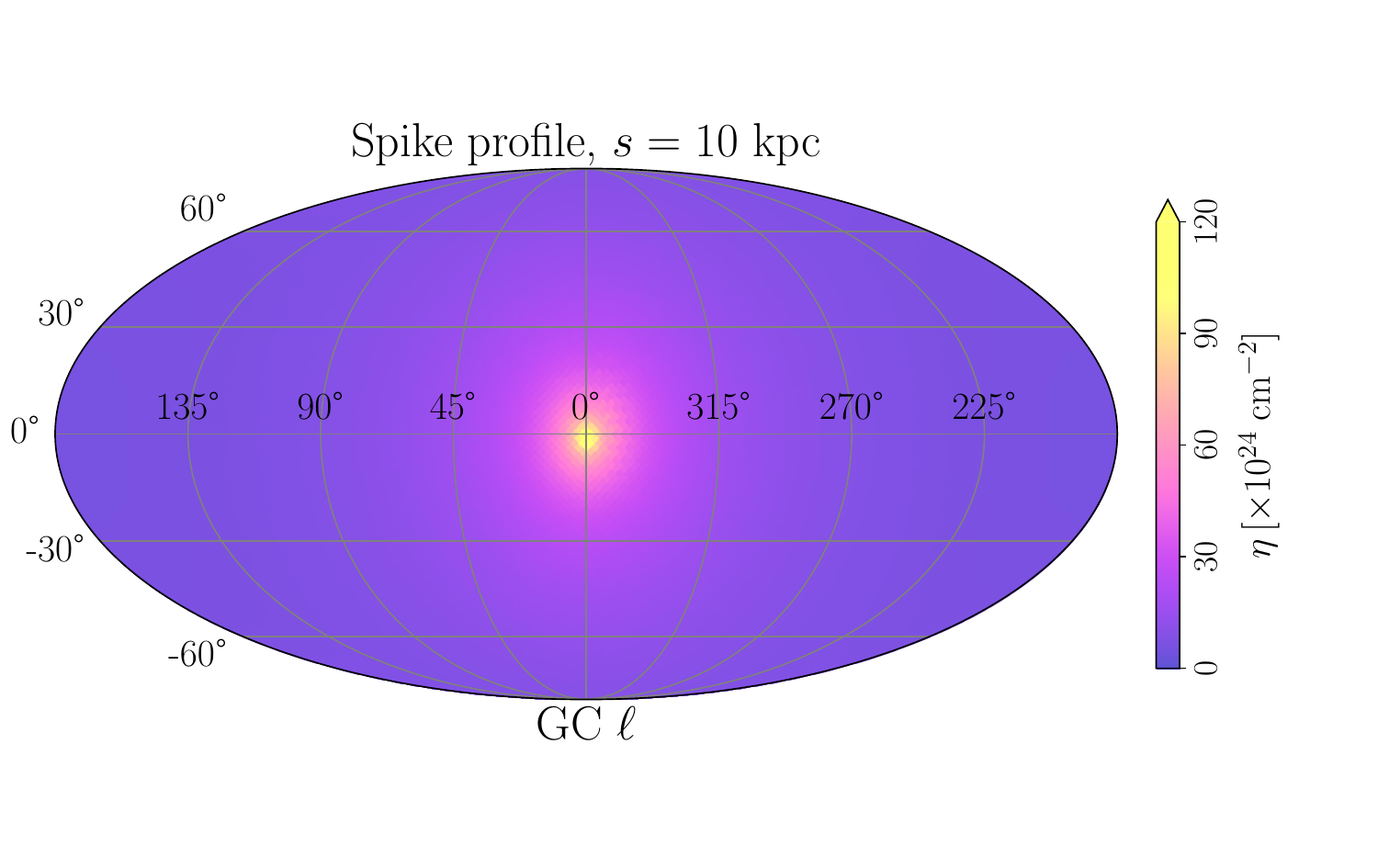}
        \caption{Spike profile skymap, $d_{\rm SN} = 10$ kpc}
        \label{subfig:SpikeSkymap}
    \end{subfigure}
     \begin{subfigure}[b]{0.39\textwidth}
        \centering
        \includegraphics[width=1\textwidth]{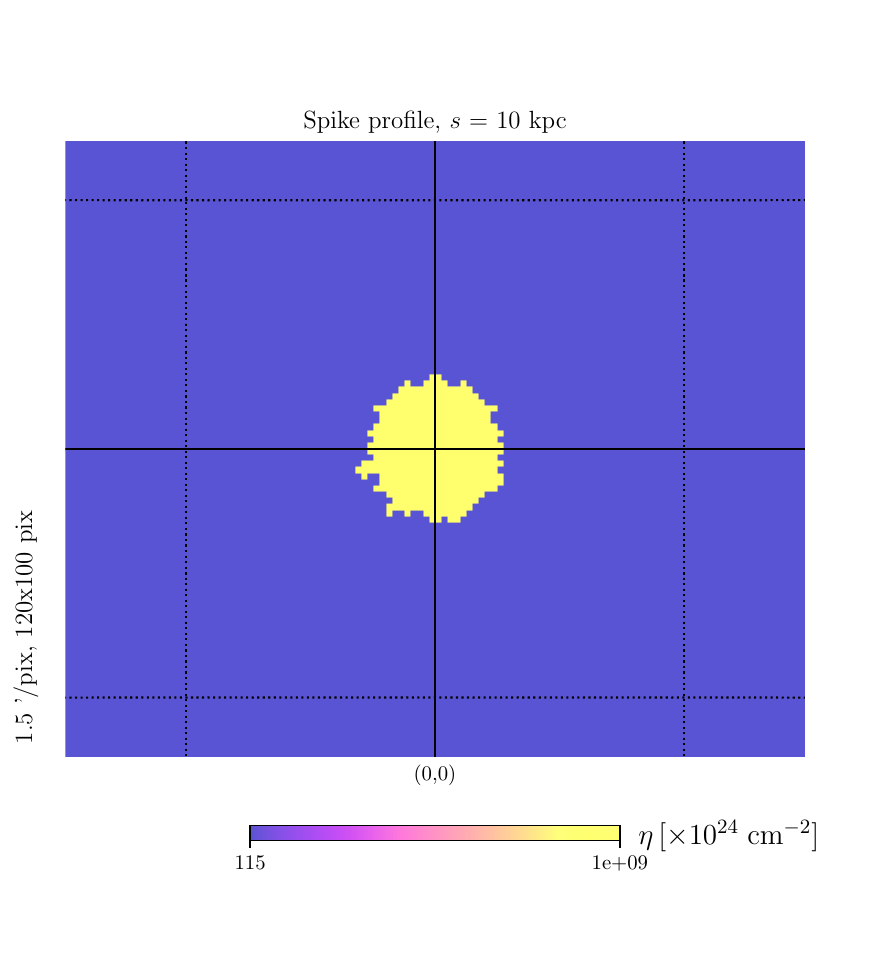}
        \caption{Spike profile zoomed}
        \label{subfig:SpikeZoom}
    \end{subfigure}
    \caption{The same as Fig.~\ref{fig:Skymaps:Einasto}, but for the spike profile.
    At the center of the spike profile, if the line-of-sight passes through the spike radius, $\eta$ can reach $\sim 10^{33}$~cm$^{-2}$ for 1 MeV DM. The spike radius is $\sim 40$ pc. The spacing between grid lines on the zoomed plot in panel (b) corresponds to 1 degree. In panel (b), there is a $10^7$ enhancement in the column density inside the spike radius.
    }
    \label{fig:Skymaps:spike}
\end{figure}

Once a DM halo profile is selected, the column density $\eta$ of DM for a given mass and supernova location is given by
\begin{equation}
    \eta \, (m_\chi;d_{\rm SN}, \ell, b) = \frac{1}{m_\chi} \int_0^{d_{\rm SN}} \rho_{\chi}\,[\,r(s,\,\ell,\,b)] \, \mathrm{d}s \,, \label{eq:ColumnDensity}
\end{equation}
where the semicolon notation indicates that $\eta$ is calculated for a specific supernova. Here the galactic radial coordinate $r$ is expressed in terms of the galactic coordinates $s$, $l$, and $b$, 
\begin{equation}
    r = \sqrt{s^2 + R_\odot^2 - 2sR_\odot \cos \ell \cos b} .
\end{equation}
As explicit examples, the skymaps of $\eta$ for the Einasto and spike profiles are displayed in Figs.~\ref{fig:Skymaps:Einasto} and \ref{fig:Skymaps:spike}, respectively. In both figures, we have set $d_{\rm SN}=10$~kpc and $m_{\chi} = 1$ MeV. 
It should be noted that, for the spike profile case, the integrated column density $\eta$ can reach as high as $10^{33}$ cm$^{-2}$ at the galactic center,\footnote{This assumes a DM mass of $m_\chi$ = 1 MeV, which is consistent with the integrated column density calculated in Ref.~\cite{Cline:2023tkp}, where $\eta = \Sigma_\chi/m_\chi \sim 6\times 10^{34}$ cm$^{-2}$ for 1 MeV DM.} much larger than that of the Einasto and other profiles.  

With the inputs for the DM profile and supernova location fixed, the opacity $\tau$ for a given model can be expressed as a function of the masses $m_\chi$, $m_{\rm med}$, and coupling $g$  (and the dimensionful parameter $\mu$ if applicable) of the model:
\begin{equation} \label{opacitydef}
    \tau(g,m_\chi, m_{\rm{med}}; E_\nu) = \sigma_{\nu\chi}(g, m_\chi,m_{\rm{med}};E_\nu) \, \eta(m_\chi; d_{\rm SN}, \ell, b).
\end{equation}
The region of parameter space where the opacity $\tau \sim \mathcal{O}(1)$ is particularly interesting. As we will discuss more quantitatively in the next subsection, the attenuation of the neutrino flux $\Phi$ reaching the Earth is qualitatively expected to follow an exponential decay, i.e. $\Phi/\Phi_0 \sim e^{-\tau}$. However, there is an additional contribution to the neutrino flux from down-scattering, where a higher-energy neutrino scatters off DM, producing a lower-energy neutrino. These two effects and their interplay can be systematically addressed through the cascade equation, which we discuss in detail in the next subsection. Solving the cascade equation is crucial to determine the modified neutrino flux as a function of the neutrino energy. The reference opacity contours of $\tau =0.1$ (i.e. where the observed neutrino flux for $E_\nu = 25$ MeV would be $\sim$90\% of the expected one, assuming negligible down-scattering) for the Einasto and spike DM halo profiles are shown in
Figs.~\ref{fig:exclusion_fermionDM_scalarMed_t}
through \ref{fig:exclusion_3}, 
which are presented as the solid and dotted blue lines, respectively. In some of those subfigures, the dip structure around $m_{\rm med} \sim 20$ MeV is due to the resonance $\nu \chi \to {\rm med} \to \nu \chi$ in $\nu$-DM scattering.  It is clear that for all the models in these figures, the parameter space for $\tau = 0.1$ and larger values for the Einasto DM profile has been excluded by existing constraints. However, if DM has the spike profile, with much larger DM density at the galactic center, there is still a wide-open parameter space that can produce $\tau = 0.1$, i.e. generating a significant attenuation effect for the neutrino fluxes from a galactic supernova explosion that can be potentially observable in future neutrino experiments.

\subsection{Cascade equation and modified supernova neutrino flux}
\label{sec:cascade}

We now describe the cascade equation that measures the change in neutrino flux as a function of neutrino energy during its interaction with the DM medium, following the procedure outlined in Refs.~\cite{Vincent:2017svp,McMullen:2021ikf}. This equation is given by:
\begin{equation}
    \frac{\mathrm{d}\Phi(E,\eta)}{\mathrm{d}\eta} = -\sigma(E) \Phi(E,\eta)+\int_E^{E_{\rm max}} \mathrm{d}\Tilde{E}\frac{\mathrm{d}\sigma(\Tilde{E},E)}{\mathrm{d}E}\Phi(\Tilde{E},\eta), \label{Eq:CascadeFull}
\end{equation}
where $E_{\rm max}$ is the maximal energy for the supernova neutrino flux, $\Phi = {\rm d} \phi/{\rm d}E$ is the differential neutrino flux, and $\eta$ is the DM column density as defined in Eq.~\eqref{eq:ColumnDensity}. 
The first term on the right-hand side of Eq.~(\ref{Eq:CascadeFull}) represents a loss, describing the rate at which neutrinos with energy $E$ undergo down-scattering with DM, where $\sigma (E)$ is the corresponding total $\nu$-DM scattering cross section. In contrast, the second term in the equation above represents a gain, describing the rate at which neutrinos with energy $\tilde{E}$ scatter off DM, producing neutrinos with energy $E$. Here, ${\rm d}\sigma(\tilde{E}, E)/{\rm d}E$ is the differential cross section with respect to the final-state neutrino energy $E$. 

A few comments should be made regarding the implications of the loss and gain terms in Eq.~(\ref{Eq:CascadeFull}). The differential cross section in the gain term, ${\rm d}\sigma(\tilde{E},E)/{\rm d}E$, implicitly includes the angular information of the outgoing neutrino. This can be easily read from Eq.~(\ref{eqn:scattering:relation}): the incoming neutrino energy $\tilde{E}$ is dictated by the outgoing neutrino energy $E$, DM mass $m_\chi$ and the $\nu$-DM scattering angle $\theta$. As a result, a down-scattered neutrino from energy $\tilde{E}$ to $E$ at a given point is not always guaranteed to reach Earth. Strictly speaking, the cascade equation is only valid in the forward-scattering regime, where the momentum exchange between the neutrino and DM is sufficiently small. However, the scenarios we consider here do not necessarily involve small momentum exchanges. Nonetheless, we argue that the contributions from the gain term are subdominant or negligible, meaning that solving the cascade equation (\ref{Eq:CascadeFull}) does not significantly deviate from the actual situation. Typically, ${\rm d}\sigma(\tilde{E},E)/{\rm d}E$ tends to favor lower energy regions. As a result, for the typical values of $E$ under consideration, $\Delta \tilde{E}\, {{\rm d}\sigma(\tilde{E},E)}/{{\rm d}E}$ remains small. Additionally, the integral is upper bounded by $E_{\max}$, which we set to be 140 MeV throughout our analysis, as the initial supernova flux diminishes well below it. Furthermore, the typical supernova neutrino flux $\Phi(\tilde{E})$ rapidly decreases with increasing $\tilde{E}$. Taking all these factors into account, one finds that the gain term is subdominant or negligible compared to the loss term. Naturally, if $E$ is too small, this argument no longer holds, and the gain term could become significant. However, our benchmark neutrino detectors have MeV-scale energy thresholds, making them insensitive to such low-energy contributions.

Beyond this mathematical reasoning, realistic phenomenological considerations also suggest that the gains from higher-energy neutrinos are subdominant or negligible. For example, even if the scattering angle is very small, the DM-scattered neutrino trajectory might also potentially lead to a sizable time delay of the neutrino signals~\cite{Murase:2019xqi,Carpio:2022sml,Ge:2024ftz,Chauhan:2025hoz}. In light of the huge distance of 10 kpc from the source to the Earth, the time delay could be significant, even up to the order of ${\cal O} ({\rm year})$~\cite{Carpio:2022sml}. In contrast, we consider a 10-second exposure in our analysis, ensuring that the detectors remain insensitive to contributions from these delayed components.
Unfortunately, the cascade equation~\eqref{Eq:CascadeFull} does not incorporate this effect. Thus, our results can be interpreted as {\it conservative} when probing the model parameter space, based on the discrepancy between the expected ``prompt" neutrino fluxes with and without $\nu$-DM interactions. 

Indeed, our numerical studies support these arguments and expectations, showing that the right-hand side of the cascade equation~(\ref{Eq:CascadeFull}) is dominated by the first term in the parameter space of interest, i.e. the loss term due to the {\it down}-scattering of neutrinos with DM, when considering neutrino energies $E_\nu \gtrsim$ 15-20 MeV at the detectors. Below 15 MeV, our results show a sizable effect from down-scattering of the neutrino flux. However, the flux itself is reduced in comparison to the peak, where the effect of attenuation is greatest. Therefore, it is a reasonable approximation to neglect the subleading contributions and effects from the gain term and delayed neutrino signals and rely on the original cascade equation~\eqref{Eq:CascadeFull} to estimate the expected attenuation near the peak of the SN neutrino flux.

The cascade equation can be solved numerically by ``vectorizing" it, converting it into a matrix equation~\cite{Vincent:2017svp}. By transforming $E \rightarrow \vec{E}$ and $\Phi \rightarrow \vec{\phi}$, the cascade equation is simplified to be
\begin{equation}
    \frac{{\rm d} \vec{\phi}}{{\rm d}\eta} = M \Vec{\phi} = -{\rm diag}( \Vec{\sigma})+C  \,,
    \label{Eq:CascadeSimple}
\end{equation}
where ``diag'' takes the diagonal elements of $\vec\sigma$, and the elements of matrix $C$ are given by
\begin{equation}
    C_{ij}=\frac{{\rm d}\Vec{\sigma}(E_i,\Tilde{E_j})}{{\rm d} E_i} {\rm d}\Tilde{E_j} \,.
\end{equation}
Converting the linearly spaced energy interval to a logarithmically spaced interval, we have ${\rm d} \Tilde{E}_j \rightarrow \Tilde{E}_j(\log \tilde{E}_{j+1}-\log \tilde{E}_j)$.  

The solution to Eq.~\eqref{Eq:CascadeSimple} can be obtained by solving for the eigenvalues $\lambda_i$ and eigenvectors $\hat{\phi}_i$ of the matrix $M$, with the eigenvectors forming a complete basis. The solution is
\begin{equation}
    \Vec{\phi} = \sum_{i=1}^{N} c_i \hat{\phi}_i e^{\lambda_i \eta}, \label{Eq:CascadeSolution}
\end{equation}
where the coefficients $c_i$ are determined by requiring the flux matches the modeled supernova flux when no DM column density is present, i.e. $\eta = 0$. To make the solution valid, in the calculations we have take the energy spacing $\Tilde{E}_j(\log \tilde{E}_{j+1}-\log \tilde{E}_j)$ smaller than the scale of changes in neutrino flux or the scattering cross section, otherwise the changes will not be observed. 

\subsection{Event numbers}

We now report the expected number of supernova-induced neutrino events in reconstructed energy deposits at three benchmark large-volume neutrino detectors, namely, DUNE, Hyper-K, and JUNO. Our analysis procedure is similar to that given in Ref.~\cite{Hajjar:2023knk}. The flavor-dependent neutrino spectra at the source point follow the best-fit parameters of \texttt{Warren20} spectra~\cite{Warren:2019lgb}, as also summarized in Table 1 of Ref.~\cite{Hajjar:2023knk}. Since neutrinos oscillate among flavors while traveling to the Earth, we account for these effects in estimating the neutrino fluxes entering the benchmark detectors, following the methods described in Ref.~\cite{Hajjar:2023knk}. For each neutrino flavor $\alpha$, this allows us to estimate the flux $\Phi_{0,\alpha}$ that would be observed without $\nu$-DM interactions. We then apply the procedures from the previous subsection to calculate the attenuated neutrino fluxes $\Phi_\alpha$. For some of the models in Table~\ref{tab:models}, the scattering cross sections of neutrinos and antineutrinos with DM are different, which have been taken into account in our calculations.

For a given neutrino flux $\Phi_\alpha$, the number of detected neutrino scattering events $N_{\rm evt}$ as a function of the true energy deposited, $E_{\rm true}$, is given by
\begin{equation}
    \frac{{\rm d} N_{\rm evt}}{{\rm d} E_{\rm true}}=\sum_{i={\rm scat~proc}} \int {\rm d} E_{\nu} \frac{{\rm d} \Phi_\alpha(E_\nu)}{{\rm d} E_\nu}\frac{{\rm d} \sigma_i (E_{\rm true}, E_\nu)}{{\rm d} E_{\rm true}} N_i \Delta t,
\end{equation}
where the reference time exposure $\Delta t$ is set to be 10 seconds, and $N_i$ is the number of targets in the detector for the $i$th scattering process. Since neutrinos undergo different scattering processes (as indicated by the summation symbol over $i$), we take into account the following relevant processes:
\begin{subequations}
    \begin{align}
    \hbox{Inverse beta decay (IBD):}& \;\; \bar{\nu}_e+p \to e^+ + n \,, \\
    \hbox{$\nu_e$ charged-current scattering (CC):}& \;\; \nu_e+N \to e^- +X \,, 
    \end{align}
\end{subequations}
where $N$ denotes the target nucleus, and $X$ represents any final-state particles other than electrons. The cross sections for these processes are taken from Refs.~\cite{Formaggio:2012cpf,Ricciardi:2022pru}. Since actual detectors have finite energy resolution and detection efficiency $\varepsilon$, we incorporate these factors into our analysis as follows:
\begin{equation}
    \frac{{\rm d} N_{\rm evt}}{{\rm d} E_{\rm rec}}=\int {\rm d} E_{\rm true} \, \varepsilon(E_{\rm true}) \,\mathcal{R}(E_{\rm rec};E_{\rm true}) \frac{{\rm d} N_{\rm evt}}{{\rm d} E_{\rm true}} \,,
\end{equation}
with $E_{\rm rec}$ the reconstructed visible energy. Although detection efficiency $\varepsilon$ generally depends on $E_{\rm true}$, for simplicity, we apply a constant value of $\varepsilon = 1$ in our analysis. The detector resolution is parameterized by $\mathcal{R}$, which is given by
\begin{equation}
    \mathcal{R}(E_{\rm rec};E_{\rm true})=\frac{1}{\sqrt{2\pi}\sigma_{\rm det}(E_{\rm true})} \exp \left\{ -\frac{(E_{\rm true}-E_{\rm rec})^2}{2\sigma_{\rm det}^2(E_{\rm true})} \right\} \,,
\end{equation}
where $\sigma_{\rm det}(E_{\rm true})$ represents the fractional detector resolution as a function of the true deposited energy.

The descriptions so far are detector-independent, but the specific parameter values depend on the individual detectors. The key parameter values of the benchmark neutrino detectors DUNE, Hyper-K, and JUNO used in our analysis are summarized in Table~\ref{tab:keyparam}. In addition, each detector has its own deposited energy resolution function~\cite{Hajjar:2023knk}\footnote{The DUNE resolution function here differs from that of Ref.~\cite{Hajjar:2023knk}. We use $0.16 E_{\rm true}$ based on \cite{Castiglioni:2020tsu}}, given by 
\begin{subequations}
\begin{align}
    \textrm{DUNE: } & \sigma_{\rm det} = 0.16 E_{\rm true} \,, \\
    \textrm{Hyper-K:  } & \sigma_{\rm det} = -0.123 + 0.0349 E_{\rm true} + 0.376\sqrt{E_{\rm true}} \,, \\
    \textrm{JUNO: } & \sigma_{\rm det}^2 = 0.0261^2 E_{\rm true} + 0.0082^2 E_{\rm true}^2 + 0.0123 \,,
\end{align}
\end{subequations}
where $\sigma_{\rm det}$ and $E_{\rm true}$ are in unit of MeV.

The DUNE threshold can in principle go down to $\sim 1.5$ MeV~\cite{GilBotella:2003sz}, or even lower~\cite{mooney}; however, we use the conservative value of 5 MeV to be consistent with the unattenuated flux calculation in Ref.~\cite{Hajjar:2023knk}, as well as the official DUNE supernova analysis~\cite{DUNE:2020zfm}. Since the DUNE detector is unable to identify neutrons, we do not consider the inverse beta decay process at DUNE. Based on the detector’s (fiducial) mass and the type of target nucleus shown in the Table, one can straightforwardly calculate the number of target electrons/protons for elastic scattering/inverse beta decay and target nuclei for charged-current interactions. For example, the DUNE detector consists of $6.03 \times 10^{32}$ target argon nuclei and $1.09 \times 10^{34}$ target electrons (protons).  

\begin{table}[t]
    \centering
    \caption{Key detector specifications and parameter values for the benchmark neutrino experiments DUNE, Hyper-K, and JUNO chosen for our analysis.  The CC processes are subdominant for Hyper-K and JUNO \cite{Hajjar:2023knk}.}
    \label{tab:keyparam}
    \begin{tabular}{c|c c c c c}
    \hline \hline
    Detector & Mass [kt] & Target nucleus & Processes & $\varepsilon$ &  $E_{\rm th}$ [MeV] \\
    \hline
    DUNE     & 40  & Argon & CC& 1 &  5 \\
    Hyper-K  & 220  & Oxygen & IBD, CC & 1 & 3\\
    JUNO     & 20  & Carbon & IBD, CC & 1  & 0.1\\
    \hline \hline
    \end{tabular}
\end{table}

\begin{figure}[t!]
     \centering    \includegraphics[width=0.85\textwidth]{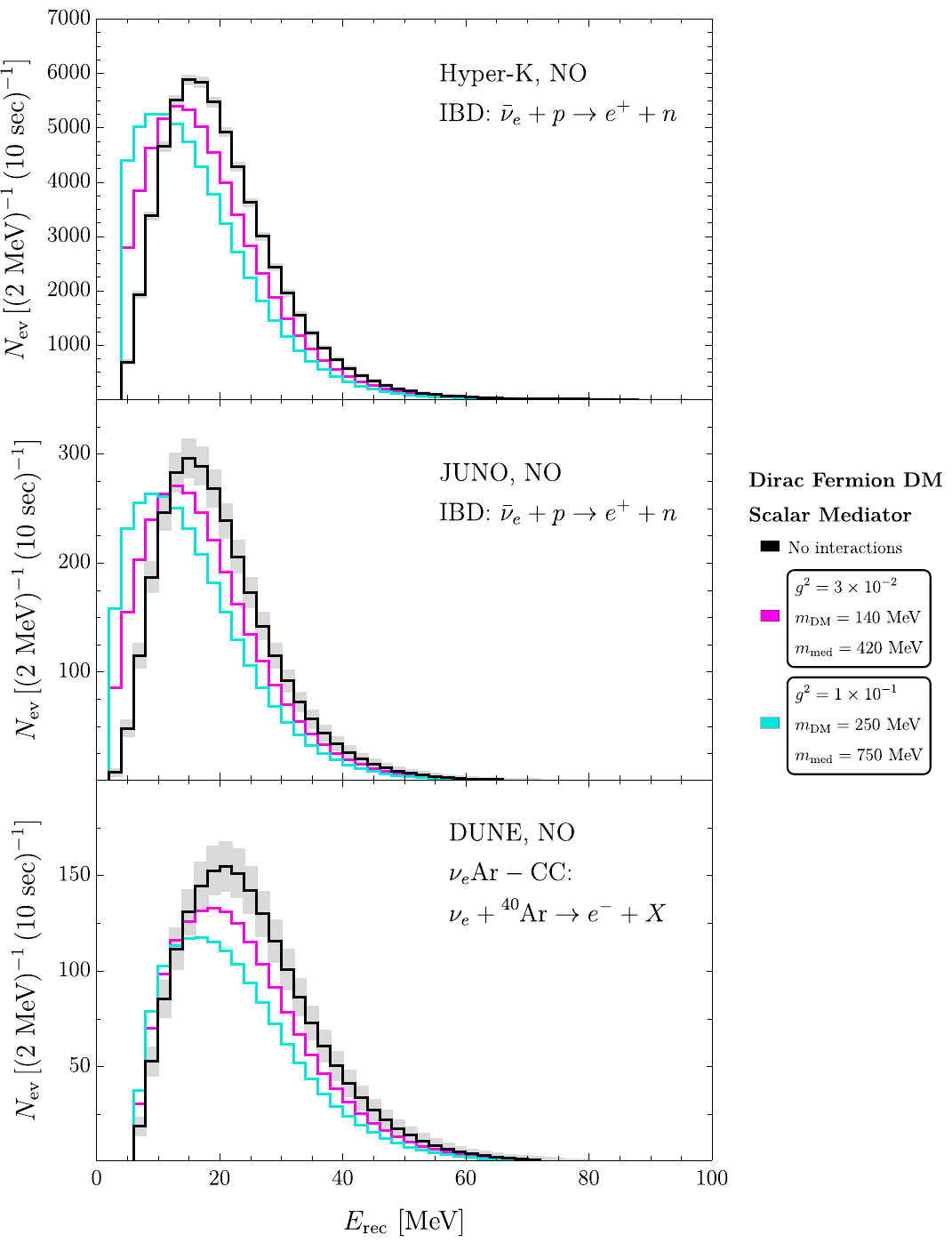}
    \caption{Impact of $\nu$-DM interactions on supernova induced neutrino event rates at Hyper-K (top), JUNO (middle), and DUNE (bottom), assuming a DM spike  profile, for a benchmark model  involving a Dirac fermion DM and a scalar mediator in Fig.~\ref{fig:exclusion_fermionDM_scalarMed_t}. 
    The magenta and cyan lines correspond to the two benchmark points in Fig.~\ref{fig:exclusion_fermionDM_scalarMed_t}, while the black lines are for the case without any $\nu$-DM interaction. 
   The inset plots of each panel represent a zoomed-in picture of the higher energy bins, from 50 MeV to 100 MeV, with the bin heights on a logarithmic scale. The shaded light gray band represents the statistical uncertainty in the unattenuated supernova neutrino flux.}
    \label{fig:Event Rate}
\end{figure}

Our results for the event distributions are presented in Fig.~\ref{fig:Event Rate}, showing the expected number of supernova neutrino events as functions of the reconstructed energy of final-state electrons at the benchmark detectors Hyper-K (top panel), JUNO (middle panel), and DUNE (bottom panel), compared to the case without $\nu$-DM interactions. As justified in the previous section, we adopt a DM spike profile to see observable effect. The two different sets of parameter choices are consistent with existing bounds, shown in Fig.~\ref{fig:exclusion_fermionDM_scalarMed_t} by the magenta and cyan dots, involving a Dirac fermion DM and a scalar mediator: 
\begin{align}
g_\nu g_\chi &= 3\times10^{-2} \,, & m_{\chi} &= 140\, {\rm MeV} \,, & m_{\rm med} &= 420\, {\rm MeV} \,, \nonumber \\
g_\nu g_\chi &= 1\times10^{-1} \,, & m_{\chi} &= 250\, {\rm MeV} \,, & m_{\rm med} &= 750\, {\rm MeV} \,.
\end{align}
The modified spectra for these two sets of parameters are shown in magenta and cyan, respectively, while the case without $\nu$-DM interactions is depicted as the black lines in Fig.~\ref{fig:Event Rate}. The results indicate that, in the presence of $\nu$-DM interactions, the expected number of supernova-induced neutrino events is lower than in the case without such interactions, and the overall energy spectrum becomes softer. Whether this effect is experimentally observable depends on many factors, including the statistical and systematic uncertainties associated with the event rate, as well as with the supernova modeling itself. Here we only show (by the gray shaded region) the statistical uncertainty on the original unattenuated flux, given by $\pm \sqrt{N_{\rm bin}}$ for a given number of events $N_{\rm bin}$ in any particular energy bin. A detailed treatment of the uncertainties associated with detector systematics and supernova modeling is beyond the scope of the present work. But we should emphasize that due to the nontrivial energy dependence of the neutrino flux by virtue of the cascade equation~\eqref{Eq:CascadeFull}, the attenuation effect at higher energies and the accumulation of events at lower energies should be distinguishable from potential astrophysical uncertainties on the flux itself.   

We find that, if the DM halo follows a spike profile, the expected supernova neutrino energy spectrum becomes significantly softer, depending on the details of the underlying DM scenario, and may become observable in future neutrino experiments. This allows us to probe a broad range of unexplored parameter space, as suggested by the results shown in Figs.~\ref{fig:exclusion_fermionDM_scalarMed_t} through \ref{fig:exclusion_3}. Turning the argument around, neutrino detectors are sensitive to the DM spike profile in this channel, meaning that if there is no significant deviation from the expected behavior of the supernova neutrino flux without $\nu$-DM interactions, we can exclude the possibility of spiked DM profiles with sizable couplings to neutrinos. Even with the Einasto DM halo profile, the energy spectra may exhibit significant anomalous behavior at higher energies, well above the detector energy threshold. Consequently, upcoming large-volume neutrino detectors such as DUNE, Hyper-K, and JUNO will allow us to explore new regions of parameter space for models involving DM interactions with neutrinos. 

\section{Discussion and conclusions}
\label{sec:con}

In this paper, we have performed a detailed study of the $\nu$-DM interactions. To this effect, we worked in a simplified model framework and wrote down the effective dimension-4 couplings of the mediators to neutrinos and DM. Here we summarize the main points of our paper. The key information about the DM and associated mediator models is collected in Table~\ref{tab:models}. The comparison of our  cross sections with those in the existing literature can also be found in this Table and in Figs.~\ref{fig:xs:comparison:1}, \ref{fig:xs:comparison:2}, and \ref{fig:xs:comparison:3}. The details of the $\nu$-DM scattering cross sections are given in Appendix~\ref{Appendix:xs:nuDM}. We have collected all relevant astrophysical, cosmological, and laboratory limits on the couplings of neutrinos and DM particles in Section~\ref{sec:bounds}. Among these limits, the new ones derived in this paper are: 
\begin{itemize}
    \item the bullet cluster limit on the self-interactions of DM applied to each model and translated into the mass-coupling plane;
    \item the collisional damping limits for all the models; 
    \item the CMB $\nu$-DM limits for all the models; 
    \item the NGC 1068 bound applied to each model; 
    \item an updated SN1987A opacity bound for each DM model; 
    \item the annihilation cross sections of DM into neutrinos and the resultant thermal relic density of DM as functions of the couplings and masses; 
    \item the updated meson and $Z$ decay limits including the 1-loop contributions and cancellations of IR divergences, including heavier $D$ and $B$ meson decays. 
\end{itemize}
Our summary plots containing all relevant constraints for the models are presented in Figs.~\ref{fig:exclusion_fermionDM_scalarMed_t}, \ref{fig:exclusion_2} and \ref{fig:exclusion_3} for a benchmark choice of $m_{\rm DM} = m_{\rm med}/3$. For all the models in this paper, it is found that the most stringent limits for the parameter space we are interested in are mainly from CMB, bullet cluster and meson/$Z$ decays. The CMB and bullet cluster limits exclude mediators lighter than the scale of ${\cal O}({\rm MeV})$, while the meson and $Z$ limits are the most constraining for ${\cal O}({\rm MeV}) \lesssim m_{\rm med} \lesssim {\cal O}({\rm GeV})$.
    
As an application of our results, we explore the observational prospects of $\nu$-DM interactions for a future galactic supernova event in large-volume neutrino experiments. The supernova neutrinos from the Milky Way Galaxy will inevitably pass through the galactic DM halo en route to Earth. Assuming a supernova explosion at the distance of 10 kpc from the Earth but on the opposite side of the galactic center such that the neutrinos pass through the galactic center en route to us, we have examined the attenuation of the neutrino flux due to $\nu$-DM scattering. As explicit examples, the attenuation effects for some benchmark points of the couplings and mass parameters in the case of Dirac fermion DM with a scalar mediator are demonstrated in Fig.~\ref{fig:Event Rate}. The important point here is that the parameters of the benchmark points chosen here satisfy all the constraints discussed in Section~\ref{sec:bounds}, as shown in Fig.~\ref{fig:exclusion_fermionDM_scalarMed_t} by the magenta and cyan dots. As exemplified in Fig.~\ref{fig:Event Rate}, the attenuation effects due to $\nu$-DM interaction could be observable in some regions of the parameter space, in particular for a spike DM profile, at future neutrino experiments, such as DUNE, Hyper-K, and JUNO. In the presence of sizable $\nu$-DM interactions, the number of observed neutrino events within the first few seconds is expected to be lower than that predicted from  supernova simulations. If no such reduction is observed in future neutrino experiments, the neutrino events could then be used to constrain the $\nu$-DM couplings.
This is true for many of the models considered in Figs.~\ref{fig:exclusion_fermionDM_scalarMed_t} through \ref{fig:exclusion_3}, where the neutrino flux attenuation effect is measured by the opacity $\tau$, which is chosen to be 0.1 in these figures. It turns out that there is still a large parameter space for which the attenuation effect could be significant for the spike DM profile. But in the absence of a DM spike, the viable parameter space to probe $\nu$-DM interactions in future experiments is rather limited, as indicated by the opacity contours for the Einasto DM profile.

We conclude with some further remarks and possible future directions:   
\begin{itemize}
    \item In this paper, we have only considered the effective couplings of neutrinos with DM. Some potential UV-completions and underlying theories can be found in Appendix~\ref{sec:UV}; however, there is scope for concrete model building efforts, especially for light DM and light mediators. 
    The minimal set of astrophysical, cosmological, and laboratory limits discussed in this paper are readily applicable to these UV-complete models, but they  might be subject to additional limits such as those from the charged lepton sector, which need further scrutiny. 
    \item We have not considered any specific DM production mechanism in this work. Although we show the thermal relic density curve due to DM annihilation into neutrinos in our summary plots, there exists other possibilities for getting the correct relic density, e.g. in non-thermal and freeze-in dark matter scenarios. Moreover, some of the cosmological limits discussed here can be avoided or at least relaxed in non-standard cosmological scenarios. In those situations, the complementarity between the cosmological and other bounds can be important in finding viable parameter space for observable $\nu$-DM interactions.   
    \item We only discussed one potential observational aspect of the $\nu$-DM interaction effect in  galactic supernova neutrinos. Other possible effects of $\nu$-DM interactions include boosting the DM particles to relativistic energies by supernova neutrinos~\cite{Das:2021lcr,Lin:2022dbl,Lin:2023nsm,Lin:2024vzy,Das:2024ghw} or by the cosmic relic neutrino background~\cite{Cho:2022axv}. In the former case, the DM direct detection experiments such as XENONnT~\cite{XENON:2023cxc}, LUX-ZEPLIN~\cite{LZ:2018qzl}, and PandaX-4T~\cite{PandaX-4T:2021bab} can also be used to set limits on the $\nu$-DM interactions, whereas in the latter case, neutrino experiments may even be able to detect the C$\nu$B flux due to its interaction with DM. This is much like two sides of the same coin. The presence of $\nu$-DM interactions could make both DM and neutrino experiments complementary to each other in probing the unknown sectors of neutrino and dark matter more efficiently. 
\end{itemize}

\section*{Acknowledgments}
We thank Kaladi Babu, Brian Batell, Christopher Cappiello, Dibya Chattopadhyay, André de Gouvêa, Bhaskar Dutta, Subhajit Ghosh, Matheus Hostert, Alejandro Ibarra, Stephan Meighen-Berger, Shmuel Nussinov, Jong-Chul Park, Amol Patwardhan, Dan Pershey, Yago Porto, Seodong Shin, Anna Suliga, and Anil Thapa for many illuminating discussions.
We also thank Satoshi Shirai and Aaron Vincent for communications on Refs.~\cite{Heston:2024ljf} and \cite{Arguelles:2017atb}, respectively. Finally, we thank Saurabh Kadam for creating the schematic of neutrino-DM interactions. 
The work of BD is supported by the U.S. Department of Energy grant No.~DE-SC 0017987, and by a Humboldt Fellowship from the Alexander von Humboldt Foundation. The work of DK is supported in part by the U.S. Department of Energy Grant DE-SC0010813. The work of DS is partly supported by NSF Grant No.~PHY-2210361 and U.S. DOE Grant DE-SC0010813, by the Maryland Center for Fundamental Physics, and by the Mitchell Institute for Fundamental Physics and Astronomy. KS is supported in part by the National Science Foundation under Award No. PHY-2412671. YZ is supported by the National Natural Science Foundation of China under grant No. 12175039, the 2021 Jiangsu Shuangchuang (Mass Innovation and Entrepreneurship) Talent Program No.~JSSCBS20210144, the State Key Laboratory of Dark Matter Physics, and the “Fundamental Research Funds for the Central Universities”. BD, DK, DS, and KS also wish to acknowledge the Center for Theoretical Underground Physics and Related Areas (CETUP*) and the Institute for Underground Science at SURF for hospitality and for providing a stimulating environment. KS would like to thank the Aspen Center for Physics, which is supported by National Science Foundation grant PHY-2210452, for hospitality during the final stages of this work.

\appendix

\section{UV-complete models} 
\label{sec:UV}

In this Appendix, we provide some UV-complete models for each of the DM-mediator types discussed in Section~\ref{sec:nuDMInteractions}. See also Ref.~\cite{Babu:2025czb} for a detailed discussion. This is by no means an exhaustive list. Our goal is to simply illustrate that it is possible to have gauge-invariant, UV-complete models where the DM dominantly couples to neutrinos at leading order, whereas the couplings of the DM to charged leptons and other SM particles tend to be smaller, generated only at loop level.  Nevertheless, the extra ingredients in the UV-complete models might induce additional constraints on the parameter space shown in the main text. In this sense, the constraints shown in the main text should be treated as the conservative (or bare minimum) ones, when we talk about $\nu$-DM interactions.\footnote{In some cases, additional interactions could also alleviate some constraints; for instance, thermal dark matter having non-negligible interactions with both electrons and
neutrinos are more elusive to CMB observations than purely neutrinophilic
ones~\cite{Escudero:2018mvt}. }

In general, the UV-completion of the mediator to neutrinos and DM can be classified into the following groups: (i) med-$\nu$-$\nu$ and med-DM-DM couplings, and (ii) med-$\nu$-DM couplings.

Let us first consider the scalar couplings to neutrinos and DM particles in the form of  $\phi$-$\nu$-$\nu$ and $\phi$-DM-DM. 
\begin{itemize}
    \item One of the well-motivated UV-completions for the couplings to neutrinos is the Majoron particle from spontaneous breaking of the global leptonic symmetry~\cite{Chikashige:1980qk,Chikashige:1980ui,Gelmini:1980re,Schechter:1981cv,Aulakh:1982yn}. The couplings of Majoron to the charged leptons and other SM particles can be generated at the 1-loop or 2-loop levels~\cite{Chikashige:1980ui,Garcia-Cely:2017oco,Heeck:2019guh}.

    \item Another well-motivated example of a neutrinophilic mediator can be constructed from the type-I seesaw mechanism~\cite{Minkowski:1977sc,Mohapatra:1979ia,Yanagida:1979as,Gell-Mann:1979vob,Glashow:1979nm} augmented by a scalar $\phi$, where the SM-singlet fermions dominantly couple to $\phi$ which in turn couple to the dark sector, and their interactions with SM neutrinos are only induced by the active-sterile mixing~\cite{Xu:2023xva, Das:2025asx}.  
    
    \item The scalar $\phi$ can be identified as the neutral component of the $SU(2)_R$-breaking triplet in the left-right symmetric model~\cite{Pati:1974yy,Mohapatra:1974hk,Senjanovic:1975rk} (see also Refs.~\cite{Maiezza:2015lza,Nemevsek:2016enw,BhupalDev:2016nfr,Dev:2017dui}). The coupling of $\phi$ to the charged leptons can be generated radiatively at the 1-loop order~\cite{Nemevsek:2016enw}. Note that if $\phi$ is very light (keV scale), it can also be a good DM candidate itself~\cite{Nemevsek:2012cd, Nemevsek:2023yjl, Dev:2025fcv}.  
    
    \item $\phi$ can also be the scalar in the neutrinophilic two-Higgs-doublet model, which is in charge of the tiny neutrino masses~\cite{Wang:2006jy,Gabriel:2006ns}. 
    
    \item The couplings of $\phi$ to neutrinos can be generated from dimension-6 operators such as $(LH)(LH)\phi / \Lambda^2$, with the active neutrinos being of Dirac or Majorana particles~\cite{Berryman:2018ogk,Kelly:2019wow,deGouvea:2019qaz}. Such operators can originate from the type-II seesaw model with the $SU(2)_L$ triplet $\Delta$~\cite{Magg:1980ut,Schechter:1980gr,Cheng:1980qt,Mohapatra:1980yp,Lazarides:1980nt} extended by a singlet $\phi$, with scalar couplings in the form of $\phi H \Delta^\dagger H$. Then the dimension-6 operator can be obtained by integrating out the triplet $\Delta$. The phenomenological details of such a UV-completion can be found in Ref.~\cite{Dev:2021axj}. One may also generate such dimension-6 operators by involving the vector-like fermions, which is similar to the type-I seesaw~\cite{Minkowski:1977sc,Mohapatra:1979ia,Yanagida:1979as,Gell-Mann:1979vob,Glashow:1979nm}, or the $SU(2)_L$-triplet fermions in the type-III seesaw~\cite{Foot:1988aq}. More details can be found in e.g. Ref.~\cite{deGouvea:2019qaz}. 
    \item The leptonic scalar $\phi$ can also play the role of a portal to the dark sector, and the DM particle is stabilized, e.g. by introducing a discrete $Z_2$ symmetry~\cite{Kelly:2019wow}. 
\end{itemize}

The couplings of gauge boson mediator to neutrinos and DM in the form of $Z'$-$\nu$-$\nu$ and $Z'$-DM-DM can originate from the following UV-completions.
\begin{itemize}
    \item The couplings of $Z'$ to neutrinos can be realized e.g. in the neutrinophilic two-Higgs-doublet model with the $U(1)$ gauge symmetry~\cite{Nomura:2017wxf}.
    
    \item The couplings of $Z'$ boson to neutrinos and DM can also be realized in neutrino portal DM with a vector mediator~\cite{Cherry:2014xra, Farzan:2016wym,Farzan:2017xzy, Blennow:2019fhy,Abdallah:2021npg}. Both DM and heavy neutrinos are charged under the gauge group. The heavy neutrinos couple to the gauge boson, then the heavy-light neutrino mixing will induce the couplings of the gauge boson $Z'$ to the active neutrinos. In such models, the couplings of $Z'$ to the charged leptons are generated at the 1-loop order~\cite{Chauhan:2020mgv, Chauhan:2022iuh}.
\end{itemize}

For the coupling in the form of $\phi/Z'$-$\nu$-DM, we can have the following motivations and UV-completions:
\begin{itemize}
    \item A representative UV-completion of this type of operator appears in the radiative generation of neutrino masses at 1-loop or 2-loop level~\cite{Tao:1996vb,Ma:2006km,Farzan:2012ev,Arhrib:2015dez,Cai:2017jrq, Hagedorn:2018spx,Babu:2019mfe,Alvarado:2021fbw,Herms:2023cyy}. In such models, the active neutrinos couple to the scalar $\phi$ and the fermion $\chi$ in the form of $\phi \nu \chi$. Either $\phi$ or $\chi$ can play the role of DM, and the other one is then the mediator. Here the scalar $\phi$ may also be a vector boson $Z'$. The stabilization of DM in such frameworks can be realized by either introducing an ad hoc $Z_2$ symmetry, or by introducing a chiral $U(1)$ gauge symmetry acting on a dark sector which protects the masses of the dark sector fermions,
and also guarantees the stability of the dark matter particle by virtue of an unbroken discrete $Z_N$ gauge symmetry~\cite{Babu:2024zoe}.
    
    \item The DM particle can couple to the heavy neutrino $N$ via the term $\phi \chi N$. Then the heavy-light neutrino mixing will induce the coupling of $\phi$ with neutrinos and DM, which is dubbed as neutrino portal DM~\cite{Batell:2017cmf,Batell:2017rol,Folgado:2018qlv,Becker:2018rve,Blennow:2019fhy,Lamprea:2019qet, Abdelrahim:2025fiz}. In such scenarios, the couplings of DM particle to the charged leptons and other SM particles can be generated at the 1-loop level~\cite{Becker:2018rve,Blennow:2019fhy,Lamprea:2019qet}.
    
    \item The couplings in the form of $\phi \nu \chi$ can be generated by dimension-5 operators e.g. $(H L) (\phi \chi)/\Lambda$. After the spontaneous symmetry breaking of the SM Higgs, we will obtain the effective coupling of $\phi$ with neutrino and DM $\chi$~\cite{Falkowski:2009yz,Ghosh:2017jdy,Zhang:2020nis,Orlofsky:2021mmy}. Such dimension-5 operators can originate from ${\cal N} = 1$ super-Yang-Mills theories such as supersymmetry~\cite{Falkowski:2009yz}.
    \item Another UV-completion of the $\phi\chi\nu$ vertex comes from a simple extension of the 2HDM by adding one complex scalar singlet and right-handed neutrinos~\cite{Dutta:2020scq, Dev:2025sah}. 
\end{itemize}

\section{$\nu$-DM scattering cross sections}
\label{Appendix:xs:nuDM}

In this Appendix, we collect all the $\nu$-DM scattering cross sections for the models in Section~\ref{sec:nuDMInteractions}. 
The differential cross section for $\nu$-DM scattering is given by~\cite{Arguelles:2017atb} 
\begin{eqnarray}
\frac{{\rm d}\sigma}{{\rm d}\cos\theta} = \frac{1}{8\pi} \frac{E_\nu^{\prime\, 2}}{4m_\chi^2 E_\nu^2} \overline{|{\cal M}|^2} \,,
\end{eqnarray}
where $\overline{|{\cal M}|^2}$ is the usual spin sum-averaged amplitude square, and $m_\chi$, $E_\nu$, $E'_\nu$, and $\theta$ denote the DM mass, the incoming neutrino energy, the outgoing neutrino energy, and the scattering angle of the outgoing neutrino with respect to the momentum direction of the incoming neutrino, respectively. The quantities $E_\nu$, $E'_\nu$, and $\cos\theta$ are all measured in the laboratory frame, and are correlated  via
\begin{eqnarray} 
\label{eqn:scattering:relation}
\frac{1}{E'_\nu} = \frac{1}{E_\nu} + \frac{1-\cos\theta}{m_\chi} \,. \label{eq:relEvEv}
\end{eqnarray}
Here the simple correlation of $E'_\nu$ with the scattering angle $\theta$, as well as $E_\nu$ and $m_\chi$, has far-reaching implications for the kinetics of scattered neutrinos and the resultant time delay of supernova neutrino signals in the terrestrial experiments (cf. Section~\ref{sec:cascade}).  
In the non-relativistic limit of DM velocity, i.e., $v_{\chi} \ll 1$ hence $p_{\chi}^\mu\approx(m_\chi, \vec{0})$, the Mandelstam variables are further simplified to
\begin{subequations}
\begin{align}
\label{eq:s}
s &= (p_{\chi}+p_\nu)^2 = m_\chi^2 + 2 m_\chi E_\nu \,, \\
t &= (p'_\nu - p_\nu)^2 = -2 E_\nu E'_\nu (1-\cos\theta) \,, \\
\label{eq:u}
u &= (p'_\nu - p_{\chi})^2 = m_\chi^2 - 2m_\chi E'_\nu \,,
\end{align}
\end{subequations}
which satisfy the usual identity $s + t + u = 2 m_\chi^2$ in combination with Eq.~\eqref{eq:relEvEv}. In the calculations of the amplitude square for $s$-channel processes, we will need the widths of the mediators, which can be found in Appendix~\ref{appendix:width}.

Let us first consider the DM scenarios with a real scalar mediator $\phi$. For the complex scalar DM case with Lagrangian given in Eq.~(\ref{eqn:L:scalar:scalar}), the spin sum-averaged amplitude square is given by
\begin{eqnarray}
\label{eqn:ampsq:scalar:scalar}
\overline{|{\cal M}|^2} =
- 2 \left( g_{\nu\, s}^2 + g_{\nu\, p}^2 \right) \mu^2 \frac{t}{(t-m_\phi)^2} \,.
\end{eqnarray}
For the case of Dirac fermion DM $\chi$ with couplings given in Eq.~(\ref{eqn:L:scalar:Dirac}), the spin sum-averaged amplitude squares for neutrino and antineutrino are, respectively
\begin{subequations}
\begin{align}
\label{eqn:ampsq:scalar:Dirac:u}
\overline{ \sum |{\cal M}|_{\chi\nu}^2} & \ = \ 
\frac12 (g_s^2+g_p^2)^2 \frac{(u-m_\chi^2)^2}{(u-m_\phi)^2} \,, \\
\label{eqn:ampsq:scalar:Dirac:s}
\overline{ \sum |{\cal M}|_{\chi\bar\nu}^2} & \ = \ 
\frac12 (g_s^2+g_p^2)^2 \frac{(s-m_\chi^2)^2}{(s-m_\phi)^2 + m_\phi^2 \Gamma_\phi^2} \,.
\end{align}
\end{subequations}
If the DM is a Majorana fermion, the amplitude square is
\begin{eqnarray}
\label{eqn:ampsq:scalar:Majorana}
\overline{ \sum |{\cal M}|_{}^2} &=& 
\frac12 (g_s^2+g_p^2)^2
\left[ \frac{(s-m_\chi^2)^2}{(s-m_\phi)^2 + m_\phi^2 \Gamma_\phi^2} + \frac{(u-m_\chi^2)^2}{(u-m_\phi^2)^2} \right. \nonumber \\
&& \left. +
\frac{2(s-m_\chi^2)}{(s-m_\phi)^2 + m_\phi^2 \Gamma_\phi^2} \frac{(u-m_\chi^2)^2}{u-m_\phi^2} \right] \,,
\end{eqnarray}
where the last term in the bracket is the interference term. If the couplings of mediator $\phi$ to neutrino and DM are in the form of Eq.~(\ref{eqn:L:scalar:Dirac:t}), the amplitude square for the Dirac fermion DM case is then
\begin{eqnarray}
\label{eqn:ampsq:scalar:Dirac:t}
\overline{ \sum |{\cal M}|_{}^2} = 
2 \left( g_{\nu\, s}^2 + g_{\nu\, p}^2 \right) 
\frac{t \left[ g_{\chi\, s}^2 (t-4m_\chi^2) + g_{\chi\, p}^2 t \right]}{(t-m_\phi)^2} \,.
\end{eqnarray}
For the case of complex vector DM with couplings in Eq.~(\ref{eqn:L:scalar:vector}), the spin sum-averaged amplitude square is
\begin{eqnarray}
\label{eqn:ampsq:scalar:vector}
\overline{\sum |{\cal M}|_{}^2} = -
 \frac{2\left( g_{\nu\, s}^2 + g_{\nu\, p}^2 \right)  \mu^2}{(t-m_\phi)^2} \frac{t}{m_\chi^4} 
\Big[ t^2 - 4 m_\chi^2 t + 12 m_\chi^4 \Big] \,.
\end{eqnarray}

For the scenarios with a fermion mediator $N$, if DM $\chi$ is a real scalar, the couplings are in the form of Eq.~(\ref{eqn:L:fermion:scalar_complex}), and the corresponding amplitude square is
\begin{eqnarray}
\label{eqn:ampsq:fermion:scalar_real}
\overline{\sum |{\cal M}|_{}^2} &=& 
 (g_s^2 + g_p^2)^2 (m_\chi^4-su) 
\left[ \frac{1}{(s-m_N)^2 + m_N^2 \Gamma_N^2} + \frac{1}{(u-m_N)^2} \right. \nonumber \\
&& \left. - \frac{2(s-m_N^2)}{(s-m_N)^2 + m_N^2 \Gamma_N^2} \frac{1}{u-m_N^2} \right] \,.
\end{eqnarray}
For the case of complex scalar DM, the $\nu$-DM and $\bar\nu$-DM scatterings proceed in the $s$- and $u$-channels, respectively, and the corresponding amplitude squares are:
\begin{subequations}
\label{eqn:ampsq:fermion:scalar_complex:su}
\begin{align}
\label{eqn:ampsq:fermion:scalar_complex:s}
\overline{\sum |{\cal M}|_{\chi\nu}^2} & \ =  \ 
(g_s^2 + g_p^2)^2
\frac{m_\chi^4-su}{(s-m_{N}^2)^2 + m_N^2 \Gamma_N^2} \,, \\
\label{eqn:ampsq:fermion:scalar_complex:u}
\overline{ \sum |{\cal M}|_{\chi\bar\nu}^2 } & \ = \ 
(g_s^2 + g_p^2)^2
\frac{m_\chi^4-su}{(u-m_{N}^2)^2} \,.
\end{align}
\end{subequations}
For the case of vector DM with couplings in Eq.~(\ref{eqn:L:fermion:vector}), the amplitude square can be written as
\begin{equation}
\label{eqn:ampsq:fermion:vector}
\overline{ \sum |{\cal M}|_{}^2} = \frac{g^4}{3} 
\left[ \frac{f_{\rm VF}^{(s)}}{(s-m_N^2)^2 + m_N^2 \Gamma_N^2} 
+ \frac{f_{\rm VF}^{(u)}}{(u-m_N^2)^2}
+ \frac{2(s-m_N^2)f_{\rm VF}^{(su)}}{\left[ (s-m_N^2)^2+m_N^2 \Gamma_N^2 \right](u-m_N^2)} \right] \,, 
\end{equation}
with the functions
\begin{subequations}
\begin{align}
f_{\rm VF}^{(s)} &= 4m_\chi^4 - s (7s+4u) + \frac{4s^2(s+u)}{m_\chi^2} - \frac{s^3 u}{m_\chi^4} \,, \\
f_{\rm VF}^{(u)} &= 4m_\chi^4 - u (4s+7u) + \frac{4u^2(s+u)}{m_\chi^2} - \frac{s u^3}{m_\chi^4} \,, \\
f_{\rm VF}^{(su)} &= -16m_\chi^4 + 8m_\chi^2 (s+u) + 7su - \frac{4su(s+u)}{m_\chi^2} + \frac{s^2u^2}{m_\chi^4} \,.
\end{align}
\end{subequations}

Let us now consider the scenarios with a vector mediator $Z'$. For the case of complex scalar DM with couplings in Eq.~(\ref{eqn:L:vector:scalar_complex:t}), the amplitude square is
\begin{eqnarray}
\label{eqn:ampsq:vector:scalar_complex:t}
\overline{ \sum |{\cal M}|^2} = 
\frac{ g_\nu^2 g_\chi^2}{(t-m_{Z'}^2)^2} \left[ 
t (4m_\chi^2-t) + (s-u)^2 \right] \,.
\end{eqnarray}
For the cases of Dirac fermion DM with couplings in Eq.~(\ref{eqn:L:vector:Dirac:t}) and Majorana fermion DM with couplings in Eq.~(\ref{eqn:L:vector:Majorana:t}), the amplitude squares are, respectively,
\begin{eqnarray}
\label{eqn:ampsq:vector:Dirac:t}
\overline{ \sum |{\cal M}|_{\rm D}^2} &=& 
\frac{2g_\nu^2}{(t-m_{Z'}^2)^2}  
\Big[ g_{\chi L}^2 (s-m_\chi^2)^2 
+ 2 g_{\chi L} g_{\chi R} m_\chi^2 t
+ g_{\chi R}^2 (u-m_\chi^2)^2 \Big]  \,, \\
\label{eqn:ampsq:vector:Majorana:t}
\overline{ \sum |{\cal M}|_{\rm M}^2} &=& 
\frac{ 2g_\nu^2 g_\chi^2}{(t-m_{Z'}^2)^2} \Big[ 
s^2 + u^2 - 2 m_\chi^4 \Big] \,.
\end{eqnarray}
For the $s$- and $u$-channel processes involving a Dirac fermion DM with couplings in Eq.~(\ref{eqn:L:vector:Dirac:u}), the amplitude squares for the neutrino and antineutrino are, respectively,
\begin{subequations}
\begin{align}
\label{eqn:ampsq:vector:Dirac:u}
\overline{ \sum |{\cal M}|_{\chi \nu}^2} &= 
\frac{2g^4}{(u-m_{Z'}^2)^2} \left[ (u-m_\chi^2)^2 + \frac{m_\chi^4 t}{m_{Z'}^2} + \frac{m_\chi^4 (s-m_\chi^2)^2}{4m_{Z'}^4} \right]  \,, \\
 \label{eqn:ampsq:vector:Dirac:s}
\overline{ \sum |{\cal M}|_{\chi \bar\nu}^2} &= 
\frac{2g^4}{(s-m_{Z'}^2)^2 + m_{Z'}^2 \Gamma_{Z'}^2} \left[ (s-m_\chi^2)^2 + \frac{m_\chi^4 t}{m_{Z'}^2} + \frac{m_\chi^4 (u-m_\chi^2)^2}{4m_{Z'}^4} \right]  \,.
\end{align}
\end{subequations}
For the case of Majorana fermion DM, there are both the $s$- and $u$-channel contributions, and the amplitude square is
\begin{eqnarray}
\label{eqn:ampsq:vector:Majorana:su}
\overline{ \sum |{\cal M}|_{}^2} &=& 2g^4 \left\{
\frac{1}{(s-m_{Z'}^2)^2} \left[ (s-m_\chi^2)^2 + \frac{m_\chi^4 t}{m_{Z'}^2} + \frac{m_\chi^4 (u-m_\chi^2)^2}{4m_{Z'}^4} \right] \right. \nonumber \\
&& + \frac{1}{(u-m_{Z'}^2)^2} \left[ (u-m_\chi^2)^2 + \frac{m_\chi^4 t}{m_{Z'}^2} + \frac{m_\chi^4 (s-m_\chi^2)^2}{4m_{Z'}^4} \right] \nonumber \\
&& + \frac{2(s-m_{Z'}^2)}{\left[ (s-m_{Z'}^2)^2 + m_{Z'}^2 \Gamma_{Z'}^2 \right] (u-m_{Z'}^2)} \Bigg[ -2 (u-m_\chi^2)^2 \nonumber \\
&& \left. + \frac{(m_\chi^2-u)(-2 m_\chi^4 + m_\chi^2 s + su)}{m_{Z'}^2} + \frac{m_\chi^4 (m_\chi^4-su)}{2m_{Z'}^4} \Bigg] \right\} \,,
\end{eqnarray}
where the last term is the cross term between the $s$- and $u$-channel diagrams.

For the case of real and complex vector DM with couplings in Eqs.~(\ref{eqn:L:vector:vector_real:t}) and (\ref{eqn:L:vector:vector_complex:t}), the corresponding amplitude squares are, respectively,
\begin{subequations}
\begin{align}
\label{eqn:ampsq:vector:vector_real:t}
\overline{ \sum |{\cal M}|_{\rm r}^2} &= 
\frac{ g_\nu^2 g_\chi^2}{3(t-m_{Z'}^2)^2} \left[ 
2m_\chi^2 t
- \frac{t(s^2-t^2+u^2)}{m_{\chi}^2} - \frac{t^2 (m_\chi^2-s)(m_\chi^2-u)}{m_{\chi}^4} \right] \,, \\
\label{eqn:ampsq:vector:vector_complex:t}
\overline{ \sum |{\cal M}|_{\rm c}^2} &= 
\frac{g_\nu^2 g_\chi^2}{3(t-m_{Z'}^2)^2} \left[ 
4 (3m_\chi^4 - 3su - t^2 )
+ \frac{t(2t^2-(s-u)^2)}{m_{\chi}^2} + \frac{t^2 (m_\chi^2-s)(u-m_\chi^2)}{m_{\chi}^4} \right] \,. 
\end{align}
\end{subequations}

\section{DM self-scattering cross sections}
\label{app:dmself}

This Appendix is for the DM self-interaction cross sections, which are relevant to the limits of bullet clusters on SIDM in Section~\ref{subsection:bulletcluster}. 
For the scattering of DM particles ${\rm DM} + {\rm DM} \, (\overline{\rm DM}) \to {\rm DM} + {\rm DM} \,(\overline{\rm DM})$, the Mandelstam variables are
\begin{subequations}
\begin{align}
s & =  4 m^2 (1+v^2+v^4) \,, \\
t & =  -2 m^2 (v^2+v^4) (1-\cos\theta) \,, \\
u & =  -2 m^2 (v^2+v^4) (1+\cos\theta) \,,     
\end{align}
\end{subequations}
with $m$ and $v$ the mass and velocity of DM, respectively, and $\cos\theta$ the DM scattering angle. Then, in the non-relativistic limit, the cross section times DM velocity is
\begin{eqnarray}
\sigma v = \frac{1}{16\pi s} \overline{\sum |{\cal M}|^2} \,.
\end{eqnarray}
The widths of mediators can be found in Appendix~\ref{appendix:width}, whenever needed.

Let us first consider the case of scalar DM. For the scattering of real DM $\chi \chi \to \chi \chi$ mediated by a scalar mediator $\phi$ with the Lagrangian ${\cal L} = - \mu \phi \chi \chi/2$: 
\begin{eqnarray}
\sigma v (\chi \chi) &\simeq& \frac{\mu^4}{128\pi m^2 m_\phi^4} \frac{(8m^2-3m_\phi^2)^2}{(4m^2-m_\phi^2)^2 + m_\phi^2 \Gamma_\phi^2} \,.
\end{eqnarray}
For the case of complex scalar DM $\chi \chi^\dagger \to \chi \chi^\dagger$ with a scalar mediator, with the coupling in Eq.~(\ref{eqn:L:scalar:scalar}),
\begin{eqnarray}
\sigma v (\chi \chi^\dagger) &\simeq& \frac{\mu^4}{16\pi m^2 m_\phi^4} \frac{(2m^2-m_\phi^2)^2}{(4m^2-m_\phi^2)^2 + m_\phi^2 \Gamma_\phi^2} \,.
\end{eqnarray}
For the case of a vector mediator $V^\mu$ with coupling in Eq.~(\ref{eqn:L:vector:scalar_complex:t}),
\begin{eqnarray}
\sigma v (\chi \chi^\dagger) &\simeq& \frac{g_\chi^2 m^2}{4\pi m_V^4} \,.
\end{eqnarray}

For the scattering of complex scalar DM particles $\phi \phi \to \phi \phi$ applying to asymmetric DM cases, with a scalar or vector mediator, the cross sections are, respectively,
\begin{eqnarray}
\sigma v_\phi (\chi\chi) &\simeq& \frac{\mu^4}{16\pi m^2 m_\phi^4} \,, \\
\sigma v_V (\phi\phi) &\simeq& \frac{g_\chi^2 m^2}{\pi m_V^4} \,.
\end{eqnarray}

For the case of Dirac fermion DM scattering $\chi \bar\chi \to \chi \bar\chi$ mediated by a scalar $\phi$ with the couplings in Eq.~(\ref{eqn:L:scalar:Dirac:t}), we have 
\begin{subequations}
\begin{align}
\sigma v_s (\chi\bar\chi) &\simeq \frac{g_{\chi\,s}^4 m^2}{4\pi m_\phi^4} \,, \\
\sigma v_p (\chi\bar\chi) &\simeq \frac{g_{\chi\,p}^4 m^2}{4\pi \left[ (4m^2-m_\phi^2)^2 + m_\phi^2 \Gamma_\phi^2 \right]} \,.
\end{align}
\end{subequations}
For the vector mediator case, the couplings are given in Eq.~(\ref{eqn:L:vector:Dirac:t}). For the left-handed, right-handed, vector, and axial-vector couplings,
\begin{subequations}
\begin{align}
\sigma v_{L,\,R} (\chi\bar\chi) &\simeq \frac{g_{\chi L,\,\chi R}^4 m^2}{4\pi m_V^4} \frac{7m^4 - 5m^2 m_V^2 + m_V^4}{(4m^2-m_V^2)^2 + m_V^2 \Gamma_V^2} \,, \\
\sigma v_V (\chi \bar\chi) &\simeq \frac{g_{\chi V}^4 m^2}{4\pi m_V^4} \frac{16m^4 - 20m^2 m_V^2 + 7m_V^4}{(4m^2-m_V^2)^2 + m_V^2 \Gamma_V^2} \,, \\
\sigma v_A (\chi \bar\chi) &\simeq \frac{7g_{\chi A}^4 m^2}{4\pi m_V^4}  \,.
\end{align}
\end{subequations}
For the case of Dirac fermion DM-DM scattering $\chi \chi \to \chi \chi$, which applies to asymmetric DM scenarios,
\begin{subequations}
\begin{align}
\sigma v_s (\chi\chi) &\simeq \frac{g_{\chi\,s}^4 m^2}{4\pi m_\phi^4} \,, \\
\sigma v_p (\chi\chi) &\simeq \frac{5g_{\chi\,p}^4 m^2 v^4}{24\pi m_\phi^4} \,, \\
\sigma v_{L,\,R,\, V} (\chi\chi) &\simeq \frac{g_{\chi L,\, \chi R,\, \chi V}^4 m^2}{4\pi m_V^4}  \,, \\
\sigma v_A (\chi\chi) &\simeq \frac{9g_{\chi A}^4 m^2}{4\pi m_V^4}  \,.
\end{align}
\end{subequations}

For the scattering of real vector DM $\chi \chi \to \chi \chi$ mediated by a scalar mediator $\phi$ or a vector mediator $V^\mu$, with couplings given in Eqs.~(\ref{eqn:L:scalar:vector}) and (\ref{eqn:L:vector:vector_real:t}), respectively,  
\begin{eqnarray}
\sigma v_\phi (\chi\chi) &\simeq& \frac{\mu^4}{384\pi m^2 m_\phi^4} \frac{128 m^4 - 80 m^2 m_\phi^2 + 15 m_\phi^2}{(4m^2-m_\phi^2)^2 + m_\phi^2 \Gamma_\phi^2} \,, \\
\sigma v_V (\chi\chi) &\simeq& \frac{g_\chi^4 m^2}{3\pi m_V^4} \,.
\end{eqnarray}
For the case of a complex vector DM, with couplings in Eqs~(\ref{eqn:L:scalar:vector}) and (\ref{eqn:L:vector:vector_complex:t}), the cross sections for the scattering $\chi \chi^\dagger \to \chi\chi^\dagger$ are, respectively,
\begin{eqnarray}
\sigma v_\phi (\chi\chi^\dagger) &\simeq& \frac{\mu^4}{24\pi m^2 m_\phi^4} \frac{6 m^4 - 4 m^2 m_\phi^2 + m_\phi^4}{(4m^2-m_\phi^2)^2 + m_\phi^2 \Gamma_\phi^2} \,, \\
\sigma v_V (\chi\chi^\dagger) &\simeq& \frac{g_\chi^4 m^2}{3\pi m_V^4} \,.
\end{eqnarray}
For the scattering $\chi \chi \to \chi \chi$,
\begin{eqnarray}
\sigma v_\phi (\chi\chi) &\simeq& \frac{\mu^4}{24\pi m^2 m_\phi^4}  \,, \\
\sigma v_V (\chi\chi) &\simeq& \frac{2g_\chi^4 m^2}{3\pi m_V^4} \,.
\end{eqnarray}

\section{Neutrino self-scattering cross sections}
\label{Appendix:xs:nu-nu}

This Appendix is on the cross sections of $\nu$-$\nu$ self-scattering, which are relevant to the astrophysical and cosmological limits on neutrino self-interactions in Section~\ref{sec:bounds}, e.g. those from SN1987A, IceCube data, and CMB. Although the calculations in this Appendix are not used for our limits in Section~\ref{sec:bounds}, they may be of interest for future studies or to readers exploring related contexts.

For Dirac neutrinos with a scalar mediator $\phi$, the couplings can be written in the form of 
\begin{eqnarray}
{\cal L} = - g_{ij} \phi \bar\nu_i \nu_j \,,
\end{eqnarray}
with $i,\,j$ the mass indices. Following Ref.~\cite{Esteban:2021tub}, for the scattering $\nu_i + \bar\nu_j \to \nu_k + \bar\nu_l$ and $\nu_i + \nu_j \to \nu_k + \nu_l$, the differential cross section is
\begin{eqnarray}
\frac{d\sigma}{dt} = \left( 1-\frac12 \delta_{kl} \right) \frac{1}{16\pi s^2} \overline{\sum |{\cal M}|^2} \,,
\end{eqnarray}
where $\delta_{kl} = 1 (0)$ for $k=l$ ($k\neq l$) is the symmetry factor for identical particles in the final state, and $t \in [-s, 0]$ for relativistic neutrinos. For concreteness, we have neglected neutrino masses. For the $\nu - \bar\nu$ and $\nu - \nu$ scatterings, the total cross sections are, respectively, 
\begin{subequations}
\begin{align}
\sigma_{ij}^{\rm D} (\nu\bar\nu, s) \ = \ & \frac{1}{16\pi} \sum_{k,\,l}
\left[ g_{ij}^2 g_{kl}^2 \frac{s}{(s-m_\phi^2)^2 + m_\phi^2 \Gamma_\phi^2} 
+ g_{ik}^2 g_{jl}^2 \left( \frac{s+2m_\phi^2} {s(s+m_\phi^2)} + \frac{2m_\phi^2}{s^2} \log \frac{m_\phi^2}{s+m_\phi^2} \right) \right. \nonumber \\ & \left. + g_{ij} g_{kl} g_{ik} g_{jl} \left( 1 - \frac{m_\phi^2}{s} \right) \frac{s + m_\phi^2 \log \frac{m_\phi^2}{s+m_\phi^2}}{(s-m_\phi^2)^2 + m_\phi^2 \Gamma_\phi^2} \right] \,, \\
\sigma_{ij}^{\rm D} (\nu\nu, s) \ = \ & \frac{1}{16\pi} \sum_{k,\,l}
\left( 1-\frac12 \delta_{kl} \right) \left[  \left( g_{ik}^2 g_{jl}^2 + g_{il}^2 g_{jk}^2 \right) \left( \frac{s+2m_\phi^2} {s(s+m_\phi^2)} + \frac{2m_\phi^2}{s^2} \log \frac{m_\phi^2}{s+m_\phi^2} \right) \right. \nonumber \\ 
& \left. + g_{ik} g_{jl} g_{il} g_{jk} \left( \frac{1}{s} + \frac{2m_\phi^2 (s+m_\phi^2)}{s^2 (s+2m_\phi^2)} \log \frac{m_\phi^2}{s+m_\phi^2} \right)  \right] \,,
\end{align}
\end{subequations}
where $\Gamma_\phi$ is the width of scalar $\phi$ [cf. Eq.~(\ref{eqn:width:phi:nu})]. For Majorana neutrinos,
\begin{eqnarray}
\sigma_{ij}^{\rm M} (s) &=& \frac{1}{16\pi} \sum_{k,\,l}
\left( 1-\frac12 \delta_{kl} \right) \left[ g_{ij}^2 g_{kl}^2 \frac{s}{(s-m_\phi^2)^2 + m_\phi^2 \Gamma_\phi^2} \right. \nonumber \\
&& + \left( g_{ik}^2 g_{jl}^2 + g_{il}^2 g_{kl}^2 \right) \left( \frac{s+2m_\phi^2} {s(s+m_\phi^2)} + \frac{2m_\phi^2}{s^2} \log \frac{m_\phi^2}{s+m_\phi^2} \right) \nonumber \\ 
&& + g_{ij} g_{kl} \left( g_{ik} g_{jl} + g_{il} g_{jk} \right) \left( 1 - \frac{m_\phi^2}{s} \right) \frac{s + m_\phi^2 \log \frac{m_\phi^2}{s+m_\phi^2}}{(s-m_\phi^2)^2 + m_\phi^2 \Gamma_\phi^2} \nonumber \\
&&\left. + g_{ik} g_{jl} g_{il} g_{jk} \left( \frac{1}{s} + \frac{2m_\phi^2 (s+m_\phi^2)}{s^2 (s+2m_\phi^2)} \log \frac{m_\phi^2}{s+m_\phi^2} \right)  \right] \,.
\end{eqnarray}
If the couplings are of the form $g \phi \bar\nu \gamma_5 \nu$, the results are the same.

For the case of a vector mediator $V^\mu$, if the couplings are of the form of $g_{ij} V^\mu \bar\nu_i \gamma_\mu P_{L,\,R} \nu_j$, the cross sections are, for the case of Dirac neutrinos,
\begin{subequations}
\begin{align}
\sigma_{ij}^{} (\nu\bar\nu, s) \ = \ & \frac{1}{16\pi} \sum_{k,\,l}
\left[ \frac13 g_{ij}^2 g_{kl}^2 \frac{s}{(s-m_V^2)^2 + m_V^2 \Gamma_V^2} \right. \nonumber \\
&  + g_{ik}^2 g_{jl}^2 \left( \frac{1}{s} \left( 2 + \frac{s}{m_V^2} \right) + \frac{2}{s} \left( 1 + \frac{m_V^2}{s} \right) \log \frac{m_V^2}{s+m_V^2} \right) \nonumber \\ 
& \left. + g_{ij} g_{kl} g_{ik} g_{jl} \left( 3 + \frac{2m_V^2}{s} + 2\left( 1 + \frac{m_V^2}{s} \right)^2 \log \frac{m_V^2}{s+ m_V^2} \right) \frac{s-m_V^2}{(s-m_V^2)^2 + m_V^2 \Gamma_V^2} \right] \,, \\ 
\sigma_{ij}^{} (\nu\nu, s) \ = \ & \frac{1}{16\pi} \sum_{k,\,l}
\left( 1-\frac12 \delta_{kl} \right) \left[  \left( g_{ik}^2 g_{jl}^2 + g_{il}^2 g_{jk}^2 \right)  \frac{s} {m_V^2(s+m_V^2)}  \right. \nonumber \\ 
& \left. - g_{ik} g_{jl} g_{il} g_{jk} \left( \frac{4}{s+2m_V^2} \log \frac{m_V^2}{s+m_V^2} \right)  \right] \,.
\end{align}
\end{subequations}
For the vector couplings of the form $g_{ij} V^\mu \bar\nu_i \gamma_\mu \nu_j$ or axial-vector couplings $g_{ij} V^\mu \bar\nu_i \gamma_\mu \gamma_5 \nu_j$, 
\begin{subequations}
\begin{align}
\sigma_{ij}^{} (\nu\bar\nu, s) \ = \ & \frac{1}{16\pi} \sum_{k,\,l}
\left[ \frac43 g_{ij}^2 g_{kl}^2 \frac{s}{(s-m_V^2)^2 + m_V^2 \Gamma_V^2} \right. \nonumber \\
&  + g_{ik}^2 g_{jl}^2 \left( \frac{2(2m_V^4+3sm_V^2+2s^2)}{sm_V^2(s+m_V^2)} + \frac{4}{s} \left( 1 + \frac{m_V^2}{s} \right) \log \frac{m_V^2}{s+m_V^2} \right) \nonumber \\ 
& + 2 g_{ij} g_{kl} g_{ik} g_{jl} \left( 3 + \frac{2m_V^2}{s} + 2\left( 1 + \frac{m_V^2}{s} \right)^2 \log \frac{m_V^2}{s+ m_V^2} \right) \nonumber \\
& \left. \times \frac{s-m_V^2}{(s-m_V^2)^2 + m_V^2 \Gamma_V^2} \right] \,, \\
\sigma_{ij}^{} (\nu\nu, s) \ = \ & \frac{1}{16\pi} \sum_{k,\,l} \left( 1-\frac12 \delta_{kl} \right)
\left[ g_{ik}^2 g_{jl}^2 \left( \frac{2(2m_V^4+3sm_V^2+2s^2)}{sm_V^2(s+m_V^2)} \right. \right. \nonumber \\
& \left. \left. + \frac{4}{s} \left( 1 + \frac{m_V^2}{s} \right) \log \frac{m_V^2}{s+m_V^2} \right) \right. \nonumber \\ 
& + g_{il}^2 g_{jk}^2 \left( \frac{2}{s} \left( \frac{m_V^2}{s+m_V^2} + \frac{2s}{m_V^2} + 1 \right) + \frac{4}{s} \left( 1 + \frac{m_V^2}{s} \right) \log \frac{m_V^2}{s+m_V^2} \right) \nonumber \\
& \left. - g_{ik} g_{jl} g_{il} g_{jk} \left( \frac{8}{s+2m_V^2} \log \frac{m_V^2}{s+m_V^2} \right)  \right] \,.
\end{align}
\end{subequations}

\section{Annihilation cross sections of DM into neutrinos}
\label{app:thermalrelic}

In this Appendix, we collect all the annihilation cross sections of thermal DM particles into neutrinos, which are relevant to the relic density of DM particles in Section~\ref{subsubsec:thermalrelic}.  The relevant widths of the mediators can be found in Appendix~\ref{appendix:width}.

For the case of scalar mediator $\phi$, if DM is a scalar or vector particle, with couplings respectively in the form of Eqs.~(\ref{eqn:L:scalar:scalar}) and (\ref{eqn:L:scalar:vector}), the leading-order annihilation cross sections are respectively 
\begin{eqnarray}
\sigma v_{\text{S}} &=& \frac{\left( g_{\nu \, s}^2 + g_{\nu\, p}^2 \right) \mu^2}{4\pi} \frac{1}{(s-m_\phi^2)^2 + m_\phi^2 \Gamma_\phi^2} \nonumber \\
&\simeq& \frac{\left( g_{\nu \, s}^2 + g_{\nu\, p}^2 \right) \mu^2}{4\pi} \frac{1}{(4m^2-m_\phi^2)^2 + m_\phi^2 \Gamma_\phi^2} \,, \\
\sigma v_{\text{V}} &=& \frac{ \left( g_{\nu \, s}^2 + g_{\nu \, p}^2 \right) \mu^2 }{144\pi \xi^2} \frac{1 - 4 \xi + 12 \xi^2}{(s-m_\phi^2)^2 + m_\phi^2 \Gamma_\phi^2} \nonumber \\
&\simeq& \frac{ \left( g_{\nu \, s}^2 + g_{\nu \, p}^2 \right) \mu^2 }{12\pi} \frac{1}{(4m^2-m_\phi^2)^2 + m_\phi^2 \Gamma_\phi^2} \,.
\end{eqnarray}
We have defined the dimensionless parameter $\xi \equiv m^2/s$. 
For Dirac and Majorana fermion DM with couplings in Eq.~(\ref{eqn:L:scalar:Dirac}), the annihilation cross sections are, respectively,
\begin{subequations}
\begin{align}
\sigma v_{\rm D} &= \frac{ \left( g_{s}^2 + g_{p}^2 \right)^2 }{32\pi s} \left[ \frac{\eta + 2(\xi-\eta)^2}{\eta + (\xi-\eta)^2} + \frac{4(\xi-\eta)}{\sqrt{1-4\xi}} {\rm arctanh} \frac{\sqrt{1-4\xi}}{1-2(\xi-\eta)} \right] \nonumber \\
&\simeq \frac{ \left( g_{s}^2 + g_{p}^2 \right)^2 }{32\pi} \frac{m^2}{(m^2+m_\phi^2)^2} \,, \\
\sigma v_{\rm M} &= \frac{ \left( g_{s}^2 + g_{p}^2 \right)^2 }{16\pi s} \left[ 2 + \frac{(\xi-\eta)^2}{\eta + (\xi-\eta)^2} - \frac{4 \left(3 (\xi -\eta )^2+2 \eta -\xi \right)}{\sqrt{1-4 \xi } (1-2(\xi-\eta))} {\rm arctanh} \frac{\sqrt{1-4\xi}}{1-2(\xi-\eta)} \right] \nonumber \\
&\simeq \frac{ \left( g_{s}^2 + g_{p}^2 \right)^2 }{16\pi} \frac{m^2}{(m^2+m_\phi^2)^2} \,,
\end{align} 
\end{subequations}
where the dimensionless parameter $\eta \equiv m_{\rm med}^2/s = m_\phi^2/s$.
If the couplings are in the form of Eq.~(\ref{eqn:L:scalar:Dirac:t}), then
\begin{eqnarray}
\sigma v &=& \frac{ g_{\nu \, s}^2 + g_{\nu\, p}^2 }{8\pi} \frac{s \left[ g_{\chi\,s}^2 (1-4\xi) + g_{\chi\,p}^2 \right]}{(s-m_\phi^2)^2 + m_\phi^2 \Gamma_\phi^2} \nonumber \\
&\simeq& \frac{ g_{\nu \, s}^2 + g_{\nu\, p}^2 }{2\pi} \left( g_{\chi\,p}^2 + \frac{1}{4} g_{\chi\,s}^2 v^2 \right) \frac{m^2}{(4m^2-m_\phi^2)^2 + m_\phi^2 \Gamma_\phi^2} \,.
\end{eqnarray}

For the case of fermion mediator $N$, if the DM particle is a real or complex scalar $\phi$ with coupling in Eq.~(\ref{eqn:L:fermion:scalar_complex}), the annihilation cross sections are, respectively, 
\begin{subequations}
\begin{align}
\sigma v_{\rm r} &= \frac{ \left( g_{s}^2 + g_{p}^2 \right)^2}{4\pi s} \left[ -3 + \frac{2 \left(1-2 (2\xi-3\eta) + 6 (\xi-\eta)^2 \right)}{\sqrt{1-4 \xi } (1-2 (\xi-\eta))} {\rm arctanh} \frac{\sqrt{1-4 \xi }}{1-2 (\xi-\eta)} \right] \nonumber \\
&\simeq \frac{ \left( g_{s}^2 + g_{p}^2 \right)^2 v^4 }{60\pi} \frac{m^6}{(m^2 + m_N^2)^4} \,, \\
\sigma v_{\rm c} &= \frac{ \left( g_{s}^2 + g_{p}^2 \right)^2 }{4\pi s} \left[ -1 + \frac{1-2 (\xi-\eta)}{\sqrt{1-4 \xi }} {\rm arctanh} \frac{\sqrt{1-4 \xi }}{1-2 (\xi-\eta)} \right]  \nonumber \\
&\simeq \frac{ \left( g_{s}^2 + g_{p}^2 \right)^2 v^2 }{48\pi} \frac{m^2}{(m^2 + m_N^2)^2} \,.
\end{align} 
\end{subequations}
For the case of real scalar DM, there are both $t$- and $u$-channel contributions, and they cancel out at the order of $v^2$. For the case of vector DM with coupling in Eq.~(\ref{eqn:L:fermion:vector}), the cross section is 
\begin{eqnarray}
\sigma v &=& \frac{g^4}{36\pi \xi^2 s} \left[ -8 \xi ^2 +4 \xi \eta -3 \eta ^2  -\frac{4 \xi \eta  (\xi-\eta)^2}{\eta + (\xi-\eta)^2} \right. \nonumber \\
&& \left. + \frac{2 \left( 4 \xi ^2+\eta ^2 + 2 \eta  \left(8 \xi^2 -6 \eta  \xi + 3 \eta ^2 \right)+2 (\xi-\eta)^2 \left(8
   \xi^2+3 \eta ^2\right)\right)}{\sqrt{1-4 \xi } (1-2 (\xi-\eta))} {\rm arctanh} \frac{\sqrt{1-4 \xi }}{1-2 (\xi-\eta)}  \right] \nonumber \\
&\simeq& \frac{2g_{}^4 }{9\pi} \frac{m^2}{(m^2 + m_N^2)^2} \,.
\end{eqnarray}

For the case of vector mediator, if the DM particle is a complex scalar with coupling in Eq.~(\ref{eqn:L:vector:scalar_complex:t}), the annihilation cross section is 
\begin{eqnarray}
\sigma v &=&  \frac{g_{\nu}^2 g_\chi^2 }{12\pi} \frac{s(1-4\xi)}{(s - m_{Z'}^2)^2 + m_{Z'}^2 \Gamma_{Z'}^2} \nonumber \\ 
&\simeq& \frac{g_{\nu}^2 g_\chi^2 v^2 }{12\pi} \frac{m^2}{(4m^2 - m_{Z'}^2)^2 + m_{Z'}^2 \Gamma_{Z'}^2} \,.
\end{eqnarray}
For the Dirac and Majorana fermion DM with couplings in Eqs.~(\ref{eqn:L:vector:Dirac:t}) and (\ref{eqn:L:vector:Majorana:t}), the cross sections are, respectively, 
\begin{subequations}
\begin{align}
\sigma v_{\rm D} &= \frac{g_{\nu}^2}{24\pi} \frac{ s \left[ ( g_{\chi L}^2 + g_{\chi R}^2) (1-\xi) + 6 g_{\chi L} g_{\chi R} \xi \right]}{(s - m_{Z'}^2)^2 + m_{Z'}^2 \Gamma_{Z'}^2} \nonumber \\
&\simeq \frac{g_{\nu}^2 (g_{\chi L} + g_{\chi R})^2}{8\pi} \frac{m^2}{(4m^2 - m_{Z'}^2)^2 + m_{Z'}^2 \Gamma_{Z'}^2} \,, \\
\sigma v_{\rm M} &=  \frac{g_{\nu}^2 g_\chi^2 }{12\pi} \frac{s(1-4\xi)}{(s - m_{Z'}^2)^2 + m_{Z'}^2 \Gamma_{Z'}^2} \nonumber \\ 
&\simeq \frac{g_{\nu}^2 g_\chi^2 v^2}{12\pi} \frac{m^2}{(4m^2 - m_{Z'}^2)^2 + m_{Z'}^2 \Gamma_{Z'}^2} \,.
\end{align} 
\end{subequations}
If the couplings of fermion DM are in the form of Eq.~(\ref{eqn:L:vector:Dirac:u}), the cross sections for Dirac and Majorana fermion DM particles are, respectively,
\begin{subequations}
\begin{align}
\sigma v_{\rm D} =& \ \frac{g_{}^4}{32\pi\eta^2 s} \left[ \frac{ 2 \xi ^4 +\eta \xi ^2  (5-4 \xi ) +2 \eta ^2 \left(2-4 \xi +5 \xi ^2\right) +4 \eta ^3 (3-4 \xi)  + 8 \eta ^4 }{\eta + (\xi-\eta)^2} \right. \nonumber \\
& \left. + \frac{4 \left(\xi ^3 -\eta  \xi ^2 -4 \eta ^2 (1-\xi) -4 \eta ^3 \right)}{\sqrt{1-4 \xi }} {\rm arctanh} \frac{\sqrt{1-4
   \xi }}{1 - 2 (\xi-\eta)} \right] \nonumber \\
\simeq & \ \frac{g_{}^4 }{32\pi} \frac{m^2 (m^2 + 2m_{Z'}^2)^2}{m_{Z'}^4 (m^2 + m_{Z'}^2)^2} \,, \\
\sigma v_{\rm M} =& \frac{g^4}{16\pi \eta^2 s} \ \left[ (\xi+2 \eta)^2 + \frac{\xi^4 +2 \eta \xi^2 (2-\xi) +\eta ^2 \left(4-8\xi + 5 \xi^2\right) +8 \eta ^3 (1-\xi) + 4 \eta ^4}{\eta + (\xi-\eta)^2} \right. \nonumber \\
& \left. - \frac{4 \left( 2 \xi^4 +\eta \xi (2-3 \xi) +2 \eta^2 \left(2-2\xi + \xi^2\right) +12 \eta ^3 (1-\xi) + 8 \eta^4 \right)}{\sqrt{1-4 \xi } (1-2(\xi-\eta))} {\rm arctanh} \frac{\sqrt{1-4
   \xi }}{1 - 2 (\xi-\eta)} \right] \nonumber \\
\simeq & \ \frac{g_{}^4 v^2}{48\pi} \frac{m^2 (m^8 + 4 m^6 m_{Z'}^2 + 13 m^4 m_{Z'}^4 + 12 m^2m_{Z'}^6 + 4m_{Z'}^8)}{m_{Z'}^4 (m^2 + m_{Z'}^2)^4} \,.
\end{align} 
\end{subequations}
For the Dirac case, there are only $t$-channel diagrams, while for the Majorana fermion DM there are both $t$- and $u$-channel contributions, which cancel out at the level of $v^0$. With a real or complex vector DM with couplings in Eqs.~(\ref{eqn:L:vector:vector_real:t}) and 
(\ref{eqn:L:vector:vector_complex:t}), the annihilation cross sections are, respectively, 
\begin{subequations}
\begin{align}
\sigma v_{\rm r} &= \frac{g_{\nu}^2 g_\chi^2 }{432\pi \xi^2} \frac{s (1-16\xi^2)}{(4m^2 - m_{Z'}^2)^2 + m_{Z'}^2 \Gamma_{Z'}^2} \nonumber \\
&\simeq \frac{2g_{\nu}^2 g_\chi^2 v^2 }{27\pi} \frac{m^2}{(4m^2 - m_{Z'}^2)^2 + m_{Z'}^2 \Gamma_{Z'}^2} \,, \\
\label{eqn:DM:annihilation:vector:vector:c}
\sigma v_{\rm c} &= \frac{g_{\nu}^2 g_\chi^2 }{432\pi \xi^2} \frac{s (1-4\xi) (1-4\xi + 12\xi^2)}{(4m^2 - m_{Z'}^2)^2 + m_{Z'}^2 \Gamma_{Z'}^2} \nonumber \\
&\simeq \frac{g_{\nu}^2 g_\chi^2 v^2 }{36\pi} \frac{m^2}{(4m^2 - m_{Z'}^2)^2 + m_{Z'}^2 \Gamma_{Z'}^2} \,.
\end{align} 
\end{subequations}

\section{Annihilation cross sections of neutrinos into DM}
\label{app:nu2DM}

In this Appendix, we list all the cross sections of neutrino-antineutrino scattering into DM particles, which are relevant to the BBN limits in Section~\ref{subsubsec:BBN}. All relevant widths of the mediators can be found in Appendix~\ref{appendix:width}.

Let us first consider the cases with a scalar mediator $\phi$. If DM is a complex scalar or vector, with couplings respectively in the form of Eqs.~(\ref{eqn:L:scalar:scalar}) and (\ref{eqn:L:scalar:vector}), the cross sections for the scattering process $\nu \bar\nu \to \chi \chi^\dagger$ are respectively 
\begin{eqnarray}
\sigma_{\text{S}} &=& \frac{\mu^2 (g_{\nu \, s}^2 + g_{\nu \, p}^2)}{8\pi} \frac{\sqrt{1-4\xi}}{(s-m_\phi^2)^2 + m_\phi^2 \Gamma_\phi^2} \,, \\
\sigma_{\text{V}} &=& \frac{\mu^2 (g_{\nu\,s}^2 + g_{\nu\,p}^2)}{32\pi \xi^2} \frac{1-4\xi + 12 \xi^2}{(s-m_\phi^2)^2 + m_\phi^2 \Gamma_\phi^2} \sqrt{1-4\xi} \,.
\end{eqnarray}
For the case of a Dirac or Majorana fermion DM with couplings in Eq.~(\ref{eqn:L:scalar:Dirac:t}), the neutrino-antineutrino annihilation process proceeds in the $s$-channel, and the corresponding cross section is
\begin{eqnarray}
\sigma_{\text{F}}^{(s)} &=& \frac{g_{\nu \, s}^2 + g_{\nu \, p}^2}{4\pi} \frac{s\left( g_{\chi \, s}^2 (1-4\xi) + g_{\chi \, p}^2 \right)}{(s-m_\phi^2)^2 + m_\phi^2 \Gamma_\phi^2} \sqrt{1-4\xi} \,.
\end{eqnarray}
If the couplings are in the form of Eq.~(\ref{eqn:L:scalar:Dirac}), the processes are, respectively, in the $t$-channel for Dirac fermion DM, and $t$- and $u$-channels for the Majorana fermion DM, with the cross sections
\begin{subequations}
\begin{align}
\sigma_{\text{D}}^{(t)} =& \ \frac{(g_{s}^2 + g_{p}^2)^2}{16\pi s} \left[ \frac{\sqrt{1-4 \xi } \left(\eta + 2 (\xi -\eta )^2 \right)}{\eta + (\xi -\eta )^2} + 4 (\xi - \eta) {\rm arctanh} \frac{\sqrt{1-4 \xi }}{1-2 (\xi-\eta)} \right] \,, \\
\sigma_{\text{M}}^{(tu)} =& \ \frac{(g_{s}^2 + g_{p}^2)^2}{8\pi s} \left[ \sqrt{1-4\xi} \left( 2 + \frac{(\xi -\eta )^2}{(\xi -\eta )^2+\eta } \right) \right. \nonumber \\ 
& \left. + \frac{4 \left(2 \eta -\xi + 3 (\xi -\eta )^2 \right)}{1 - 2 (\xi-\eta)} {\rm arctanh} \frac{\sqrt{1-4 \xi }}{1+2 (\xi-\eta)} \right]  \,.
\end{align}
\end{subequations}

Let us now move on to the fermion mediator cases. The couplings of real or complex scalar DM are given in Eq.~(\ref{eqn:L:fermion:scalar_complex}), and the resultant cross sections are, respectively,
\begin{subequations}
\begin{align}
\sigma_{\text{R}}^{(tu)} =& \ \frac{(g_{s}^2 + g_{p}^2)^2}{16\pi s (1-2(\xi-\eta))} \bigg[ -3 \sqrt{1-4 \xi } \Big( 1 - 2 (\xi-\eta) \Big) \nonumber \\ 
& + 2 \Big( (1-4\xi) + 6 \left(\eta + (\xi-\eta)^2 \right) \Big) {\rm arctanh} \frac{\sqrt{1-4 \xi }}{1-2 (\xi-\eta)} \bigg] \,, \\
\sigma_{\text{C}}^{(t)} =& \ \frac{(g_{s}^2 + g_{p}^2)^2}{8\pi s} \bigg[ -\sqrt{1-4 \xi }  + \Big( 1-2 (\xi-\eta) \Big) {\rm arctanh} \frac{\sqrt{1-4 \xi }}{1-2 (\xi-\eta)} \bigg] \,. 
\end{align}
\end{subequations}
For the case of a vector DM, with couplings given in Eq.~(\ref{eqn:L:fermion:vector}), 
\begin{eqnarray}
\sigma_{\text{V}}^{} &=& \frac{g^4}{8\pi \xi^2 s} \bigg[ - \sqrt{1-4 \xi } \left( 8 \xi ^2 -4 \eta  \xi + 3 \eta ^2  + \frac{4 \eta  \xi  (\xi -\eta )^2}{\eta + (\xi -\eta )^2} \right) \nonumber \\ 
&& + \frac{2 \left(4 \xi ^2 + \eta ^2 + 2 \eta  \left(8 \xi ^2 -6 \eta  \xi + 3 \eta ^2\right)+2 (\xi -\eta )^2 \left(8\xi ^2+3 \eta ^2\right) \right)}{1-2 (\xi- \eta)} {\rm arctanh} \frac{\sqrt{1-4 \xi }}{1-2 (\xi-\eta)} \bigg] \,. \nonumber \\ && 
\end{eqnarray}

For the case of vector mediator, for the complex scalar DM with coupling in Eq.~(\ref{eqn:L:vector:scalar_complex:t}), the neutrino-antineutrino annihilation cross section is 
\begin{eqnarray}
\sigma_{\text{S}} &=& \frac{g_{\nu}^2 g_\chi^2}{24\pi} \frac{s (1-4\xi)^{3/2}}{(s-m_V^2)^2 + m_V^2 \Gamma_V^2} \,.
\end{eqnarray}
For the case of Dirac and Majorana fermion DM with couplings in Eqs.~(\ref{eqn:L:vector:Dirac:t}) and (\ref{eqn:L:vector:Majorana:t}), the $\nu \bar\nu \to \chi \bar\chi$ process proceeds in the $s$-channel, and the corresponding cross sections are, respectively,
\begin{subequations}
\begin{align}
\label{eqn:xs:nunu_to_DMDM:vector:Dirac:s}
\sigma_{\text{D}}^{(s)} =& \ \frac{g_{\nu}^2}{12\pi} \frac{s \left[ (g_{\chi\,L}^2 + g_{\chi\, R}^2) (1-\xi) + 6 \xi g_{\chi\,L} g_{\chi\,R} \right]}{(s-m_V^2)^2 + m_V^2 \Gamma_V^2} \sqrt{1-4\xi} \,, \\
\sigma_{\text{M}}^{(s)} =& \ \frac{g_{\nu}^2 g_\chi^2}{6\pi} \frac{s (1-4\xi)^{3/2}}{(s-m_V^2)^2 + m_V^2 \Gamma_V^2} \,.
\end{align}
\end{subequations}
For the case of Dirac fermion DM with vector coupling to the vector mediator, i.g. $g_{\chi\,L} = g_{\chi\,R}$, our cross section in Eq.~(\ref{eqn:xs:nunu_to_DMDM:vector:Dirac:s}) is consistent with that in the equation below Eq.~(B2) of Ref.~\cite{Manzari:2023gkt}.
For couplings in the form of Eq.~(\ref{eqn:L:vector:Dirac:u}), the neutrino annihilation process are in the $t$-channel for Dirac fermion DM, and $t$- and $u$-channels for the Majorana type, and the corresponding cross sections turn out to be
\begin{subequations}
\begin{align}
\sigma_{\text{D}}^{(t)} =& \ \frac{g^4}{16\pi \eta^2 s}  
\bigg[ \frac{\sqrt{1-4 \xi}}{\eta + (\xi-\eta)^2} \Big( 2 \xi ^4 +\eta \xi^2 (5-4 \xi ) +2 \eta ^2 (2-\xi  (4-5 \xi))+ 4 \eta ^3 (3-4 \xi ) + 8 \eta ^4 \Big) \nonumber \\
& +4 \Big( \xi^3 - \eta \xi^2 - 4 \eta ^2 (1-\xi) -4 \eta ^3 \Big) {\rm arctanh} \frac{\sqrt{1-4 \xi}}{1-2 (\xi -\eta)} \bigg] \,, \\
\sigma_{\text{M}}^{(tu)} =& \ \frac{g^4}{8\pi \eta^2 s}  \bigg[ \frac{\sqrt{1-4 \xi }}{\eta + (\xi-\eta)^2} \Big( \xi^4 + 4 \eta \xi^2  (1-\xi) +\eta ^2 (4-\xi  (10-13 \xi)) +6 \eta ^3 (2-3 \xi ) + 8 \eta ^4 \Big) \nonumber \\
& + \frac{2 \left( \xi^3 (1-2 \xi) -2 \eta \xi^2 (3-4 \xi ) -2\eta ^2 \left(13 \xi^2-10\xi+4\right) - 12 \eta^3 (2-3\xi ) - 16 \eta^4  \right)}{1-2 (\xi-\eta)} \nonumber \\
& \times {\rm arctanh} \frac{\sqrt{1-4 \xi}}{1-2 (\xi -\eta)} \bigg] \,.
\end{align}
\end{subequations}
For the case of a real or complex vector DM with couplings in Eqs.~(\ref{eqn:L:vector:vector_real:t}) and 
(\ref{eqn:L:vector:vector_complex:t}), the corresponding cross sections are respectively
\begin{subequations}
\begin{align}
\sigma_{\rm R} =& \ \frac{g_\nu^2 g_\chi^2}{96\pi \xi^2} \frac{s (1+4\xi) (1-4\xi)^{3/2}}{(s - m_{V}^2)^2 + m_{V}^2 \Gamma_{V}^2} \,, \\
\label{eqn:xs:nunu_to_DMDM}
\sigma_{\rm C} =& \ \frac{g_\nu^2 g_\chi^2}{96\pi \xi^2} \frac{s(1-4\xi)^{3/2} (1 - 4\xi + 12 \xi^2)}{(s - m_{V}^2)^2 + m_{V}^2 \Gamma_{V}^2} \,.
\end{align}
\end{subequations}

\section{Mediator decay widths}
\label{appendix:width}

In this Appendix we collect all the decay widths of the mediators to neutrinos and DM particles, which are relevant to the calculations of scattering cross sections in Appendices~\ref{Appendix:xs:nuDM}, \ref{app:dmself}, \ref{Appendix:xs:nu-nu} and \ref{app:thermalrelic} above.

Let us start with the cases of mediator decaying into a $\nu \bar\nu$ pair. For scalar mediator $\phi$ or vector mediator $Z'$, with couplings to neutrinos in the form of Eqs.~(\ref{eqn:L:scalar:scalar}) and (\ref{eqn:L:vector:scalar_complex:t}), the partial widths are, respectively, 
\begin{eqnarray}
\label{eqn:width:phi:nu}
\Gamma (\phi \to \nu \bar\nu) &=& 3 \times \frac{ \left( g_{\nu\, s}^2 + g_{\nu\, p}^2 \right) m_\phi }{8\pi} \,, \\
\Gamma (Z' \to \nu \bar\nu) &=& 3 \times \frac{ g_\nu^2 m_{Z'} }{24\pi} \,,
\end{eqnarray}
where the factor of $3$ is for three generations of neutrinos, assuming flavor-universal couplings. 

We now consider the decays of the mediator to a pair of DM particles. For the coupling of scalar mediator $\phi$ to complex scalar $\chi$ in the form of Eq.~(\ref{eqn:L:scalar:scalar}), the partial width is 
\begin{equation}
\label{eqn:width:phi:DM}
\Gamma (\phi \to \chi \chi^\dagger) = \frac{\mu^2}{16\pi m_\phi} (1-4x)^{1/2} \,,
\end{equation}
where we have defined the dimensionless ratio $x \equiv m_{\rm DM}^2 / m_{\rm med}^2$. For the case of real scalar DM $\chi$ with the coupling $\mu \phi \chi\chi/2$, there is an extra symmetry factor of $1/2$ for identical particles in the final state. For the case of Dirac fermion DM $\chi$, with the couplings in Eq.~(\ref{eqn:L:scalar:Dirac:t}), the partial width is 
\begin{equation}
\Gamma (\phi \to \chi \chi^\dagger) = \frac{m_\phi}{8\pi} \left[ g_{\chi\,s}^2 \left( 1-4x \right)^{1/2} + g_{\chi\,p}^2 \left( 1-4x \right)^{3/2} \right]\,.
\end{equation}
For the case of complex vector DM $\chi$ with coupling in Eq~(\ref{eqn:L:scalar:vector}), the partial width is 
\begin{equation}
\Gamma (\phi \to \chi \chi^\dagger) = \frac{\mu^2}{64\pi x^2 m_\phi} (1-4x)^{1/2} (1-4x+12x^2) \,.
\end{equation}
If DM is a real vector, there is an extra factor of $1/2$.

For the case of vector mediator, if the DM is a complex scalar with coupling in Eq.~(\ref{eqn:L:vector:scalar_complex:t}), the partial width is 
\begin{equation}
\Gamma (V^\mu \to \chi \chi^\dagger) = \frac{g_\chi^2 m_V}{48\pi} (1-4x)^{3/2} \,.
\end{equation}
For Dirac and Majorana fermion DM with couplings in Eqs.~(\ref{eqn:L:vector:Dirac:t}) and (\ref{eqn:L:vector:Majorana:t}), the widths are, respectively,
\begin{subequations}
\begin{align}
\label{eqn:width:vector:Dirac}
\Gamma_{\rm D} (V^\mu \to \chi \bar\chi) &= \frac{m_{V}}{24\pi} \left[ (g_{\chi L}^2 + g_{\chi R}^2) (1-x) + 6 g_{\chi L} g_{\chi R} x \right] (1-4x)^{1/2} \,, \\
\Gamma_{\rm M} (V^\mu \to \chi \bar\chi) &= \frac{g_\chi^2 m_{V}}{12\pi} (1-4x)^{3/2} \,.
\end{align}
\end{subequations}
For a complex vector DM with coupling in Eq.~(\ref{eqn:L:vector:vector_complex:t}), the width is 
\begin{equation}
\label{eqn:width:MedV:DMV}
\Gamma ( V^\mu \to \chi \chi^\dagger) = \frac{g_\chi^2 m_V}{192\pi x^2} (1-4x)^{1/2} (1-8x+28x^2-48x^3) \,.
\end{equation}
For the case of a real vector, there is an extra factor of $1/2$.

Let us now consider the decays of the mediator particle into a neutrino and DM. For the case of a real scalar mediator and Dirac fermion DM with coupling in Eq.~(\ref{eqn:L:scalar:Dirac}), the partial width is 
\begin{equation}
\Gamma ( \phi \to \nu \bar\chi + \bar\nu \chi) = 3 \times \frac{(g_s^2+g_p^2) m_\phi}{8\pi} (1-x)^2 \,.
\end{equation}
Here we take into account both neutrinos and antineutrinos in the final state, and the factor of 3 is for three generations of neutrinos. For the case of Dirac fermion mediator $N$ and complex scalar DM $\phi$ or vector DM $Z'$ with couplings in Eqs.~(\ref{eqn:L:fermion:scalar_complex}) and (\ref{eqn:L:fermion:vector}), the widths are, respectively, 
\begin{eqnarray}
\Gamma ( N \to \nu \phi ) &=& 3\times \frac{ (g_s^2+g_p^2) m_N}{32\pi} (1-x) \,, \\
\Gamma ( N \to \nu Z' ) &=& 3\times \frac{g^2 m_N}{32\pi x} (1-x) \,.
\end{eqnarray} 
Given a vector mediator $Z'$ and Dirac fermion DM $\chi$ with coupling in Eq.~(\ref{eqn:L:vector:Dirac:u}), the partial width is 
\begin{equation}
\Gamma ( Z' \to \nu \bar\chi + \bar\nu \chi) = 3\times \frac{g^2 m_{Z'}}{24\pi} (1-x) (2-x) \,.
\end{equation}

\bibliography{ref}

@article{Agashe:2024owh,
    author = "Agashe, Kaustubh and Airen, Sagar and Franceschini, Roberto and Kim, Doojin and Kotwal, Ashutosh V. and Ricci, Lorenzo and Sathyan, Deepak",
    title = "{\textquotedblleft{}Unification\textquotedblright{} of BSM searches and SM measurements: the case of lepton+[inline-graphic not available: see fulltext] and m$_{W}$}",
    eprint = "2404.17574",
    archivePrefix = "arXiv",
    primaryClass = "hep-ph",
    doi = "10.1007/JHEP02(2025)139",
    journal = "JHEP",
    volume = "02",
    pages = "139",
    year = "2025"
}

@article{Sun:2025gyj,
    author = "Sun, Jun-Wei and Wu, Lei and Xu, Yan-Hao and Zhu, Bin",
    title = "{Probing supernova neutrino boosted dark matter with collective excitations}",
    eprint = "2501.07591",
    archivePrefix = "arXiv",
    primaryClass = "hep-ph",
    doi = "10.1103/xhl3-v17n",
    journal = "Phys. Rev. D",
    volume = "112",
    number = "1",
    pages = "015014",
    year = "2025"
}

@article{Arguelles:2019ouk,
    author = {Arg{\"u}elles, Carlos A. and Diaz, Alejandro and Kheirandish, Ali and Olivares-Del-Campo, Andr{\'e}s and Safa, Ibrahim and Vincent, Aaron C.},
    title = "{Dark matter annihilation to neutrinos}",
    eprint = "1912.09486",
    archivePrefix = "arXiv",
    primaryClass = "hep-ph",
    doi = "10.1103/RevModPhys.93.035007",
    journal = "Rev. Mod. Phys.",
    volume = "93",
    number = "3",
    pages = "035007",
    year = "2021"
}

@article{BetancourtKamenetskaia:2025rwk,
    author = "Betancourt Kamenetskaia, Boris and Park, Jong-Chul and Reichard, Merlin and Tomar, Gaurav",
    title = "{Investigating sub-MeV dark matter annihilation to neutrinos using direct detection experiments}",
    eprint = "2505.07403",
    archivePrefix = "arXiv",
    primaryClass = "hep-ph",
    doi = "10.1103/tzv8-53rz",
    journal = "Phys. Rev. D",
    volume = "112",
    number = "5",
    pages = "055019",
    year = "2025"
}

@article{Abdelrahim:2025fiz,
    author = "Abdelrahim, Amro E. B. and Batell, Brian and Berger, Joshua and McKeen, David and Shams Es Haghi, Barmak",
    title = "{Cosmological Histories in Neutrino Portal Dark Matter}",
    eprint = "2506.09137",
    archivePrefix = "arXiv",
    primaryClass = "hep-ph",
    reportNumber = "PITT-PACC-2502, UTWI-15-2025",
    month = "6",
    year = "2025"
}

@article{Pal:2024yom,
    author = "Pal, Sourav and Samanta, Rickmoy and Pal, Supratik",
    title = "{Exploring neutrino interactions in light of present and upcoming galaxy surveys}",
    eprint = "2409.03712",
    archivePrefix = "arXiv",
    primaryClass = "astro-ph.CO",
    doi = "10.1088/1475-7516/2025/03/047",
    journal = "JCAP",
    volume = "03",
    pages = "047",
    year = "2025"
}

@article{Xu:2023xva,
    author = "Xu, Xun-Jie and Zhou, Siyu and Zhu, Junyu",
    title = "{The $\nu_{R}$-philic scalar dark matter}",
    eprint = "2310.16346",
    archivePrefix = "arXiv",
    primaryClass = "hep-ph",
    doi = "10.1088/1475-7516/2024/04/012",
    journal = "JCAP",
    volume = "04",
    pages = "012",
    year = "2024"
}

@article{Wang:2023csv,
    author = "Wang, Isaac R. and Xu, Xun-Jie",
    title = "{Imprints of light dark matter on the evolution of cosmic neutrinos}",
    eprint = "2312.17151",
    archivePrefix = "arXiv",
    primaryClass = "hep-ph",
    reportNumber = "FERMILAB-PUB-23-0829-T",
    doi = "10.1088/1475-7516/2024/05/050",
    journal = "JCAP",
    volume = "05",
    pages = "050",
    year = "2024"
}

@article{Pandey:2018wvh,
    author = "Pandey, Sujata and Karmakar, Siddhartha and Rakshit, Subhendu",
    title = "{Interactions of astrophysical neutrinos with dark matter: a model building perspective}",
    eprint = "1810.04203",
    archivePrefix = "arXiv",
    primaryClass = "hep-ph",
    doi = "10.1007/JHEP11(2021)215",
    journal = "JHEP",
    volume = "01",
    pages = "095",
    year = "2019",
    note = "[Erratum: JHEP 11, 215 (2021)]"
}

@article{Lambiase:2025twn,
    author = "Lambiase, Gaetano and Poddar, Tanmay Kumar and Visinelli, Luca",
    title = "{Impact of the cosmic neutrino background on black hole superradiance}",
    eprint = "2503.02940",
    archivePrefix = "arXiv",
    primaryClass = "hep-ph",
    reportNumber = "CA21106; CA21136",
    doi = "10.1103/yfwv-37ss",
    journal = "Phys. Rev. D",
    volume = "112",
    number = "1",
    pages = "016010",
    year = "2025"
}

@article{Cheek:2025kks,
    author = "Cheek, Andrew and Visinelli, Luca and Zhang, Hong-Yi",
    title = "{Testing the Dark Origin of Neutrino Masses with Oscillation Experiments}",
    eprint = "2503.08439",
    archivePrefix = "arXiv",
    primaryClass = "hep-ph",
    doi = "10.1103/wyns-m4y5",
    journal = "Phys. Rev. Lett.",
    volume = "135",
    number = "3",
    pages = "031801",
    year = "2025"
}

@article{Berlin:2018ztp,
    author = "Berlin, Asher and Blinov, Nikita",
    title = "{Thermal neutrino portal to sub-MeV dark matter}",
    eprint = "1807.04282",
    archivePrefix = "arXiv",
    primaryClass = "hep-ph",
    reportNumber = "SLAC-PUB-17278",
    doi = "10.1103/PhysRevD.99.095030",
    journal = "Phys. Rev. D",
    volume = "99",
    number = "9",
    pages = "095030",
    year = "2019"
}

@article{Akita:2023yga,
    author = "Akita, Kensuke and Ando, Shin'ichiro",
    title = "{Constraints on dark matter-neutrino scattering from the Milky-Way satellites and subhalo modeling for dark acoustic oscillations}",
    eprint = "2305.01913",
    archivePrefix = "arXiv",
    primaryClass = "astro-ph.CO",
    reportNumber = "CTPU-PTC-23-16",
    doi = "10.1088/1475-7516/2023/11/037",
    journal = "JCAP",
    volume = "11",
    pages = "037",
    year = "2023"
}

@article{Wilkinson:2014ksa,
    author = "Wilkinson, Ryan J. and Boehm, Celine and Lesgourgues, Julien",
    title = "{Constraining Dark Matter-Neutrino Interactions using the CMB and Large-Scale Structure}",
    eprint = "1401.7597",
    archivePrefix = "arXiv",
    primaryClass = "astro-ph.CO",
    reportNumber = "IPPP-14-03, DCPT-14-06, CERN-PH-TH-2014-013, LAPTH-006-14",
    doi = "10.1088/1475-7516/2014/05/011",
    journal = "JCAP",
    volume = "05",
    pages = "011",
    year = "2014"
}

@article{Giare:2023qqn,
    author = "Giar{\`e}, William and G{\'o}mez-Valent, Adri{\`a} and Di Valentino, Eleonora and van de Bruck, Carsten",
    title = "{Hints of neutrino dark matter scattering in the CMB? Constraints from the marginalized and profile distributions}",
    eprint = "2311.09116",
    archivePrefix = "arXiv",
    primaryClass = "astro-ph.CO",
    doi = "10.1103/PhysRevD.109.063516",
    journal = "Phys. Rev. D",
    volume = "109",
    number = "6",
    pages = "063516",
    year = "2024"
}

@article{Crumrine:2024sdn,
    author = "Crumrine, Wendy and Nadler, Ethan O. and An, Rui and Gluscevic, Vera",
    title = "{Dark matter coupled to radiation: Limits from the Milky~Way satellites}",
    eprint = "2406.19458",
    archivePrefix = "arXiv",
    primaryClass = "astro-ph.CO",
    doi = "10.1103/PhysRevD.111.023530",
    journal = "Phys. Rev. D",
    volume = "111",
    number = "2",
    pages = "023530",
    year = "2025"
}

@article{Manzari:2023gkt,
    author = "Manzari, Claudio Andrea and Martin Camalich, Jorge and Spinner, Jonas and Ziegler, Robert",
    title = "{Supernova limits on muonic dark forces}",
    eprint = "2307.03143",
    archivePrefix = "arXiv",
    primaryClass = "hep-ph",
    reportNumber = "CERN-TH-2023-134",
    doi = "10.1103/PhysRevD.108.103020",
    journal = "Phys. Rev. D",
    volume = "108",
    number = "10",
    pages = "103020",
    year = "2023"
}

@article{Chauhan:2025hoz,
    author = "Chauhan, Garv and Gustafson, R. Andrew and Herrera, Gonzalo and Johnson, Taj and Shoemaker, Ian",
    title = "{The Dark Matter Diffused Supernova Neutrino Background}",
    eprint = "2505.03882",
    archivePrefix = "arXiv",
    primaryClass = "hep-ph",
    month = "5",
    year = "2025"
}

@article{Cappiello:2025tws,
    author = "Cappiello, Christopher V. and Dev, P. S. Bhupal and Patwardhan, Amol V.",
    title = "{New Supernova Constraints on Neutrinophilic Dark Sector}",
    eprint = "2503.09691",
    archivePrefix = "arXiv",
    primaryClass = "hep-ph",
    month = "3",
    year = "2025"
}

@article{Fujiwara:2023lsv,
    author = "Fujiwara, Motoko and Herrera, Gonzalo",
    title = "{Tidal disruption events and dark matter scatterings with neutrinos and photons}",
    eprint = "2312.11670",
    archivePrefix = "arXiv",
    primaryClass = "hep-ph",
    doi = "10.1016/j.physletb.2024.138573",
    journal = "Phys. Lett. B",
    volume = "851",
    pages = "138573",
    year = "2024"
}

@article{Bertolez-Martinez:2025trs,
    author = "Bert{\'o}lez-Mart{\'\i}nez, Toni and Herrera, Gonzalo and Mart{\'\i}nez-Mirav{\'e}, Pablo and Terol Calvo, Jorge",
    title = "{The Highest-Energy Neutrino Event Constrains Dark Matter-Neutrino Interactions}",
    eprint = "2506.08993",
    archivePrefix = "arXiv",
    primaryClass = "hep-ph",
    month = "6",
    year = "2025"
}

@article{Mondol:2025uuw,
    author = "Mondol, Ranjini and Bouri, Subhadip and Saha, Akash Kumar and Laha, Ranjan",
    title = "{Road through Dark$\nu$ess: Probing dark matter-neutrino interactions using KM3-230213A}",
    eprint = "2506.19910",
    archivePrefix = "arXiv",
    primaryClass = "hep-ph",
    month = "6",
    year = "2025"
}

@article{Ricciardi:2022pru,
    author = "Ricciardi, Giulia and Vignaroli, Natascia and Vissani, Francesco",
    title = "{An accurate evaluation of electron (anti-)neutrino scattering on nucleons}",
    eprint = "2206.05567",
    archivePrefix = "arXiv",
    primaryClass = "hep-ph",
    doi = "10.1007/JHEP08(2022)212",
    journal = "JHEP",
    volume = "08",
    pages = "212",
    year = "2022"
}

@article{Tulin:2017ara,
    author = "Tulin, Sean and Yu, Hai-Bo",
    title = "{Dark Matter Self-interactions and Small Scale Structure}",
    eprint = "1705.02358",
    archivePrefix = "arXiv",
    primaryClass = "hep-ph",
    doi = "10.1016/j.physrep.2017.11.004",
    journal = "Phys. Rept.",
    volume = "730",
    pages = "1--57",
    year = "2018"
}

@article{Agashe:2023itp,
    author = "Agashe, Kaustubh and Airen, Sagar and Franceschini, Roberto and Kim, Doojin and Kotwal, Ashutosh V. and Ricci, Lorenzo and Sathyan, Deepak",
    title = "{A new purpose for the W-boson mass measurement: Searching for New Physics in lepton+MET}",
    eprint = "2310.13687",
    archivePrefix = "arXiv",
    primaryClass = "hep-ph",
    reportNumber = "UMD-PP-023-04, MI-HET-817",
    doi = "10.1016/j.physletb.2024.138774",
    journal = "Phys. Lett. B",
    volume = "855",
    pages = "138774",
    year = "2024"
}

@article{Olivares-DelCampo:2017feq,
    author = "Olivares-Del Campo, Andr\'es and B\oe{}hm, C\'eline and Palomares-Ruiz, Sergio and Pascoli, Silvia",
    title = "{Dark matter-neutrino interactions through the lens of their cosmological implications}",
    eprint = "1711.05283",
    archivePrefix = "arXiv",
    primaryClass = "hep-ph",
    reportNumber = "IFIC-17-54, IPPP-17-84",
    doi = "10.1103/PhysRevD.97.075039",
    journal = "Phys. Rev. D",
    volume = "97",
    number = "7",
    pages = "075039",
    year = "2018"
}

@article{Arguelles:2017atb,
    author = {Arg\"uelles, Carlos A. and Kheirandish, Ali and Vincent, Aaron C.},
    title = "{Imaging Galactic Dark Matter with High-Energy Cosmic Neutrinos}",
    eprint = "1703.00451",
    archivePrefix = "arXiv",
    primaryClass = "hep-ph",
    doi = "10.1103/PhysRevLett.119.201801",
    journal = "Phys. Rev. Lett.",
    volume = "119",
    number = "20",
    pages = "201801",
    year = "2017"
}

@article{Telalovic:2024cot,
    author = "Telalovic, Bernanda and Fiorillo, Damiano F. G. and Mart{\'\i}nez-Mirav{\'e}, Pablo and Vitagliano, Edoardo and Bustamante, Mauricio",
    title = "{The next galactic supernova can uncover mass and couplings of particles decaying to neutrinos}",
    eprint = "2406.15506",
    archivePrefix = "arXiv",
    primaryClass = "hep-ph",
    doi = "10.1088/1475-7516/2024/11/011",
    journal = "JCAP",
    volume = "11",
    pages = "011",
    year = "2024"
}

@article{Navarro:1995iw,
    author = "Navarro, Julio F. and Frenk, Carlos S. and White, Simon D. M.",
    title = "{The Structure of cold dark matter halos}",
    eprint = "astro-ph/9508025",
    archivePrefix = "arXiv",
    doi = "10.1086/177173",
    journal = "Astrophys. J.",
    volume = "462",
    pages = "563--575",
    year = "1996"
}

@article{Navarro:1996gj,
    author = "Navarro, Julio F. and Frenk, Carlos S. and White, Simon D. M.",
    title = "{A Universal density profile from hierarchical clustering}",
    eprint = "astro-ph/9611107",
    archivePrefix = "arXiv",
    doi = "10.1086/304888",
    journal = "Astrophys. J.",
    volume = "490",
    pages = "493--508",
    year = "1997"
}

@article{Franarin:2018gfk,
    author = "Franarin, Tarso and Fairbairn, Malcolm and Davis, Jonathan H.",
    title = "{JUNO Sensitivity to Resonant Absorption of Galactic Supernova Neutrinos by Dark Matter}",
    eprint = "1806.05015",
    archivePrefix = "arXiv",
    primaryClass = "hep-ph",
    reportNumber = "KCL-PH-TH/2018-25, KCL-PH-TH-2018-25",
    month = "6",
    year = "2018"
}

@article{Blinov:2019gcj,
    author = "Blinov, Nikita and Kelly, Kevin James and Krnjaic, Gordan Z and McDermott, Samuel D",
    title = "{Constraining the Self-Interacting Neutrino Interpretation of the Hubble Tension}",
    eprint = "1905.02727",
    archivePrefix = "arXiv",
    primaryClass = "astro-ph.CO",
    reportNumber = "FERMILAB-PUB-19-175-A-T",
    doi = "10.1103/PhysRevLett.123.191102",
    journal = "Phys. Rev. Lett.",
    volume = "123",
    number = "19",
    pages = "191102",
    year = "2019"
}

@article{Escudero:2019gzq,
    author = "Escudero, Miguel and Hooper, Dan and Krnjaic, Gordan and Pierre, Mathias",
    title = "{Cosmology with A Very Light L$_{\mu}$ - L$_{\tau}$ Gauge Boson}",
    eprint = "1901.02010",
    archivePrefix = "arXiv",
    primaryClass = "hep-ph",
    reportNumber = "FERMILAB-PUB-19-001-A, LPT-Orsay-18-15, IFIC-19-02, KCL-19-01,
  IFT-UAM/CSIC-19-7, KCL-19-01",
    doi = "10.1007/JHEP03(2019)071",
    journal = "JHEP",
    volume = "03",
    pages = "071",
    year = "2019"
}

@article{Carpio:2022sml,
    author = "Carpio, Jose Alonso and Kheirandish, Ali and Murase, Kohta",
    title = "{Time-delayed neutrino emission from supernovae as a probe of dark matter-neutrino interactions}",
    eprint = "2204.09650",
    archivePrefix = "arXiv",
    primaryClass = "hep-ph",
    doi = "10.1088/1475-7516/2023/04/019",
    journal = "JCAP",
    volume = "04",
    pages = "019",
    year = "2023"
}

@article{Esteban:2021tub,
    author = "Esteban, Ivan and Pandey, Sujata and Brdar, Vedran and Beacom, John F.",
    title = "{Probing secret interactions of astrophysical neutrinos in the high-statistics era}",
    eprint = "2107.13568",
    archivePrefix = "arXiv",
    primaryClass = "hep-ph",
    reportNumber = "FERMILAB-PUB-21-328-T, nuhep-th/21-06",
    doi = "10.1103/PhysRevD.104.123014",
    journal = "Phys. Rev. D",
    volume = "104",
    number = "12",
    pages = "123014",
    year = "2021"
}

@article{Blennow:2019fhy,
    author = "Blennow, M. and Fernandez-Martinez, E. and Olivares-Del Campo, A. and Pascoli, S. and Rosauro-Alcaraz, S. and Titov, A. V.",
    title = "{Neutrino Portals to Dark Matter}",
    eprint = "1903.00006",
    archivePrefix = "arXiv",
    primaryClass = "hep-ph",
    reportNumber = "FTUAM-19-5, IFT-UAM/CSIC-19-19, IPPP/19/17",
    doi = "10.1140/epjc/s10052-019-7060-5",
    journal = "Eur. Phys. J. C",
    volume = "79",
    number = "7",
    pages = "555",
    year = "2019"
}

@article{Fayet:2006sa,
    author = "Fayet, Pierre and Hooper, Dan and Sigl, Gunter",
    title = "{Constraints on light dark matter from core-collapse supernovae}",
    eprint = "hep-ph/0602169",
    archivePrefix = "arXiv",
    reportNumber = "FERMILAB-PUB-06-001-A, LPTENS-06-04",
    doi = "10.1103/PhysRevLett.96.211302",
    journal = "Phys. Rev. Lett.",
    volume = "96",
    pages = "211302",
    year = "2006"
}

@article{Bertoni:2014mva,
    author = "Bertoni, Bridget and Ipek, Seyda and McKeen, David and Nelson, Ann E.",
    title = "{Constraints and consequences of reducing small scale structure via large dark matter-neutrino interactions}",
    eprint = "1412.3113",
    archivePrefix = "arXiv",
    primaryClass = "hep-ph",
    doi = "10.1007/JHEP04(2015)170",
    journal = "JHEP",
    volume = "04",
    pages = "170",
    year = "2015"
}

@article{Mangano:2006mp,
    author = "Mangano, Gianpiero and Melchiorri, Alessandro and Serra, Paolo and Cooray, Asantha and Kamionkowski, Marc",
    title = "{Cosmological bounds on dark matter-neutrino interactions}",
    eprint = "astro-ph/0606190",
    archivePrefix = "arXiv",
    doi = "10.1103/PhysRevD.74.043517",
    journal = "Phys. Rev. D",
    volume = "74",
    pages = "043517",
    year = "2006"
}

@article{Ferrer:2022kei,
    author = "Ferrer, Francesc and Herrera, Gonzalo and Ibarra, Alejandro",
    title = "{New constraints on the dark matter-neutrino and dark matter-photon scattering cross sections from TXS 0506+056}",
    eprint = "2209.06339",
    archivePrefix = "arXiv",
    primaryClass = "hep-ph",
    reportNumber = "TUM-HEP 1418/22",
    doi = "10.1088/1475-7516/2023/05/057",
    journal = "JCAP",
    volume = "05",
    pages = "057",
    year = "2023"
}

@article{Cline:2022qld,
    author = "Cline, James M. and Gao, Shan and Guo, Fangyi and Lin, Zhongan and Liu, Shiyan and Puel, Matteo and Todd, Phillip and Xiao, Tianzhuo",
    title = "{Blazar Constraints on Neutrino-Dark Matter Scattering}",
    eprint = "2209.02713",
    archivePrefix = "arXiv",
    primaryClass = "hep-ph",
    doi = "10.1103/PhysRevLett.130.091402",
    journal = "Phys. Rev. Lett.",
    volume = "130",
    number = "9",
    pages = "091402",
    year = "2023"
}

@article{Cline:2023tkp,
    author = "Cline, James M. and Puel, Matteo",
    title = "{NGC 1068 constraints on neutrino-dark matter scattering}",
    eprint = "2301.08756",
    archivePrefix = "arXiv",
    primaryClass = "hep-ph",
    doi = "10.1088/1475-7516/2023/06/004",
    journal = "JCAP",
    volume = "06",
    pages = "004",
    year = "2023"
}

@article{Bhattiprolu:2023akk,
    author = "Bhattiprolu, Prudhvi N. and McGehee, Robert and Pierce, Aaron",
    title = "{Dark sink enhances the direct detection of freeze-in dark matter}",
    eprint = "2312.14152",
    archivePrefix = "arXiv",
    primaryClass = "hep-ph",
    reportNumber = "FTPI-MINN-23-24, LCTP-23-17",
    doi = "10.1103/PhysRevD.110.L031702",
    journal = "Phys. Rev. D",
    volume = "110",
    number = "3",
    pages = "L031702",
    year = "2024"
}

@article{Berryman:2022hds,
    author = "Berryman, Jeffrey M. and others",
    title = "{Neutrino self-interactions: A white paper}",
    eprint = "2203.01955",
    archivePrefix = "arXiv",
    primaryClass = "hep-ph",
    reportNumber = "CERN-TH-2022-024, DESY-22-035, FERMILAB-PUB-22-099-T",
    doi = "10.1016/j.dark.2023.101267",
    journal = "Phys. Dark Univ.",
    volume = "42",
    pages = "101267",
    year = "2023"
}

@article{deGouvea:2019qaz,
    author = "de Gouv\^ea, Andr\'e and Dev, P. S. Bhupal and Dutta, Bhaskar and Ghosh, Tathagata and Han, Tao and Zhang, Yongchao",
    title = "{Leptonic Scalars at the LHC}",
    eprint = "1910.01132",
    archivePrefix = "arXiv",
    primaryClass = "hep-ph",
    reportNumber = "PITT-PACC 1909, MI-TH-1936",
    doi = "10.1007/JHEP07(2020)142",
    journal = "JHEP",
    volume = "07",
    pages = "142",
    year = "2020"
}

@article{Vincent:2017svp,
    author = {Vincent, Aaron C. and Arg\"uelles, Carlos A. and Kheirandish, Ali},
    title = "{High-energy neutrino attenuation in the Earth and its associated uncertainties}",
    eprint = "1706.09895",
    archivePrefix = "arXiv",
    primaryClass = "hep-ph",
    doi = "10.1088/1475-7516/2017/11/012",
    journal = "JCAP",
    volume = "11",
    pages = "012",
    year = "2017"
}

@article{McMullen:2021ikf,
    author = "McMullen, Adam and Vincent, Aaron and Arguelles, Carlos and Schneider, Austin",
    collaboration = "IceCube",
    title = "{Dark matter neutrino scattering in the galactic centre with IceCube}",
    eprint = "2107.11491",
    archivePrefix = "arXiv",
    primaryClass = "astro-ph.HE",
    reportNumber = "PoS-ICRC2021-569",
    doi = "10.1088/1748-0221/16/08/C08001",
    journal = "JINST",
    volume = "16",
    number = "08",
    pages = "C08001",
    year = "2021"
}

@article{Hajjar:2023knk,
    author = "Hajjar, Rasmi and Mena, Olga and Palomares-Ruiz, Sergio",
    title = "{Earth tomography with supernova neutrinos at future neutrino detectors}",
    eprint = "2303.09369",
    archivePrefix = "arXiv",
    primaryClass = "hep-ph",
    doi = "10.1103/PhysRevD.108.083011",
    journal = "Phys. Rev. D",
    volume = "108",
    number = "8",
    pages = "083011",
    year = "2023"
}

@article{Escudero:2018mvt,
    author = "Escudero, Miguel",
    title = "{Neutrino decoupling beyond the Standard Model: CMB constraints on the Dark Matter mass with a fast and precise $N_{\rm eff}$ evaluation}",
    eprint = "1812.05605",
    archivePrefix = "arXiv",
    primaryClass = "hep-ph",
    reportNumber = "KCL-2018-76",
    doi = "10.1088/1475-7516/2019/02/007",
    journal = "JCAP",
    volume = "02",
    pages = "007",
    year = "2019"
}

@article{Nemevsek:2012cd,
    author = "Nemevsek, Miha and Senjanovic, Goran and Zhang, Yue",
    title = "{Warm Dark Matter in Low Scale Left-Right Theory}",
    eprint = "1205.0844",
    archivePrefix = "arXiv",
    primaryClass = "hep-ph",
    doi = "10.1088/1475-7516/2012/07/006",
    journal = "JCAP",
    volume = "07",
    pages = "006",
    year = "2012"
}

@article{Nemevsek:2023yjl,
    author = "Nemev{\v{s}}ek, Miha and Zhang, Yue",
    title = "{Anatomy of diluted dark matter in the minimal left-right symmetric model}",
    eprint = "2312.00129",
    archivePrefix = "arXiv",
    primaryClass = "hep-ph",
    doi = "10.1103/PhysRevD.109.056021",
    journal = "Phys. Rev. D",
    volume = "109",
    number = "5",
    pages = "056021",
    year = "2024"
}

@article{Chauhan:2022iuh,
    author = "Chauhan, Garv and Dev, P. S. Bhupal and Xu, Xun-Jie",
    title = "{Probing the {\ensuremath{\nu}}R-philic Z' at DUNE near detectors}",
    eprint = "2204.11876",
    archivePrefix = "arXiv",
    primaryClass = "hep-ph",
    doi = "10.1016/j.physletb.2023.137907",
    journal = "Phys. Lett. B",
    volume = "841",
    pages = "137907",
    year = "2023"
}

@article{Chauhan:2020mgv,
    author = "Chauhan, Garv and Xu, Xun-Jie",
    title = "{How dark is the $\nu_R$-philic dark photon?}",
    eprint = "2012.09980",
    archivePrefix = "arXiv",
    primaryClass = "hep-ph",
    doi = "10.1007/JHEP04(2021)003",
    journal = "JHEP",
    volume = "04",
    pages = "003",
    year = "2021"
}

@article{Dev:2025fcv,
    author = "Dev, P. S. Bhupal and Heeck, Julian and Thapa, Anil",
    title = "{Decaying scalar dark matter in the minimal left-right symmetric model}",
    eprint = "2501.14669",
    archivePrefix = "arXiv",
    primaryClass = "hep-ph",
    month = "1",
    year = "2025"
}

@article{Dutta:2024kuj,
    author = "Dutta, Bhaskar and Huang, Wei-Chih and Kim, Doojin and Newstead, Jayden L. and Park, Jong-Chul and Ali, Iman Shaukat",
    title = "{Prospects for Light Dark Matter Searches at Large-Volume Neutrino Detectors}",
    eprint = "2402.04184",
    archivePrefix = "arXiv",
    primaryClass = "hep-ph",
    doi = "10.1103/PhysRevLett.133.161801",
    journal = "Phys. Rev. Lett.",
    volume = "133",
    number = "16",
    pages = "161801",
    year = "2024"
}

@article{Heston:2024ljf,
    author = "Heston, Sean and Horiuchi, Shunsaku and Shirai, Satoshi",
    title = "{Constraining neutrino-DM interactions with Milky~Way dwarf spheroidals and supernova neutrinos}",
    eprint = "2402.08718",
    archivePrefix = "arXiv",
    primaryClass = "hep-ph",
    doi = "10.1103/PhysRevD.110.023004",
    journal = "Phys. Rev. D",
    volume = "110",
    number = "2",
    pages = "023004",
    year = "2024"
}

@article{Sabti:2019mhn,
    author = "Sabti, Nashwan and Alvey, James and Escudero, Miguel and Fairbairn, Malcolm and Blas, Diego",
    title = "{Refined Bounds on MeV-scale Thermal Dark Sectors from BBN and the CMB}",
    eprint = "1910.01649",
    archivePrefix = "arXiv",
    primaryClass = "hep-ph",
    reportNumber = "KCL-2019-75",
    doi = "10.1088/1475-7516/2020/01/004",
    journal = "JCAP",
    volume = "01",
    pages = "004",
    year = "2020"
}

@article{Choi:2019zxy,
    author = "Choi, Ki-Young and Chun, Eung Jin and Kim, Jongkuk",
    title = "{Neutrino Oscillations in Dark Matter}",
    eprint = "1909.10478",
    archivePrefix = "arXiv",
    primaryClass = "hep-ph",
    doi = "10.1016/j.dark.2020.100606",
    journal = "Phys. Dark Univ.",
    volume = "30",
    pages = "100606",
    year = "2020"
}

@article{Boehm:2013jpa,
    author = "Boehm, C\'eline and Dolan, Matthew J. and McCabe, Christopher",
    title = "{A Lower Bound on the Mass of Cold Thermal Dark Matter from Planck}",
    eprint = "1303.6270",
    archivePrefix = "arXiv",
    primaryClass = "hep-ph",
    reportNumber = "IPPP-13-18, DCPT-13-36",
    doi = "10.1088/1475-7516/2013/08/041",
    journal = "JCAP",
    volume = "08",
    pages = "041",
    year = "2013"
}

@article{Mosbech:2020ahp,
    author = "Mosbech, Markus R. and Boehm, Celine and Hannestad, Steen and Mena, Olga and Stadler, Julia and Wong, Yvonne Y. Y.",
    title = "{The full Boltzmann hierarchy for dark matter-massive neutrino interactions}",
    eprint = "2011.04206",
    archivePrefix = "arXiv",
    primaryClass = "astro-ph.CO",
    doi = "10.1088/1475-7516/2021/03/066",
    journal = "JCAP",
    volume = "03",
    pages = "066",
    year = "2021"
}

@article{Hooper:2021rjc,
    author = "Hooper, Deanna C. and Lucca, Matteo",
    title = "{Hints of dark matter-neutrino interactions in Lyman-\ensuremath{\alpha} data}",
    eprint = "2110.04024",
    archivePrefix = "arXiv",
    primaryClass = "astro-ph.CO",
    reportNumber = "ULB-TH/21-15",
    doi = "10.1103/PhysRevD.105.103504",
    journal = "Phys. Rev. D",
    volume = "105",
    number = "10",
    pages = "103504",
    year = "2022"
}

@article{Lim:2023lss,
    author = "Lim, Sung Hak and Putney, Eric and Buckley, Matthew R. and Shih, David",
    title = "{Mapping dark matter in the Milky Way using normalizing flows and Gaia DR3}",
    eprint = "2305.13358",
    archivePrefix = "arXiv",
    primaryClass = "astro-ph.GA",
    reportNumber = "CTPU-PTC-25-01",
    doi = "10.1088/1475-7516/2025/01/021",
    journal = "JCAP",
    volume = "01",
    pages = "021",
    year = "2025"
}

@article{Serpico:2004nm,
    author = "Serpico, Pasquale Dario and Raffelt, Georg G.",
    title = "{MeV-mass dark matter and primordial nucleosynthesis}",
    eprint = "astro-ph/0403417",
    archivePrefix = "arXiv",
    reportNumber = "MPP-2004-34",
    doi = "10.1103/PhysRevD.70.043526",
    journal = "Phys. Rev. D",
    volume = "70",
    pages = "043526",
    year = "2004"
}

@article{Koren:2019wwi,
    author = "Koren, Seth",
    title = "{Neutrino -- Dark Matter Scattering and Coincident Detections of UHE Neutrinos with EM Sources}",
    eprint = "1903.05096",
    archivePrefix = "arXiv",
    primaryClass = "hep-ph",
    doi = "10.1088/1475-7516/2019/09/013",
    journal = "JCAP",
    volume = "09",
    pages = "013",
    year = "2019"
}

@article{Giovanetti:2024orj,
    author = "Giovanetti, Cara and Schmaltz, Martin and Weiner, Neal",
    title = "{Neutrino-dark sector equilibration and primordial element abundances}",
    eprint = "2402.10264",
    archivePrefix = "arXiv",
    primaryClass = "hep-ph",
    doi = "10.1103/PhysRevD.111.043526",
    journal = "Phys. Rev. D",
    volume = "111",
    number = "4",
    pages = "043526",
    year = "2025"
}

@article{Murase:2019xqi,
    author = "Murase, Kohta and Shoemaker, Ian M.",
    title = "{Neutrino Echoes from Multimessenger Transient Sources}",
    eprint = "1903.08607",
    archivePrefix = "arXiv",
    primaryClass = "hep-ph",
    doi = "10.1103/PhysRevLett.123.241102",
    journal = "Phys. Rev. Lett.",
    volume = "123",
    number = "24",
    pages = "241102",
    year = "2019"
}

@article{Kelly:2018tyg,
    author = "Kelly, Kevin J. and Machado, Pedro A. N.",
    title = "{Multimessenger Astronomy and New Neutrino Physics}",
    eprint = "1808.02889",
    archivePrefix = "arXiv",
    primaryClass = "hep-ph",
    reportNumber = "FERMILAB-PUB-18-374-T, NUHEP-TH/18-07",
    doi = "10.1088/1475-7516/2018/10/048",
    journal = "JCAP",
    volume = "10",
    pages = "048",
    year = "2018"
}

@article{Lessa:2007up,
    author = "Lessa, A. P. and Peres, O. L. G.",
    title = "{Revising limits on neutrino-Majoron couplings}",
    eprint = "hep-ph/0701068",
    archivePrefix = "arXiv",
    doi = "10.1103/PhysRevD.75.094001",
    journal = "Phys. Rev. D",
    volume = "75",
    pages = "094001",
    year = "2007"
}

@article{Berryman:2018ogk,
    author = "Berryman, Jeffrey M. and De Gouv\^ea, Andr\'e and Kelly, Kevin J. and Zhang, Yue",
    title = "{Lepton-Number-Charged Scalars and Neutrino Beamstrahlung}",
    eprint = "1802.00009",
    archivePrefix = "arXiv",
    primaryClass = "hep-ph",
    reportNumber = "NUHEP-TH-18-03, FERMILAB-PUB-18-020-T",
    doi = "10.1103/PhysRevD.97.075030",
    journal = "Phys. Rev. D",
    volume = "97",
    number = "7",
    pages = "075030",
    year = "2018"
}

@article{deSalas:2016svi,
    author = "de Salas, P. F. and Lineros, R. A. and T\'ortola, M.",
    title = "{Neutrino propagation in the galactic dark matter halo}",
    eprint = "1601.05798",
    archivePrefix = "arXiv",
    primaryClass = "astro-ph.HE",
    reportNumber = "IFIC-16-01",
    doi = "10.1103/PhysRevD.94.123001",
    journal = "Phys. Rev. D",
    volume = "94",
    number = "12",
    pages = "123001",
    year = "2016"
}

@article{Reynoso:2016hjr,
    author = "Reynoso, Mat\'\i{}as M. and Sampayo, Oscar A.",
    title = "{Propagation of high-energy neutrinos in a background of ultralight scalar dark matter}",
    eprint = "1605.09671",
    archivePrefix = "arXiv",
    primaryClass = "hep-ph",
    doi = "10.1016/j.astropartphys.2016.05.004",
    journal = "Astropart. Phys.",
    volume = "82",
    pages = "10--20",
    year = "2016"
}

@article{Karmakar:2020yzn,
    author = "Karmakar, Siddhartha and Pandey, Sujata and Rakshit, Subhendu",
    title = "{Astronomy with energy dependent flavour ratios of extragalactic neutrinos}",
    eprint = "2010.07336",
    archivePrefix = "arXiv",
    primaryClass = "hep-ph",
    doi = "10.1007/JHEP10(2021)004",
    journal = "JHEP",
    volume = "10",
    pages = "004",
    year = "2021"
}

@article{vandenAarssen:2012vpm,
    author = "van den Aarssen, Laura G. and Bringmann, Torsten and Pfrommer, Christoph",
    title = "{Is dark matter with long-range interactions a solution to all small-scale problems of \textbackslash{}Lambda CDM cosmology?}",
    eprint = "1205.5809",
    archivePrefix = "arXiv",
    primaryClass = "astro-ph.CO",
    doi = "10.1103/PhysRevLett.109.231301",
    journal = "Phys. Rev. Lett.",
    volume = "109",
    pages = "231301",
    year = "2012"
}

@article{Boehm:2000gq,
    author = "Boehm, C. and Fayet, Pierre and Schaeffer, R.",
    title = "{Constraining dark matter candidates from structure formation}",
    eprint = "astro-ph/0012504",
    archivePrefix = "arXiv",
    doi = "10.1016/S0370-2693(01)01060-7",
    journal = "Phys. Lett. B",
    volume = "518",
    pages = "8--14",
    year = "2001"
}

@article{Hooper:2007tu,
    author = "Hooper, Dan and Kaplinghat, Manoj and Strigari, Louis E. and Zurek, Kathryn M.",
    title = "{MeV Dark Matter and Small Scale Structure}",
    eprint = "0704.2558",
    archivePrefix = "arXiv",
    primaryClass = "astro-ph",
    reportNumber = "FERMILAB-PUB-07-064-A, MADPH-07-1482",
    doi = "10.1103/PhysRevD.76.103515",
    journal = "Phys. Rev. D",
    volume = "76",
    pages = "103515",
    year = "2007"
}

@article{Diacoumis:2017hff,
    author = "Diacoumis, James A. D. and Wong, Yvonne Y. Y.",
    title = "{Using CMB spectral distortions to distinguish between dark matter solutions to the small-scale crisis}",
    eprint = "1707.07050",
    archivePrefix = "arXiv",
    primaryClass = "astro-ph.CO",
    doi = "10.1088/1475-7516/2017/09/011",
    journal = "JCAP",
    volume = "09",
    pages = "011",
    year = "2017"
}

@article{DiValentino:2017oaw,
    author = "Di Valentino, Eleonora and B\o{}ehm, C\'eline and Hivon, Eric and Bouchet, Fran\c{c}ois R.",
    title = "{Reducing the $H_0$ and $\sigma_8$ tensions with Dark Matter-neutrino interactions}",
    eprint = "1710.02559",
    archivePrefix = "arXiv",
    primaryClass = "astro-ph.CO",
    doi = "10.1103/PhysRevD.97.043513",
    journal = "Phys. Rev. D",
    volume = "97",
    number = "4",
    pages = "043513",
    year = "2018"
}

@article{Ghosh:2017jdy,
    author = "Ghosh, Subhajit and Khatri, Rishi and Roy, Tuhin S.",
    title = "{Dark neutrino interactions make gravitational waves blue}",
    eprint = "1711.09929",
    archivePrefix = "arXiv",
    primaryClass = "astro-ph.CO",
    reportNumber = "TIFR-TH-17-43",
    doi = "10.1103/PhysRevD.97.063529",
    journal = "Phys. Rev. D",
    volume = "97",
    number = "6",
    pages = "063529",
    year = "2018"
}

@article{Escudero:2015yka,
    author = "Escudero, Miguel and Mena, Olga and Vincent, Aaron C. and Wilkinson, Ryan J. and B\oe{}hm, C\'eline",
    title = "{Exploring dark matter microphysics with galaxy surveys}",
    eprint = "1505.06735",
    archivePrefix = "arXiv",
    primaryClass = "astro-ph.CO",
    reportNumber = "IFIC-15-32, IPPP-15-29, DCPT-15-58",
    doi = "10.1088/1475-7516/2015/9/034",
    journal = "JCAP",
    volume = "09",
    pages = "034",
    year = "2015"
}

@article{Dey:2022ini,
    author = "Dey, Antara and Paul, Arnab and Pal, Supratik",
    title = "{Constraints on dark matter\textendash{}neutrino interaction from 21-cm cosmology and forecasts on SKA1-Low}",
    eprint = "2207.02451",
    archivePrefix = "arXiv",
    primaryClass = "astro-ph.CO",
    doi = "10.1093/mnras/stad1838",
    journal = "Mon. Not. Roy. Astron. Soc.",
    volume = "524",
    number = "1",
    pages = "100--107",
    year = "2023"
}

@article{Mosbech:2022uud,
    author = "Mosbech, Markus R. and Boehm, Celine and Wong, Yvonne Y. Y.",
    title = "{Probing dark matter interactions with 21cm observations}",
    eprint = "2207.03107",
    archivePrefix = "arXiv",
    primaryClass = "astro-ph.CO",
    reportNumber = "CPPC-2022-06",
    doi = "10.1088/1475-7516/2023/03/047",
    journal = "JCAP",
    volume = "03",
    pages = "047",
    year = "2023"
}

@article{Brax:2023tvn,
    author = "Brax, Philippe and van de Bruck, Carsten and Di Valentino, Eleonora and Giar\`e, William and Trojanowski, Sebastian",
    title = "{Extended analysis of neutrino-dark matter interactions with small-scale CMB experiments}",
    eprint = "2305.01383",
    archivePrefix = "arXiv",
    primaryClass = "astro-ph.CO",
    doi = "10.1016/j.dark.2023.101321",
    journal = "Phys. Dark Univ.",
    volume = "42",
    pages = "101321",
    year = "2023"
}

@article{Farzan:2014gza,
    author = "Farzan, Yasaman and Palomares-Ruiz, Sergio",
    title = "{Dips in the Diffuse Supernova Neutrino Background}",
    eprint = "1401.7019",
    archivePrefix = "arXiv",
    primaryClass = "hep-ph",
    reportNumber = "IFIC-14-02",
    doi = "10.1088/1475-7516/2014/06/014",
    journal = "JCAP",
    volume = "06",
    pages = "014",
    year = "2014"
}

@article{Yuksel:2007ac,
    author = "Yuksel, Hasan and Horiuchi, Shunsaku and Beacom, John F. and Ando, Shin'ichiro",
    title = "{Neutrino Constraints on the Dark Matter Total Annihilation Cross Section}",
    eprint = "0707.0196",
    archivePrefix = "arXiv",
    primaryClass = "astro-ph",
    doi = "10.1103/PhysRevD.76.123506",
    journal = "Phys. Rev. D",
    volume = "76",
    pages = "123506",
    year = "2007"
}

@article{Beacom:2006tt,
    author = "Beacom, John F. and Bell, Nicole F. and Mack, Gregory D.",
    title = "{General Upper Bound on the Dark Matter Total Annihilation Cross Section}",
    eprint = "astro-ph/0608090",
    archivePrefix = "arXiv",
    reportNumber = "KRL-MAP-322",
    doi = "10.1103/PhysRevLett.99.231301",
    journal = "Phys. Rev. Lett.",
    volume = "99",
    pages = "231301",
    year = "2007"
}

@article{Palomares-Ruiz:2007trf,
    author = "Palomares-Ruiz, Sergio and Pascoli, Silvia",
    title = "{Testing MeV dark matter with neutrino detectors}",
    eprint = "0710.5420",
    archivePrefix = "arXiv",
    primaryClass = "astro-ph",
    reportNumber = "IPPP-07-81, DCPT-07-162",
    doi = "10.1103/PhysRevD.77.025025",
    journal = "Phys. Rev. D",
    volume = "77",
    pages = "025025",
    year = "2008"
}

@article{Chattopadhyay:2026kbm,
    author = "Chattopadhyay, Dibya S. and Dev, P. S. Bhupal and Porto, Yago",
    title = "{Ruling Out Spiky WIMP Dark Matter using Indirect Searches}",
    eprint = "2602.23348",
    archivePrefix = "arXiv",
    primaryClass = "hep-ph",
    reportNumber = "CETUP-2025-019",
    month = "2",
    year = "2026"
}

@article{Babu:2025czb,
    author = "Babu, K. S. and Dev, P. S. Bhupal and Thapa, Anil",
    title = "{Large Neutrino-Dark Matter Interactions: From Effective Field Theory to Ultraviolet Completions}",
    eprint = "2512.25035",
    archivePrefix = "arXiv",
    primaryClass = "hep-ph",
    month = "12",
    year = "2025"
}

@article{Palomares-Ruiz:2007egs,
    author = "Palomares-Ruiz, Sergio",
    title = "{Model-independent bound on the dark matter lifetime}",
    eprint = "0712.1937",
    archivePrefix = "arXiv",
    primaryClass = "astro-ph",
    reportNumber = "IPPP-07-96, DCPT-07-192",
    doi = "10.1016/j.physletb.2008.05.040",
    journal = "Phys. Lett. B",
    volume = "665",
    pages = "50--53",
    year = "2008"
}

@article{Covi:2009xn,
    author = "Covi, Laura and Grefe, Michael and Ibarra, Alejandro and Tran, David",
    title = "{Neutrino Signals from Dark Matter Decay}",
    eprint = "0912.3521",
    archivePrefix = "arXiv",
    primaryClass = "hep-ph",
    reportNumber = "DESY-09-164, TUM-HEP-743-09",
    doi = "10.1088/1475-7516/2010/04/017",
    journal = "JCAP",
    volume = "04",
    pages = "017",
    year = "2010"
}

@article{IceCube:2023ies,
    author = "Abbasi, R. and others",
    collaboration = "IceCube",
    title = "{Search for neutrino lines from dark matter annihilation and decay with IceCube}",
    eprint = "2303.13663",
    archivePrefix = "arXiv",
    primaryClass = "astro-ph.HE",
    doi = "10.1103/PhysRevD.108.102004",
    journal = "Phys. Rev. D",
    volume = "108",
    number = "10",
    pages = "102004",
    year = "2023"
}

@article{Arguelles:2022nbl,
    author = {Arg\"uelles, Carlos A. and Delgado, Diyaselis and Friedlander, Avi and Kheirandish, Ali and Safa, Ibrahim and Vincent, Aaron C. and White, Henry},
    title = "{Dark matter decay to neutrinos}",
    eprint = "2210.01303",
    archivePrefix = "arXiv",
    primaryClass = "hep-ph",
    doi = "10.1103/PhysRevD.108.123021",
    journal = "Phys. Rev. D",
    volume = "108",
    number = "12",
    pages = "123021",
    year = "2023"
}

@article{ElAisati:2017ppn,
    author = "El Aisati, Chaimae and Garcia-Cely, Camilo and Hambye, Thomas and Vanderheyden, Laurent",
    title = "{Prospects for discovering a neutrino line induced by dark matter annihilation}",
    eprint = "1706.06600",
    archivePrefix = "arXiv",
    primaryClass = "hep-ph",
    doi = "10.1088/1475-7516/2017/10/021",
    journal = "JCAP",
    volume = "10",
    pages = "021",
    year = "2017"
}

@article{Moline:2014xua,
    author = "Molin\'e, \'A and Ibarra, Alejandro and Palomares-Ruiz, Sergio",
    title = "{Future sensitivity of neutrino telescopes to dark matter annihilations from the cosmic diffuse neutrino signal}",
    eprint = "1412.4308",
    archivePrefix = "arXiv",
    primaryClass = "astro-ph.CO",
    reportNumber = "CFTP-14-019, TUM-HEP-970-14, IFIC-14-78",
    doi = "10.1088/1475-7516/2015/06/005",
    journal = "JCAP",
    volume = "06",
    pages = "005",
    year = "2015"
}

@article{Boehm:2003xr,
    author = "Boehm, C. and Mathis, H. and Devriendt, J. and Silk, J.",
    title = "{Non-linear evolution of suppressed dark matter primordial power spectra}",
    eprint = "astro-ph/0309652",
    archivePrefix = "arXiv",
    doi = "10.1111/j.1365-2966.2005.09032.x",
    journal = "Mon. Not. Roy. Astron. Soc.",
    volume = "360",
    pages = "282--287",
    year = "2005"
}

@article{Boehm:2014vja,
    author = "Boehm, C. and Schewtschenko, J. A. and Wilkinson, R. J. and Baugh, C. M. and Pascoli, S.",
    title = "{Using the Milky Way satellites to study interactions between cold dark matter and radiation}",
    eprint = "1404.7012",
    archivePrefix = "arXiv",
    primaryClass = "astro-ph.CO",
    reportNumber = "IPPP-14-11, DCPT-14-22, DURAST-2014-0020, DURAST-2014-0004",
    doi = "10.1093/mnrasl/slu115",
    journal = "Mon. Not. Roy. Astron. Soc.",
    volume = "445",
    pages = "L31--L35",
    year = "2014"
}

@article{Schewtschenko:2014fca,
    author = "Schewtschenko, J. A. and Wilkinson, R. J. and Baugh, C. M. and B\oe{}hm, C. and Pascoli, S.",
    title = "{Dark matter\textendash{}radiation interactions: the impact on dark matter haloes}",
    eprint = "1412.4905",
    archivePrefix = "arXiv",
    primaryClass = "astro-ph.CO",
    reportNumber = "IPPP-14-106, DCPT-14-212",
    doi = "10.1093/mnras/stv431",
    journal = "Mon. Not. Roy. Astron. Soc.",
    volume = "449",
    number = "4",
    pages = "3587--3596",
    year = "2015"
}

@article{Schewtschenko:2015rno,
    author = "Schewtschenko, J. A. and Baugh, C. M. and Wilkinson, R. J. and B\oe{}hm, C. and Pascoli, S. and Sawala, T.",
    title = "{Dark matter\textendash{}radiation interactions: the structure of Milky Way satellite galaxies}",
    eprint = "1512.06774",
    archivePrefix = "arXiv",
    primaryClass = "astro-ph.CO",
    reportNumber = "DURAST-2016-0001, IPPP-16-03",
    doi = "10.1093/mnras/stw1078",
    journal = "Mon. Not. Roy. Astron. Soc.",
    volume = "461",
    number = "3",
    pages = "2282--2287",
    year = "2016"
}

@article{Boehm:2004th,
    author = "Boehm, Celine and Schaeffer, Richard",
    title = "{Constraints on dark matter interactions from structure formation: Damping lengths}",
    eprint = "astro-ph/0410591",
    archivePrefix = "arXiv",
    doi = "10.1051/0004-6361:20042238",
    journal = "Astron. Astrophys.",
    volume = "438",
    pages = "419--442",
    year = "2005"
}

@article{Adhikari:2022sbh,
    author = "Adhikari, Susmita and others",
    title = "{Astrophysical Tests of Dark Matter Self-Interactions}",
    eprint = "2207.10638",
    archivePrefix = "arXiv",
    primaryClass = "astro-ph.CO",
    month = "7",
    year = "2022"
}

@article{Asatrian:2012tp,
    author = "Asatrian, H. M. and Hovhannisyan, A. and Yeghiazaryan, A.",
    title = "{The phase space analysis for three and four massive particles in final states}",
    eprint = "1210.7939",
    archivePrefix = "arXiv",
    primaryClass = "hep-ph",
    doi = "10.1103/PhysRevD.86.114023",
    journal = "Phys. Rev. D",
    volume = "86",
    pages = "114023",
    year = "2012"
}

@article{Navarro:2003ew,
    author = "Navarro, Julio F. and Hayashi, Eric and Power, Chris and Jenkins, Adrian and Frenk, Carlos S. and White, Simon D. M. and Springel, Volker and Stadel, Joachim and Quinn, Thomas R.",
    title = "{The Inner structure of Lambda-CDM halos 3: Universality and asymptotic slopes}",
    eprint = "astro-ph/0311231",
    archivePrefix = "arXiv",
    doi = "10.1111/j.1365-2966.2004.07586.x",
    journal = "Mon. Not. Roy. Astron. Soc.",
    volume = "349",
    pages = "1039",
    year = "2004"
}

@article{Lin:2019yux,
    author = "Lin, Hai-Nan and Li, Xin",
    title = "{The Dark Matter Profiles in the Milky Way}",
    eprint = "1906.08419",
    archivePrefix = "arXiv",
    primaryClass = "astro-ph.GA",
    doi = "10.1093/mnras/stz1698",
    journal = "Mon. Not. Roy. Astron. Soc.",
    volume = "487",
    number = "4",
    pages = "5679--5684",
    year = "2019"
}

@article{Gao:2024irf,
    author = "Gao, Jiansong and Hochberg, Yonit and Lehmann, Benjamin V. and Nam, Sae Woo and Szypryt, Paul and Vissers, Michael R. and Xu, Tao",
    title = "{Detecting Light Dark Matter with Kinetic Inductance Detectors}",
    eprint = "2403.19739",
    archivePrefix = "arXiv",
    primaryClass = "hep-ph",
    reportNumber = "MIT-CTP/5654",
    month = "3",
    year = "2024"
}

@article{DUNE:2020zfm,
    author = "Abi, B. and others",
    collaboration = "DUNE",
    title = "{Supernova neutrino burst detection with the Deep Underground Neutrino Experiment}",
    eprint = "2008.06647",
    archivePrefix = "arXiv",
    primaryClass = "hep-ex",
    reportNumber = "FERMILAB-PUB-20-380-LBNF, FERMILAB-PUB-20-380-LBNF",
    doi = "10.1140/epjc/s10052-021-09166-w",
    journal = "Eur. Phys. J. C",
    volume = "81",
    number = "5",
    pages = "423",
    year = "2021"
}

@article{Ullio:2001fb,
    author = "Ullio, Piero and Zhao, HongSheng and Kamionkowski, Marc",
    title = "{A Dark matter spike at the galactic center?}",
    eprint = "astro-ph/0101481",
    archivePrefix = "arXiv",
    doi = "10.1103/PhysRevD.64.043504",
    journal = "Phys. Rev. D",
    volume = "64",
    pages = "043504",
    year = "2001"
}

@article{Ou:2023adg,
    author = "Ou, Xiaowei and Eilers, Anna-Christina and Necib, Lina and Frebel, Anna",
    title = "{The dark matter profile of the Milky Way inferred from its circular velocity curve}",
    eprint = "2303.12838",
    archivePrefix = "arXiv",
    primaryClass = "astro-ph.GA",
    doi = "10.1093/mnras/stae034",
    journal = "Mon. Not. Roy. Astron. Soc.",
    volume = "528",
    number = "1",
    pages = "693--710",
    year = "2024"
}

@article{Hooper:2010mq,
    author = "Hooper, Dan and Goodenough, Lisa",
    title = "{Dark Matter Annihilation in The Galactic Center As Seen by the Fermi Gamma Ray Space Telescope}",
    eprint = "1010.2752",
    archivePrefix = "arXiv",
    primaryClass = "hep-ph",
    reportNumber = "FERMILAB-PUB-10-414-A",
    doi = "10.1016/j.physletb.2011.02.029",
    journal = "Phys. Lett. B",
    volume = "697",
    pages = "412--428",
    year = "2011"
}

@article{Georgi:1981pg,
    author = "Georgi, Howard M. and Glashow, Sheldon Lee and Nussinov, Shmuel",
    title = "{Unconventional Model of Neutrino Masses}",
    reportNumber = "HUTP-81/A026",
    doi = "10.1016/0550-3213(81)90336-9",
    journal = "Nucl. Phys. B",
    volume = "193",
    pages = "297--316",
    year = "1981"
}

@article{Brune:2018sab,
    author = {Brune, Tim and P\"as, Heinrich},
    title = "{Massive Majorons and constraints on the Majoron-neutrino coupling}",
    eprint = "1808.08158",
    archivePrefix = "arXiv",
    primaryClass = "hep-ph",
    reportNumber = "DO-TH 18/23",
    doi = "10.1103/PhysRevD.99.096005",
    journal = "Phys. Rev. D",
    volume = "99",
    number = "9",
    pages = "096005",
    year = "2019"
}

@article{Deppisch:2020sqh,
    author = "Deppisch, Frank F. and Graf, Lukas and Rodejohann, Werner and Xu, Xun-Jie",
    title = "{Neutrino Self-Interactions and Double Beta Decay}",
    eprint = "2004.11919",
    archivePrefix = "arXiv",
    primaryClass = "hep-ph",
    doi = "10.1103/PhysRevD.102.051701",
    journal = "Phys. Rev. D",
    volume = "102",
    number = "5",
    pages = "051701",
    year = "2020"
}

@article{Pasquini:2015fjv,
    author = "Pasquini, P. S. and Peres, O. L. G.",
    title = "{Bounds on Neutrino-Scalar Yukawa Coupling}",
    eprint = "1511.01811",
    archivePrefix = "arXiv",
    primaryClass = "hep-ph",
    doi = "10.1103/PhysRevD.93.053007",
    journal = "Phys. Rev. D",
    volume = "93",
    number = "5",
    pages = "053007",
    year = "2016",
    note = "[Erratum: Phys.Rev.D 93, 079902 (2016)]"
}

@article{Oldengott:2017fhy,
    author = "Oldengott, Isabel M. and Tram, Thomas and Rampf, Cornelius and Wong, Yvonne Y. Y.",
    title = "{Interacting neutrinos in cosmology: exact description and constraints}",
    eprint = "1706.02123",
    archivePrefix = "arXiv",
    primaryClass = "astro-ph.CO",
    doi = "10.1088/1475-7516/2017/11/027",
    journal = "JCAP",
    volume = "11",
    pages = "027",
    year = "2017"
}

@article{Kreisch:2019yzn,
    author = "Kreisch, Christina D. and Cyr-Racine, Francis-Yan and Dor\'e, Olivier",
    title = "{Neutrino puzzle: Anomalies, interactions, and cosmological tensions}",
    eprint = "1902.00534",
    archivePrefix = "arXiv",
    primaryClass = "astro-ph.CO",
    doi = "10.1103/PhysRevD.101.123505",
    journal = "Phys. Rev. D",
    volume = "101",
    number = "12",
    pages = "123505",
    year = "2020"
}

@article{RoyChoudhury:2020dmd,
    author = "Roy Choudhury, Shouvik and Hannestad, Steen and Tram, Thomas",
    title = "{Updated constraints on massive neutrino self-interactions from cosmology in light of the $H_0$ tension}",
    eprint = "2012.07519",
    archivePrefix = "arXiv",
    primaryClass = "astro-ph.CO",
    doi = "10.1088/1475-7516/2021/03/084",
    journal = "JCAP",
    volume = "03",
    pages = "084",
    year = "2021"
}

@article{Brinckmann:2020bcn,
    author = "Brinckmann, Thejs and Chang, Jae Hyeok and LoVerde, Marilena",
    title = "{Self-interacting neutrinos, the Hubble parameter tension, and the cosmic microwave background}",
    eprint = "2012.11830",
    archivePrefix = "arXiv",
    primaryClass = "astro-ph.CO",
    reportNumber = "YITP-SB-2020-40",
    doi = "10.1103/PhysRevD.104.063523",
    journal = "Phys. Rev. D",
    volume = "104",
    number = "6",
    pages = "063523",
    year = "2021"
}

@article{Taule:2022jrz,
    author = "Taule, Petter and Escudero, Miguel and Garny, Mathias",
    title = "{Global view of neutrino interactions in cosmology: The free streaming window as seen by Planck}",
    eprint = "2207.04062",
    archivePrefix = "arXiv",
    primaryClass = "astro-ph.CO",
    reportNumber = "TUM-HEP-1406/22",
    doi = "10.1103/PhysRevD.106.063539",
    journal = "Phys. Rev. D",
    volume = "106",
    number = "6",
    pages = "063539",
    year = "2022"
}

@article{Camarena:2024zck,
    author = "Camarena, David and Cyr-Racine, Francis-Yan",
    title = "{Strong constraints on a simple self-interacting neutrino cosmology}",
    eprint = "2403.05496",
    archivePrefix = "arXiv",
    primaryClass = "astro-ph.CO",
    doi = "10.1103/PhysRevD.111.023504",
    journal = "Phys. Rev. D",
    volume = "111",
    number = "2",
    pages = "023504",
    year = "2025"
}

@article{Archidiacono:2013dua,
    author = "Archidiacono, Maria and Hannestad, Steen",
    title = "{Updated constraints on non-standard neutrino interactions from Planck}",
    eprint = "1311.3873",
    archivePrefix = "arXiv",
    primaryClass = "astro-ph.CO",
    doi = "10.1088/1475-7516/2014/07/046",
    journal = "JCAP",
    volume = "07",
    pages = "046",
    year = "2014"
}

@article{Blum:2018ljv,
    author = "Blum, Kfir and Nir, Yosef and Shavit, Michal",
    title = "{Neutrinoless double-beta decay with massive scalar emission}",
    eprint = "1802.08019",
    archivePrefix = "arXiv",
    primaryClass = "hep-ph",
    doi = "10.1016/j.physletb.2018.08.022",
    journal = "Phys. Lett. B",
    volume = "785",
    pages = "354--361",
    year = "2018"
}

@article{Barabash:2019nnr,
    author = "Barabash, A. S.",
    editor = "Civitarese, Osvaldo and Stekl, Ivan and Suhonen, Jouni",
    title = "{Average and recommended half-life values for two-neutrino double beta decay: upgrade-2019}",
    eprint = "1907.06887",
    archivePrefix = "arXiv",
    primaryClass = "nucl-ex",
    doi = "10.1063/1.5130963",
    journal = "AIP Conf. Proc.",
    volume = "2165",
    number = "1",
    pages = "020002",
    year = "2019"
}

@article{Laha:2013xua,
    author = "Laha, Ranjan and Dasgupta, Basudeb and Beacom, John F.",
    title = "{Constraints on New Neutrino Interactions via Light Abelian Vector Bosons}",
    eprint = "1304.3460",
    archivePrefix = "arXiv",
    primaryClass = "hep-ph",
    doi = "10.1103/PhysRevD.89.093025",
    journal = "Phys. Rev. D",
    volume = "89",
    number = "9",
    pages = "093025",
    year = "2014"
}

@article{Carlson:2012pc,
    author = "Carlson, Carl E. and Rislow, Benjamin C.",
    title = "{New Physics and the Proton Radius Problem}",
    eprint = "1206.3587",
    archivePrefix = "arXiv",
    primaryClass = "hep-ph",
    doi = "10.1103/PhysRevD.86.035013",
    journal = "Phys. Rev. D",
    volume = "86",
    pages = "035013",
    year = "2012"
}

@article{Dutta:2021cip,
    author = "Dutta, Bhaskar and Kim, Doojin and Thompson, Adrian and Thornton, Remington T. and Van de Water, Richard G.",
    title = "{Solutions to the MiniBooNE Anomaly from New Physics in Charged Meson Decays}",
    eprint = "2110.11944",
    archivePrefix = "arXiv",
    primaryClass = "hep-ph",
    reportNumber = "MI-HET-766, LA-UR-21-30532",
    doi = "10.1103/PhysRevLett.129.111803",
    journal = "Phys. Rev. Lett.",
    volume = "129",
    number = "11",
    pages = "111803",
    year = "2022"
}

@article{Dutta:2023fnl,
    author = "Dutta, Bhaskar and Karthikeyan, Aparajitha and Kim, Doojin",
    title = "{Longer-lived mediators from charged mesons and photons at neutrino experiments}",
    eprint = "2308.01491",
    archivePrefix = "arXiv",
    primaryClass = "hep-ph",
    reportNumber = "MI-HET-809",
    doi = "10.1103/PhysRevD.109.075029",
    journal = "Phys. Rev. D",
    volume = "109",
    number = "7",
    pages = "075029",
    year = "2024"
}

@article{Krnjaic:2019rsv,
    author = "Krnjaic, Gordan and Marques-Tavares, Gustavo and Redigolo, Diego and Tobioka, Kohsaku",
    title = "{Probing Muonphilic Force Carriers and Dark Matter at Kaon Factories}",
    eprint = "1902.07715",
    archivePrefix = "arXiv",
    primaryClass = "hep-ph",
    reportNumber = "FERMILAB-PUB-18-665-A, KEK-TH-2105",
    doi = "10.1103/PhysRevLett.124.041802",
    journal = "Phys. Rev. Lett.",
    volume = "124",
    number = "4",
    pages = "041802",
    year = "2020"
}

@article{Barger:1981vd,
    author = "Barger, Vernon D. and Keung, Wai-Yee and Pakvasa, S.",
    title = "{Majoron Emission by Neutrinos}",
    reportNumber = "MAD/PH/15",
    doi = "10.1103/PhysRevD.25.907",
    journal = "Phys. Rev. D",
    volume = "25",
    pages = "907",
    year = "1982"
}

@article{Brdar:2020nbj,
    author = "Brdar, Vedran and Lindner, Manfred and Vogl, Stefan and Xu, Xun-Jie",
    title = "{Revisiting neutrino self-interaction constraints from $Z$ and $\tau$ decays}",
    eprint = "2003.05339",
    archivePrefix = "arXiv",
    primaryClass = "hep-ph",
    doi = "10.1103/PhysRevD.101.115001",
    journal = "Phys. Rev. D",
    volume = "101",
    number = "11",
    pages = "115001",
    year = "2020"
}

@article{Dev:2024ygx,
    author = "Dev, P. S. Bhupal and Kim, Doojin and Sathyan, Deepak and Sinha, Kuver and Zhang, Yongchao",
    title = "{New laboratory constraints on neutrinophilic mediators}",
    eprint = "2407.12738",
    archivePrefix = "arXiv",
    primaryClass = "hep-ph",
    reportNumber = "CETUP-2024-005",
    doi = "10.1016/j.physletb.2025.139765",
    journal = "Phys. Lett. B",
    volume = "868",
    pages = "139765",
    year = "2025"
}

@article{Rai:2021vvq,
    author = "Rai, Mudit and Chen, Lisong and Boyanovsky, Daniel",
    title = "{Infrared dressing in real time: emergence of anomalous dimensions}",
    eprint = "2105.06572",
    archivePrefix = "arXiv",
    primaryClass = "hep-ph",
    doi = "10.1103/PhysRevD.104.085021",
    journal = "Phys. Rev. D",
    volume = "104",
    pages = "085021",
    year = "2021"
}

@book{Peskin:1995ev,
    author = "Peskin, Michael E. and Schroeder, Daniel V.",
    title = "{An Introduction to quantum field theory}",
    doi = "10.1201/9780429503559",
    isbn = "978-0-201-50397-5, 978-0-429-50355-9, 978-0-429-49417-8",
    publisher = "Addison-Wesley",
    address = "Reading, USA",
    year = "1995"
}

@article{Agarwal:2021ais,
    author = "Agarwal, Neelima and Magnea, Lorenzo and Signorile-Signorile, Chiara and Tripathi, Anurag",
    title = "{The infrared structure of perturbative gauge theories}",
    eprint = "2112.07099",
    archivePrefix = "arXiv",
    primaryClass = "hep-ph",
    reportNumber = "CERN-TH-2021-145",
    doi = "10.1016/j.physrep.2022.10.001",
    journal = "Phys. Rept.",
    volume = "994",
    pages = "1--120",
    year = "2023"
}

@article{Kinoshita:1962ur,
    author = "Kinoshita, T.",
    title = "{Mass singularities of Feynman amplitudes}",
    doi = "10.1063/1.1724268",
    journal = "J. Math. Phys.",
    volume = "3",
    pages = "650--677",
    year = "1962"
}

@article{Lee:1964is,
    author = "Lee, T. D. and Nauenberg, M.",
    editor = "Feinberg, G.",
    title = "{Degenerate Systems and Mass Singularities}",
    doi = "10.1103/PhysRev.133.B1549",
    journal = "Phys. Rev.",
    volume = "133",
    pages = "B1549--B1562",
    year = "1964"
}

@article{PIENU:2021clt,
    author = "Aguilar-Arevalo, A. and others",
    collaboration = "PIENU",
    title = "{Search for three body pion decays ${\pi}^+{\to}l^+{\nu}X$}",
    eprint = "2101.07381",
    archivePrefix = "arXiv",
    primaryClass = "hep-ex",
    doi = "10.1103/PhysRevD.103.052006",
    journal = "Phys. Rev. D",
    volume = "103",
    number = "5",
    pages = "052006",
    year = "2021"
}

@article{NA62:2021bji,
    author = "Cortina Gil, Eduardo and others",
    collaboration = "NA62",
    title = "{Search for $K^+$ decays to a muon and invisible particles}",
    eprint = "2101.12304",
    archivePrefix = "arXiv",
    primaryClass = "hep-ex",
    reportNumber = "CERN-EP-2021-018",
    doi = "10.1016/j.physletb.2021.136259",
    journal = "Phys. Lett. B",
    volume = "816",
    pages = "136259",
    year = "2021"
}

@article{Barger:2011mt,
    author = "Barger, Vernon and Chiang, Cheng-Wei and Keung, Wai-Yee and Marfatia, Danny",
    title = "{Constraint on parity-violating muonic forces}",
    eprint = "1109.6652",
    archivePrefix = "arXiv",
    primaryClass = "hep-ph",
    doi = "10.1103/PhysRevLett.108.081802",
    journal = "Phys. Rev. Lett.",
    volume = "108",
    pages = "081802",
    year = "2012"
}

@article{Bakhti:2017jhm,
    author = "Bakhti, Pouya and Farzan, Yasaman",
    title = "{Constraining secret gauge interactions of neutrinos by meson decays}",
    eprint = "1702.04187",
    archivePrefix = "arXiv",
    primaryClass = "hep-ph",
    doi = "10.1103/PhysRevD.95.095008",
    journal = "Phys. Rev. D",
    volume = "95",
    number = "9",
    pages = "095008",
    year = "2017"
}

@article{Huang:2024tbo,
    author = "Huang, Jihong and Zhou, Shun",
    title = "{Helicity-changing decays of cosmological relic neutrinos}",
    eprint = "2407.04932",
    archivePrefix = "arXiv",
    primaryClass = "hep-ph",
    doi = "10.1088/1475-7516/2024/09/067",
    journal = "JCAP",
    volume = "09",
    pages = "067",
    year = "2024"
}

@article{Bickendorf:2022buy,
    author = "Bickendorf, Gerrit and Drees, Manuel",
    title = "{Constraints on light leptophilic dark matter mediators from decay experiments}",
    eprint = "2206.05038",
    archivePrefix = "arXiv",
    primaryClass = "hep-ph",
    doi = "10.1140/epjc/s10052-022-11128-9",
    journal = "Eur. Phys. J. C",
    volume = "82",
    number = "12",
    pages = "1163",
    year = "2022"
}

@article{Smirnov:2022sfo,
    author = "Smirnov, Alexei Yu. and Xu, Xun-Jie",
    title = "{Neutrino bound states and bound systems}",
    eprint = "2201.00939",
    archivePrefix = "arXiv",
    primaryClass = "hep-ph",
    doi = "10.1007/JHEP08(2022)170",
    journal = "JHEP",
    volume = "08",
    pages = "170",
    year = "2022"
}

@article{Kelly:2021mcd,
    author = "Kelly, Kevin J. and Kling, Felix and Tuckler, Douglas and Zhang, Yue",
    title = "{Probing neutrino-portal dark matter at the Forward Physics Facility}",
    eprint = "2111.05868",
    archivePrefix = "arXiv",
    primaryClass = "hep-ph",
    reportNumber = "FERMILAB-PUB-21-588-T, CERN-TH-2021-186, DESY 21-188",
    doi = "10.1103/PhysRevD.105.075026",
    journal = "Phys. Rev. D",
    volume = "105",
    number = "7",
    pages = "075026",
    year = "2022"
}

@article{Batell:2021blf,
    author = "Batell, Brian and Feng, Jonathan L. and Trojanowski, Sebastian",
    title = "{Detecting Dark Matter with Far-Forward Emulsion and Liquid Argon Detectors at the LHC}",
    eprint = "2101.10338",
    archivePrefix = "arXiv",
    primaryClass = "hep-ph",
    reportNumber = "PITT-PACC-2101, UCI-TR-2021-01",
    doi = "10.1103/PhysRevD.103.075023",
    journal = "Phys. Rev. D",
    volume = "103",
    number = "7",
    pages = "075023",
    year = "2021"
}

@article{Chikashige:1980ui,
    author = "Chikashige, Y. and Mohapatra, Rabindra N. and Peccei, R. D.",
    title = "{Are There Real Goldstone Bosons Associated with Broken Lepton Number?}",
    reportNumber = "MPI-PAE-PTH-36-80",
    doi = "10.1016/0370-2693(81)90011-3",
    journal = "Phys. Lett. B",
    volume = "98",
    pages = "265--268",
    year = "1981"
}

@article{Borexino:2008gab,
    author = "Alimonti, G. and others",
    collaboration = "Borexino",
    title = "{The Borexino detector at the Laboratori Nazionali del Gran Sasso}",
    eprint = "0806.2400",
    archivePrefix = "arXiv",
    primaryClass = "physics.ins-det",
    doi = "10.1016/j.nima.2008.11.076",
    journal = "Nucl. Instrum. Meth. A",
    volume = "600",
    pages = "568--593",
    year = "2009"
}

@article{COHERENT:2017ipa,
    author = "Akimov, D. and others",
    collaboration = "COHERENT",
    title = "{Observation of Coherent Elastic Neutrino-Nucleus Scattering}",
    eprint = "1708.01294",
    archivePrefix = "arXiv",
    primaryClass = "nucl-ex",
    doi = "10.1126/science.aao0990",
    journal = "Science",
    volume = "357",
    number = "6356",
    pages = "1123--1126",
    year = "2017"
}

@article{Bauer:2020itv,
    author = "Bauer, Martin and Foldenauer, Patrick and Mosny, Martin",
    title = "{Flavor structure of anomaly-free hidden photon models}",
    eprint = "2011.12973",
    archivePrefix = "arXiv",
    primaryClass = "hep-ph",
    reportNumber = "IPPP/20/59",
    doi = "10.1103/PhysRevD.103.075024",
    journal = "Phys. Rev. D",
    volume = "103",
    number = "7",
    pages = "075024",
    year = "2021"
}

@article{Rodejohann:2019quz,
    author = "Rodejohann, Werner and Xu, Xun-Jie",
    title = "{Loop-enhanced rate of neutrinoless double beta decay}",
    eprint = "1907.12478",
    archivePrefix = "arXiv",
    primaryClass = "hep-ph",
    doi = "10.1007/JHEP11(2019)029",
    journal = "JHEP",
    volume = "11",
    pages = "029",
    year = "2019"
}

@article{Park:2019ibn,
    author = "Park, Minsu and Kreisch, Christina D. and Dunkley, Jo and Hadzhiyska, Boryana and Cyr-Racine, Francis-Yan",
    title = "{$\Lambda$CDM or self-interacting neutrinos: How CMB data can tell the two models apart}",
    eprint = "1904.02625",
    archivePrefix = "arXiv",
    primaryClass = "astro-ph.CO",
    doi = "10.1103/PhysRevD.100.063524",
    journal = "Phys. Rev. D",
    volume = "100",
    number = "6",
    pages = "063524",
    year = "2019"
}

@article{Kelly:2019wow,
    author = "Kelly, Kevin J. and Zhang, Yue",
    title = "{Mononeutrino at DUNE: New Signals from Neutrinophilic Thermal Dark Matter}",
    eprint = "1901.01259",
    archivePrefix = "arXiv",
    primaryClass = "hep-ph",
    reportNumber = "FERMILAB-PUB-19-002-T",
    doi = "10.1103/PhysRevD.99.055034",
    journal = "Phys. Rev. D",
    volume = "99",
    number = "5",
    pages = "055034",
    year = "2019"
}

@article{Ge:2021lur,
    author = "Ge, Shao-Feng and Pasquini, Pedro",
    title = "{Probing light mediators in the radiative emission of neutrino pair}",
    eprint = "2110.03510",
    archivePrefix = "arXiv",
    primaryClass = "hep-ph",
    doi = "10.1140/epjc/s10052-022-10131-4",
    journal = "Eur. Phys. J. C",
    volume = "82",
    number = "3",
    pages = "208",
    year = "2022"
}

@article{Song:2015xaa,
    author = "Song, Ningqiang and Boyero Garcia, R. and Gomez-Cadenas, J. J. and Gonzalez-Garcia, M. C. and Peralta Conde, A. and Taron, Josep",
    title = "{Conditions for Statistical Determination of the Neutrino Mass Spectrum in Radiative Emission of Neutrino Pairs in Atoms}",
    eprint = "1510.00421",
    archivePrefix = "arXiv",
    primaryClass = "hep-ph",
    reportNumber = "YITP-SB-15-38",
    doi = "10.1103/PhysRevD.93.013020",
    journal = "Phys. Rev. D",
    volume = "93",
    number = "1",
    pages = "013020",
    year = "2016"
}

@article{Choubey:2017eyg,
    author = "Choubey, Sandhya and Goswami, Srubabati and Gupta, Chandan and Lakshmi, S. M. and Thakore, Tarak",
    title = "{Sensitivity to neutrino decay with atmospheric neutrinos at the INO-ICAL detector}",
    eprint = "1709.10376",
    archivePrefix = "arXiv",
    primaryClass = "hep-ph",
    doi = "10.1103/PhysRevD.97.033005",
    journal = "Phys. Rev. D",
    volume = "97",
    number = "3",
    pages = "033005",
    year = "2018"
}

@article{Lancaster:2017ksf,
    author = "Lancaster, Lachlan and Cyr-Racine, Francis-Yan and Knox, Lloyd and Pan, Zhen",
    title = "{A tale of two modes: Neutrino free-streaming in the early universe}",
    eprint = "1704.06657",
    archivePrefix = "arXiv",
    primaryClass = "astro-ph.CO",
    doi = "10.1088/1475-7516/2017/07/033",
    journal = "JCAP",
    volume = "07",
    pages = "033",
    year = "2017"
}

@article{Abrahao:2015rba,
    author = "Abrah\~ao, Thamys and Minakata, Hisakazu and Nunokawa, Hiroshi and Quiroga, Alexander A.",
    title = "{Constraint on Neutrino Decay with Medium-Baseline Reactor Neutrino Oscillation Experiments}",
    eprint = "1506.02314",
    archivePrefix = "arXiv",
    primaryClass = "hep-ph",
    reportNumber = "INT-PUB-15-024",
    doi = "10.1007/JHEP11(2015)001",
    journal = "JHEP",
    volume = "11",
    pages = "001",
    year = "2015"
}

@article{Ibe:2014pja,
    author = "Ibe, Masahiro and Kaneta, Kunio",
    title = "{Cosmic neutrino background absorption line in the neutrino spectrum at IceCube}",
    eprint = "1407.2848",
    archivePrefix = "arXiv",
    primaryClass = "hep-ph",
    reportNumber = "IPMU-14-0160, ICRR-REPROT-687-2014-13",
    doi = "10.1103/PhysRevD.90.053011",
    journal = "Phys. Rev. D",
    volume = "90",
    number = "5",
    pages = "053011",
    year = "2014"
}

@article{Ioka:2014kca,
    author = "Ioka, Kunihto and Murase, Kohta",
    title = "{IceCube PeV\textendash{}EeV neutrinos and secret interactions of neutrinos}",
    eprint = "1404.2279",
    archivePrefix = "arXiv",
    primaryClass = "astro-ph.HE",
    reportNumber = "KEK-TH-1723, KEK-COSMO-141",
    doi = "10.1093/ptep/ptu090",
    journal = "PTEP",
    volume = "2014",
    number = "6",
    pages = "061E01",
    year = "2014"
}

@article{Ng:2014pca,
    author = "Ng, Kenny C. Y. and Beacom, John F.",
    title = "{Cosmic neutrino cascades from secret neutrino interactions}",
    eprint = "1404.2288",
    archivePrefix = "arXiv",
    primaryClass = "astro-ph.HE",
    doi = "10.1103/PhysRevD.90.065035",
    journal = "Phys. Rev. D",
    volume = "90",
    number = "6",
    pages = "065035",
    year = "2014",
    note = "[Erratum: Phys.Rev.D 90, 089904 (2014)]"
}

@article{Cyr-Racine:2013jua,
    author = "Cyr-Racine, Francis-Yan and Sigurdson, Kris",
    title = "{Limits on Neutrino-Neutrino Scattering in the Early Universe}",
    eprint = "1306.1536",
    archivePrefix = "arXiv",
    primaryClass = "astro-ph.CO",
    doi = "10.1103/PhysRevD.90.123533",
    journal = "Phys. Rev. D",
    volume = "90",
    number = "12",
    pages = "123533",
    year = "2014"
}

@article{Bell:2005dr,
    author = "Bell, Nicole F. and Pierpaoli, Elena and Sigurdson, Kris",
    title = "{Cosmological signatures of interacting neutrinos}",
    eprint = "astro-ph/0511410",
    archivePrefix = "arXiv",
    reportNumber = "KRL-MAP-309",
    doi = "10.1103/PhysRevD.73.063523",
    journal = "Phys. Rev. D",
    volume = "73",
    pages = "063523",
    year = "2006"
}

@article{Beacom:2004yd,
    author = "Beacom, John F. and Bell, Nicole F. and Dodelson, Scott",
    title = "{Neutrinoless universe}",
    eprint = "astro-ph/0404585",
    archivePrefix = "arXiv",
    reportNumber = "FERMILAB-PUB-04-050-A",
    doi = "10.1103/PhysRevLett.93.121302",
    journal = "Phys. Rev. Lett.",
    volume = "93",
    pages = "121302",
    year = "2004"
}

@article{Farzan:2002wx,
    author = "Farzan, Yasaman",
    title = "{Bounds on the coupling of the Majoron to light neutrinos from supernova cooling}",
    eprint = "hep-ph/0211375",
    archivePrefix = "arXiv",
    reportNumber = "SLAC-PUB-9543, SISSA-69-2002-EP",
    doi = "10.1103/PhysRevD.67.073015",
    journal = "Phys. Rev. D",
    volume = "67",
    pages = "073015",
    year = "2003"
}

@article{Beacom:2002cb,
    author = "Beacom, John F. and Bell, Nicole F.",
    title = "{Do Solar Neutrinos Decay?}",
    eprint = "hep-ph/0204111",
    archivePrefix = "arXiv",
    reportNumber = "FERMILAB-PUB-02-061-A",
    doi = "10.1103/PhysRevD.65.113009",
    journal = "Phys. Rev. D",
    volume = "65",
    pages = "113009",
    year = "2002"
}

@article{Ansarifard:2024zxm,
    author = "Ansarifard, Saeed and Gonzalez-Garcia, M. C. and Maltoni, Michele and Pinheiro, Joao Paulo",
    title = "{Solar neutrinos and leptonic spin forces}",
    eprint = "2405.05340",
    archivePrefix = "arXiv",
    primaryClass = "hep-ph",
    reportNumber = "IFT-UAM/CSIC-24-67, YITP-SB-2024-08",
    doi = "10.1007/JHEP07(2024)172",
    journal = "JHEP",
    volume = "07",
    pages = "172",
    year = "2024"
}

@article{Lindner:2001th,
    author = "Lindner, Manfred and Ohlsson, Tommy and Winter, Walter",
    title = "{Decays of supernova neutrinos}",
    eprint = "astro-ph/0105309",
    archivePrefix = "arXiv",
    reportNumber = "TUM-HEP-417-01",
    doi = "10.1016/S0550-3213(01)00603-4",
    journal = "Nucl. Phys. B",
    volume = "622",
    pages = "429--456",
    year = "2002"
}

@article{Kachelriess:2000qc,
    author = "Kachelriess, M. and Tomas, R. and Valle, J. W. F.",
    title = "{Supernova bounds on Majoron emitting decays of light neutrinos}",
    eprint = "hep-ph/0001039",
    archivePrefix = "arXiv",
    reportNumber = "CERN-TH-2000-002, FTUV-99-71, IFIC-99-74",
    doi = "10.1103/PhysRevD.62.023004",
    journal = "Phys. Rev. D",
    volume = "62",
    pages = "023004",
    year = "2000"
}

@inproceedings{Bilenky:1999dn,
    author = "Bilenky, Mikhail S. and Santamaria, Arcadi",
    title = "{'Secret' neutrino interactions}",
    booktitle = "{Neutrino Mixing: Meeting in Honor of Samoil Bilenky's 70th Birthday}",
    eprint = "hep-ph/9908272",
    archivePrefix = "arXiv",
    reportNumber = "FTUV-99-57, IFIC-99-60",
    pages = "50--61",
    month = "8",
    year = "1999"
}

@article{Dicus:1988jh,
    author = "Dicus, Duane A. and Nussinov, S. and Pal, Palash B. and Teplitz, Vigdor L.",
    title = "{Implications of Relativistic Gas Dynamics for Neutrino-neutrino Cross-sections}",
    reportNumber = "DOE-ER40200-161",
    doi = "10.1016/0370-2693(89)90480-2",
    journal = "Phys. Lett. B",
    volume = "218",
    pages = "84--90",
    year = "1989"
}

@article{Manohar:1987ec,
    author = "Manohar, Aneesh",
    title = "{A Limit on the Neutrino-neutrino Scattering Cross-section From the Supernova}",
    reportNumber = "MIT-CTP-1461",
    doi = "10.1016/0370-2693(87)91171-3",
    journal = "Phys. Lett. B",
    volume = "192",
    pages = "217",
    year = "1987"
}

@article{Bardin:1970wq,
    author = "Bardin, D. Yu. and Bilenky, Samoil M. and Pontecorvo, B.",
    title = "{On the nu - nu interaction}",
    doi = "10.1016/0370-2693(70)90602-7",
    journal = "Phys. Lett. B",
    volume = "32",
    pages = "121--124",
    year = "1970"
}

@article{Camarena:2023cku,
    author = "Camarena, David and Cyr-Racine, Francis-Yan and Houghteling, John",
    title = "{Confronting self-interacting neutrinos with the full shape of the galaxy power spectrum}",
    eprint = "2309.03941",
    archivePrefix = "arXiv",
    primaryClass = "astro-ph.CO",
    doi = "10.1103/PhysRevD.108.103535",
    journal = "Phys. Rev. D",
    volume = "108",
    number = "10",
    pages = "103535",
    year = "2023"
}

@article{Suliga:2020jfa,
    author = "Suliga, Anna M. and Tamborra, Irene",
    title = "{Astrophysical constraints on nonstandard coherent neutrino-nucleus scattering}",
    eprint = "2010.14545",
    archivePrefix = "arXiv",
    primaryClass = "hep-ph",
    doi = "10.1103/PhysRevD.103.083002",
    journal = "Phys. Rev. D",
    volume = "103",
    number = "8",
    pages = "083002",
    year = "2021"
}

@article{He:2023oke,
    author = "He, Adam and An, Rui and Ivanov, Mikhail M. and Gluscevic, Vera",
    title = "{Self-interacting neutrinos in light of large-scale structure data}",
    eprint = "2309.03956",
    archivePrefix = "arXiv",
    primaryClass = "astro-ph.CO",
    reportNumber = "MIT-CTP/5608",
    doi = "10.1103/PhysRevD.109.103527",
    journal = "Phys. Rev. D",
    volume = "109",
    number = "10",
    pages = "103527",
    year = "2024"
}

@article{Kreisch:2022zxp,
    author = "Kreisch, Christina D. and others",
    title = "{Atacama Cosmology Telescope: The persistence of neutrino self-interaction in cosmological measurements}",
    eprint = "2207.03164",
    archivePrefix = "arXiv",
    primaryClass = "astro-ph.CO",
    doi = "10.1103/PhysRevD.109.043501",
    journal = "Phys. Rev. D",
    volume = "109",
    number = "4",
    pages = "043501",
    year = "2024"
}

@article{Forastieri:2019cuf,
    author = "Forastieri, Francesco and Lattanzi, Massimiliano and Natoli, Paolo",
    title = "{Cosmological constraints on neutrino self-interactions with a light mediator}",
    eprint = "1904.07810",
    archivePrefix = "arXiv",
    primaryClass = "astro-ph.CO",
    doi = "10.1103/PhysRevD.100.103526",
    journal = "Phys. Rev. D",
    volume = "100",
    number = "10",
    pages = "103526",
    year = "2019"
}

@article{Yang:2018yvk,
    author = "Yang, Yue and Kneller, James P.",
    title = "{Neutrino flavor transformation in supernovae as a probe for nonstandard neutrino-scalar interactions}",
    eprint = "1803.04504",
    archivePrefix = "arXiv",
    primaryClass = "astro-ph.HE",
    doi = "10.1103/PhysRevD.97.103018",
    journal = "Phys. Rev. D",
    volume = "97",
    number = "10",
    pages = "103018",
    year = "2018"
}

@article{Dighe:2017sur,
    author = "Dighe, Amol and Sen, Manibrata",
    title = "{Nonstandard neutrino self-interactions in a supernova and fast flavor conversions}",
    eprint = "1709.06858",
    archivePrefix = "arXiv",
    primaryClass = "hep-ph",
    reportNumber = "TIFR-TH-17-33",
    doi = "10.1103/PhysRevD.97.043011",
    journal = "Phys. Rev. D",
    volume = "97",
    number = "4",
    pages = "043011",
    year = "2018"
}

@article{Das:2017iuj,
    author = "Das, Anirban and Dighe, Amol and Sen, Manibrata",
    title = "{New effects of non-standard self-interactions of neutrinos in a supernova}",
    eprint = "1705.00468",
    archivePrefix = "arXiv",
    primaryClass = "hep-ph",
    reportNumber = "TIFR-TH-17-18",
    doi = "10.1088/1475-7516/2017/05/051",
    journal = "JCAP",
    volume = "05",
    pages = "051",
    year = "2017"
}

@article{Arhrib:2015dez,
    author = "Arhrib, Abdesslam and B\oe{}hm, C\'eline and Ma, Ernest and Yuan, Tzu-Chiang",
    title = "{Radiative Model of Neutrino Mass with Neutrino Interacting MeV Dark Matter}",
    eprint = "1512.08796",
    archivePrefix = "arXiv",
    primaryClass = "hep-ph",
    doi = "10.1088/1475-7516/2016/04/049",
    journal = "JCAP",
    volume = "04",
    pages = "049",
    year = "2016"
}

@article{Araki:2015mya,
    author = "Araki, Takeshi and Kaneko, Fumihiro and Ota, Toshihiko and Sato, Joe and Shimomura, Takashi",
    title = "{MeV scale leptonic force for cosmic neutrino spectrum and muon anomalous magnetic moment}",
    eprint = "1508.07471",
    archivePrefix = "arXiv",
    primaryClass = "hep-ph",
    reportNumber = "UME-PP-002, STUPP-15-223",
    doi = "10.1103/PhysRevD.93.013014",
    journal = "Phys. Rev. D",
    volume = "93",
    number = "1",
    pages = "013014",
    year = "2016"
}

@article{Araki:2014ona,
    author = "Araki, Takeshi and Kaneko, Fumihiro and Konishi, Yasufumi and Ota, Toshihiko and Sato, Joe and Shimomura, Takashi",
    title = "{Cosmic neutrino spectrum and the muon anomalous magnetic moment in the gauged $L_\mu-L_\tau$ model}",
    eprint = "1409.4180",
    archivePrefix = "arXiv",
    primaryClass = "hep-ph",
    reportNumber = "STUPP-14-219",
    doi = "10.1103/PhysRevD.91.037301",
    journal = "Phys. Rev. D",
    volume = "91",
    number = "3",
    pages = "037301",
    year = "2015"
}

@article{Blennow:2008er,
    author = "Blennow, Mattias and Mirizzi, Alessandro and Serpico, Pasquale D.",
    title = "{Nonstandard neutrino-neutrino refractive effects in dense neutrino gases}",
    eprint = "0810.2297",
    archivePrefix = "arXiv",
    primaryClass = "hep-ph",
    reportNumber = "FERMILAB-PUB-08-532-A, MPP-2008-134",
    doi = "10.1103/PhysRevD.78.113004",
    journal = "Phys. Rev. D",
    volume = "78",
    pages = "113004",
    year = "2008"
}

@article{Gavela:2008ra,
    author = "Gavela, M. B. and Hernandez, D. and Ota, T. and Winter, W.",
    title = "{Large gauge invariant non-standard neutrino interactions}",
    eprint = "0809.3451",
    archivePrefix = "arXiv",
    primaryClass = "hep-ph",
    reportNumber = "FTUAM-08-15, IFT-UAM-CSIC-08-53, LPT-ORSAY-08-75",
    doi = "10.1103/PhysRevD.79.013007",
    journal = "Phys. Rev. D",
    volume = "79",
    pages = "013007",
    year = "2009"
}

@article{Kolb:1987qy,
    author = "Kolb, Edward W. and Turner, Michael S.",
    title = "{Supernova SN 1987a and the Secret Interactions of Neutrinos}",
    reportNumber = "FERMILAB-PUB-87-110-A",
    doi = "10.1103/PhysRevD.36.2895",
    journal = "Phys. Rev. D",
    volume = "36",
    pages = "2895",
    year = "1987"
}

@article{Heurtier:2016otg,
    author = "Heurtier, Lucien and Zhang, Yongchao",
    title = "{Supernova Constraints on Massive (Pseudo)Scalar Coupling to Neutrinos}",
    eprint = "1609.05882",
    archivePrefix = "arXiv",
    primaryClass = "hep-ph",
    reportNumber = "ULB-TH-16-16",
    doi = "10.1088/1475-7516/2017/02/042",
    journal = "JCAP",
    volume = "02",
    pages = "042",
    year = "2017"
}

@article{Chang:2022aas,
    author = "Chang, Po-Wen and Esteban, Ivan and Beacom, John F. and Thompson, Todd A. and Hirata, Christopher M.",
    title = "{Toward Powerful Probes of Neutrino Self-Interactions in Supernovae}",
    eprint = "2206.12426",
    archivePrefix = "arXiv",
    primaryClass = "hep-ph",
    doi = "10.1103/PhysRevLett.131.071002",
    journal = "Phys. Rev. Lett.",
    volume = "131",
    number = "7",
    pages = "071002",
    year = "2023"
}

@article{Akita:2022etk,
    author = "Akita, Kensuke and Im, Sang Hui and Masud, Mehedi",
    title = "{Probing non-standard neutrino interactions with a light boson from next galactic and diffuse supernova neutrinos}",
    eprint = "2206.06852",
    archivePrefix = "arXiv",
    primaryClass = "hep-ph",
    reportNumber = "CTPU-PTC-22-13",
    doi = "10.1007/JHEP12(2022)050",
    journal = "JHEP",
    volume = "12",
    pages = "050",
    year = "2022"
}

@article{Reddy:2021rln,
    author = "Reddy, Sanjay and Zhou, Dake",
    title = "{Dark lepton superfluid in protoneutron stars}",
    eprint = "2107.06279",
    archivePrefix = "arXiv",
    primaryClass = "hep-ph",
    reportNumber = "INT-PUB-21-016",
    doi = "10.1103/PhysRevD.105.023026",
    journal = "Phys. Rev. D",
    volume = "105",
    number = "2",
    pages = "023026",
    year = "2022"
}

@article{Shalgar:2019rqe,
    author = "Shalgar, Shashank and Tamborra, Irene and Bustamante, Mauricio",
    title = "{Core-collapse supernovae stymie secret neutrino interactions}",
    eprint = "1912.09115",
    archivePrefix = "arXiv",
    primaryClass = "astro-ph.HE",
    doi = "10.1103/PhysRevD.103.123008",
    journal = "Phys. Rev. D",
    volume = "103",
    number = "12",
    pages = "123008",
    year = "2021"
}

@article{Davoudiasl:2005fd,
    author = "Davoudiasl, Hooman and Huber, Patrick",
    title = "{Probing the origins of neutrino mass with supernova data}",
    eprint = "hep-ph/0504265",
    archivePrefix = "arXiv",
    reportNumber = "MADPH-05-1424",
    doi = "10.1103/PhysRevLett.95.191302",
    journal = "Phys. Rev. Lett.",
    volume = "95",
    pages = "191302",
    year = "2005"
}

@article{Choi:1989hi,
    author = "Choi, Kiwoon and Santamaria, A.",
    title = "{Majorons and Supernova Cooling}",
    reportNumber = "CMU-HEP89-23, MPI-PAE-PTH-75-89",
    doi = "10.1103/PhysRevD.42.293",
    journal = "Phys. Rev. D",
    volume = "42",
    pages = "293--306",
    year = "1990"
}

@article{Zhou:2011rc,
    author = "Zhou, Shun",
    title = "{Comment on astrophysical consequences of a neutrinophilic 2HDM}",
    eprint = "1106.3880",
    archivePrefix = "arXiv",
    primaryClass = "hep-ph",
    reportNumber = "MPP-2011-70",
    doi = "10.1103/PhysRevD.84.038701",
    journal = "Phys. Rev. D",
    volume = "84",
    pages = "038701",
    year = "2011"
}

@article{Galais:2011jh,
    author = "Galais, S\'ebastien and Kneller, James and Volpe, Cristina",
    title = "{The neutrino-neutrino interaction effects in supernovae: the point of view from the matter basis}",
    eprint = "1102.1471",
    archivePrefix = "arXiv",
    primaryClass = "astro-ph.SR",
    doi = "10.1088/0954-3899/39/3/035201",
    journal = "J. Phys. G",
    volume = "39",
    pages = "035201",
    year = "2012"
}

@article{Sher:2011mx,
    author = "Sher, Marc and Triola, Christopher",
    title = "{Astrophysical Consequences of a Neutrinophilic Two-Higgs-Doublet Model}",
    eprint = "1105.4844",
    archivePrefix = "arXiv",
    primaryClass = "hep-ph",
    doi = "10.1103/PhysRevD.83.117702",
    journal = "Phys. Rev. D",
    volume = "83",
    pages = "117702",
    year = "2011"
}

@article{Fiorillo:2022cdq,
    author = "Fiorillo, Damiano F. G. and Raffelt, Georg G. and Vitagliano, Edoardo",
    title = "{Strong Supernova 1987A Constraints on Bosons Decaying to Neutrinos}",
    eprint = "2209.11773",
    archivePrefix = "arXiv",
    primaryClass = "hep-ph",
    doi = "10.1103/PhysRevLett.131.021001",
    journal = "Phys. Rev. Lett.",
    volume = "131",
    number = "2",
    pages = "021001",
    year = "2023"
}

@article{Fiorillo:2023ytr,
    author = "Fiorillo, Damiano F. G. and Raffelt, Georg G. and Vitagliano, Edoardo",
    title = "{Large Neutrino Secret Interactions Have a Small Impact on Supernovae}",
    eprint = "2307.15115",
    archivePrefix = "arXiv",
    primaryClass = "hep-ph",
    doi = "10.1103/PhysRevLett.132.021002",
    journal = "Phys. Rev. Lett.",
    volume = "132",
    number = "2",
    pages = "021002",
    year = "2024"
}

@article{Fiorillo:2023cas,
    author = "Fiorillo, Damiano F. G. and Raffelt, Georg G. and Vitagliano, Edoardo",
    title = "{Supernova emission of secretly interacting neutrino fluid: Theoretical foundations}",
    eprint = "2307.15122",
    archivePrefix = "arXiv",
    primaryClass = "hep-ph",
    doi = "10.1103/PhysRevD.109.023017",
    journal = "Phys. Rev. D",
    volume = "109",
    number = "2",
    pages = "023017",
    year = "2024"
}

@article{Fiorillo:2024upk,
    author = "Fiorillo, Damiano F. G. and Vitagliano, Edoardo",
    title = "{Self-Interacting Dark Sectors in Supernovae Can Behave as a Relativistic Fluid}",
    eprint = "2404.07714",
    archivePrefix = "arXiv",
    primaryClass = "hep-ph",
    doi = "10.1103/PhysRevLett.133.251004",
    journal = "Phys. Rev. Lett.",
    volume = "133",
    number = "25",
    pages = "251004",
    year = "2024"
}

@article{Barger:1999bg,
    author = "Barger, Vernon D. and Learned, J. G. and Lipari, P. and Lusignoli, Maurizio and Pakvasa, S. and Weiler, Thomas J.",
    title = "{Neutrino decay and atmospheric neutrinos}",
    eprint = "hep-ph/9907421",
    archivePrefix = "arXiv",
    reportNumber = "MADPH-99-1127, UH-511-931-99",
    doi = "10.1016/S0370-2693(99)00887-4",
    journal = "Phys. Lett. B",
    volume = "462",
    pages = "109--114",
    year = "1999"
}

@article{Barger:1998xk,
    author = "Barger, Vernon D. and Learned, J. G. and Pakvasa, S. and Weiler, Thomas J.",
    title = "{Neutrino decay as an explanation of atmospheric neutrino observations}",
    eprint = "astro-ph/9810121",
    archivePrefix = "arXiv",
    reportNumber = "MADPH-98-1077, UH-511-916-98, VAND-TH-98-17, FERMILAB-PUB-98-331-T",
    doi = "10.1103/PhysRevLett.82.2640",
    journal = "Phys. Rev. Lett.",
    volume = "82",
    pages = "2640--2643",
    year = "1999"
}

@article{Gelmini:1982rr,
    author = "Gelmini, Graciela B. and Nussinov, Shmuel and Roncadelli, Marco",
    title = "{Bounds and Prospects for the Majoron Model of Left-handed Neutrino Masses}",
    reportNumber = "MPI-PAE-PTH-37-82",
    doi = "10.1016/0550-3213(82)90107-9",
    journal = "Nucl. Phys. B",
    volume = "209",
    pages = "157--173",
    year = "1982"
}

@article{Goldman:1982bd,
    author = "Goldman, J. Terrance and Kolb, Edward W. and Stephenson, Jr., G. J.",
    title = "{Majorons and Muon Decay}",
    reportNumber = "LA-UR-82-1401",
    doi = "10.1103/PhysRevD.26.2503",
    journal = "Phys. Rev. D",
    volume = "26",
    pages = "2503",
    year = "1982"
}

@article{Stephenson:1993rx,
    author = "Stephenson, Jr., G. J. and Goldman, J. Terrance",
    title = "{Observable consequences of a scalar boson coupled only to neutrinos}",
    eprint = "hep-ph/9309308",
    archivePrefix = "arXiv",
    reportNumber = "LA-UR-93-3348",
    month = "9",
    year = "1993"
}

@article{Bilenky:1992xn,
    author = "Bilenky, Mikhail S. and Bilenky, Samoil M. and Santamaria, A.",
    title = "{Invisible width of the Z boson and 'secret' neutrino-neutrino interactions}",
    reportNumber = "FTUV-92-40, BI-TP-92-36",
    doi = "10.1016/0370-2693(93)90703-K",
    journal = "Phys. Lett. B",
    volume = "301",
    pages = "287--291",
    year = "1993"
}

@article{Santamaria:1985xa,
    author = "Santamaria, A. and Pich, A. and Bernabeu, J.",
    title = "{Single Majoron Emission in $\mu$ Decay}",
    reportNumber = "Print-85-0243 (VALENCIA)",
    doi = "10.1103/PhysRevD.32.2461",
    journal = "Phys. Rev. D",
    volume = "32",
    pages = "2461",
    year = "1985"
}

@article{Gelmini:1983ea,
    author = "Gelmini, G. B. and Valle, J. W. F.",
    title = "{Fast Invisible Neutrino Decays}",
    reportNumber = "CERN-TH-3791",
    doi = "10.1016/0370-2693(84)91258-9",
    journal = "Phys. Lett. B",
    volume = "142",
    pages = "181--187",
    year = "1984"
}

@article{Calatayud-Cadenillas:2024wdw,
    author = "Calatayud-Cadenillas, A. and P{\'e}rez-G, A. and Gago, A. M.",
    title = "{Distinguishing beyond-standard model effects in neutrino oscillation}",
    eprint = "2408.04234",
    archivePrefix = "arXiv",
    primaryClass = "hep-ph",
    doi = "10.1016/j.physletb.2025.139377",
    journal = "Phys. Lett. B",
    volume = "863",
    pages = "139377",
    year = "2025"
}

@article{Berezhiani:1987gf,
    author = "Berezhiani, Z. G. and Vysotsky, M. I.",
    title = "{Neutrino Decay in Matter}",
    reportNumber = "ITEP-44-1987",
    doi = "10.1016/0370-2693(87)91375-X",
    journal = "Phys. Lett. B",
    volume = "199",
    pages = "281",
    year = "1987"
}

@article{Mazumdar:2020ibx,
    author = "Mazumdar, Arindam and Mohanty, Subhendra and Parashari, Priyank",
    title = "{Flavour specific neutrino self-interaction: H $_{0}$ tension and IceCube}",
    eprint = "2011.13685",
    archivePrefix = "arXiv",
    primaryClass = "hep-ph",
    doi = "10.1088/1475-7516/2022/10/011",
    journal = "JCAP",
    volume = "10",
    pages = "011",
    year = "2022"
}

@article{Das:2020xke,
    author = "Das, Anirban and Ghosh, Subhajit",
    title = "{Flavor-specific interaction favors strong neutrino self-coupling in the early universe}",
    eprint = "2011.12315",
    archivePrefix = "arXiv",
    primaryClass = "astro-ph.CO",
    reportNumber = "SLAC-PUB-17547",
    doi = "10.1088/1475-7516/2021/07/038",
    journal = "JCAP",
    volume = "07",
    pages = "038",
    year = "2021"
}

@article{Lyu:2020lps,
    author = "Lyu, Kun-Feng and Stamou, Emmanuel and Wang, Lian-Tao",
    title = "{Self-interacting neutrinos: Solution to Hubble tension versus experimental constraints}",
    eprint = "2004.10868",
    archivePrefix = "arXiv",
    primaryClass = "hep-ph",
    doi = "10.1103/PhysRevD.103.015004",
    journal = "Phys. Rev. D",
    volume = "103",
    number = "1",
    pages = "015004",
    year = "2021"
}

@article{Escudero:2019gvw,
    author = "Escudero, Miguel and Witte, Samuel J.",
    title = "{A CMB search for the neutrino mass mechanism and its relation to the Hubble tension}",
    eprint = "1909.04044",
    archivePrefix = "arXiv",
    primaryClass = "astro-ph.CO",
    reportNumber = "KCL-2019-71",
    doi = "10.1140/epjc/s10052-020-7854-5",
    journal = "Eur. Phys. J. C",
    volume = "80",
    number = "4",
    pages = "294",
    year = "2020"
}

@article{Huang:2017egl,
    author = "Huang, Guo-yuan and Ohlsson, Tommy and Zhou, Shun",
    title = "{Observational Constraints on Secret Neutrino Interactions from Big Bang Nucleosynthesis}",
    eprint = "1712.04792",
    archivePrefix = "arXiv",
    primaryClass = "hep-ph",
    doi = "10.1103/PhysRevD.97.075009",
    journal = "Phys. Rev. D",
    volume = "97",
    number = "7",
    pages = "075009",
    year = "2018"
}

@article{Venzor:2020ova,
    author = "Venzor, Jorge and P\'erez-Lorenzana, Abdel and De-Santiago, Josue",
    title = "{Bounds on neutrino-scalar nonstandard interactions from big bang nucleosynthesis}",
    eprint = "2009.08104",
    archivePrefix = "arXiv",
    primaryClass = "hep-ph",
    doi = "10.1103/PhysRevD.103.043534",
    journal = "Phys. Rev. D",
    volume = "103",
    number = "4",
    pages = "043534",
    year = "2021"
}

@article{Grohs:2020xxd,
    author = "Grohs, E. and Fuller, George M. and Sen, Manibrata",
    title = "{Consequences of neutrino self interactions for weak decoupling and big bang nucleosynthesis}",
    eprint = "2002.08557",
    archivePrefix = "arXiv",
    primaryClass = "astro-ph.CO",
    doi = "10.1088/1475-7516/2020/07/001",
    journal = "JCAP",
    volume = "07",
    pages = "001",
    year = "2020"
}

@article{Huang:2021dba,
    author = "Huang, Guo-yuan and Rodejohann, Werner",
    title = "{Solving the Hubble tension without spoiling Big Bang Nucleosynthesis}",
    eprint = "2102.04280",
    archivePrefix = "arXiv",
    primaryClass = "hep-ph",
    doi = "10.1103/PhysRevD.103.123007",
    journal = "Phys. Rev. D",
    volume = "103",
    pages = "123007",
    year = "2021"
}

@article{Hyde:2023eph,
    author = "Hyde, Jeffrey M.",
    title = "{Constraints on Neutrino Self-Interactions from IceCube Observation of NGC 1068}",
    eprint = "2307.02361",
    archivePrefix = "arXiv",
    primaryClass = "hep-ph",
    month = "7",
    year = "2023"
}

@article{Bustamante:2020mep,
    author = "Bustamante, Mauricio and Rosenstr\o{}m, Charlotte and Shalgar, Shashank and Tamborra, Irene",
    title = "{Bounds on secret neutrino interactions from high-energy astrophysical neutrinos}",
    eprint = "2001.04994",
    archivePrefix = "arXiv",
    primaryClass = "astro-ph.HE",
    doi = "10.1103/PhysRevD.101.123024",
    journal = "Phys. Rev. D",
    volume = "101",
    number = "12",
    pages = "123024",
    year = "2020"
}

@article{Doring:2023vmk,
    author = {D\"oring, Christian and Vogl, Stefan},
    title = "{Testing secret interaction with astrophysical neutrino point sources}",
    eprint = "2304.08533",
    archivePrefix = "arXiv",
    primaryClass = "hep-ph",
    reportNumber = "ULB-TH/23-05",
    doi = "10.1088/1475-7516/2024/07/015",
    journal = "JCAP",
    volume = "07",
    pages = "015",
    year = "2024"
}

@article{Creque-Sarbinowski:2020qhz,
    author = "Creque-Sarbinowski, Cyril and Hyde, Jeffrey and Kamionkowski, Marc",
    title = "{Resonant neutrino self-interactions}",
    eprint = "2005.05332",
    archivePrefix = "arXiv",
    primaryClass = "hep-ph",
    doi = "10.1103/PhysRevD.103.023527",
    journal = "Phys. Rev. D",
    volume = "103",
    number = "2",
    pages = "023527",
    year = "2021"
}

@article{Bustamante:2020niz,
    author = "Bustamante, Mauricio",
    title = "{New limits on neutrino decay from the Glashow resonance of high-energy cosmic neutrinos}",
    eprint = "2004.06844",
    archivePrefix = "arXiv",
    primaryClass = "astro-ph.HE",
    month = "4",
    year = "2020"
}

@article{Arcadi:2018xdd,
    author = "Arcadi, Giorgio and Heeck, Julian and Heizmann, Florian and Mertens, Susanne and Queiroz, Farinaldo S. and Rodejohann, Werner and Slez\'ak, Martin and Valerius, Kathrin",
    title = "{Tritium beta decay with additional emission of new light bosons}",
    eprint = "1811.03530",
    archivePrefix = "arXiv",
    primaryClass = "hep-ph",
    reportNumber = "ULB-TH/18-12, UCI-TR-2018-14",
    doi = "10.1007/JHEP01(2019)206",
    journal = "JHEP",
    volume = "01",
    pages = "206",
    year = "2019"
}

@article{Huang:2018nxj,
    author = "Huang, Guo-Yuan and Zhou, Shun",
    title = "{Constraining Neutrino Lifetimes and Magnetic Moments via Solar Neutrinos in the Large Xenon Detectors}",
    eprint = "1810.03877",
    archivePrefix = "arXiv",
    primaryClass = "hep-ph",
    doi = "10.1088/1475-7516/2019/02/024",
    journal = "JCAP",
    volume = "02",
    pages = "024",
    year = "2019"
}

@article{Bustamante:2016ciw,
    author = "Bustamante, Mauricio and Beacom, John F. and Murase, Kohta",
    title = "{Testing decay of astrophysical neutrinos with incomplete information}",
    eprint = "1610.02096",
    archivePrefix = "arXiv",
    primaryClass = "astro-ph.HE",
    doi = "10.1103/PhysRevD.95.063013",
    journal = "Phys. Rev. D",
    volume = "95",
    number = "6",
    pages = "063013",
    year = "2017"
}

@article{Choi:1987sd,
    author = "Choi, Kiwoon and Kim, C. W. and Kim, Jewan and Lam, W. P.",
    title = "{Constraints on the Majoron Interactions From the Supernova {SN1987A}}",
    reportNumber = "JHU-TIPAC-8722",
    doi = "10.1103/PhysRevD.37.3225",
    journal = "Phys. Rev. D",
    volume = "37",
    pages = "3225",
    year = "1988"
}

@article{Santamaria:1986kg,
    author = "Santamaria, A. and Bernabeu, J. and Pich, A.",
    title = "{Neutrino Masses, Majorons and Muon Decay}",
    reportNumber = "Print-86-0907 (VALENCIA)",
    doi = "10.1103/PhysRevD.36.1408",
    journal = "Phys. Rev. D",
    volume = "36",
    pages = "1408",
    year = "1987"
}

@article{Glashow:1985cm,
    author = "Glashow, S. L. and Manohar, Aneesh",
    title = "{MAJORONS REVISITED}",
    reportNumber = "HUTP-85/A031",
    doi = "10.1103/PhysRevLett.54.2306",
    journal = "Phys. Rev. Lett.",
    volume = "54",
    pages = "2306",
    year = "1985"
}

@article{Berezhiani:1989za,
    author = "Berezhiani, Z. G. and Smirnov, A. Yu.",
    title = "{Matter Induced Neutrino Decay and Supernova {SN1987A}}",
    doi = "10.1016/0370-2693(89)90052-X",
    journal = "Phys. Lett. B",
    volume = "220",
    pages = "279--284",
    year = "1989"
}

@article{Cowsik:1977vz,
    author = "Cowsik, R.",
    title = "{Limits on the Radiative Decay of Neutrinos}",
    doi = "10.1103/PhysRevLett.39.784",
    journal = "Phys. Rev. Lett.",
    volume = "39",
    pages = "784--787",
    year = "1977",
    note = "[Erratum: Phys.Rev.Lett. 40, 201 (1978)]"
}

@article{Frieman:1987as,
    author = "Frieman, Joshua A. and Haber, Howard E. and Freese, Katherine",
    title = "{Neutrino Mixing, Decays and Supernova Sn1987a}",
    reportNumber = "SLAC-PUB-4261, SCIPP-87-90, NSF-ITP-87-53",
    doi = "10.1016/0370-2693(88)91120-3",
    journal = "Phys. Lett. B",
    volume = "200",
    pages = "115--121",
    year = "1988"
}

@article{Baerwald:2012kc,
    author = "Baerwald, Philipp and Bustamante, Mauricio and Winter, Walter",
    title = "{Neutrino Decays over Cosmological Distances and the Implications for Neutrino Telescopes}",
    eprint = "1208.4600",
    archivePrefix = "arXiv",
    primaryClass = "astro-ph.CO",
    doi = "10.1088/1475-7516/2012/10/020",
    journal = "JCAP",
    volume = "10",
    pages = "020",
    year = "2012"
}

@article{Farzan:2011tz,
    author = "Farzan, Yasaman",
    title = "{Strategies to link tiny neutrino masses with huge missing mass of the Universe}",
    eprint = "1106.2948",
    archivePrefix = "arXiv",
    primaryClass = "hep-ph",
    doi = "10.1142/S0217751X11053572",
    journal = "Int. J. Mod. Phys. A",
    volume = "26",
    pages = "2461--2485",
    year = "2011"
}

@article{Tomas:2001dh,
    author = "Tomas, R. and Pas, H. and Valle, J. W. F.",
    title = "{Generalized bounds on Majoron - neutrino couplings}",
    eprint = "hep-ph/0103017",
    archivePrefix = "arXiv",
    doi = "10.1103/PhysRevD.64.095005",
    journal = "Phys. Rev. D",
    volume = "64",
    pages = "095005",
    year = "2001"
}

@article{Robertson:2016xjh,
    author = "Robertson, Andrew and Massey, Richard and Eke, Vincent",
    title = "{What does the Bullet Cluster tell us about self-interacting dark matter?}",
    eprint = "1605.04307",
    archivePrefix = "arXiv",
    primaryClass = "astro-ph.CO",
    doi = "10.1093/mnras/stw2670",
    journal = "Mon. Not. Roy. Astron. Soc.",
    volume = "465",
    number = "1",
    pages = "569--587",
    year = "2017"
}

@article{Markevitch:2003at,
    author = "Markevitch, Maxim and Gonzalez, A. H. and Clowe, D. and Vikhlinin, A. and David, L. and Forman, W. and Jones, C. and Murray, S. and Tucker, W.",
    title = "{Direct constraints on the dark matter self-interaction cross-section from the merging galaxy cluster 1E0657-56}",
    eprint = "astro-ph/0309303",
    archivePrefix = "arXiv",
    doi = "10.1086/383178",
    journal = "Astrophys. J.",
    volume = "606",
    pages = "819--824",
    year = "2004"
}

@article{Bradac:2006er,
    author = "Bradac, Marusa and Clowe, Douglas and Gonzalez, Anthony H. and Marshall, Phil and Forman, William and Jones, Christine and Markevitch, Maxim and Randall, Scott and Schrabback, Tim and Zaritsky, Dennis",
    title = "{Strong and weak lensing united. 3. Measuring the mass distribution of the merging galaxy cluster 1E0657-56}",
    eprint = "astro-ph/0608408",
    archivePrefix = "arXiv",
    reportNumber = "SLAC-PUB-12077",
    doi = "10.1086/508601",
    journal = "Astrophys. J.",
    volume = "652",
    pages = "937--947",
    year = "2006"
}

@article{Froustey:2020mcq,
    author = "Froustey, Julien and Pitrou, Cyril and Volpe, Maria Cristina",
    title = "{Neutrino decoupling including flavour oscillations and primordial nucleosynthesis}",
    eprint = "2008.01074",
    archivePrefix = "arXiv",
    primaryClass = "hep-ph",
    doi = "10.1088/1475-7516/2020/12/015",
    journal = "JCAP",
    volume = "12",
    pages = "015",
    year = "2020"
}

@article{Bennett:2020zkv,
    author = "Bennett, Jack J. and Buldgen, Gilles and De Salas, Pablo F. and Drewes, Marco and Gariazzo, Stefano and Pastor, Sergio and Wong, Yvonne Y. Y.",
    title = "{Towards a precision calculation of $N_{\rm eff}$ in the Standard Model II: Neutrino decoupling in the presence of flavour oscillations and finite-temperature QED}",
    eprint = "2012.02726",
    archivePrefix = "arXiv",
    primaryClass = "hep-ph",
    reportNumber = "CPPC-2020-10",
    doi = "10.1088/1475-7516/2021/04/073",
    journal = "JCAP",
    volume = "04",
    pages = "073",
    year = "2021"
}

@article{Planck:2018vyg,
    author = "Aghanim, N. and others",
    collaboration = "Planck",
    title = "{Planck 2018 results. VI. Cosmological parameters}",
    eprint = "1807.06209",
    archivePrefix = "arXiv",
    primaryClass = "astro-ph.CO",
    doi = "10.1051/0004-6361/201833910",
    journal = "Astron. Astrophys.",
    volume = "641",
    pages = "A6",
    year = "2020",
    note = "[Erratum: Astron.Astrophys. 652, C4 (2021)]"
}

@article{Steigman:2012nb,
    author = "Steigman, Gary and Dasgupta, Basudeb and Beacom, John F.",
    title = "{Precise Relic WIMP Abundance and its Impact on Searches for Dark Matter Annihilation}",
    eprint = "1204.3622",
    archivePrefix = "arXiv",
    primaryClass = "hep-ph",
    doi = "10.1103/PhysRevD.86.023506",
    journal = "Phys. Rev. D",
    volume = "86",
    pages = "023506",
    year = "2012"
}

@article{Boehm:2001hm,
    author = "Boehm, Celine and Riazuelo, Alain and Hansen, Steen H. and Schaeffer, Richard",
    title = "{Interacting dark matter disguised as warm dark matter}",
    eprint = "astro-ph/0112522",
    archivePrefix = "arXiv",
    reportNumber = "SACLAY-SPH-T-01-147",
    doi = "10.1103/PhysRevD.66.083505",
    journal = "Phys. Rev. D",
    volume = "66",
    pages = "083505",
    year = "2002"
}

@article{DUNE:2015lol,
    author = "Acciarri, R. and others",
    collaboration = "DUNE",
    title = "{Long-Baseline Neutrino Facility (LBNF) and Deep Underground Neutrino Experiment (DUNE)}: {Conceptual Design Report, Volume 2: The Physics Program for DUNE at LBNF}",
    eprint = "1512.06148",
    archivePrefix = "arXiv",
    primaryClass = "physics.ins-det",
    reportNumber = "FERMILAB-DESIGN-2016-02",
    month = "12",
    year = "2015"
}

@article{Berbig:2020wve,
    author = "Berbig, Maximilian and Jana, Sudip and Trautner, Andreas",
    title = "{The Hubble tension and a renormalizable model of gauged neutrino self-interactions}",
    eprint = "2004.13039",
    archivePrefix = "arXiv",
    primaryClass = "hep-ph",
    doi = "10.1103/PhysRevD.102.115008",
    journal = "Phys. Rev. D",
    volume = "102",
    number = "11",
    pages = "115008",
    year = "2020"
}

@article{Hall:2009bx,
    author = "Hall, Lawrence J. and Jedamzik, Karsten and March-Russell, John and West, Stephen M.",
    title = "{Freeze-In Production of FIMP Dark Matter}",
    eprint = "0911.1120",
    archivePrefix = "arXiv",
    primaryClass = "hep-ph",
    reportNumber = "OUTP-09-18-P, UCB-PTH-09-32",
    doi = "10.1007/JHEP03(2010)080",
    journal = "JHEP",
    volume = "03",
    pages = "080",
    year = "2010"
}

@article{Hui:2016ltb,
    author = "Hui, Lam and Ostriker, Jeremiah P. and Tremaine, Scott and Witten, Edward",
    title = "{Ultralight scalars as cosmological dark matter}",
    eprint = "1610.08297",
    archivePrefix = "arXiv",
    primaryClass = "astro-ph.CO",
    doi = "10.1103/PhysRevD.95.043541",
    journal = "Phys. Rev. D",
    volume = "95",
    number = "4",
    pages = "043541",
    year = "2017"
}

@article{Wang:2006jy,
    author = "Wang, Fei and Wang, Wenyu and Yang, Jin Min",
    title = "{Split two-Higgs-doublet model and neutrino condensation}",
    eprint = "hep-ph/0601018",
    archivePrefix = "arXiv",
    doi = "10.1209/epl/i2006-10293-3",
    journal = "EPL",
    volume = "76",
    pages = "388--394",
    year = "2006"
}

@article{Gabriel:2006ns,
    author = "Gabriel, S. and Nandi, S.",
    title = "{A New two Higgs doublet model}",
    eprint = "hep-ph/0610253",
    archivePrefix = "arXiv",
    reportNumber = "OSU-HEP-06-6",
    doi = "10.1016/j.physletb.2007.04.062",
    journal = "Phys. Lett. B",
    volume = "655",
    pages = "141--147",
    year = "2007"
}

@article{Ma:2006km,
    author = "Ma, Ernest",
    title = "{Verifiable radiative seesaw mechanism of neutrino mass and dark matter}",
    eprint = "hep-ph/0601225",
    archivePrefix = "arXiv",
    reportNumber = "UCRHEP-T403",
    doi = "10.1103/PhysRevD.73.077301",
    journal = "Phys. Rev. D",
    volume = "73",
    pages = "077301",
    year = "2006"
}

@article{Nomura:2017wxf,
    author = "Nomura, Takaaki and Okada, Hiroshi",
    title = "{Hidden $U(1)$ gauge symmetry realizing a neutrinophilic two-Higgs-doublet model with dark matter}",
    eprint = "1709.06406",
    archivePrefix = "arXiv",
    primaryClass = "hep-ph",
    reportNumber = "KIAS-P17069",
    doi = "10.1103/PhysRevD.97.075038",
    journal = "Phys. Rev. D",
    volume = "97",
    number = "7",
    pages = "075038",
    year = "2018"
}

@article{Dev:2021axj,
    author = "Dev, P. S. Bhupal and Dutta, Bhaskar and Ghosh, Tathagata and Han, Tao and Qin, Han and Zhang, Yongchao",
    title = "{Leptonic scalars and collider signatures in a UV-complete model}",
    eprint = "2109.04490",
    archivePrefix = "arXiv",
    primaryClass = "hep-ph",
    reportNumber = "PITT-PACC-2108, MI-TH-2112, HRI-RECAPP-2021-007",
    doi = "10.1007/JHEP03(2022)068",
    journal = "JHEP",
    volume = "03",
    pages = "068",
    year = "2022"
}

@article{Schechter:1980gr,
    author = "Schechter, J. and Valle, J. W. F.",
    title = "{Neutrino Masses in SU(2) x U(1) Theories}",
    reportNumber = "SU-4217-167, COO-3533-167",
    doi = "10.1103/PhysRevD.22.2227",
    journal = "Phys. Rev. D",
    volume = "22",
    pages = "2227",
    year = "1980"
}

@article{Cheng:1980qt,
    author = "Cheng, T. P. and Li, Ling-Fong",
    title = "{Neutrino Masses, Mixings and Oscillations in SU(2) x U(1) Models of Electroweak Interactions}",
    reportNumber = "PRINT-80-0511 (CARNEGIE-MELLON), COO-3066-152",
    doi = "10.1103/PhysRevD.22.2860",
    journal = "Phys. Rev. D",
    volume = "22",
    pages = "2860",
    year = "1980"
}

@article{Mohapatra:1980yp,
    author = "Mohapatra, Rabindra N. and Senjanovic, Goran",
    title = "{Neutrino Masses and Mixings in Gauge Models with Spontaneous Parity Violation}",
    reportNumber = "FERMILAB-PUB-80-061-THY, FERMILAB-PUB-80-061-T",
    doi = "10.1103/PhysRevD.23.165",
    journal = "Phys. Rev. D",
    volume = "23",
    pages = "165",
    year = "1981"
}

@article{Lazarides:1980nt,
    author = "Lazarides, George and Shafi, Q. and Wetterich, C.",
    title = "{Proton Lifetime and Fermion Masses in an SO(10) Model}",
    reportNumber = "FREIBURG-THEP-80-2",
    doi = "10.1016/0550-3213(81)90354-0",
    journal = "Nucl. Phys. B",
    volume = "181",
    pages = "287--300",
    year = "1981"
}

@article{Magg:1980ut,
    author = "Magg, M. and Wetterich, C.",
    title = "{Neutrino Mass Problem and Gauge Hierarchy}",
    reportNumber = "CERN-TH-2829",
    doi = "10.1016/0370-2693(80)90825-4",
    journal = "Phys. Lett. B",
    volume = "94",
    pages = "61--64",
    year = "1980"
}

@article{Foot:1988aq,
    author = "Foot, Robert and Lew, H. and He, X. G. and Joshi, Girish C.",
    title = "{Seesaw Neutrino Masses Induced by a Triplet of Leptons}",
    reportNumber = "UM-P-88/89, OZ-P-88/7",
    doi = "10.1007/BF01415558",
    journal = "Z. Phys. C",
    volume = "44",
    pages = "441",
    year = "1989"
}

@article{Minkowski:1977sc,
    author = "Minkowski, Peter",
    title = "{$\mu \to e\gamma$ at a Rate of One Out of $10^{9}$ Muon Decays?}",
    reportNumber = "Print-77-0182 (BERN)",
    doi = "10.1016/0370-2693(77)90435-X",
    journal = "Phys. Lett. B",
    volume = "67",
    pages = "421--428",
    year = "1977"
}

@article{Mohapatra:1979ia,
    author = "Mohapatra, Rabindra N. and Senjanovic, Goran",
    title = "{Neutrino Mass and Spontaneous Parity Nonconservation}",
    reportNumber = "MDDP-TR-80-060, MDDP-PP-80-105, CCNY-HEP-79-10",
    doi = "10.1103/PhysRevLett.44.912",
    journal = "Phys. Rev. Lett.",
    volume = "44",
    pages = "912",
    year = "1980"
}

@article{Yanagida:1979as,
    author = "Yanagida, Tsutomu",
    editor = "Sawada, Osamu and Sugamoto, Akio",
    title = "{Horizontal gauge symmetry and masses of neutrinos}",
    reportNumber = "KEK-79-18-95",
    journal = "Conf. Proc. C",
    volume = "7902131",
    pages = "95--99",
    year = "1979"
}

@article{Gell-Mann:1979vob,
    author = "Gell-Mann, Murray and Ramond, Pierre and Slansky, Richard",
    title = "{Complex Spinors and Unified Theories}",
    eprint = "1306.4669",
    archivePrefix = "arXiv",
    primaryClass = "hep-th",
    reportNumber = "PRINT-80-0576",
    journal = "Conf. Proc. C",
    volume = "790927",
    pages = "315--321",
    year = "1979"
}

@article{Glashow:1979nm,
    author = "Glashow, S. L.",
    editor = "L\'evy, Maurice and Basdevant, Jean-Louis and Speiser, David and Weyers, Jacques and Gastmans, Raymond and Jacob, Maurice",
    title = "{The Future of Elementary Particle Physics}",
    reportNumber = "HUTP-79-A059",
    doi = "10.1007/978-1-4684-7197-7_15",
    journal = "NATO Sci. Ser. B",
    volume = "61",
    pages = "687",
    year = "1980"
}

@article{Chikashige:1980qk,
    author = "Chikashige, Y. and Mohapatra, Rabindra N. and Peccei, R. D.",
    title = "{Spontaneously Broken Lepton Number and Cosmological Constraints on the Neutrino Mass Spectrum}",
    reportNumber = "MPI-PAE/PTh 40/80",
    doi = "10.1103/PhysRevLett.45.1926",
    journal = "Phys. Rev. Lett.",
    volume = "45",
    pages = "1926",
    year = "1980"
}

@article{Aulakh:1982yn,
    author = "Aulakh, C. S. and Mohapatra, Rabindra N.",
    title = "{Neutrino as the Supersymmetric Partner of the Majoron}",
    reportNumber = "CCNY-HEP-82-9-REV, CCNY-HEP-82-9",
    doi = "10.1016/0370-2693(82)90262-3",
    journal = "Phys. Lett. B",
    volume = "119",
    pages = "136--140",
    year = "1982"
}

@article{Gelmini:1980re,
    author = "Gelmini, G. B. and Roncadelli, M.",
    title = "{Left-Handed Neutrino Mass Scale and Spontaneously Broken Lepton Number}",
    reportNumber = "MPI-PAE-PTH-50-80",
    doi = "10.1016/0370-2693(81)90559-1",
    journal = "Phys. Lett. B",
    volume = "99",
    pages = "411--415",
    year = "1981"
}

@article{Oldengott:2014qra,
    author = "Oldengott, Isabel M. and Rampf, Cornelius and Wong, Yvonne Y. Y.",
    title = "{Boltzmann hierarchy for interacting neutrinos I: formalism}",
    eprint = "1409.1577",
    archivePrefix = "arXiv",
    primaryClass = "astro-ph.CO",
    doi = "10.1088/1475-7516/2015/04/016",
    journal = "JCAP",
    volume = "04",
    pages = "016",
    year = "2015"
}

@article{Barenboim:2019tux,
    author = "Barenboim, Gabriela and Denton, Peter B. and Oldengott, Isabel M.",
    title = "{Constraints on inflation with an extended neutrino sector}",
    eprint = "1903.02036",
    archivePrefix = "arXiv",
    primaryClass = "astro-ph.CO",
    doi = "10.1103/PhysRevD.99.083515",
    journal = "Phys. Rev. D",
    volume = "99",
    number = "8",
    pages = "083515",
    year = "2019"
}

@article{Ghosh:2019tab,
    author = "Ghosh, Subhajit and Khatri, Rishi and Roy, Tuhin S.",
    title = "{Can dark neutrino interactions phase out the Hubble tension?}",
    eprint = "1908.09843",
    archivePrefix = "arXiv",
    primaryClass = "hep-ph",
    reportNumber = "TIFR/TH/19-31",
    doi = "10.1103/PhysRevD.102.123544",
    journal = "Phys. Rev. D",
    volume = "102",
    number = "12",
    pages = "123544",
    year = "2020"
}

@article{RoyChoudhury:2022rva,
    author = "Roy Choudhury, Shouvik and Hannestad, Steen and Tram, Thomas",
    title = "{Massive neutrino self-interactions and inflation}",
    eprint = "2207.07142",
    archivePrefix = "arXiv",
    primaryClass = "astro-ph.CO",
    doi = "10.1088/1475-7516/2022/10/018",
    journal = "JCAP",
    volume = "10",
    pages = "018",
    year = "2022"
}

@article{Das:2023npl,
    author = "Das, Anirban and Ghosh, Subhajit",
    title = "{The magnificent ACT of flavor-specific neutrino self-interaction}",
    eprint = "2303.08843",
    archivePrefix = "arXiv",
    primaryClass = "astro-ph.CO",
    reportNumber = "SLAC-PUB-17708",
    doi = "10.1088/1475-7516/2023/09/042",
    journal = "JCAP",
    volume = "09",
    pages = "042",
    year = "2023"
}

@article{Venzor:2023aka,
    author = "Venzor, Jorge and Garcia-Arroyo, Gabriela and De-Santiago, Josue and P\'erez-Lorenzana, Abdel",
    title = "{Resonant neutrino self-interactions and the H0 tension}",
    eprint = "2303.12792",
    archivePrefix = "arXiv",
    primaryClass = "astro-ph.CO",
    doi = "10.1103/PhysRevD.108.043536",
    journal = "Phys. Rev. D",
    volume = "108",
    number = "4",
    pages = "043536",
    year = "2023"
}

@article{Poudou:2025qcx,
    author = "Poudou, Ad\`ele and Simon, Th\'eo and Montandon, Thomas and Teixeira, Elsa M. and Poulin, Vivian",
    title = "{Self-interacting neutrinos in light of recent CMB and LSS data}",
    eprint = "2503.10485",
    archivePrefix = "arXiv",
    primaryClass = "astro-ph.CO",
    month = "3",
    year = "2025"
}

@article{He:2025jwp,
    author = "He, Adam and Ivanov, Mikhail M. and Bird, Simeon and An, Rui and Gluscevic, Vera",
    title = "{A Fresh Look at Neutrino Self-Interactions With the Lyman-$\alpha$ Forest: Constraints from EFT and PRIYA}",
    eprint = "2503.15592",
    archivePrefix = "arXiv",
    primaryClass = "astro-ph.CO",
    reportNumber = "MIT-CTP/5854",
    month = "3",
    year = "2025"
}

@article{Das:2025asx,
    author = "Das, Anirban and Dev, P. S. Bhupal and Gao, Christina and Ghosh, Subhajit and Kim, Taegyun",
    title = "{Impostor Among $\nu$s: Dark Radiation Masquerading as Self-Interacting Neutrinos}",
    eprint = "2506.08085",
    archivePrefix = "arXiv",
    primaryClass = "hep-ph",
    month = "6",
    year = "2025"
}

@article{Dev:2025sah,
    author = "Dev, P. S. Bhupal and Dutta, Bhaskar and Goswami, Srubabati and Tang, Jianrong Paul and Ramachandran, Aaroodd Ujjayini",
    title = "{Opening up New Parameter Space for Sterile Neutrino Dark Matter}",
    eprint = "2505.22463",
    archivePrefix = "arXiv",
    primaryClass = "hep-ph",
    reportNumber = "MI-HET-858",
    month = "5",
    year = "2025"
}

@article{Cherry:2014xra,
    author = "Cherry, John F. and Friedland, Alexander and Shoemaker, Ian M.",
    title = "{Neutrino Portal Dark Matter: From Dwarf Galaxies to IceCube}",
    eprint = "1411.1071",
    archivePrefix = "arXiv",
    primaryClass = "hep-ph",
    reportNumber = "CP3-Origins-2014-034, DIAS-2014-34, LA-UR-14-28339",
    month = "11",
    year = "2014"
}

@article{Farzan:2016wym,
    author = "Farzan, Yasaman and Heeck, Julian",
    title = "{Neutrinophilic nonstandard interactions}",
    eprint = "1607.07616",
    archivePrefix = "arXiv",
    primaryClass = "hep-ph",
    reportNumber = "ULB-TH-16-12",
    doi = "10.1103/PhysRevD.94.053010",
    journal = "Phys. Rev. D",
    volume = "94",
    number = "5",
    pages = "053010",
    year = "2016"
}

@article{Farzan:2017xzy,
    author = "Farzan, Y. and Tortola, M.",
    title = "{Neutrino oscillations and Non-Standard Interactions}",
    eprint = "1710.09360",
    archivePrefix = "arXiv",
    primaryClass = "hep-ph",
    doi = "10.3389/fphy.2018.00010",
    journal = "Front. in Phys.",
    volume = "6",
    pages = "10",
    year = "2018"
}

@article{Abdallah:2021npg,
    author = "Abdallah, Waleed and Barik, Anjan Kumar and Rai, Santosh Kumar and Samui, Tousik",
    title = "{Search for a light Z' at LHC in a neutrinophilic U(1) model}",
    eprint = "2106.01362",
    archivePrefix = "arXiv",
    primaryClass = "hep-ph",
    reportNumber = "HRI-RECAPP-2021-006",
    doi = "10.1103/PhysRevD.104.095031",
    journal = "Phys. Rev. D",
    volume = "104",
    number = "9",
    pages = "095031",
    year = "2021"
}

@article{Babu:2024zoe,
    author = "Babu, K. S. and Chakdar, Shreyashi and P. K., Vishnu",
    title = "{Chiral dark matter and radiative neutrino masses from gauged U(1) symmetry}",
    eprint = "2409.09008",
    archivePrefix = "arXiv",
    primaryClass = "hep-ph",
    reportNumber = "MS-TP-24-24",
    doi = "10.1088/1475-7516/2025/02/010",
    journal = "JCAP",
    volume = "02",
    pages = "010",
    year = "2025"
}

@article{Cai:2017jrq,
    author = "Cai, Yi and Herrero-Garc{\'\i}a, Juan and Schmidt, Michael A. and Vicente, Avelino and Volkas, Raymond R.",
    title = "{From the trees to the forest: a review of radiative neutrino mass models}",
    eprint = "1706.08524",
    archivePrefix = "arXiv",
    primaryClass = "hep-ph",
    reportNumber = "ADP-17-29-T1035",
    doi = "10.3389/fphy.2017.00063",
    journal = "Front. in Phys.",
    volume = "5",
    pages = "63",
    year = "2017"
}

@article{Babu:2019mfe,
    author = "Babu, K. S. and Dev, P. S. Bhupal and Jana, Sudip and Thapa, Anil",
    title = "{Non-Standard Interactions in Radiative Neutrino Mass Models}",
    eprint = "1907.09498",
    archivePrefix = "arXiv",
    primaryClass = "hep-ph",
    reportNumber = "FERMILAB-PUB-19-304-T, OSU-HEP-19-04",
    doi = "10.1007/JHEP03(2020)006",
    journal = "JHEP",
    volume = "03",
    pages = "006",
    year = "2020"
}

@article{Dutta:2020scq,
    author = "Dutta, Bhaskar and Ghosh, Sumit and Li, Tianjun",
    title = "{Explaining $(g-2)_{\mu,e}$, the KOTO anomaly and the MiniBooNE excess in an extended Higgs model with sterile neutrinos}",
    eprint = "2006.01319",
    archivePrefix = "arXiv",
    primaryClass = "hep-ph",
    reportNumber = "MI-TH-2012",
    doi = "10.1103/PhysRevD.102.055017",
    journal = "Phys. Rev. D",
    volume = "102",
    number = "5",
    pages = "055017",
    year = "2020"
}

@article{Schechter:1981cv,
    author = "Schechter, J. and Valle, J. W. F.",
    title = "{Neutrino Decay and Spontaneous Violation of Lepton Number}",
    reportNumber = "SU-4217-203, COO-3533-203",
    doi = "10.1103/PhysRevD.25.774",
    journal = "Phys. Rev. D",
    volume = "25",
    pages = "774",
    year = "1982"
}

@article{Pati:1974yy,
    author = "Pati, Jogesh C. and Salam, Abdus",
    title = "{Lepton Number as the Fourth Color}",
    reportNumber = "IC-74-7",
    doi = "10.1103/PhysRevD.10.275",
    journal = "Phys. Rev. D",
    volume = "10",
    pages = "275--289",
    year = "1974",
    note = "[Erratum: Phys.Rev.D 11, 703--703 (1975)]"
}

@article{Mohapatra:1974hk,
    author = "Mohapatra, Rabindra N. and Pati, Jogesh C.",
    title = "{Left-Right Gauge Symmetry and an Isoconjugate Model of CP Violation}",
    reportNumber = "MDDP-TR-74-085",
    doi = "10.1103/PhysRevD.11.566",
    journal = "Phys. Rev. D",
    volume = "11",
    pages = "566--571",
    year = "1975"
}

@article{Senjanovic:1975rk,
    author = "Senjanovic, G. and Mohapatra, Rabindra N.",
    title = "{Exact Left-Right Symmetry and Spontaneous Violation of Parity}",
    reportNumber = "CCNY-HEP-75-5",
    doi = "10.1103/PhysRevD.12.1502",
    journal = "Phys. Rev. D",
    volume = "12",
    pages = "1502",
    year = "1975"
}

@article{Dev:2017dui,
    author = "Dev, P. S. Bhupal and Mohapatra, Rabindra N. and Zhang, Yongchao",
    title = "{Long Lived Light Scalars as Probe of Low Scale Seesaw Models}",
    eprint = "1703.02471",
    archivePrefix = "arXiv",
    primaryClass = "hep-ph",
    reportNumber = "ULB-TH-17-05, UMD-PP-017-21",
    doi = "10.1016/j.nuclphysb.2017.07.021",
    journal = "Nucl. Phys. B",
    volume = "923",
    pages = "179--221",
    year = "2017"
}

@article{BhupalDev:2016nfr,
    author = "Dev, P. S. Bhupal and Mohapatra, Rabindra N. and Zhang, Yongchao",
    title = "{Displaced photon signal from a possible light scalar in minimal left-right seesaw model}",
    eprint = "1612.09587",
    archivePrefix = "arXiv",
    primaryClass = "hep-ph",
    reportNumber = "ULB-TH-17-11",
    doi = "10.1103/PhysRevD.95.115001",
    journal = "Phys. Rev. D",
    volume = "95",
    number = "11",
    pages = "115001",
    year = "2017"
}

@article{Maiezza:2015lza,
    author = "Maiezza, Alessio and Nemev\v{s}ek, Miha and Nesti, Fabrizio",
    title = "{Lepton Number Violation in Higgs Decay at LHC}",
    eprint = "1503.06834",
    archivePrefix = "arXiv",
    primaryClass = "hep-ph",
    doi = "10.1103/PhysRevLett.115.081802",
    journal = "Phys. Rev. Lett.",
    volume = "115",
    pages = "081802",
    year = "2015"
}

@article{Nemevsek:2016enw,
    author = "Nemev\v{s}ek, Miha and Nesti, Fabrizio and Vasquez, Juan Carlos",
    title = "{Majorana Higgses at colliders}",
    eprint = "1612.06840",
    archivePrefix = "arXiv",
    primaryClass = "hep-ph",
    doi = "10.1007/JHEP04(2017)114",
    journal = "JHEP",
    volume = "04",
    pages = "114",
    year = "2017"
}

@article{Garcia-Cely:2017oco,
    author = "Garcia-Cely, Camilo and Heeck, Julian",
    title = "{Neutrino Lines from Majoron Dark Matter}",
    eprint = "1701.07209",
    archivePrefix = "arXiv",
    primaryClass = "hep-ph",
    reportNumber = "ULB-TH-17-01",
    doi = "10.1007/JHEP05(2017)102",
    journal = "JHEP",
    volume = "05",
    pages = "102",
    year = "2017"
}

@article{Heeck:2019guh,
    author = "Heeck, Julian and Patel, Hiren H.",
    title = "{Majoron at two loops}",
    eprint = "1909.02029",
    archivePrefix = "arXiv",
    primaryClass = "hep-ph",
    reportNumber = "UCI-TR-2019-23",
    doi = "10.1103/PhysRevD.100.095015",
    journal = "Phys. Rev. D",
    volume = "100",
    number = "9",
    pages = "095015",
    year = "2019"
}

@article{Farzan:2012ev,
    author = "Farzan, Yasaman and Pascoli, Silvia and Schmidt, Michael A.",
    title = "{Recipes and Ingredients for Neutrino Mass at Loop Level}",
    eprint = "1208.2732",
    archivePrefix = "arXiv",
    primaryClass = "hep-ph",
    reportNumber = "IPM-P-2012-027, IPPP-12-64, DCPT-12-128",
    doi = "10.1007/JHEP03(2013)107",
    journal = "JHEP",
    volume = "03",
    pages = "107",
    year = "2013"
}

@article{Hagedorn:2018spx,
    author = "Hagedorn, Claudia and Herrero-Garc\'\i{}a, Juan and Molinaro, Emiliano and Schmidt, Michael A.",
    title = "{Phenomenology of the Generalised Scotogenic Model with Fermionic Dark Matter}",
    eprint = "1804.04117",
    archivePrefix = "arXiv",
    primaryClass = "hep-ph",
    reportNumber = "ADP-18-9-T1057, CP3-ORIGINS-2018-012, ADP-18-9/T1057, CP3-Origins-2018-012-DNRF90",
    doi = "10.1007/JHEP11(2018)103",
    journal = "JHEP",
    volume = "11",
    pages = "103",
    year = "2018"
}

@article{Batell:2017cmf,
    author = "Batell, Brian and Han, Tao and McKeen, David and Shams Es Haghi, Barmak",
    title = "{Thermal Dark Matter Through the Dirac Neutrino Portal}",
    eprint = "1709.07001",
    archivePrefix = "arXiv",
    primaryClass = "hep-ph",
    reportNumber = "PITT-PACC-1710",
    doi = "10.1103/PhysRevD.97.075016",
    journal = "Phys. Rev. D",
    volume = "97",
    number = "7",
    pages = "075016",
    year = "2018"
}

@article{Orlofsky:2021mmy,
    author = "Orlofsky, Nicholas and Zhang, Yue",
    title = "{Neutrino as the dark force}",
    eprint = "2106.08339",
    archivePrefix = "arXiv",
    primaryClass = "hep-ph",
    doi = "10.1103/PhysRevD.104.075010",
    journal = "Phys. Rev. D",
    volume = "104",
    number = "7",
    pages = "075010",
    year = "2021"
}

@article{Falkowski:2009yz,
    author = "Falkowski, Adam and Juknevich, Jose and Shelton, Jessie",
    title = "{Dark Matter Through the Neutrino Portal}",
    eprint = "0908.1790",
    archivePrefix = "arXiv",
    primaryClass = "hep-ph",
    reportNumber = "RU-NHETC-09-15",
    month = "8",
    year = "2009"
}

@article{Batell:2017rol,
    author = "Batell, Brian and Han, Tao and Shams Es Haghi, Barmak",
    title = "{Indirect Detection of Neutrino Portal Dark Matter}",
    eprint = "1704.08708",
    archivePrefix = "arXiv",
    primaryClass = "hep-ph",
    reportNumber = "PITT-PACC-1620",
    doi = "10.1103/PhysRevD.97.095020",
    journal = "Phys. Rev. D",
    volume = "97",
    number = "9",
    pages = "095020",
    year = "2018"
}

@article{Becker:2018rve,
    author = "Becker, Mathias",
    title = "{Dark Matter from Freeze-In via the Neutrino Portal}",
    eprint = "1806.08579",
    archivePrefix = "arXiv",
    primaryClass = "hep-ph",
    reportNumber = "DO-TH 18/13, DO-TH-18-13",
    doi = "10.1140/epjc/s10052-019-7095-7",
    journal = "Eur. Phys. J. C",
    volume = "79",
    number = "7",
    pages = "611",
    year = "2019"
}

@article{Folgado:2018qlv,
    author = "Folgado, Miguel G. and G\'omez-Vargas, Germ\'an A. and Rius, Nuria and Ruiz De Austri, Roberto",
    title = "{Probing the sterile neutrino portal to Dark Matter with $\gamma$ rays}",
    eprint = "1803.08934",
    archivePrefix = "arXiv",
    primaryClass = "hep-ph",
    reportNumber = "FTUV-18-0226.9011, IFIC/18-10, IFIC-18-10",
    doi = "10.1088/1475-7516/2018/08/002",
    journal = "JCAP",
    volume = "08",
    pages = "002",
    year = "2018"
}
\bibliographystyle{JHEP}
\end{document}